\documentclass[extra,mreferee]{gji} 
\usepackage{timet}
\usepackage{amsmath,amssymb,amsfonts,latexsym}[mathlines] 
\usepackage{mathrsfs}
\usepackage{graphicx}
\usepackage{changepage}
\usepackage{color}
\usepackage{todonotes}

\usepackage{lineno}
\let\oldequation\equation
\let\oldendequation\endequation
\renewenvironment{equation}
  {\linenomathNonumbers\oldequation}
  {\oldendequation\endlinenomath}

\usepackage{mathtools} 
\newcommand\ggll{\mathbin{\begin{smallmatrix} \gg \\ \ll \end{smallmatrix}}}
\newcommand\llgg{\mathbin{\begin{smallmatrix} \ll \\ \gg \end{smallmatrix}}}

\usepackage{soul}

\usepackage[finalnew]{trackchanges}

\title[
Reconciling aging law and slip law of rate-and-state friction
]
  {
  Reconciling aging and slip state evolutions from laboratory-derived canons of rate-and-state friction
}
\author[D. Sato, M. Nakatani, \& R. Ando
]
  {Dye SK Sato$^1$, Masao Nakatani$^{2}$, \& Ryosuke Ando$^{3}$
\\
  $^1$ Japan Agency for Marine-Earth Science and Technology, Kanagawa, Japan
\\
  $^2$ Earthquake Research Institute, The University of Tokyo, Tokyo, Japan\\
  $^3$ Graduate School of Science, The University of Tokyo, Tokyo, Japan
  }
\date{Received 20xx Xxxx xx; in original form 20xx Xxxx xx}
\pagerange{\pageref{firstpage}--\pageref{lastpage}}
\volume{xxx}
\pubyear{xxxx}

\begin{document}

\label{firstpage}

\maketitle

\begin{summary}
The aging law and the slip law are two representative evolution laws of the rate- and state-dependent friction (RSF) law, based on canonical behaviors in three types of laboratory experiments: slide-hold-slide (SHS), velocity-step (VS), and steady-state (SS) tests. The aging law explains the SHS canon but contradicts the VS canon, and vice versa for the slip law. The later proposed composite law, which switches these two laws according to the slip rate $V$, explains both canons but contradicts the SS canon. The present study constructs evolution laws satisfying all three canons throughout the range of variables where experiments have confirmed the canons. By recompiling the three canons, we have derived constraints on the evolution law and found that the evolution rates in the strengthening phases of the SHS and VS canons are so different that complete reconciliation throughout the entire range of variables is mathematically impossible. However, for the limited range of variables probed by experiments so far, we have found that the SHS and VS canons can be reconciled without violating the SS canon by switching the evolution function according to $\Omega$, the ratio of the state $\theta$ to its steady-state value $\theta_{\rm SS}$ for the instantaneous slip rate. We could generally show that, as long as the state evolution rate $\dot \theta$ depends only on the instantaneous values of $V$ and $\theta$, simultaneous reproduction of the three canons, throughout the experimentally confirmed range, requires the aging-law-like evolution for $\Omega$ sufficiently below a threshold $\beta$ and the slip-law-like evolution for $\Omega$ sufficiently above $\beta$. The validity of the canons in existing experiments suggests $\beta \lesssim 0.01$.
\end{summary}


\section{Introduction}
\label{sec:intro}

The rate- and state-dependent friction (RSF) law is based on the frictional behaviors observed in three standard types of laboratory experiments: (1) velocity-step (VS) test, which examines the change of frictional stress upon and shortly following an imposed stepwise change of the slip rate, (2) slide-hold-slide (SHS) test, which examines the dependence of the static friction on the time of (quasi-)stationary contact, and (3) steady-state (SS) test, which examines the slip-rate dependence of the dynamic friction during steady-state sliding~\citep{dieterich1979modeling,ruina1983slip}. 

These experiments have established the following behaviors of frictional strength (see \S\ref{subsec:RSFseparation} for the distinction between frictional stress and frictional strength), which shall be formulated in \S\ref{sec:3}. 
(1) VS tests show that the frictional strength evolves exponentially with the slip displacement towards a new steady-state value corresponding to the renewed slip rate; the characteristic slip distance is independent of the magnitude and polarity of the imposed velocity step~\citep{dieterich1979modeling,ruina1983slip,kilgore1993velocity}. (2) SHS tests show that the frictional strength increases logarithmically with the time of (quasi-)stationary contact~\citep{dieterich1972time,dieterich1978time,beeler1994roles,nakatani1996effects}, referred to as `log-t healing' in the present paper. 
(3) SS tests show that the frictional strength during steady-state sliding has a negative, linear dependence on the log velocity~\citep[e.g.,][]{scholz1976role,dieterich1978time,blanpied1998effects}. 
In an attempt to capture these three canonical behaviors under the framework of the RSF, different versions of state-evolution laws have been proposed~\citep[e.g.,][]{ruina1983slip,perrin1995self,kato2001composite,nagata2012revised}, but infamously, none of them have succeeded in simultaneously reproducing all of the above three canons~\citep{beeler1994roles,marone1998laboratory,nakatani2001conceptual,bhattacharya2015critical,bhattacharya2017does}. 

In the present study, we first clarify the constraints required for the evolution law to reproduce the above canonical behaviors (\S\ref{sec:3}). 
Although with some presumptions, we try to formulate the constraints in maximally general forms. 
Then, we develop new evolution laws that simultaneously reproduce all three canons throughout the range of law's variables where experiments have confirmed the cannons (\S\ref{sec:development}). We propose a general class of such laws, followed by an example employing a specific functional form (eq.~\ref{RSFeq:modcomplaw}). 

\section*{A guide and a roadmap}

Formulas that describe the consensual frictional behaviors in three representative types of experiments (eq.~\ref{eq:req4ss}, eq.~\ref{eq:realVScanonical}, and eqs.~\ref{RSFeq:PhiSHSNSCrawtc} and \ref{RSFeq:coincidenceofBs}) have been pursued since \citet{dieterich1979modeling}. Dieterich's formulation (1979) is primarily motivated by SS and SHS tests by \citet{dieterich1972time} and \citet{dieterich1978time}. It was reorganized into a differential equation (eq.~\ref{eq:defofaginglaw}) by \citet{ruina1983slip}, called the aging law (Dieterich law) today. However, the aging law contradicts VS tests. By linear expansion of the aging law around the steady states, \citet{ruina1983slip} proposed another differential equation (eq.~\ref{eq:defofsliplaw}) called the slip law (Ruina law) today. \citet{ruina1983slip} pointed out that the slip law is a straightforward translation of VS and SS behaviors into a differential equation. However, the slip law fails to describe the log-time healing observed under low shear stresses where slip is negligible, although it explains another, more popular type of SHS (the stationary-loadpoint SHS, \S\ref{subsubsec:331}) where shear stress remains pretty high, allowing for significant slip to continue in its quasi-static hold period. Thus, neither law has succeeded in compiling the different types of experiments. This long-standing discrepancy, the aging vs. slip problem~\citep{beeler1994roles,ampuero2008earthquake}, is the background of the present paper. Before starting the lengthy mathematical derivation, we summarize the requirements for this quest and draw a roadmap to obtain an evolution law satisfying all of them.
Our nomenclature is shown in Tables \ref{table:nom1} and \ref{table:nom2}.

Several papers have claimed the successful reproduction of the three experimentally established behaviors mentioned above, but always within a limited range of variables; their complete reconciliation is impossible (\S\ref{sec:canonicalandside}). As mentioned above, the aging law and the slip law coincide around the steady states, so reconciling them is effortless around steady states. Nontrivial is the regime far from steady states, and a few studies succeeded in explaining far-from-steady-state behaviors. One outstanding proposal is the composite law (eq.~\ref{eq:defofcompositelaw}) by \citet{kato2001composite}. 
However, the coincidences between those laws and experimental behaviors are all heuristic. Namely, they are not deduced from those experimental requirements, unlike the aging law and the slip law. Since the experimental requirement from VS and SS tests is exactly the slip law, which is besides supported by the stationary-loadpoint SHS, those heuristic laws do not withstand the questions whether their predictions differing from the slip law are the case. Existing observations in VS and stationary-loadpoint SHS tests consistently support the slip law, even at slip rates almost the same as plate convergence rates~\citep{bhattacharya2022evolution}. In other words, in the aging vs. slip problem context, the true importance lies in the constraint on reasonable deviations from the slip law: the constraint on how to joint frictional slip strengthening/weakening (the slip law) and the aging of stationary contacts (the aging law). Heuristics proposing many hows do not answer this question of determining actual how, regardless of the quality of fit. 

Instead of heuristics, the present paper pursues the deduction of constraints from the experimental behaviors. Only such deduced constraints allow us to start an empirically meaningful discussion of deviations from the aging and slip laws, as such deviations are logically required to reconcile the two laws. The principles of our deductions are the consensual behaviors, termed canonical behaviors, in the three representative types of experiments mentioned earlier, which exemplify the existence of frictional changes in response to time and slip increments. Thus, the SHS canon is set by constant-$\tau_{\rm hold}$ SHS tests (\S\ref{subsubsec:331}), which evidences the existence of truly time-dependent log-t healing that the slip law cannot explain at all. The stationary-loadpoint SHS is not involved in the three canons in our study. In fact, we can show that this type of SHS test only probes the variable regime covered by VS tests, so including it or not does not affect the derived constraints on evolution law (Appendix~\ref{sec:slip_law_pred_lph}).

With the above understanding, we start the derivation of a good evolution law. The foothold is a suitable formulation of the RSF (\S\ref{subsec:RSFseparation}) provided by \citet{nakatani2001conceptual}, where strength is differentiated from stress, and constitutive law is separated from evolution law. This formulation is mathematically equivalent to the conventional formulation of the RSF, which we included considering readers' familiarity. 
Then, we iterate the three canons of SS, VS, and SHS tests (\S\ref{sec:3}). We then translate each of the three canons (eq.~\ref{eq:req4ss}, eq.~\ref{eq:realVScanonical}, and eqs.~\ref{RSFeq:PhiSHSNSCrawtc} and \ref{RSFeq:coincidenceofBs}) into the constraints on evolution laws (eqs.~\ref{RSFeq:steadystateconstraint}, \ref{RSFeq:dotthetaVS}, 
and \ref{eq:ftNSC_constrained}, then simplified to eqs.~\ref{eq:constraintforsteadystate}--\ref{eq:simplifiedtimedepofthetadot}), getting ready for derivation. 

In \S\ref{sec:canonicalandside}, we prove that simultaneous satisfaction of the three canons is mathematically impossible. This incompatibility among the canons is the fundamental answer to why the aging and slip laws conflict. From this point forward, we pursue satisfying the three canons in the variable range where experiments have confirmed the canons and prove that this is possible if the aging law and the slip law are switched according to the value of the state relative to its steady-state value (eq.~\ref{eq:constraintonftgen}). Although we first introduce eq.~(\ref{eq:constraintonftgen}) heuristically in \S\ref{sec:heuristicderivation} (eq.~\ref{eq:betaisVSSHSboundary}), we later rederive it deductively in \S\ref{sec:generalizingheuristicderivation} and again in \S\ref{subsec:discussion1} (eq.~\ref{eq:constraintonftgen_withoutTF}) with further generalization. When the state evolution law is premised to be a function of only the current slip rate and the current interface state, the evolution law satisfying eq.~(\ref{eq:constraintonftgen_withoutTF}) is the unique choice to satisfy the three canons. A few examples of the concrete function form are presented in \S\ref{RSF33}. Lastly, we extend the present results for plausible minor modifications of the canons (\S\ref{sec:extensions_DcoverL} and \S\ref{sec:extensions}).

In \S\ref{sec:discussion}, we discuss the generality of our derivation, physical interpretations, and the possibility of experimental scrutiny. Although the goodness of experimental data fits is important, it is not the present paper's point. Instead, we emphasize the following outcomes. (I) The fundamental reason for the aging vs. slip problem is the presently proven inherent incompatibility among the three canons used in preceding studies in constructing the RSF law. (II) As long as all three canons are correct under their confirmed range of experimental conditions, the possible form of evolution law is narrowly constrained; under an additional ansatz that the evolution function only depends on the rate and state, the only possibility left is the behaviors that asymptotically resemble eq.~(\ref{RSFeq:modcomplaw}).

\begin{table}
\caption{Nomenclature. The numbers indicate the equations where the associated parameters and functions appear.}
 \label{table:nom1}
 \begin{center}
  \begin{tabular}{lr|lr}
  \multicolumn{3}{c}{Mechanical variables of the frictional interface}\\
  \hline
    $\delta$ &[m] & slip &(\ref{RSFeq:thermodynamicformPhi})\\
    $t$ &[s] & time  &(\ref{RSFeq:evolutionlawtheta})\\
    $V$ &[m/s] & slip rate &(\ref{RSFeq:constitutivetheta})\\
    $\tau$ &[N/m${}^2$]& shear stress (frictional resistance)&(\ref{RSFeq:constitutivetheta})\\
    $\sigma$ &[N/m${}^2$]&normal stress\\
    \hline
    \multicolumn{3}{c}{State variables and empirical parameters of the RSF}\\
    \hline
    $\theta$ &[s]& conventional state variable &(\ref{RSFeq:constitutivetheta})\\
    $\Phi$ &[N/m${}^2$]& frictional strength; serves as a state variable &(\ref{RSFeq:conversionofthetatophinocut_taustar}--\ref{RSFeq:constitutivePhiB})\\
    $\Omega$ && ratio of instantaneous $\theta$ to $\theta_{\rm SS}$  &(\ref{RSFeq:defofOmega_steadystate})\\
    $A$ &[N/m${}^2$]& magnitude of the direct effect &(\ref{RSFeq:constitutivetheta})\\
    $B$ &[N/m${}^2$]& magnitude of the state (evolution) effect &(\ref{RSFeq:constitutivetheta})\\
    $D$&[m]& characteristic length scale of the RSF&(\ref{eq:defofaginglaw})\\
    $L$ &[m]& $V \theta $ at steady states
    &(\ref{RSFeq:Vthetaconst})\\
    $D_{\rm c}$ &[m]& characteristic slip weakening distance&(\ref{eq:realVScanonical})
    \\
    $t_{\rm c}$ &[s]& cutoff time of log-time healing&(\ref{RSFeq:PhiSHSNSCrawtc})\\
    $B_{\rm heal}$ &[N/m${}^2$]& slope of log-time healing &(\ref{RSFeq:PhiSHSNSCrawtc})\\
    \hline
    \multicolumn{3}{c}{Hypothetical functions and parameters of the RSF}\\
    \hline
    $f$ &&functional representation of $\dot \theta$&(\ref{RSFeq:evolutionlawtheta})
    \\
    $f_t$ &&functional representation of $\partial\theta/\partial t$& (\ref{RSFeq:defofsmallft})
    \\
    $f_\delta$ &[s/m]&functional representation of $\partial\theta/\partial \delta$& (\ref{RSFeq:defofsmallfdelta})
    \\
    $F$ &[N/(m${}^2$s)]& functional representation of $\dot \Phi$&(\ref{RSFeq:evolutionlawPhi})
    \\
    $F_t$ &[N/(m${}^2$s)]& functional representation of $\partial \Phi/\partial t$&(\ref{RSFeq:defofFt})
    \\
    $F_\delta$ &[N/(m${}^3$)]& functional representation of $\partial\Phi/\partial \delta$&(\ref{RSFeq:defofFdelta})
    \\
    $V_{\rm c}$ &[m/s]& possible $V$ value s.t. $V\ggll V_{\rm c}\Leftrightarrow f_t\llgg f_\delta V$ &(\ref{eq:defofcompositelaw})
    \\
    $\beta$ && possible $\Omega$ value s.t. $\Omega\ggll \beta\Leftrightarrow f_t\llgg f_\delta V$ & (\ref{RSFeq:reqforpsic})
    \\   
    $\tilde \theta$ &[s]& $\theta+\theta_{\rm X}$&(\ref{eq:thetatilde})
    \\
    $\tilde \Omega$ &&ratio of instantaneous $\tilde \theta$ to $\theta_{\rm SS}$&(\ref{eq:Omegatilde})
    \\
    \hline
  \end{tabular}
 \end{center}
\end{table}

\begin{table}
\caption{Nomenclature (continued).}
 \label{table:nom2}
 \begin{center}
  \begin{tabular}{lr|lr}
    \multicolumn{3}{c}{Subscripts}\\
  \hline
  SS && steady-state values\\
  &&(for instantaneous $V$, unless otherwise noted)\\
  $|$SHS && in slide-hold-slide tests\\
  $|$VS && in velocity step tests\\
  \_init && initial values\\
  \_$*$ && reference steady state\\
  \_$X$ && values associated with strength minimum\\
  $|$NSC && under the negligible-slip condition\\
  \_end && at the end of the preceding experimental phase&
  \\
  \_peak && at which $\tau$ takes peak values in SHS tests&
  \\
  \hline
    \multicolumn{3}{c}{Imposed experimental conditions}\\
  \hline
  $k$ &[N/m${}^3$]& stiffness of spring slider models&\\
  $\delta_{\rm m}$&[m]& load-point displacement\\
  $V_{\rm m}$ &[m/s]& load-point velocity&\\
  $V_{\rm before}$ &[m/s]&  $V_{\rm m}$ before velocity steps
  \\
  $V_{\rm after}$ &[m/s]&  $V_{\rm m}$ after velocity steps
  \\
  $t_{\rm h}$&[s]& hold-phase duration in SHS tests &
  \\
  $V_{\rm prior}$ &[m/s]& $V_{\rm m}$ at initial sliding in SHS tests
  \\
  $\tau_{\rm hold}$ &[N/m${}^2$]& $\tau$ during the hold of constant-$\tau_{\rm hold}$ SHS tests
  \\
  $\Delta \tau$ &[N/m${}^2$]& $\tau_{\rm end}-\tau_{\rm hold}$
  \\
  \hline
  \end{tabular}
 \end{center}
\end{table}

\section{Structure of the RSF law}
\subsection{Conventional view}
\label{subsec:RSFconventional}

Suppose a frictional interface between two elastic blocks under shear stress. 
The present study assumes mechanical equilibrium, neglecting inertial effects.
Sliding or not, the stress continuity in the continuum host media proximate to the interface guarantees that the shear stress applied to the interface always equals frictional shear resistance. 
Thus, we use the same variable $\tau$ to express the magnitude of these two physical quantities. 
When the distinction is relevant in context, we refer to it as the applied shear stress or frictional resistance. Note, as customary in the field, we may refer to the frictional resistance as the frictional stress. The present paper concerns only the situations where the shear stress and slip rate directions are invariant; $\tau$ and $V$ denote their absolute values. Also, we assume normal stress $\sigma$ is constant.

The RSF law~\citep{dieterich1979modeling,ruina1983slip} has various mathematically equivalent formulations. We may say the most conventional formulation at present is eq.~(\ref{RSFeq:constitutivetheta}) below~\citep{marone1998laboratory}, where the secondary variation in frictional resistance $\tau$ from its base level is ascribed to the instant effect of the slip rate (called `direct effect') and the effect of the interface's internal state (represented by the state variable $\theta$) acquired as a consequence of past slip history. The latter effect is called the `state effect,' `memory effect,' or `evolution effect.' 
\begin{equation}
\tau=\tau_* + A \ln( V/V_*) + B \ln(\theta/\theta_*) 
\label{RSFeq:constitutivetheta}
\end{equation}
Here, $A$ and $B$ are empirical parameters representing the magnitude of the direct and state effects, respectively. They are positive parameters with stress dimension. $V_*$ is the reference slip rate, which can be set to any positive value without loss of generality. $\tau_*$ and $\theta_*$ are the reference values of $\tau$ and $\theta$, respectively. Usually, they are set at $\tau_{\rm SS}(V_*)$ and $\theta_{\rm SS}(V_*)$, the values of $\tau$ and $\theta$ at steady-state sliding at $V_*$. Throughout the present paper, the subscript ${\rm SS}$ denotes the value at a steady state. In the nomenclature of classical tribology, before the RSF began with \citet{dieterich1979modeling}, $\tau_*$ is the dynamic friction at $V_*$. 

The state variable $\theta$ in eq.~(\ref{RSFeq:constitutivetheta}) is usually a non-negative variable of the time dimension~\citep[e.g.,][]{dieterich1978time, dieterich1979modeling}, which is also the case in the present paper. Note that $\theta$ has constant-factor arbitrariness, that is, eq.~(\ref{RSFeq:constitutivetheta}) remains the same against the variable conversion:
\begin{equation}
    \theta \to \theta^\prime := c \theta,
    \label{eq:constantfactorarbitrarinessintheta}
\end{equation}
where $c$ is an arbitrary positive constant. 
The significance of constant-factor arbitrariness of $\theta$ shall be discussed again when we consider the normalization of $\dot \theta$ in \S\ref{subsec:SHSrequirement}. 
The RSF law, including the present modification, is offered as a phenomenological description of experimental facts. However, we keep in mind a representative physical interpretation of $\theta$: the mean age of asperity contacts that appear and disappear on the sliding interface~\citep[e.g.,][]{dieterich1978time,dieterich1979modeling}.

The time evolution of the state variable $\theta$ may be described by a differential equation: 
\begin{equation}
\dot \theta=f(V,\theta,...),
\label{RSFeq:evolutionlawtheta}
\end{equation}
which is termed the evolution law (of $\theta$). 
The argument abbreviation $...$ in eq.~(\ref{RSFeq:evolutionlawtheta}) implies that $V$ and $\theta$ may not be the only variables that affect the state evolution, 
though the present study mainly considers the $f(V,\theta)$ form that depends only on $V$ and $\theta$, 
following many of the previous studies since \citet{ruina1983slip}. 
The two most widely used formulae~\citep{ruina1983slip} of $f$ are
\begin{equation}
f= 1 - V\theta/D \hspace{5pt} (\mbox{Aging Law})
\label{eq:defofaginglaw}
\end{equation}
and
\begin{equation}
f =  - (V\theta/D)\ln(V\theta/D) \hspace{5pt} (\mbox{Slip Law}),
\label{eq:defofsliplaw}
\end{equation}
where $D$ is an empirical constant of length dimension. 

When \citet{dieterich1979modeling} started the RSF, the state evolution was implemented by the direct substitution of observed $\theta$ variations in respective standard tests: $\theta_{\rm |VS} (\delta; \theta_{\rm init})$, $\theta_{\rm |SS}(V)$, and $\theta_{\rm |SHS}(t_{\rm h}; \theta_{\rm init})$, being functions of the slip  $\delta=\int dt V$, slip rate $V$, and hold time $t_{\rm h}$,  respectively ($\theta_{\rm |VS}$ and $\theta_{\rm |SHS}$ also depend on the initial $\theta=\theta_{\rm init}$), whereas underlying mechanisms common to those canonical behaviors were already anticipated. 
Later, \citet{ruina1980friction,ruina1983slip} and \citet{kosloff1980reformulation} came up with the idea of using a differential law (i.e., eq.~\ref{RSFeq:evolutionlawtheta}) that yields canonical behaviors as its special solutions. 
This time differential approach has enabled predictions of frictional behaviors under general loading history, not limited to the standard tests associated with the canonical behaviors. 
Since then, the RSF has been applied to many aspects of fault mechanics~\citep{scholz2019mechanics}, becoming a de facto standard of fault constitutive laws. 
However, neither the aging nor slip law can reproduce the three canons simultaneously. Since this was pointed out~\citep{beeler1994roles,marone1998laboratory}, several alternative evolution laws have been proposed, but none has achieved complete success, as mentioned in \S\ref{sec:intro}. 

Among the existing evolution laws, the aging, slip, and composite~\citep[a hybrid of the aging and slip laws,][]{kato2001composite} laws are deeply relevant to the present derivation of new evolution laws. We will explain their workings in light of the canonical behaviors in \S\ref{subsec:VSrequirement}, \S\ref{subsec:SHSrequirement}, and \S\ref{sec:canonicalandside}. 

Logically, there is no a priori reason to doubt the evolution law only: eq.~(\ref{RSFeq:constitutivetheta}) may be at fault as well~\citep{nakatani2003physicochemical,baumberger2006solid,barbot2019modulation}. However, as previous authors did, we proceed by premising eq.~(\ref{RSFeq:constitutivetheta}). Although we add a minor modification of eq.~(\ref{RSFeq:constitutivetheta}) to deal with the lower limit of the frictional strength (\S\ref{subsec:RSFminimumStrength} and \S\ref{sec:extensions}), it is irrelevant to our goal of simultaneous reproduction of the three canons. 

\subsection{Separation of the constitutive law and the state-evolution law}
\label{subsec:RSFseparation}

In the conventional view of eq.~(\ref{RSFeq:constitutivetheta}) as a function that returns the frictional resistance $\tau$ under various conditions, the RSF law seems to merely recognize secondary $\tau$ variations due to the slip-rate $V$ and interface-state $\theta$ perturbations, to be added to the base level $\tau_*$, the dynamic friction in the classical friction laws since Amontons-Coulomb to \citet{dieterich1978time}. 
However, the involvement of the direct effect term $A \ln(V/V_*)$ has made eq.~(\ref{RSFeq:constitutivetheta}) a
rheological constitutive law that describes the instantaneous relationship between the mechanical variables $\tau$ and $V$. 
This reading of eq.~(\ref{RSFeq:constitutivetheta}) also tells that the frictional rheological constitutive relation is affected by the auxiliary variable $\theta$, the state of the frictional interface at the moment---these insights from eq.~(\ref{RSFeq:constitutivetheta}) lead to another notation of the RSF law that separates the constitutive law and the evolution law~\citep{nakatani2001conceptual} clearly, which is the foothold of the present development of the evolution law, as elaborated below. 

If a variable functions as the state variable of the RSF law, any variable that has one-to-one correspondence with it is also fully functional as a state variable~\citep{ruina1983slip}. \citet{nakatani2001conceptual} introduced a new state variable $\Phi$ such that
\begin{equation}
\Phi=\tau_*+ B \ln (\theta /\theta_*)
\label{RSFeq:conversionofthetatophinocut_taustar}
\end{equation}
and rewrote eq.~(\ref{RSFeq:constitutivetheta}) into 
\begin{equation}
\tau=A\ln (V/V_*)+\Phi.
\label{RSFeq:constitutivePhiA}
\end{equation}
When solved for $V$, eq.~(\ref{RSFeq:constitutivePhiA}) becomes
\begin{equation}
V=V_* e^{(\tau-\Phi)/A},
\label{RSFeq:constitutivePhiB}
\end{equation}
which means that the new state variable $\Phi$ is the shear stress required to slide the interface of the given internal state at a reference slip rate of $V_*$. 
Thus, $\Phi$ can be regarded as a natural extension of the classical concept of frictional strength, namely the shear stress required to slide the interface (at any nonzero $V$), to the RSF world, where finite $\tau$ necessarily induces finite $V$.
[Although eqs.~(\ref{RSFeq:constitutivetheta}), (\ref{RSFeq:constitutivePhiA}), and (\ref{RSFeq:constitutivePhiB}), all mathematically equivalent to each other, give finite positive $V$ for zero or even negative $\tau$, this obvious flaw has been fixed by replacing the $\exp(\cdot)$ in eq.~(\ref{RSFeq:constitutivePhiB}) with $2\sinh(\cdot)$~\citep{rice2001rate} implied by a physical interpretation of eq.~(\ref{RSFeq:constitutivePhiB})~\citep{heslot1994creep,nakatani2001conceptual} as the flow law of thermally activated shear creep of frictional junction (collection of atomic bonds) formed at real contacts of the interface.] 
In the present paper, the state variable $\Phi$ is often called the frictional strength, or simply, strength. 
For later convenience, let us define the reference value of $\Phi$, denoted by $\Phi_*$. As it is the frictional strength, a natural choice is the dynamic friction at the reference velocity $V_*$, that is, $\Phi_*:=\Phi_{\rm SS}(V_*)$, the same as the definition of $\tau_*$ in eq.~(\ref{RSFeq:constitutivetheta}):
\begin{equation}
    \Phi_*=\tau_*. 
    \label{eq:steadystatestrengthstressequality}
\end{equation}
Hereafter, we will replace $\tau_*$ with $\Phi_*$. 

With the above understood, we can see eq.~(\ref{RSFeq:constitutivePhiB}) is a refined version of the unwritten constitutive law implicit in classical tribology before the RSF: the interface does not slip when $\tau$ is less than the threshold strength and slips when $\tau$ is equal to or above. Indeed, under the conditions where the applicability of the RSF has been experimentally confirmed, $A$ is far smaller than $\tau$ and $\Phi$, so that eq.~(\ref{RSFeq:constitutivePhiB}) tells that $V$ increases very rapidly as $\tau$ approaches and surpasses $\Phi$, quite close to the threshold-yielding supposed in the implicit constitutive law of classical tribology~\citep[cf. Fig.~1 of][]{nakatani2001conceptual}. 
By contrast, the explicitly stated part of the classical friction laws (e.g., normal-stress proportionality, static friction higher than dynamic friction, slight $V$ dependence of the dyanmic friction) is a `strength law,' which describes the threshold strength, playing the same logical role as the evolution law (eq.~\ref{RSFeq:evolutionlawtheta}) of the RSF.

In addition to having the intuitive meaning of the frictional strength, the $\Phi$-notation has another big advantage for us to seek the correct evolution law from experimental requirements: 
the state variable $\Phi$ is substantially an observable; at any moment in the experiments where both $\tau$ and $V$ are measured, the value of $\Phi$ is given as
\begin{equation}
\Phi = \tau - A\ln (V/V_*).
\label{eq:N1}
\end{equation}
Therefore, experimental results observed as $\tau(t)$ are straightforwardly convertible to $\Phi(t)$. 
Hence, we can describe the experimental requirements in terms of $\Phi$, i.e., the canonical behaviors of $\Phi$ (\S\ref{sec:3}). 
Furthermore, $\Phi$ has a unique value excepting the offset $A\ln V_*$ (eq.~\ref{eq:N1}), associated with the free choice of reference slip rate $V_*$. Therefore, the variation in $\Phi$ and its evolution law are defined without ambiguity, contrasting with those of $\theta$ defined with constant-factor arbitrariness. 
Note that the above eq.~(\ref{eq:N1}) can be regarded as the definition of $\Phi$, an alternative to the original installation of eq.~(\ref{RSFeq:conversionofthetatophinocut_taustar}) by \citet{nakatani2001conceptual} via variable transformation from $\theta$. 
With the definition of $\Phi$ by eq.~(\ref{eq:N1}), 
one can reformulate eq.~(\ref{RSFeq:constitutivetheta}) into eq.~(\ref{RSFeq:constitutivePhiA}) or (\ref{RSFeq:constitutivePhiB}), without invoking the conventional state variable $\theta$. 
From our standpoint of using the $\Phi$-notation as a primary representation of the RSF, 
eq.~(\ref{RSFeq:conversionofthetatophinocut_taustar}) is considered a variable conversion rule from $\theta$ to $\Phi$. 
Since $\Phi_*=\tau_*$ (eq.~\ref{eq:steadystatestrengthstressequality}), the $\theta\to \Phi $ conversion is also written as 
\begin{equation}
\Phi=\Phi_*+ B \ln (\theta /\theta_*).
\label{RSFeq:conversionofthetatophinocut}
\end{equation}
We use eq.~(\ref{RSFeq:conversionofthetatophinocut}) rather than eq.~(\ref{RSFeq:conversionofthetatophinocut_taustar}).
We also have the inverse conversion $\Phi \to\theta$:  
\begin{equation}
\theta =\theta_*e^{(\Phi-\Phi_*)/B}.
\label{RSFeq:conversionofphitothetanocut}
\end{equation}
When the time evolution of $\theta$ can be described by a differential equation (eq.~\ref{RSFeq:evolutionlawtheta}), so can the time evolution of $\Phi$, the frictional strength. We denote the evolution of $\Phi$ as 
\begin{equation}
\dot \Phi= F(V,\Phi,...).
\label{RSFeq:evolutionlawPhi}
\end{equation}

Thus, in the notation with $\Phi$, the RSF law separates into the rheological constitutive law (eq.~\ref{RSFeq:constitutivePhiA}, or equivalently, \ref{RSFeq:constitutivePhiB}) describing the instantaneous relationship between $\tau$ and $V$, conditioned on the parameter $\Phi$ (frictional strength of the interface at the moment), and the equation describing the time evolution of the frictional strength $\Phi$ (eq.~\ref{RSFeq:evolutionlawPhi}). 
This set is mathematically equivalent to the conventional set of eqs.~(\ref{RSFeq:constitutivetheta}) and (\ref{RSFeq:evolutionlawtheta}) through the variable conversion (eqs.~\ref{RSFeq:conversionofthetatophinocut} and \ref{RSFeq:conversionofphitothetanocut}). 
However, we need to be cautious of the fact that 
the state variable $\theta$ appears in eq.~(\ref{RSFeq:constitutivetheta}) in the form of $B \ln \theta$, 
where $B\ln(\cdot)$ is tied to some experimental facts regarding the evolution of $\Phi$. 
In consequence of using $\theta$-notation, the information on how the variation in the interface strength $B\ln(\theta/\theta_*)$ occurs spans the two equations of eqs.~(\ref{RSFeq:constitutivetheta}) and (\ref{RSFeq:evolutionlawtheta}). 
The constant $B$, appearing in eq.~(\ref{RSFeq:constitutivetheta}), is the quantity rooted in the two observations of log-V dependence of steady-state dynamic friction (\S\ref{subsec:SSrequirement}) and log-t healing (\S\ref{subsec:SHSrequirement}), meaning that eq.~(\ref{RSFeq:constitutivetheta}) already premises specific knowledge of the time evolution of strength. 
Thus, eq.~(\ref{RSFeq:constitutivetheta}) in the $\theta$-notation 
is customarily bound to specific notions about the time evolution of strength. 
For us to pursue the phenomenological development of the strength evolution law without resorting to assumptions about the underlying physics, unsuited is the $\theta$-notation, which was introduced through the physical intuition that ascribes frictional strength variations to log-t healing and its erasure by slip~\citep{dieterich1979modeling}. By contrast, in the rewritten RSF (the set of eqs.~\ref{RSFeq:constitutivePhiA} and \ref{RSFeq:evolutionlawPhi}), where the frictional strength $\Phi$ serves as the state variable, the present subject of how the frictional interface evolves is solely taken care of by the evolution law (eq.~\ref{RSFeq:evolutionlawPhi}) of $\Phi$. 

Thus, our derivation (\S\ref{sec:3} and \S\ref{sec:development}) of the evolution law shall start with describing the experimental requirements in the $\Phi$-notation, where the evolution law (eq.~\ref{RSFeq:evolutionlawPhi}) well separates from the constitutive law (eq.~\ref{RSFeq:constitutivePhiA}). 
Naturally, the derived equations and findings are valid in any RSF formulations, including the $\theta$-notation (the set of eqs.~\ref{RSFeq:constitutivetheta} and \ref{RSFeq:evolutionlawtheta}) convertible via eqs.~(\ref{RSFeq:conversionofthetatophinocut}) and (\ref{RSFeq:conversionofphitothetanocut}). We also show derived results in the $\theta$-notation to secure usability in this widely used notation. 

The dominant factor governing the frictional strength $\Phi$ is the proportionality to the normal stress $\sigma$. Hence, classical tribology defines the frictional coefficient, $\mu_{\rm clf}:= \Phi/\sigma$, the $\sigma$-normalized frictional strength. 
Since the introduction of the RSF by~\citet{dieterich1979modeling}, however, the frictional coefficient is often defined as $\tau$ (the equal magnitude of the frictional resistance and applied shear stress, as explained at the beginning of \S\ref{subsec:RSFconventional}) normalized by $\sigma$, i.e., $\mu_{\rm RSF}:= \tau/\sigma$. 
This redefinition would be acceptable, given that the RSF is a flow law. However, it is wrong and harmful that quite a few authors refer to $\mu_{\rm RSF}$ (and/or $\tau$) as ``frictional strength.'' Suppose a commonplace situation where an interface is subjected to shear stress $\tau$ far below the classical frictional strength only to induce minuscule slip rate $V$---this $\tau$ far below the strength is not the strength! 
(Note, however, referring to $\tau$ or $\mu_{\rm RSF}$ as `frictional resistance' or `frictional stress' is perfectly correct.) In the present paper, we intentionally avoid the custom of the friction industry, where quantities of stress dimensions (e.g., $\tau$, $\Phi$, $\Phi_*$, $A$, $B$) are denoted by $\sigma$-normalized quantities. (We have set forth this remark here because the conceptual distinction of strength from stress is vital in the present development of the evolution law.)

At the end of this subsection, we summarize the physical interpretation of $\Phi$. Although the present development of the new evolution law is solely based on macroscopic observations of the canonical behaviors, not resorting to specific microphysical interpretations, those insights would still help follow the present development, described phenomenologically with a long series of mathematical formulae. 
The representative interpretation of $\Phi$ is the adhesion theory of friction~\citep[ATF,][]{bowden1964friction}, where $\Phi \propto$ the real contact area $A_{\rm r}$ of the frictional interface of two rough surfaces. 
For multi-contact interfaces of rough surfaces of various solids, including rocks, the ATF assumption of $\Phi(t) \propto A_{\rm r}$ has been roughly confirmed by experiments employing electrical, optical, or acoustic techniques to monitor $A_{\rm r}(t)$~\citep{bowden1964friction,dieterich1994direct,schoenberg1980elastic,nagata2008monitoring,nagata2014high}, while additional contributions to $\Phi(t)$ due to the change of contact quality (strength per unit $A_{\rm r}$) have been recently recognized in single-contact experiments~\citep{li2011frictional}. In any case, $\Phi(t)$ corresponds to the number of atomic bonds per unit nominal interface area. Then, through the conversion eq.~(\ref{RSFeq:conversionofthetatophinocut}), the conventional state variable $\theta$ is interpreted as the effective contact time required to earn the number of atomic bonds for the given $\Phi(t)$ via log-t healing, consistent with the contact-time interpretation of $\theta$ mentioned in \S\ref{subsec:RSFconventional}.

\subsection{Lower limit of frictional strength}
\label{subsec:RSFminimumStrength}
The negative log-V dependence of the steady-state strength $\Phi_{\rm SS}(V)$, one of the canonical behaviors mentioned in \S\ref{sec:intro}, must disappear at velocities above some cutoff velocity because if it lasted to however high $V$, strength would go negative. Indeed, such a high-V cutoff has been observed around $0.1$ mm/s, $d\Phi_{\rm SS}/d\ln V$ approaches zero from above~\citep{okubo1986state,weeks1993constitutive}.  
Note that this cutoff velocity for the RSF's state effect occurs below the velocity level where recently recognized drastic dynamic weakening~\citep[e.g.,][]{di2011fault}, likely caused by thermal effects such as flash heating~\citep{rice2006heating}, comes into play. 

When the interface is kept sliding at a fixed slip rate, frictional strength converges to a steady-state level $\Phi_{\rm SS}(V)$ specific to the slip rate. 
This convergence to a steady state has been well observed over various conditions in laboratories, thus premised in the present study. This observation 
implies $\dot \Phi \gtrless 0$ when $\Phi \lessgtr \Phi_{\rm SS}(V)$, which in turn leads to $\dot \Phi >0$
whenever $\Phi$ is less than $\min_{\rm V}\Phi_{\rm SS}$, the lower limit of the steady-state strength. Therefore, we see that $\min_{\rm V}\Phi_{\rm SS}$ (denoted as $\Phi_{\rm X}$ hereafter) is
also the lower limit of the frictional strength as long as we consider situations under fixed normal stress, as we declared at the beginning of \S\ref{subsec:RSFconventional}. Namely,
\begin{equation}
\Phi\geq \Phi_{\rm X}.
\label{RSFeq:xstatePhi}
\end{equation}
Since it has been defined by a steady-state strength, $\Phi_{\rm X}$ is a time-invariant property of the interface. 

\begin{figure*}
   \includegraphics[width=80mm]{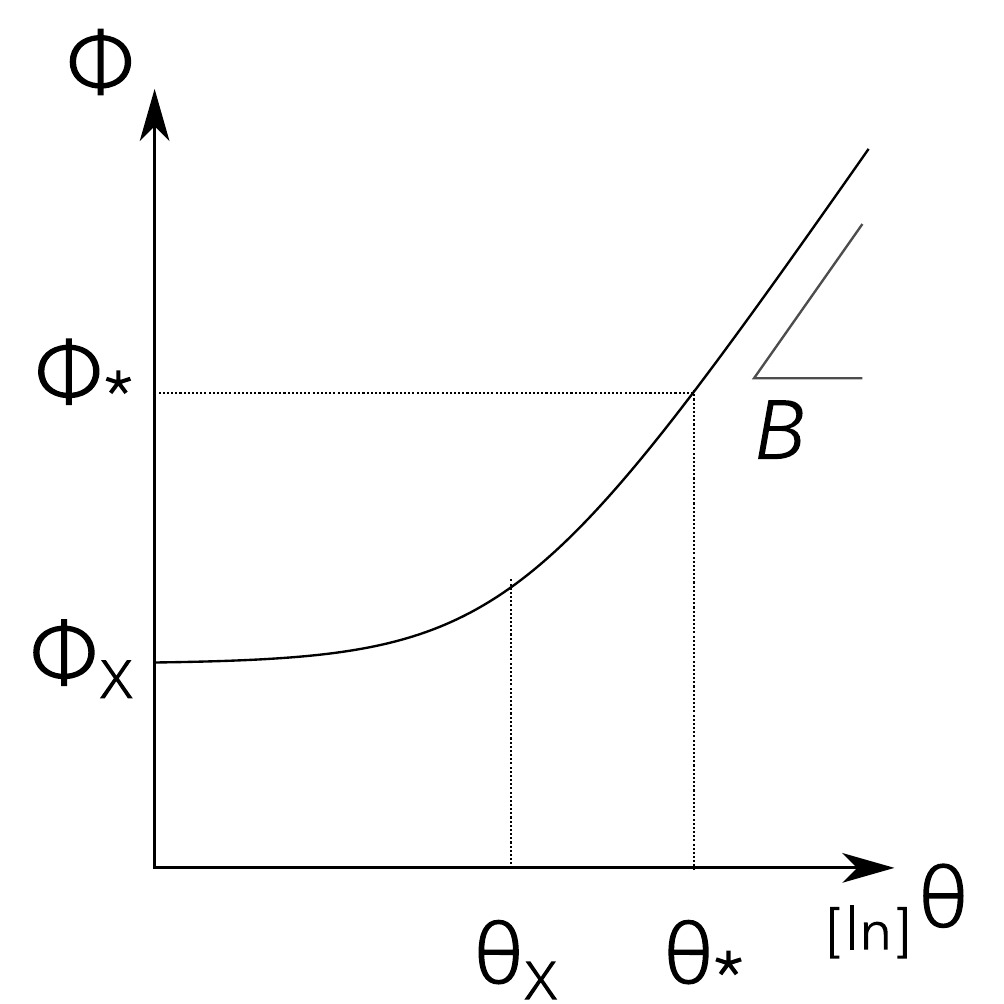}
  \caption{
  Correspondence between the strength $\Phi$ and the conventional state variable $\theta$. $\Phi_{\rm X}$ and $\theta_{\rm X}$ are state values for the X-state. $\Phi_*$ and $\theta_*$ are state values for the reference state. For $\theta\gg\theta_{\rm X}$, $\Phi$ increases linearly with $\ln \theta$, at the slope $B$.
}
  \label{fig:Xstate}
\end{figure*}

The lower limit of frictional strength can be incorporated into the RSF law in the conventional $\theta$-notation~\citep[e.g.,][]{okubo1986state,weeks1993constitutive, nakatani2006intrinsic} by redefining $\theta$ using an adjusted version of eq.~(\ref{RSFeq:conversionofthetatophinocut}): 
\begin{equation}
\Phi=\Phi_{\rm X}+ B \ln (\theta /\theta_{\rm X}+1),
\label{RSFeq:conversionofthetatophi}
\end{equation}
where $\theta_{\rm X}$ denotes the level of $\theta$ below which $\Phi$ shows little logarithmic dependence on $\theta$. 
According to the contact-age interpretation of $\theta$, $\theta_{\rm X}$ represents the short-time cutoff of the contact-age dependence of $\Phi$, 
and $\Phi_{\rm X}$ represents the minimum $\Phi$ value achieved at the short-time limit of the contact time $\theta \searrow0$. 
Figure~\ref{fig:Xstate} illustrates eq.~(\ref{RSFeq:conversionofthetatophi}); $\Phi$ loses the $\theta$ dependence for $\theta\ll\theta_{\rm X}$ and equals $\Phi_{\rm X}$ at $\theta = 0$. 
Equation~(\ref{RSFeq:conversionofthetatophi}) introduces the strength minimum while retaining the definition range of $\theta$ ($0\leq\theta<\infty$) in the conventional evolution laws, such as the aging and slip laws.

Upon replacing eq.~(\ref{RSFeq:conversionofthetatophinocut}) with eq.~(\ref{RSFeq:conversionofthetatophi}), eq.~(\ref{RSFeq:conversionofphitothetanocut}), the reverse conversion of eq.~(\ref{RSFeq:conversionofthetatophinocut}), must be replaced by
\begin{equation}
\theta =\theta_{\rm X}\left[e^{(\Phi-\Phi_{\rm X})/B}-1\right].
\label{RSFeq:conversionofphitotheta}
\end{equation}
Accordingly, eq.~(\ref{RSFeq:constitutivetheta}) is modified to
\begin{equation}
\tau = \Phi_{\rm X}+ A\ln(V/V_*)+B \ln(\theta/\theta_{\rm X}+1),
\label{RSFeq:constitutivethetacutoff}
\end{equation}
which explicitly involves $\theta_{\rm X}$ and $\Phi_{\rm X}$~\citep[e.g.,][]{okubo1986state,weeks1993constitutive, nakatani2006intrinsic}. 
We do not assume a high-V cutoff for the direct effect, following \citet{weeks1993constitutive}, who has concluded that experiments are better described if a high-V cutoff exists only for the state effect, though \citet{okubo1986state} assumed a high-V cutoff also exists for the direct effect. 

For $\theta\gg\theta_{\rm X}$, the ``1'' within the logarithm in eq.~(\ref{RSFeq:constitutivethetacutoff}) is negligible so that eq.~(\ref{RSFeq:constitutivethetacutoff}) approaches to eq.~(\ref{RSFeq:constitutivetheta}). 
Hence, when $\theta\gg\theta_{\rm X}$, 
$\Phi_*$ (equal to $\tau_*$ from eq.~\ref{eq:steadystatestrengthstressequality}) has the approximate expression: 
\begin{equation}
\Phi_*= \Phi_{\rm X}+ B\ln(\theta_*/\theta_{\rm X}+1).
\label{eq:Xstate2starstate}
\end{equation}

\citet{okubo1986state} derived eq.~(\ref{RSFeq:conversionofthetatophi}) from the original RSF law of \citet{dieterich1979modeling}, which fits the healing experiments of \citet{dieterich1978time}, where a short-time cutoff is observed. At that point, $\theta_{\rm X}$ was purely a phenomenological fitting parameter. Later, \citet{nakatani2006intrinsic} have indicated, without resorting to any microphysical interpretations, that eq.~(\ref{RSFeq:conversionofthetatophi}) is the unique function to represent the lower limit of strength in the canonical log-t healing (dealt with in \S\ref{subsec:SHSrequirement}). Here, we reiterate their theoretical findings relevant to the present development, especially concerning the canonical log-t healing behavior: The differential equation to produce log-t healing of strength $\Phi(t) =\Phi(0)+ B\ln(t/t_{\rm c} + 1)$
is $\dot\Phi\propto \exp (-\Phi/B)$, where the strengthening rate has a negative exponential dependency on the instantaneous strength. This differential equation leads to the cutoff time $t_{\rm c}$ of the log-t healing being an increasing function of the initial strength. Then, $\theta_{\rm X}$ turns out to be the $t_{\rm c}$ value for the healing starting with the minimum strength $\Phi_{\rm X}$, which is the possible minimum $t_{\rm c}$. 

One mechanism to produce the above-mentioned differential equation governing $\Phi(t)$ is the quantitative model of \citet{brechet1994effect}, where asperity contacts are being squashed to increase the real contact area via the exponential creep of the contacting asperities driven by the high local normal stress acting on the contacts. 
The secondary variations in $\Phi$ under constant normal stress in the three standard experiments (\S\ref{sec:intro}), taken as canonical behaviors to be reproduced in the present development, are often interpreted as the logarithmic dependence of frictional strength on the average contact time~\citep[e.g.,][]{dieterich1978time,dieterich1979modeling}; this corresponds to the $\theta \to \Phi$ conversion eq.~(\ref{RSFeq:conversionofthetatophinocut}) if we accept the contact-time interpretation of $\theta$ mentioned in \S\ref{subsec:RSFconventional}.

In the rest of the paper, the internal physical state corresponding to the lower-limit strength $\Phi_{\rm X}$ and the associated state-variable value ($\theta = 0$) shall be referred to as X-state. The above representative interpretation of $\theta$, taken together with ATF~\citep{bowden1964friction}, suggests that the X-state is the real contact area instantly attained upon applying the normal load, then providing the minimum strength value as eq.~(\ref{RSFeq:xstatePhi}). 

When treating the minimum strength, we suppose the X-state exists as in eq.~(\ref{RSFeq:constitutivethetacutoff}). Hence, in the $\theta$-notation, the RSF framework becomes the set of eqs.~(\ref{RSFeq:constitutivethetacutoff}) and (\ref{RSFeq:evolutionlawtheta}), in place of the set of eqs.~(\ref{RSFeq:constitutivetheta}) and (\ref{RSFeq:evolutionlawtheta}). Since the former framework includes the latter as its special case of $\theta\gg\theta_{\rm X}$ (cf. eq.~\ref{eq:Xstate2starstate}), only treatment under the former set suffices. 
By contrast, in the $\Phi$-notation, the framework consisting of eqs.~(\ref{RSFeq:constitutivePhiA}) and (\ref{RSFeq:evolutionlawPhi}) can be used as is, just by regarding eq.~(\ref{RSFeq:xstatePhi}) as the definition range of $\Phi$. 
In the $\Phi$-notation, it is clear that the lower strength limit is also a requirement to the evolution law eq.~(\ref{RSFeq:evolutionlawPhi}).

\section{Experimental requirements and empirical constraints on evolution laws}
\label{sec:3}

In \S\ref{subsec:SSrequirement}--\ref{subsec:SHSrequirement}, we formulate each canonical behavior of the frictional strength $\Phi$ in SS, VS, and SHS tests~\citep{ruina1983slip,dieterich1979modeling} to derive constraints on evolution laws as empirical formulae, which we use in the present development of better evolution laws (\S\ref{sec:development}). 
Substantial consensus exists in the field on the canonical behaviors we formulate in this section. These behaviors have been the target of earlier works of the same purpose, i.e., finding an evolution law capable of reproducing all those canonical behaviors simultaneously; the present paper is trying to take part in this quest. 

Following \citet{beeler1997roles}, we limit (except in \S\ref{subsec:discussion1}) the scope of the present development to the state evolution described by the following thermodynamic formulation (TDF) in terms of slip $\delta$ and time $t$: 
\begin{equation}
d\Phi = 
\left( \frac{\partial \Phi}{\partial t} 
\right)_\delta dt
+
\left( \frac{\partial \Phi}{\partial \delta} 
\right)_t  d\delta.
\label{RSFeq:thermodynamicformPhi}
\end{equation}
This is a presumption, but many existing evolution laws, including the aging, slip, and composite laws, fit the above total differential form. We denote the purely time-dependent and slip-driven parts of eq.~(\ref{RSFeq:thermodynamicformPhi}) with
\begin{flalign}
F_t&:=
\left( \frac{\partial \Phi}{\partial t} \right)_\delta 
\label{RSFeq:defofFt}
\\
F_\delta&:=
\left( \frac{\partial \Phi}{\partial \delta} \right)_t  
\label{RSFeq:defofFdelta}
\end{flalign}
Dividing eq.~(\ref{RSFeq:thermodynamicformPhi}) by $dt$, we see that the evolution function $F$ (eq.~\ref{RSFeq:evolutionlawPhi}) of $\Phi$ is expressed as
\begin{equation}
F=F_{t} +F_{\delta} V.
\label{eq:Phievolutiondecomposition}
\end{equation}

We are aware that $\Phi$ can evolve also in response to the changes in normal stress $\sigma$ and shear stress $\tau$, but the present work does not consider them. The shear stress effect~\citep{nagata2012revised} does not seem directly related to the difficulty of reproducing the evolutions in VS and SHS tests simultaneously, the main issue the present paper attempts to resolve. 
Although \citet{nagata2012revised} noted that their evolution law, where the explicit effect of $\tau$ has been added to the aging law, approximately reproduced the symmetric evolutions following the positive and negative velocity steps, this success is limited to small-magnitude steps---\citet{bhattacharya2015critical} have demonstrated that the Nagata law, with the laboratory-derived magnitude of the shear stress effect~\citep{nagata2012revised}, predicts an asymmetric response if the velocity is stepped by as large as three orders of magnitude. Meanwhile, the normal stress effect~\citep{linker1992effects} is beyond the scope of the present work. 
In any case, these stress effects reported so far are represented as additional stress-dependent evolution terms to be added to eq.~(\ref{RSFeq:thermodynamicformPhi}). Hence, the presently developed evolution law, which has the form of eq.~(\ref{RSFeq:thermodynamicformPhi}), can be extended to include those stress effects as additional terms. 

Under the premise of the TDF, the function $f$ describing the evolution of $\theta$ can be expressed in a form parallel to eq.~(\ref{eq:Phievolutiondecomposition}). That is, when we define the purely time-dependent part $f_t$ and the slip-driven part $f_\delta$ of the $\theta$ evolution as
\begin{flalign}
f_t
&:=\left(
\frac{\partial \theta}{\partial t}
\right)_\delta
\label{RSFeq:defofsmallft}
\\
f_\delta&:=
\left(
\frac{\partial \theta}{\partial \delta}
\right)_t
\label{RSFeq:defofsmallfdelta}
\end{flalign}
then $f$ can be written as
\begin{equation}
f=f_t+f_\delta V.
\label{eq:thermodynamicalf}
\end{equation}
Note that $f$, $f_t$, and $f_\delta$ are connected with $F$, $F_t$, and $F_\delta$, respectively, 
via the variable conversion formulae eqs.~(\ref{RSFeq:conversionofthetatophi}) and (\ref{RSFeq:conversionofphitotheta}).

To start formulating canons, we introduce idealizations with respect to the experimental settings and their reasoning. 
The friction law describes a nature of the frictional interface, thus determined by the properties of the interface and free from the experimental apparatus. The RSF premises this natural expectation, known as the spatial locality of physical laws, and the RSF users accounts for the influence of machine stiffness and unwanted machine motions by using spring-slider models when fitting the RSF to experimental data. Conversely, since the friction law itselef is independent of anything outside the interface, here we are allowed to use the idealized machine property and idealized machine motions when formulating laws, without loss of applicability to actual experiments. We follow these basic concepts of the RSF, and in formulating the canons, we suppose that ram velocity and force changes can be stepwise, much faster than the time scale of interest, but within the quasistatic regime. In formulating the VS canon, we further assume the stiffness is sufficiently high. Meanwhile, 
note that log-t healing during stationary-loadpoint SHS is affected by finite stiffness $k$ as $\tau(t)$ and $V(t)$ depend on $k$, while they are not affected by $k$ during constant-$\tau$ holds (\S\ref{subsubsec:331}).

\subsection{SS tests at various velocities} 
\label{subsec:SSrequirement}

Suppose a slip rate $V$ sufficiently low so that the high-velocity cutoff of the state effect can be ignored, that is, $\exp[(\Phi_{\rm SS}(V)-\Phi_{\rm X})/B]\gg 1$ (e.g., $\Phi_{\rm SS}(V)-\Phi_{\rm X}\geq 3B$). 
For such $V$, SS tests show the following negative logarithmic velocity dependence of $\Phi_{\rm SS}$~\citep[e.g.,][]{dieterich1979modeling}: 
\begin{equation}
\frac{d \Phi_{\rm SS}(V)}{d \ln V} \simeq -B
\hspace{10pt}(\mbox{Requirement on the steady state}).
\label{eq:req4ss}
\end{equation}


Then, we derive constraints on the evolution laws to reproduce the above canonical behavior. We use the $\theta$-notation here in consideration of readers' familiarity. For the concerned range of $V$ s.t. $\exp[\Phi_{\rm SS}(V)-\Phi_{\rm X})/B]\gg 1$, eq.~(\ref{RSFeq:conversionofphitotheta}) leads to $d\Phi_{\rm SS}/d \ln\theta_{\rm SS}=B$. Comparing this with the requirement eq.~(\ref{eq:req4ss}), we have
\begin{equation}
\frac{d\ln\theta_{\rm SS}}{d\ln V}=-1,
\end{equation}
which leads to 
$\theta_{\rm SS}(V)\propto 1/V$, that is,
\begin{equation}
V\theta_{\rm SS}(V)=const. =:L, 
\label{RSFeq:Vthetaconst}
\end{equation}
or equivalently,
\begin{equation}
\theta_{\rm SS}(V)=L/V. 
\label{eq:steadystateform_theta}
\end{equation}
Here, $L$ is the constant value of $V\theta_{\rm SS}$, which has a length dimension. Equation~(\ref{eq:steadystateform_theta}) is consistent with the representative interpretation of $\theta$ ($\theta_{\rm SS}$) as the average contact time~\citep{dieterich1978time}. 

The evolution function $f$ must satisfy $f(\theta_{\rm SS}(V), V) = 0$ for $\theta_{\rm SS}(V)$ defined by eq.~(\ref{eq:steadystateform_theta}). 
We use eq.~(\ref{eq:steadystateform_theta}) [its essence is eq.~(\ref{RSFeq:Vthetaconst})] to constrain the appropriate evolution law in terms of the steady state.

For convenience in subsequent derivation (\S\ref{sec:development}), we here introduce a variable $\Omega$~\citep{rubin2005earthquake}, which is $\theta$ normalized by $\theta_{\rm SS}(V)$: 
\begin{equation}
\Omega := V\theta/L.
\label{RSFeq:defofOmega_steadystate}
\end{equation}
$\Omega$ can be regarded as another state variable. Its steady-state value is unity: 
\begin{equation}
\Omega_{\rm SS} = 1.
\label{RSFeq:steadystateconstraint}
\end{equation}
Equation~(\ref{RSFeq:steadystateconstraint}) is equivalent to eq.~(\ref{eq:steadystateform_theta}) and hence works as an alternative expression of the constraint. 


\subsection{Transient evolution in ideal VS tests}
\label{subsec:VSrequirement}

In ideal VS tests, the interface slip rate $V$ is abruptly changed from one value ($V_{\rm before}$) to another ($V_{\rm after}$), following steady-state sliding at $V=V_{\rm before}$, which sets the strength to $\Phi_{\rm SS}(V_{\rm before})$ as the initial condition for the post-step evolution. In laboratory experiments, such a perfect step change of $V$ is approximately realized by abruptly changing the load-point velocity $V_{\rm m}$ from $V_{\rm before}$ to $V_{\rm after}$ on a sufficiently stiff loading apparatus. Following the velocity step, the strength $\Phi$ evolves from the old steady-state value $\Phi_{\rm SS}(V_{\rm before})$ to the new value $\Phi_{\rm SS}(V_{\rm after})$, exponentially with the slip displacement $\delta$ since the moment of the step. A critical, well-established experimental observation is that the characteristic slip distance $D_{\rm c}$ in this exponential evolution does not depend on either $V_{\rm before}$ or $V_{\rm after}$~\citep{ruina1980friction,bhattacharya2015critical}: 
\begin{equation}
    \Phi_{\rm VS}(\delta)=\Phi_{\rm SS}(V_{\rm after})-[\Phi_{\rm SS}(V_{\rm after})-\Phi_{\rm SS}(V_{\rm before})]e^{-\delta /D_{\rm c}}
    \label{eq:realVScanonical}
    \hspace{10pt}(\mbox{VS requirement}),
\end{equation}
where $\Phi_{\rm VS}$ denotes the strength $\Phi$ during VS tests. Equation~(\ref{eq:realVScanonical}) is the canonical behavior in VS tests. Good evolution laws must reproduce it. 

The VS canon, idealized by using the infinite machine stiffness, clearly indicates the time-evolution law that governs the slip weakening and strengthening. 
We now derive that time-evolution law 
in a time-derivative form, which applies to
whatever $V(t)$, including the real-world VS under finite stiffness, where non-Heaviside $V(t)$ arises from the elastic interaction and the RSF
constitutive law (eq.~\ref{eq:N1}), 
reproducing the non-exponential behaviors of the frictional stress under actual VS tests.

Equation~(\ref{eq:realVScanonical}) is a solution of the following differential equation of $\Phi$ with respect to $\delta$~\citep{ruina1983slip}: 
\begin{equation}
    \frac{d\Phi}{d\delta} =- \frac{\Phi-\Phi_{\rm SS}(V_{\rm after})}{D_{\rm c}}.
    \label{eq:sliplawPhiform}
\end{equation}
Because eq.~(\ref{eq:realVScanonical}) holds for any slip rates ever tested, the above fact translates to  
\begin{equation}
d\Phi_{\rm VS} =- \frac{\Phi-\Phi_{\rm SS}(V_{\rm after})}{D_{\rm c}}d\delta
\label{RSFeq:RequirementinVS} 
\end{equation}
within the framework of the TDF (eq.~\ref{RSFeq:thermodynamicformPhi}). 
Here, $d\Phi_{\rm VS}$,  the strength variation $d\Phi$ in VS tests, derives from the increment of slip $d\delta$, not from the lapse of time $dt$. 

There is no a priori reason to think that $D_{\rm c}$ is equal to $L$ (the product of $\theta_{\rm SS}$ and $V$, eq.~\ref{RSFeq:Vthetaconst}), though conventional RSF laws usually assume $D_{\rm c} = L$ implicitly~\citep{nakatani2001conceptual,nakatani2006intrinsic}. For generality's sake, we treat $D_{\rm c}$ and $L$ as two independent parameters for the time being.

\begin{figure*}
   \includegraphics[width=150mm]{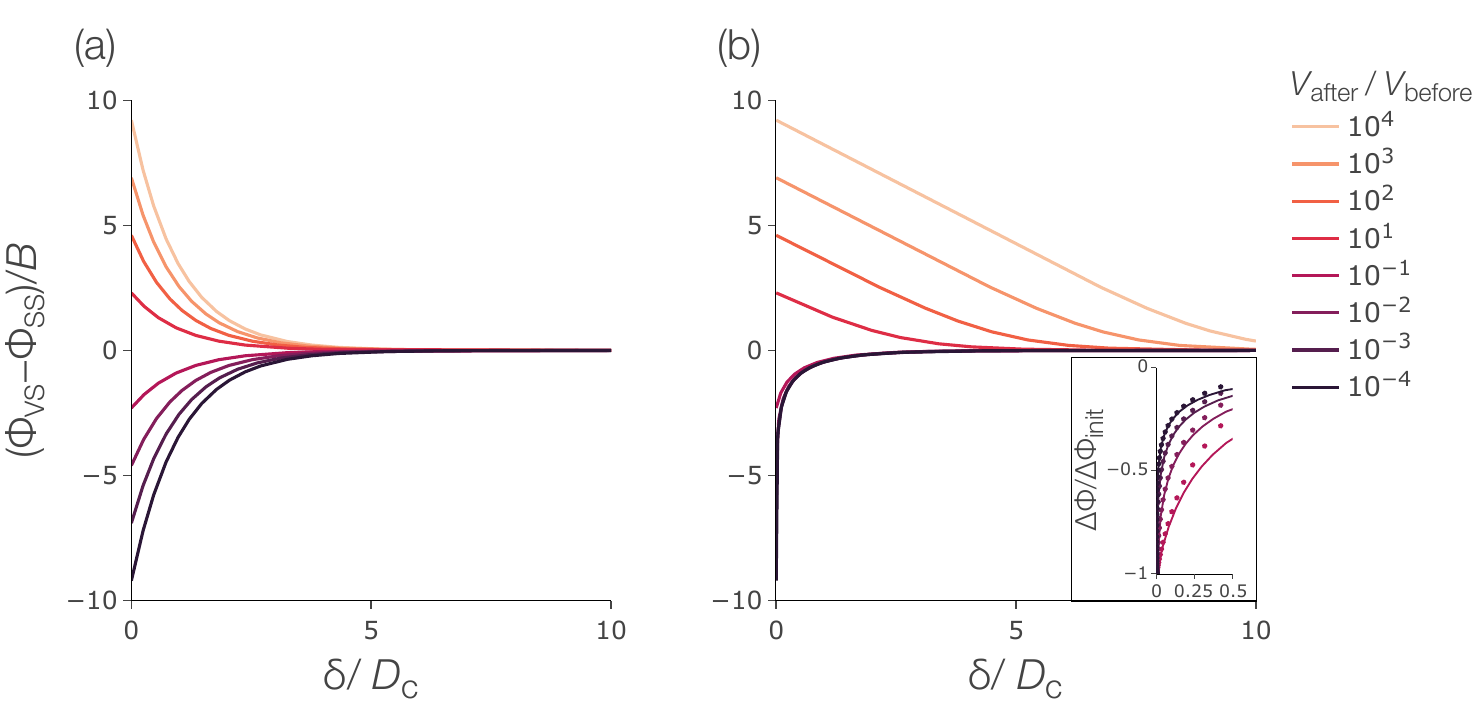}
  \caption{
Strength evolution in VS tests, predicted by (a) the slip law and (b) the aging law, simulated for ideal VS tests on an infinitely stiff apparatus, where interface slip rate is abruptly changed from $V_{\rm before}$ to $V_{\rm after}$. $D=D_{\rm c}(=L)$ is assumed. The value of $\theta_{\rm X}$ is  assumed to be so small that
$\theta\gg\theta_{\rm X}$ holds throughout the tests. Normalized frictional strength $[\Phi_{\rm VS}-\Phi_{\rm SS}(V_{\rm after})]/B$ is plotted against the normalized slip distance $\delta/D_{\rm c}$ since the moment of the velocity step. Results for different $V_{\rm after}/V_{\rm before}$ values are shown in different colors. The inset in panel (b) focuses on the early stage ($\delta<D_{\rm c}/2$) of evolution following negative ($V_{\rm after}<V_{\rm before}$) steps, showing $\Delta \Phi=\Phi_{\rm VS}-\Phi_{\rm SS}(V_{\rm after})$ normalized by its initial value $\Delta \Phi_{\rm init}$ right after the step. For comparison, the log-t healing curves, predicted by $\dot \theta=1$, are shown by the dotted lines of the same colors. 
  }
  \label{fig:AgingSlipVS}
\end{figure*}

In the context of the quest for the correct evolution law, one essential feature of eq.~(\ref{eq:realVScanonical}) is the symmetry of the transient strength evolution following the positive and negative $V$ steps, a qualitative feature repeatedly confirmed in various experiments~\citep[e.g.,][]{ruina1980friction}. 
As far as $\theta \gg \theta_{\rm X}$ is considered, the slip law (eq.~\ref{eq:defofsliplaw}) is identical to eq.~(\ref{eq:sliplawPhiform}) and exactly predicts the canonical behavior eq.~(\ref{eq:realVScanonical})~\citep[also see Fig.~\ref{fig:AgingSlipVS}a]{ruina1983slip}. 
In contrast, the other widely-used evolution law, the aging law (eq.~\ref{eq:defofaginglaw}), predicts infamously asymmetric behaviors~\citep[e.g.,][also see Fig.~\ref{fig:AgingSlipVS}b]{ampuero2008earthquake}, far different from eq.~(\ref{eq:realVScanonical}). For later comparison with the presently proposed evolution law, we derive, in Appendix~\ref{sec:VSagingderivation}, approximate analytic expressions of the aging-law prediction following the positive and negative $V$ steps. 

Equating eq.~(\ref{RSFeq:RequirementinVS}), the requirement from VS tests written in the TDF, with eq.~(\ref{RSFeq:thermodynamicformPhi}), the evolution law written in the TDF, we obtain 
\begin{equation}
F_{t|{\rm VS}} dt
+
F_{\delta|{\rm VS}}  d\delta=
- \frac{\Phi-\Phi_{\rm SS}(V_{\rm after})}{D_{\rm c}}d\delta
,
\label{eq:VSconstraintderivationfirstline}
\end{equation}
where $F_{t|{\rm VS}}$ and $F_{\delta|{\rm VS}}$ denote $F_{t}$ and $F_{\delta}$ in VS tests. 
Equating coefficients of the infinitesimals $d\delta$ and $dt$, 
\begin{flalign}
F_{t|{\rm VS}} &=0
\label{RSFeq:PhiVS_PTH} 
\\
F_{\delta|{\rm VS}} &=- \frac{1}{D_{\rm c}}[\Phi-\Phi_{\rm SS}(V)].
\label{RSFeq:PhiVS} 
\end{flalign}
The above derivation is identical to the discussion of \citet{ruina1983slip} for the evolution law of $\Theta:= (\Phi-\Phi_*)/\sigma$, and hence the evolution law derived from eqs.~(\ref{RSFeq:PhiVS_PTH}) and (\ref{RSFeq:PhiVS}) is nothing more than the slip law in the $\Theta$-notation by \citet{ruina1983slip}. 

Thus, it has been concluded that $F_t$, representing purely time-dependent healing of the strength, must be zero in VS tests (eq.~\ref{RSFeq:PhiVS_PTH}). 
This reflects that eq.~(\ref{RSFeq:RequirementinVS}) describes the strength variations in VS tests as pure responses to slip (i.e., $d\Phi_{\rm VS}\propto d\delta$), irrespective of $V_{\rm after}$ and $V_{\rm before}$ values. 
Nonetheless, we argue that such laboratory observation, which only has limited accuracy, should not be taken as the constraint that $F_{t|{\rm VS}}$ is completely null; non-zero $F_{t|{\rm VS}}$ may exist as long as the predicted evolution in VS tests still approximates the canonical exponential convergence over a length constant $D_{\rm c}$. 
Accordingly, we weaken $=$ to $\simeq$ in the canon from VS tests (eq.~\ref{eq:realVScanonical}) and relax eq.~(\ref{RSFeq:RequirementinVS}) to the following: 
\begin{equation}
d\Phi_{\rm VS} \simeq - \frac{\Phi-\Phi_{\rm SS}(V_{\rm after})}{D_{\rm c}}d\delta
\hspace{10pt}(\mbox{A weakened form of eq.~}\ref{RSFeq:RequirementinVS}).
\label{eq:relaxeddifferentialVSrequirement}
\end{equation}
Hereafter, we adopt eq.~(\ref{eq:relaxeddifferentialVSrequirement}) instead of eq.~(\ref{RSFeq:RequirementinVS}) because eq.~(\ref{RSFeq:RequirementinVS}) precludes any possibilities other than the very slip law, which severely disagrees with the healing behaviors observed in SHS tests (see \S\ref{subsec:SHSrequirement}, as well as \S\ref{sec:SHSsliplaw}). 

By the same token, we obtain a weakened form of eq.~(\ref{eq:VSconstraintderivationfirstline}): 
\begin{equation}
F_{t|{\rm VS}} dt
+
F_{\delta|{\rm VS}}  d\delta
\simeq
- \frac{\Phi-\Phi_{\rm SS}(V_{\rm after})}{D_{\rm c}}d\delta,
\end{equation}
Further regarding $\simeq$ as $=$ in the leading order, we obtain a weakened form of eq.~(\ref{RSFeq:PhiVS_PTH}):
\begin{equation}
    |F_{t|{\rm VS}}| \ll
|F_{\delta|{\rm VS}}V| .
\label{RSFeq:PhiVS_negligiblePTH}
\end{equation}
The constraint on $F_\delta$ remains eq.~(\ref{RSFeq:PhiVS}).
Equation~(\ref{RSFeq:PhiVS_negligiblePTH}) states that purely time-dependent healing $F_t$ must be sufficiently small compared to the slip-driven state evolution $F_\delta V$. We deem that eq.~(\ref{RSFeq:PhiVS_negligiblePTH}) is more suitable as a constraint imposed by laboratory observation (i.e., approx. exponential convergence with approx. constant $D_{\rm c}$).

Strictly speaking, eq.~(\ref{RSFeq:PhiVS_negligiblePTH}) is ill-defined when $\Phi$ is exactly $\Phi_{\rm SS}$ because $F_\delta V$ becomes zero as per eq.~(\ref{RSFeq:PhiVS}). Hence, right at the steady state, we need to use eq.~(\ref{RSFeq:PhiVS_PTH}) instead of eq.~(\ref{RSFeq:PhiVS_negligiblePTH}). To avoid complex presentations incurred by such conditional branching, the behaviors under $\Phi=\Phi_{\rm SS}$ shall be hereafter evaluated by considering the behaviors in the limit of $\Phi\to\Phi_{\rm SS}$. Thus, in developing the new evolution law (\S\ref{sec:development}), we use eqs.~(\ref{RSFeq:PhiVS}) and (\ref{RSFeq:PhiVS_negligiblePTH}) as the constraints from VS tests. 

Lastly, we rewrite eqs.~(\ref{RSFeq:PhiVS}) and (\ref{RSFeq:PhiVS_negligiblePTH}) into $\theta$-notation. Using the conversion formula eq.~(\ref{RSFeq:conversionofphitotheta}), eq.~(\ref{RSFeq:PhiVS}) is rewritten as
\begin{flalign}
f_{\delta|{\rm VS}} &= B^{-1}F_{\delta|{\rm VS}}\theta_{\rm X}e^{(\Phi-\Phi_{\rm X})/B}
\\
&=-\frac{\Phi-\Phi_{\rm SS}(V)}{BD_{\rm c}}\theta_{\rm X}e^{(\Phi-\Phi_{\rm X})/B},
\end{flalign}
where $f_{\delta|{\rm VS}}$ denotes $f_{\delta}$ in VS tests. 
Using eqs.~(\ref{RSFeq:conversionofthetatophi}) and
(\ref{RSFeq:conversionofphitotheta}), this is further rewritten into
\begin{equation}
f_{\delta|{\rm VS}} = -\frac{\theta+\theta_{\rm X}}{D_{\rm c}}\ln\left[
\frac{\theta+\theta_{\rm X}}{\theta_{\rm SS}(V)+\theta_{\rm X}}
\right].
\label{RSFeq:dotthetaVS}
\end{equation}
Except for the terms related to the small-$\theta$ cutoff (\S\ref{subsec:RSFminimumStrength}), eq.~(\ref{RSFeq:dotthetaVS}) coincides with eq.~(\ref{eq:defofsliplaw}), the slip law without the cutoff~\citep[e.g.,][]{ruina1983slip}. Likewise, eq.~(\ref{RSFeq:PhiVS_negligiblePTH}) can be converted into
\begin{equation}
    |f_{t|{\rm VS}}|\ll |f_{\delta|{\rm VS}}V| ,
\label{RSFeq:negligiblehealing}
\end{equation}
or equivalently,
\begin{equation}
    f_{\rm VS}\simeq f_{\delta|{\rm VS}} V,
    \label{eq:fvsisfdeltaV}
\end{equation}
where $f_{{\rm VS}}$ and $f_{t|{\rm VS}}$ denote $f$ and $f_t$ in VS tests. 
In the present development of evolution laws (\S\ref{sec:development}), eq.~(\ref{RSFeq:dotthetaVS}) serves as the main constraint from the VS canon, while eq.~(\ref{RSFeq:negligiblehealing}) is used as an auxiliary constraint.

\subsection{SHS tests under the negligible slip condition}
\label{subsec:SHSrequirement}

Log-t healing of frictional strength $\Phi$ during (quasi-)stationary contact~\citep[e.g.,][]{dieterich1972time} is another experimental requirement that good evolution laws must reproduce. The phenomenon is observed in laboratory Slide-Hold-Slide (SHS) tests (see \S\ref{subsubsec:331} for details), which evaluate the change in strength $\Delta\Phi$ that has occurred in the `Hold' period, which is meant to be the stage of stationary contact (i.e., $V=0$) lasting for a designated duration $t_{\rm h}$. Although SHS tests robustly exhibit the log-t healing, namely the linear increase of $\Delta \Phi$ with the logarithm of $t_{\rm h}$, there is a considerable complication in its interpretation, as briefed below. 

If the interface slip rate $V$ during hold is strictly zero as intended, the observed log-t healing is entirely ascribed to the time-dependent change in strength~\citep[e.g.,][]{rabinowicz1958intrinsic,dieterich1972time,dieterich1978time}. Within the TDF framework (eq.~\ref{eq:Phievolutiondecomposition}), a positive $F_t$ term expresses such purely time-dependent evolution as in the aging law~\citep[e.g.,][]{dieterich1979modeling,ruina1983slip}. However, in the majority of SHS tests, slow slip continues during the hold phase (see \S\ref{subsubsec:331} for detailed explanation) to an extent sufficient for the slip law, which lacks the purely time-dependent evolution (i.e., $F_t=0$), to reproduce the observed log-t healing as a consequence of slip-dependent evolution expressed by the $F_\delta V$ term~\citep{ruina1983slip,beeler1994roles}. 
Thus, the log-t healing observed in such SHS tests can be explained by either the slip law (as `apparent' healing) or the aging law (as `purely time-dependent' healing). Although the behaviors predicted by the two laws are not the same even for such SHS tests~\citep[e.g.,][]{beeler1994roles,bhattacharya2017does}, the difference is not significant enough to confidently decide on either laws~\citep[e.g.,][]{bhattacharya2017does}, given the limited experimental accuracy and coverage of conditions. 
In contrast, as detailed in \S\ref{subsubsec:331}, some SHS tests were done at virtually zero slip rates during hold~\citep{nakatani1996effects} or at slip rates way too low for the slip law to reproduce the log-t healing. Nevertheless, significant log-t healing was observed in these tests. For example, in the experiments of \citet{nakatani1996effects}, shear stress during hold was kept as low as approximately $0.02\sigma$, reduced by approximately $0.85\sigma$ from the preceding steady-state sliding. The interface slip rate $V$ during the hold expected from the RSF constitutive law (eq.~\ref{RSFeq:constitutivePhiB}) is so low that slip during hold does not even reach the atomic spacing in rock-forming minerals within the lifetime of human experimenters. 
An absurdly huge $F_\delta$ would be required to ascribe the observed healing to slip-driven evolution under such low $V$. Hence, their observation of significant log-t healing implies that purely time-dependent healing does operate, i.e., $F_t > 0$, at least in the holds at virtually zero slip rates.

Therefore, in developing new evolution laws, we require significant log-t healing to occur in SHS tests even under such low-V conditions that slip-driven evolution cannot produce significant evolution. Such low-V conditions can be expressed mathematically as
\begin{equation}
    |F_t|\gg |F_\delta V|,    
    \label{eq:NSCdef}
\end{equation}
which is equivalent to $|f_t|\gg |f_\delta V|$. 
This condition shall be called the `negligible slip condition' (NSC). 
Notice a logical parallel between the NSC (eq.~\ref{eq:NSCdef}) and the negligible purely time-dependent healing in VS tests (eq.~\ref{RSFeq:PhiVS_negligiblePTH}). 

We emphasize that the reproduction of log-t healing in SHS tests at the NSC (abbreviated as SHS$|$NSC tests hereafter) is an essential requirement in the quest for correct evolution law because the inability to explain the evident occurrence of log-t healing under the NSC is the only definite weak point of the slip law~\citep[e.g.,][]{beeler1994roles,bhattacharya2017does}. Accordingly, modifying the slip law so that log-t healing can be robustly reproduced even at the NSC has been a primary direction in the quest for the correct evolution law reproducing all canonical behaviors~\citep[e.g.,][]{kato2001composite}. 

As will become apparent in the derivation shown later (\S\ref{sec:development}), the canonical behaviors in SS, VS, and SHS$|$NSC tests are sufficient to constrain the evolution laws. Therefore, no room is left to employ behaviors in SHS tests that do not satisfy the NSC as another behavior that good evolution laws should reproduce. Instead, behaviors in such SHS$|$non-NSC tests would be helpful to cross-validate the presently proposed evolution laws because they shall not be tailored to reproduce the results of those tests. In addition, we note that SHS$|$non-NSC tests involve quite a wide variety of slip history during hold, so a comprehensive picture of the log-t healing in such tests has not been established: another reason why we do not adopt SHS$|$non-NSC tests as the target to be reproduced by the evolution laws we develop.

\subsubsection{Technical details of SHS tests}
\label{subsubsec:331}
\begin{figure*}
   \includegraphics[width=120mm]{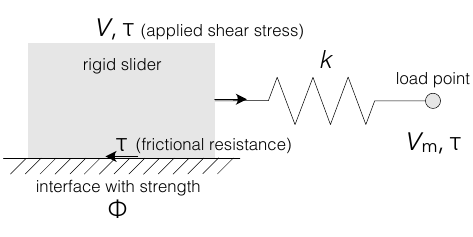}
  \caption{Schematic of the spring-slider model. Variables in the figure correspond to those defined in \S\ref{subsubsec:331}. Force directions associated with $\tau$ are indicated for conceptual clarity. 
}
  \label{fig:springslider}
\end{figure*}

Laboratory friction experiments can be modeled as a spring-slider system (Fig.~\ref{fig:springslider}). The load point is the servo-controlled ram applying the shear stress. Inertia is negligible in the usual velocity range of typical RSF studies, and the equation of motion of the rigid slider is
\begin{equation}
    \dot \tau = k(V_{\rm m}-V),
    \label{eq:springslidereq}
\end{equation}
where the interface slip rate $V$ follows the RSF constitutive law eq.~(\ref{RSFeq:constitutivePhiB}). The load-point velocity $V_{\rm m}(t)$ can be specified as desired by the experimenter. System stiffness $k$ considers the elastic deformation of the sample and apparatus but is usually dominated by the apparatus's deformation. When desired, an experimenter can specify the applied shear stress $\tau(t)$ instead of $V_{\rm m}(t)$ by switching the feedback source of the apparatus's servo-motion from the measured ram displacement to the measured shear load. 

In the first slide of SHS, $V_{\rm m}$ is held constant at $V_{\rm prior}$, and steady-state sliding at $V = V_{\rm prior}$ is achieved, setting up the initial condition ($\Phi_{\rm init}=\Phi_{\rm SS}(V_{\rm prior})$) for the subsequent `Hold' phase. Then, the experimenter starts the hold phase, meant to be the stage of (quasi-)stationary contact. There are two different ways to realize the hold phase, on which physical conditions $(\tau(t), V(t))$ during the hold depend, as explained below.

The first type is `stationary-loadpoint hold,' where the load point is halted and kept unmoved (i.e., $V_m(t) = 0$ is imposed) for $t_{\rm h}$. 
Neither $\tau(t)$ nor $V(t)$ during hold can be specified by the experimenter during this type of hold. The shear stress starts with $\tau_{\rm SS}(V_{\rm prior})$ and decreases with time as the slow slip following eq.~(\ref{RSFeq:constitutivePhiB}) continues while decelerating due to the decreasing $\tau(t)$ and the increasing $\Phi(t)$. 
Note $\tau(t)$ rarely drops much below $\tau_{\rm SS}(V_{\rm prior})$, and hence, a significant level of $V$ is maintained during the hold, enabling the slip law to explain the observed log-t healing as the slip-dependent evolution during the hold period. Detailed analysis of the stationary-loadpoint hold can be found in \citet{bhattacharya2017does}. 

The other type is `constant-$\tau$ hold,' where the load-point motion is controlled to keep the shear stress constant at a designated value $\tau_{\rm hold}$ taken to be lower than $\tau_{\rm SS}(V_{\rm prior})$ by $\Delta\tau(>0)$. 
The interface slip rate $V$ at the beginning of the hold, denoted by $V_{\rm init}$, is $V_{\rm prior}\exp(-\Delta\tau/A)$ (eq.~\ref{eq:initialV}). During the hold, $V$ usually decelerates further as $\Phi$ increases while $\tau$ is kept constant~\citep[e.g.,][]{nakatani2001conceptual}. Detailed analyses, with each of the four specific evolution laws discussed in the present paper, are given in Appendix \ref{sec:SHSHSprinciple}. Note that the system stiffness $k$ does not affect the slip history during this type of hold, unlike in the stationary-loadpoint hold, where $\tau$ changes due to the continuing slip according to eq.~(\ref{eq:springslidereq}) involving $k$.
Although the majority of the SHS tests in literature employ stationary-loadpoint hold, the present paper mainly focuses on constant-$\tau$ hold tests because, by adopting sufficiently low $\tau_{\rm hold}$~\citep[e.g.,][]{nakatani1996effects}, they can realize the NSC during hold, under which condition we require the log-t healing as the canon from SHS tests, as declared earlier. 

After the hold duration $t_{\rm h}$ has elapsed, the interface is again slid to evaluate the strength $\Phi_{\rm end}$ established by the quasi-stationary contact for the duration $t_h$. Commonly to both types of hold, this last reloading stage of SHS is realized by advancing the load point at $V_{\rm m} = V_{\rm reload}$ to find the peak shear stress $\tau_{\rm peak}$, i.e., the static friction in classical tribology. Equation~(\ref{eq:springslidereq}) guarantees that, at the peak stress (i.e., when $\dot \tau = 0$), $V$ is equal to $V_{\rm m}$ ($= V_{\rm reload}$) irrespective of the system stiffness. 
For simplicity, the present paper assumes $V_{\rm reload} = V_{\rm prior}$ as usual, so that $\tau_{\rm peak} - \tau_{\rm SS}(V_{\rm prior})$ directly gives $\Phi_{\rm peak} - \Phi_{\rm init}$, where the direct effect term has been canceled thanks to $V_{\rm reload} = V_{\rm prior}$. In ideal experiments with $k = \infty$, the slip rate $V$ is instantly brought to $V_{\rm reload}$ as soon as the reloading starts so that $\Phi_{\rm peak}$ equals $\Phi_{\rm end}$, and the strength change that has occurred during the hold ($\Delta\Phi := \Phi_{\rm end} - \Phi_{\rm init}$) is equal to the directly observable quantity of $\tau_{\rm peak} -  \tau_{\rm SS}(V_{\rm prior})$. In real-world experiments with finite $k$, however, the state changes somewhat during the reloading from the end of the hold to the peak stress, which must be corrected for in estimating $\Phi_{\rm end}$ and, in turn, $\Delta \Phi$. The correction is usually done via simulations assuming some evolution law~\citep[e.g.,][]{blanpied1998effects,nakatani2001conceptual}. Experimental requirements from SHS tests must be based on $\Delta\Phi(t_{\rm h})$ estimated after this correction. 

Meanwhile, \citet{nagata2008monitoring} have demonstrated that acoustic transmissivity across the frictional interface is linearly related to $\Phi$ (observable as per eq.~\ref{eq:N1}). With this method, $\Phi_{\rm end}$ or even the continuous record of $\Delta\Phi(t)$ throughout the hold period can be monitored in-situ, whereby the reloading operation and the above-mentioned correction become unnecessary. 
Although the linear relationship between acoustic transmissivity and $\Phi$ breaks during reloading to $\tau_{\rm peak}$~\citep[e.g.,][]{nagata2008monitoring}, which is the initial part of slip-weakening of $\Phi$, the present paper refers to the acoustic-transmissivity- $\Phi$ linearity only for SS and hold phase of SHS tests.
Furthermore, for transparent materials, optical transmissivity can play the same role~\citep{dieterich1994direct,nagata2014high}. It also supports the ATF~\citep{bowden1964friction} interpretation of $\Phi$ because optical transmissivity is proportional to the real contact area. 
These acoustic/optical results, though indirect measurements of the strength $\Phi$, may also be usable as a basis to formulate the canonical log-t healing behavior. 

Significant occurrence of log-t healing at any $\tau_{\rm hold}$ levels is seen consistently in experiments on nominally bare surfaces~\citep{nakatani1996effects,berthoud1999physical,ryan2018role}. Furthermore, the positive $\tau_{\rm hold}$-dependence of $B_{\rm heal}$, which increases by a factor of 2--2.5 from $\tau_{\rm hold}/\sigma \simeq +0$ to $\tau_{\rm hold}$ somewhat below dynamic friction, has been observed in all these experiments. 
\citet{nakatani1996effects} and \citet{ryan2018role} investigated many different levels of $\tau_{\rm hold}$ in this range and commonly found that $B(\tau_{\rm hold})$ was approximately constant for $\tau_{\rm hold} > 0.4\sigma$ and exhibited quasi-linear positive dependence for $0 < \tau_{\rm hold}/\sigma < 0.4$.
Note that \citet{ryan2018role} log-t healing was quantitatively examined only by continuous acoustic transmission monitoring during the hold phase, as they did not do SHS with many different $t_{\rm h}$. 
Strengthening at very high-$\tau_{\rm hold}$ will be apparent healing driven by the slip strengthening as in the stationary-loadpoint SHS, but frictional healing at moderately and extremely low-$\tau_{\rm hold}$ is an indication of purely time-dependent healing.

The RSF parameters, such as $A$, $B$, and $D_{\rm c}$, depend on experimental conditions. In the present development of state evolution law, we often presume a functional form $f(V, \theta)$, adopted in many RSF evolution laws. Though this may sound restrictive, the dependence of parameters on $\tau$, $V$, and $\Omega$ can be expressed even under this restriction.



\subsubsection{Canonical behavior and evolution-law constraint}
\label{subsubsec:332}

Finally, we formulate the requirements from SHS tests. The canonical log-t healing of $\Phi$ during the hold time $t_{\rm h}$ (during the hold, before reloading), spent at the NSC, is written by a function $\Phi_{\rm SHS|NSC} (t_{\rm h})$:
\begin{equation}
\Phi_{\rm SHS|NSC} (t_{\rm h})= \Phi_{\rm init}+B_{\rm heal} \ln(t_{\rm h}/t_{\rm c}+1)
\hspace{10pt}(\mbox{Requirement 1 from SHS tests}),
\label{RSFeq:PhiSHSNSCrawtc}
\end{equation}
where $\Phi_{\rm init}$ is the initial strength at the beginning of the hold period [$=\Phi_{\rm SS}(V_{\rm prior})$], 
$t_{\rm c}$ denotes the apparent cutoff time of the log-t healing~\citep{nakatani2006intrinsic}, inversely depending on $V_{\rm prior}$ as shall be explained in \S\ref{sec:34}, 
and $B_{\rm heal}$ is an empirical parameter of stress dimension representing the magnitude of strengthening per e-fold increase in $t_{\rm h}$. 
In most literature, $B_{\rm heal}$ is supposed to be equal to the magnitude of the log-V dependence of the steady-state strength~\citep[eq.~\ref{eq:req4ss}; e.g.,][]{dieterich1979modeling,okubo1986state}, which the present paper follows: 
\begin{equation}
B_{\rm heal}=B 
\hspace{10pt}(\mbox{Requirement 2 from SHS tests}).
\label{RSFeq:coincidenceofBs}
\end{equation}

However, for evolution laws having $f(V,\theta)$ form, including the aging law and the slip law (eqs.~\ref{eq:defofaginglaw} and \ref{eq:defofsliplaw}), 
it may not be appropriate to regard eq.~(\ref{RSFeq:coincidenceofBs}) as a part of the canons. 
As long as we presume that $\dot\theta$ only depends on the instantaneous $V$ and $\theta$, namely $\dot\theta=f(V,\theta)$, eq.~(\ref{RSFeq:coincidenceofBs}) follows three of the other empirical formulae, eqs.~(\ref{eq:req4ss}), (\ref{RSFeq:PhiSHSNSCrawtc}), and (\ref{eq:tcdetection}), as shown in Appendix~\ref{sec:SHSderivationdetail}. 
In this sense, eq.~(\ref{RSFeq:coincidenceofBs}) is a theoretical consequence of other empirical formulae rather than a requirement. 
Nonetheless, for the sake of prudence, we position eq.~(\ref{RSFeq:coincidenceofBs}) as a requirement from SHS tests, which would not harm the present development. 

\begin{figure*}
   \includegraphics[width=115mm]{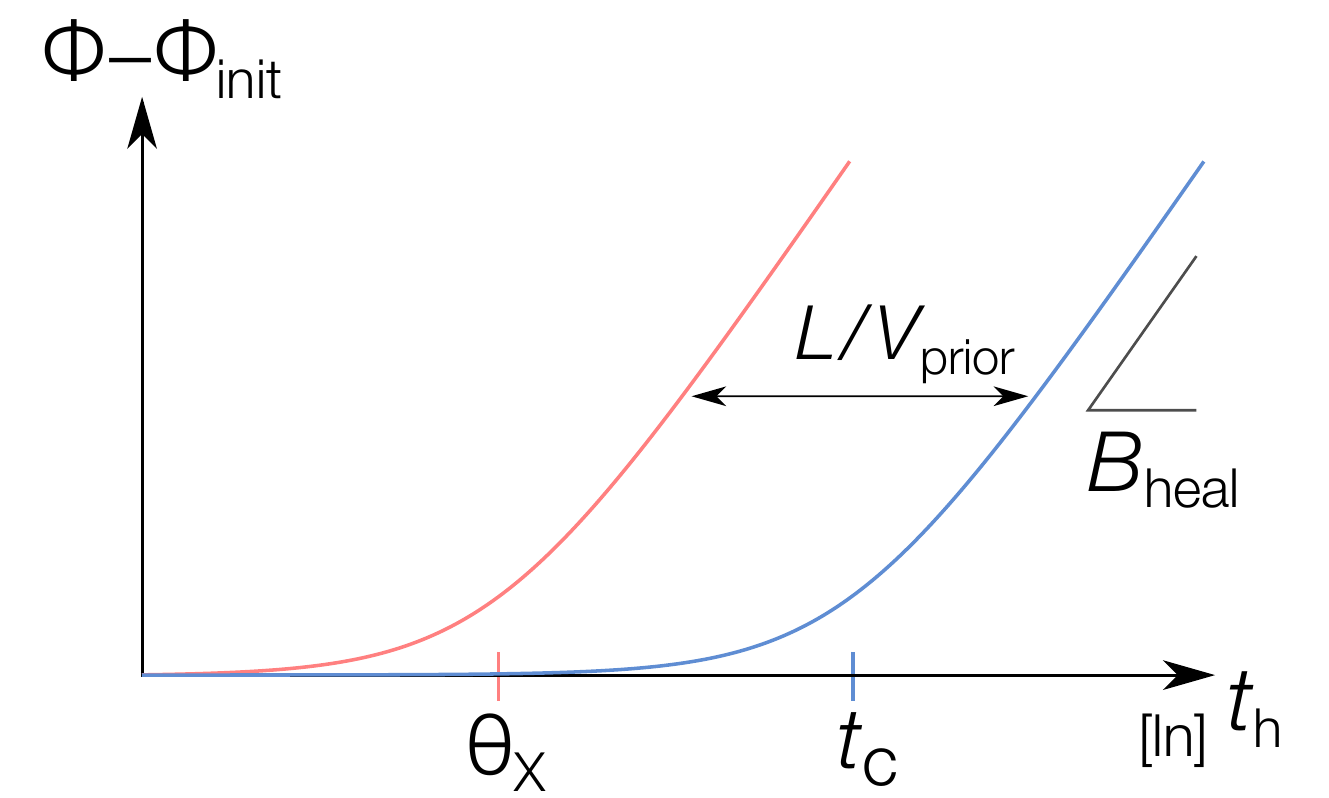}
  \caption{
  The canonical behavior of the strength $\Phi$ in SHS tests (schematic). Plotted against the hold time $t_{\rm h}$. The red curve shows the case with initial strength $\Phi_{\rm init}=\Phi_{\rm X}$, whose cutoff time $t_{\rm c} = \theta_{\rm X}$ is indicated by the red tick on the $t_{\rm h}$-axis. The blue curve shows the general case with $\Phi_{\rm init}=\Phi_{\rm SS}(V_{\rm prior})$, where $V_{\rm prior}$ is the slip rate in the preceding steady-state slide. The cutoff time is  $t_{\rm c} = \theta_{\rm X}+L/V_{\rm prior}$, indicated by the blue tick on the $t_{\rm h}$-axis. For $t_{\rm h} \gg t_{\rm c}(V_{\rm prior})$, strength increases linearly with $\ln t_{\rm h}$, at a slope $B_{\rm heal}$ independent of $\Phi_{\rm init}$. 
}
  \label{fig:LTH}
\end{figure*}

%
%
%


We now derive the constraints on the evolution law under the NSC from the SHS requirements formulated above. 
Let the evolution law during an NSC hold be denoted as $\dot\theta=f_{\rm SHS|NSC}$. Then, the value of $\theta$ during the NSC hold, denoted by $\theta_{\rm SHS|NSC}$, can be written as 
\begin{equation}
\theta_{\rm SHS|NSC}(t_{\rm h}) =\theta_{\rm init} + \int^{t_{\rm h}}_0 dt f_{\rm SHS|NSC}(t),
\label{eq:thetaSHSNSC}
\end{equation}
where $\theta_{\rm init}$ is the initial $\theta$ value converted from $\Phi_{\rm init}$ via eq.~(\ref{RSFeq:conversionofphitotheta}), that is,
\begin{equation}
    \Phi_{\rm init}= \Phi_{\rm X}+ B\ln(\theta_{\rm init}/\theta_{\rm X} + 1).
    \label{eq:Phiinitvalue}
\end{equation}
Converting $\theta_{\rm SHS|NSC}$ in eq.~(\ref{eq:thetaSHSNSC}) to $\Phi_{\rm SHS|NSC}$ via eq.~(\ref{RSFeq:conversionofthetatophi}), we have
\begin{equation}
    \Phi_{\rm SHS|NSC}(t_{\rm h}) =\Phi_{\rm X} +B \ln\left[\left(\theta_{\rm init} + \int^{t_{\rm h}}_0 dt f_{\rm SHS|NSC}\right) /\theta_{\rm X} + 1\right].
\end{equation}
Subtracting $\Phi_{\rm init}$ (eq.~\ref{eq:Phiinitvalue}) from both sides, it becomes
\begin{equation}
    \Phi_{\rm SHS|NSC}(t_{\rm h})-\Phi_{\rm init} =B \ln
    \left(\frac{\theta_{\rm init} + \int^{t_{\rm h}}_0 dt f_{\rm SHS|NSC}+\theta_{\rm X}}{\theta_{\rm init} + \theta_{\rm X}}\right),
\end{equation}
reduced to
\begin{equation}
    \Phi_{\rm SHS|NSC}(t_{\rm h})-\Phi_{\rm init} =B \ln
    \left(\frac{\int^{t_{\rm h}}_0 dt f_{\rm SHS|NSC}}{\theta_{\rm init} + \theta_{\rm X}}+1\right).
    \label{eq:PhiSHS|NSCasintegral}
\end{equation}
Comparison of eq.~(\ref{eq:PhiSHS|NSCasintegral}) with the requirement eq.~(\ref{RSFeq:PhiSHSNSCrawtc}) leads to 
\begin{equation}
    t_{\rm c}^{B_{\rm heal}/B}=c_*^{-1}(\theta_{\rm init}+\theta_{\rm X})
    \label{eq:tc_depon_Bheal_cstar}
\end{equation}
and 
\begin{equation}
    (t_{\rm h}+t_{\rm c})^{B_{\rm heal}/B}=c_*^{-1}\left[\int^{t_{\rm h}}_0 dt f_{\rm SHS|NSC}+(\theta_{\rm init}+\theta_{\rm X})\right].
    \label{eq:thetanscwithbheal}
\end{equation}
In the above, $c_*(>0)$ is an arbitrary positive constant.
Under the requirement eq.~(\ref{RSFeq:coincidenceofBs}), 
eqs.~(\ref{eq:tc_depon_Bheal_cstar}) and (\ref{eq:thetanscwithbheal}) become 
\begin{equation}
    t_{\rm c}=c_*^{-1}(\theta_{\rm init}+\theta_{\rm X})
    \label{eq:tc_depon_cstar}
\end{equation}
and
\begin{equation}
    t_{\rm h}+t_{\rm c}=c_*^{-1}\left[\int^{t_{\rm h}}_0 dt f_{\rm SHS|NSC}+(\theta_{\rm init}+\theta_{\rm X})\right].
    \label{eq:integratedftSHS_witharbitrariness}
\end{equation}
By differentiating eq.~(\ref{eq:integratedftSHS_witharbitrariness}) with respect to the hold time $t_{\rm h}$, we arrive at
\begin{equation}
    f_{\rm SHS|NSC}=c_*.
    \label{eq:ftSHS_witharbitrariness}
\end{equation}
Since the evolution law under the NSC is almost the same as $f_t$ at the NSC (denoted by $f_{t|{\rm NSC}}$), 
\begin{equation}
    f_{t|{\rm NSC}}=c_*.
    \label{eq:ftNSC_witharbitrariness}
\end{equation}

The above derivation leaves indefiniteness by a factor $c_*$ in the evolution law of $\theta$. 
Considering the limit of $\theta_{\rm init}\to 0$ in eq.~(\ref{eq:tc_depon_cstar}), we see that the value of $c_*$ is also expressed as 
\begin{equation}
c_*=\theta_{\rm X}/t_{\rm cX}, 
\label{eq:meaningofcstar}
\end{equation}
where $t_{\rm cX}$ denotes the value of $t_{\rm c}$ when the (NSC-)hold is started with the X-state (i.e., $\Phi_{\rm init}=\Phi_{\rm X}$). Note that $t_{\rm cX}$ is the lowest possible value of $t_{\rm c}$ ($t_{\rm c}\geq t_{\rm cX}$). The indefiniteness of $c_*$ is rooted in the constant-factor indefiniteness of the absolute value of $\theta$ appearing in the $B\ln (\theta/\theta_*)$ term of eq.~(\ref{RSFeq:constitutivetheta}). 
Since $c_*$ can be set to any positive value without loss of generality, we adopt the $c_*$ value such that $\theta_{\rm X}$ equals the minimum cutoff time of log-t healing, $t_{\rm cX}$. Accordingly, noting eq.~(\ref{eq:meaningofcstar}), we define 
\begin{equation}
    \theta_{\rm X}:=t_{\rm cX}
    \label{eq:normalizationoftheta}
\end{equation}
by choosing
\begin{equation}
    c_*=1.
    \label{eq:normalizationofcstar}
\end{equation}
Under this convention, eqs.~(\ref{eq:tc_depon_cstar}) and (\ref{eq:ftNSC_witharbitrariness}) reduce to
\begin{equation}
    t_{\rm c}=\theta_{\rm init}+\theta_{\rm X}
    \label{eq:tcnaturallyexpected}
\end{equation}
and
\begin{equation}
    f_{t|{\rm NSC}}=1,
    \label{eq:ftNSC_constrained}
\end{equation}
which are consistent with \citet{nakatani2006intrinsic}. 

Noticing that eq.~(\ref{eq:ftNSC_constrained}) implies $\theta$ increases by unity per unit time, we see that the convention eq.~(\ref{eq:normalizationofcstar}) defines the unit of $\theta$ consistent with the conventional interpretation that $\theta$ is the average contact time~\citep{dieterich1978time}, just as eq.~(\ref{eq:steadystateform_theta}) for the steady state. Such a natural unit system is possible only with $c_*=1$, and we adopt eq.~(\ref{eq:normalizationofcstar}) hereafter. 
Thus, eq.~(\ref{eq:ftNSC_constrained}) is the constraint on the evolution law of $\theta$ as per requirements from SHS tests.

Using the $\theta\to\Phi$ variable conversion eq.~(\ref{RSFeq:conversionofthetatophi}), eq.~(\ref{eq:ftNSC_constrained}) can be rewritten into
the expression of $F_t$ under the NSC (denoted by $F_{t|{\rm NSC}}$):
\begin{equation}
    F_{t|{\rm NSC}}=\frac{B}{t_{\rm cX}}e^{-(\Phi-\Phi_{\rm X})/B}
    \label{eq:FtNSCwithtcX}
\end{equation}
Equation~(\ref{eq:normalizationoftheta}) was used in the above derivation. 
We note that eq.~(\ref{eq:FtNSCwithtcX}) can also be obtained
from eq.~(\ref{eq:ftNSC_witharbitrariness}) via the same variable conversion (eq.~\ref{RSFeq:conversionofthetatophi}). The $c_*$ in eq.~(\ref{eq:ftNSC_witharbitrariness}) disappears after
this variable conversion without choosing specific $c_*$ values, indicating that the $c_*$-indefiniteness in the $\theta$'s evolution law (eq.~\ref{eq:ftNSC_witharbitrariness}) is merely apparent, not existent in  $\Phi$, the observable. 

Meanwhile, $\Phi$'s evolution law (eq.~\ref{eq:FtNSCwithtcX}) is the ODE expression of the purely time-dependent
log-t healing phenomena~\citep{nakatani2006intrinsic}. Hence, in addition to the two SHS requirements of eqs.~(\ref{RSFeq:PhiSHSNSCrawtc}) and (\ref{RSFeq:coincidenceofBs}), all the subsidiary features (Fig.~\ref{fig:LTH},
\S\ref{sec:34}) expected and observed in SHS tests are reproduced by eq.~(\ref{eq:FtNSCwithtcX}). They include the inverse proportionality of $t_{\rm c}$ to $V_{\rm prior}$ and the existence of the lower limit for $t_{\rm c}$: $t_{\rm c} \geq  t_{\rm cX}$ ($= \theta_{\rm X}$ under our
choice of $c_* = 1$).

\begin{figure*}
\includegraphics[width=150mm]{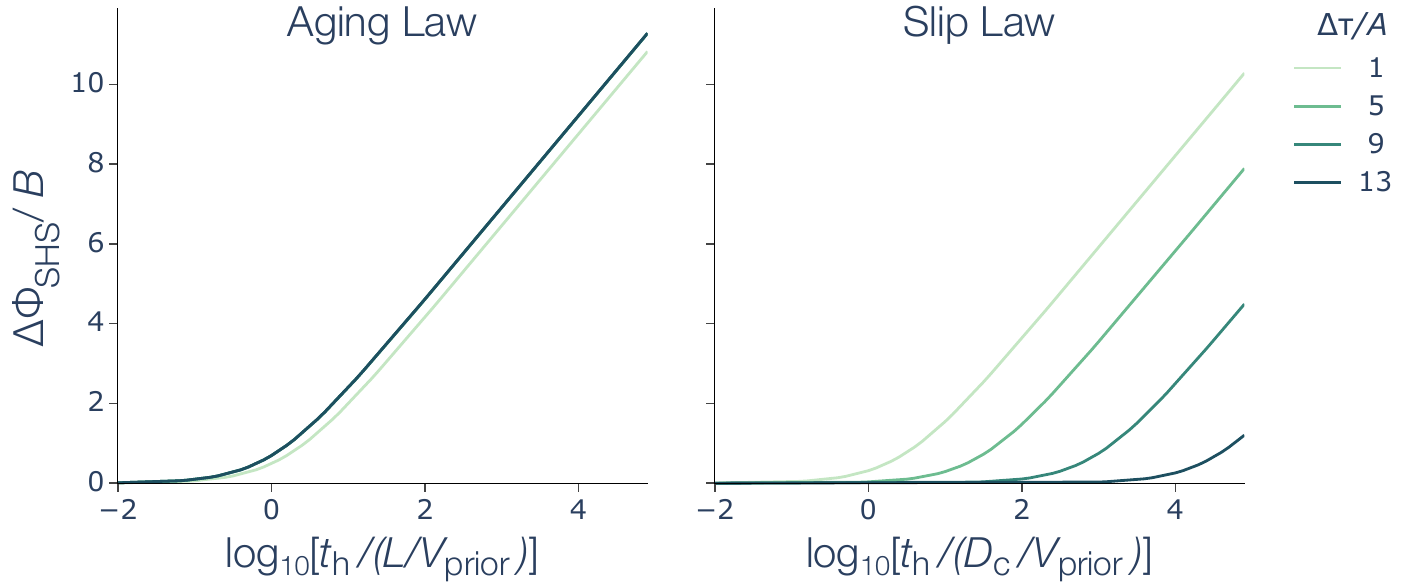}
  \caption{
Strength evolution during the hold at a constant $\tau_{\rm hold}$, predicted by the aging law (left) and the slip law (right).
Following a steady-state sliding at $V_{\rm prior}$, shear stress is abruptly reduced by $\Delta\tau = A,5A,9A,13A$ and then kept constant. $A=B$ and $\theta_{\rm X}\ll L/V_{\rm prior}$ are assumed. The predicted increase of strength $\Delta \Phi_{\rm SHS}= \Phi- \Phi_{\rm init}$, where $\Phi_{\rm init}=\Phi_{\rm SS}(V_{\rm prior})$ is the strength at the start of hold, is plotted after normalization by $B$, against the elapsed time $t_{\rm h}$ since the start of the hold after normalization by $L/V_{\rm prior}$ for the aging law and by $D_{\rm c}/V_{\rm prior}$ for the slip law, the characteristic time dimensions of the purely time-dependent log-t healing and the slip-driven apparent time healing, respectively.
  }
  \label{fig:AgingSlipSHS}
\end{figure*}

As mentioned earlier, the aging law and the slip law predict very different behaviors for SHS$|$NSC tests, the main reason we have decided to impose the significant log-t healing in SHS$|$NSC tests as the canon from SHS tests. Here, we quickly confirm the relevant differences between the predictions by the two laws using numerical simulations. Figure~\ref{fig:AgingSlipSHS} compares predictions by the two laws for constant-$\tau$ SHS tests with $\Delta\tau=A,5A,9A,13A$. See the caption for the parameter setting, which reflects typical laboratory values and commonly used implicit assumptions in RSF studies. Further details of this numerical simulation and analytic expressions of the predicted behaviors are found in Appendix~\ref{sec:SHSHSprinciple}. 

Figure~\ref{fig:AgingSlipSHS} (left) shows that the aging law predicts the canonical log-t healing, little affected by $\Delta\tau$ except the slightly smaller healing seen for the smallest $\Delta\tau$ where the slip-weakening term of the aging law becomes non-negligible and partially counters the purely time-dependent healing. The predicted $t_{\rm c}$ is almost fixed at $L/V_{\rm prior}$, as explained in \S\ref{sec:34}. In contrast, though the slip law also predicts log-t healing, the predicted $t_{\rm c}$ strongly depends on $\Delta\tau$ 
(Fig.~\ref{fig:AgingSlipSHS}, right), following eq.~(\ref{eq:tcslippertcaging}). Roughly speaking, $t_{\rm c}$ predicted by the slip law is prolonged by a factor of $\exp(\Delta \tau/A)$. Although other differences, including the slip law's prediction of $A$ln-t healing (eq.~\ref{eq:phifortime}) instead of $B$ln-t healing expected from the canon, exist, this prolonged $t_{\rm c}$ is the prediction in most striking disagreement with laboratory results of SHS$|$NSC tests, as we demonstrate below by referring to the laboratory data of \citet{nakatani1996effects}.

\citet{nakatani1996effects} did constant-$\tau$ SHS tests at various levels of $\tau_{\rm hold}$, ranging from $0.02\sigma$ to $0.84\sigma$, approximately corresponding to $\Delta \tau$ of $0.83\sigma$ to $0.01\sigma$. 
With $A = 0.01\sigma$, as typical for nominally bare granite interfaces they used, the corresponding range of $\Delta\tau/A$ is 1--83. Their healing data \citep[Figure 4a of][]{nakatani1996effects} indicate that $t_{\rm c}$ was about 3--30 ms, with no evident systematic dependence on $\tau_{\rm hold}$. For this range of $\Delta\tau/A$, the slip law predicts that $t_{\rm c}$ increases with $\Delta \tau$ by a factor of about $10^{34}$. Even if we exclude tests with $\tau_{\rm hold}<0.3\sigma$ (i.e., $\Delta \tau > 0.55 \sigma$), where their $B_{\rm heal}$ showed a  systematic dependence on $\tau_{\rm hold}$, the slip law still predicts a variation in $t_{\rm c}$ by a factor of about $10^{22}$. Thus, the slip law's prediction on $t_{\rm c}$ strongly contradicts the laboratory observation of $t_{\rm c}$ little affected by $\tau_{\rm hold}$ over its entire range from zero to $\tau_{\rm SS}(V_{\rm prior})$.

We can also argue against the slip law based on the absolute value of $t_{\rm c}$ predicted by the law. The slip law's prediction of $t_{\rm c}$ (eq.~\ref{eq:tcsliplaw}) can be rewritten into: 
$t_{\rm c|slip} = (D_{\rm c}/V_{\rm prior})\times (A/\Delta \tau)\times (A/B)\times\exp(-\Delta \tau/A)$. [Note that $D_{\rm c}/V_{\rm prior}$, instead of $L/V_{\rm prior}$, appears as the characteristic time of the apparent log-t healing predicted by the slip law, or more generally, by any apparent healing deriving from the same form of ODE ($\dot \theta = -(L/D_{\rm c})\Omega\ln\Omega$). This is because the governing ODE can be written as $\dot\theta^\prime = -(\theta^\prime V/D_{\rm c})\ln (\theta^\prime V/D_{\rm c})$ using a state variable $\theta^\prime$ whose steady-state value is $D_{\rm c}/V$.] 
According to the slip law, $D_{\rm c}$ can be directly observed in VS tests (Fig.~\ref{fig:AgingSlipVS}a). Assuming the typical value of $D_{\rm c} = 1$ $\mu$m, and $V_{\rm prior}$ of about $20$ $\mu$m/s in their tests, $D_{\rm c}/V_{\rm prior} = 50$ ms. For typical interfaces with $A\simeq B$, $t_{\rm c|slip}$ is about a century for tests with $\Delta\tau=27A$. For the lowest $\tau_{\rm hold}$ of $0.02\sigma$ in \citet{nakatani1996effects}, $\Delta \tau = 83A$, and $t_{\rm c|slip}$ predicted by eq.~(\ref{eq:tcsliplaw}) is about $10^{25}$ years, $10^{15}$ times longer than the age of our observable universe, whereas those experimenters found steadfast log-t healing (about $0.01\sigma$ strengthening per 10-fold increase in $t_{\rm h}$) for the much shorter range of $t_{\rm h}$ (30--3000 s) they tested.

As mentioned earlier, \citet{bhattacharya2017does} analyzed stationary-loadpoint SHS tests and showed that experimental results are better fitted by the slip law than the aging law when the machine stiffness is small. However, such tests are not
under the NSC. In tests under the NSC (such as constant-$\tau$ hold tests with $e^{\Delta\tau/A}\gg1$), it is clear that the slip law, where $f_t = 0$, cannot satisfy experimental requirements. Hence, as declared in \S\ref{subsubsec:331}, the present paper
basically considers constant-$\tau$ SHS tests, where the NSC can be imposed by choosing a sufficiently low $\tau_{\rm hold}$. Although constant-$\tau$ SHS tests
do have a unique methodological defect that, for interfaces with $A>B$, strength recovery saturates at very long $t_{\rm h}$ when $\Omega$ has approached $\Omega_{\rm SS}$, this is not an issue rooted in evolution laws (\S\ref{sec:SHSsaturation}). 
Fortunately, this saturation is expected to occur late enough so that constant-$\tau$ SHS tests can realize the NSC 
uniformly over a wide range of hold time since the start of the hold even for $A>B$ interfaces, allowing acquisition of healing data sufficient to constrain evolution laws. The present paper seeks evolution laws that reproduce the $B\ln t$ healing
during the hold interval before the saturation occurs, where $\Omega\ll \Omega_{\rm SS}$. 

\subsection{$V_{\rm prior}$ dependence of the cutoff time of log-t healing and its implication on the relationship between $L$ defined in SS tests and $D_{\rm c}$ defined in VS tests}
\label{sec:34}

Using Marone's (1998) laboratory data of stationary-loadpoint SHS tests with
various $V_{\rm prior}$ values, \citet{nakatani2006intrinsic} have found that $t_{\rm c}$ is
inversely proportional to $V_{\rm prior}$: 
\begin{equation}
    t_{\rm c}= L_{\rm heal}/V_{\rm prior},
    \label{eq:tcdetection}
\end{equation}
where $L_{\rm heal}$ is a proportional constant of length dimension, determined from the data. 
\citet{nakatani2006intrinsic} have interpreted this observation as a manifestation of the
dependence of $t_{\rm c}$ on the initial strength $\Phi_{\rm init} = \Phi_{\rm SS}(V_{\rm prior}$) obeying eq.~(\ref{eq:req4ss}), 
as expected of the ODE eq.~(\ref{eq:FtNSCwithtcX}).

We here interpret the same observation (eq.~\ref{eq:tcdetection}) in terms of $\theta$, which may be interpreted as the effective contact time under the natural unit system of $c_*=1$ (eq.~\ref{eq:normalizationofcstar}), as mentioned earlier. Evolution laws satisfying the SHS canon predict
$t_{\rm c}$ as eq.~(\ref{eq:tcnaturallyexpected}). Combined with the SS constraint eq.~(\ref{eq:steadystateform_theta}), eq.~(\ref{eq:tcnaturallyexpected}) yields $t_{\rm c} \simeq
L/V_{\rm prior}$ for $\theta_{\rm init}\gg \theta_{\rm X}$ (Fig.~\ref{fig:LTH}):
\begin{equation}
    t_{\rm c}= L/V_{\rm prior}+\theta_{\rm X}\simeq L/V_{\rm prior}.\label{eq:tcVpriapproxtheory}
\end{equation}
In the above derivation, we assumed $B_{\rm heal}=B$ (eq.~\ref{RSFeq:coincidenceofBs}) as declared in \S\ref{subsubsec:332}. 
In the natural unit system of $c_*=1$, the right-hand side of eq.~(\ref{eq:tcVpriapproxtheory}) is the average contact time at the beginning of the hold (Fig.~\ref{fig:LTH}). 
Comparison of eq.~(\ref{eq:tcdetection}) to eq.~(\ref{eq:tcVpriapproxtheory}) leads to
\begin{equation}
    L_{\rm heal}=L.
    \label{eq:LhealisL}
\end{equation}
Thus, we may say that \citet{nakatani2006intrinsic} have detected the length scale $L$ of eq.~(\ref{eq:steadystateform_theta}), 
which relates the effective contact time at steady states to the slip rate. Here, an additional note regarding $c_*$ may be helpful. We have introduced $c_*$ as the ratio of $\theta_{\rm X}$ to $t_{\rm cX}$ (eq.~\ref{eq:meaningofcstar}), which is observable only under $\theta_{\rm init} \lesssim \theta_{\rm X}$. For $\theta_{\rm init}\gg
\theta_{\rm X}$, eqs.~(\ref{eq:steadystateform_theta}), (\ref{eq:tc_depon_cstar}), and (\ref{eq:tcdetection}) combinedly yield $c_*=L/L_{\rm heal}$, normalizing indefinite length dimension $L$ in SS tests by the observable $L_{\rm heal}$ in SHS tests. 
As declared in \S\ref{subsubsec:332}, the present paper adopts $c_* = 1$ so that (\ref{eq:LhealisL}) holds. Hence,  we do not distinguish $L$ and
$L_{\rm heal}$ hereafter.

Furthermore, \citet{nakatani2006intrinsic} have verified that $L_{\rm heal}$ ($t_{\rm c}V_{\rm prior}$) obtained from the SHS tests of \citet{marone1998effect} is of the same order as $D_{\rm c}$ obtained as the characteristic evolution distance in the VS tests on the same interface: 
\begin{equation}
L\sim D_{\rm c}.
\label{RSFeq:LsimilartoDc}
\end{equation}

Note, however, the above argument of \citet{nakatani2006intrinsic} presumed the
NSC, which may not have been the case in the stationary-loadpoint hold phase of SHS tests by \citet{marone1998effect} they analyze. 
As shown in Appendix \ref{sec:slip_law_pred_lph}, stationary-loadpoint SHS has difficulty in probing the NSC regime. 
Rather, support for 
$t_{\rm c}= L_{\rm heal}/V_{\rm prior}$ (eq.~\ref{eq:tcdetection}) and thus $L\sim D_{\rm c}$ (eq.~\ref{RSFeq:LsimilartoDc}) comes from the constant-$\tau_{\rm hold}$ SHS tests of \citet{nakatani1996effects}, where the same order of $t_{\rm c}$ values were found in tests with low $\tau_{\rm hold}$ satisfying the NSC and tests with so high $\tau_{\rm hold}$ that cannot satisfy the NSC; the SHS$|$NSC and SS canons (i.e., the aging law under the NSC) predict $t_{\rm c}= L_{\rm heal}/V_{\rm prior}$ under the NSC ($|f_t|\gg |f_{\delta }V|$), whereas the VS and SS canons (i.e., the slip law) assert $t_{\rm c}\sim D_{\rm c}/V_{\rm prior}$ when $|f_t|\ll |f_{\delta }V|$ (Fig.~\ref{fig:AgingSlipSHS}).
Although we think that eqs.~(\ref{eq:tcdetection}) and (\ref{RSFeq:LsimilartoDc}) are likely, they are not as well-founded as thought to be. 
Nonetheless, we proceed by assuming eq.~(\ref{eq:tcdetection}) as it is tied to the presumption of $B=B_{\rm heal}$ (Appendix \ref{sec:slip_law_pred_lph}).
Meanwhile, 
regarding eq.~(\ref{RSFeq:LsimilartoDc}),
although we first assume $L\sim D_{\rm c}$ (eq.~\ref{RSFeq:LsimilartoDc}) and utilize it as an auxiliary constraint in developing evolution laws in
\S\ref{RSF33}, we consider the effects of $L/D_{\rm c}$ significantly differing from unity in \S\ref{sec:extensions_DcoverL}.

\section{Development of new evolution law}
\label{sec:development}
In \S\ref{sec:3}, we have derived constraints on the slip-dependent evolution term $f_\delta$ (equivalently $F_\delta$) and the purely time-dependent evolution term $f_t$ (equivalently $F_t$) of the evolution law to be developed, based on the canonical behaviors in ideal SS, VS, and SHS tests. In \S\ref{sec:development}, we seek evolution laws that satisfy all these constraints, 
namely eqs.~(\ref{RSFeq:steadystateconstraint}), (\ref{RSFeq:dotthetaVS}), and (\ref{eq:ftNSC_constrained}), plus an auxiliary constraint eq.~(\ref{RSFeq:negligiblehealing}). 

For the present goals of simultaneous reproduction of the three canons, we better start with evolution behaviors in the range $\theta,\theta_{\rm SS}\gg\theta_{\rm X}$, 
where experiments explicitly constrain the behavior of $\theta$. 
By contrast, for $\theta\lesssim\theta_{\rm X}$, the evolution law is mostly unconstrained; we only know that the observable quantity
$\Phi=\Phi_{\rm X}+B\ln (\theta/\theta_{\rm X}+1)$ little depends on $\theta$ so that the development path is less evident. Note that the conventional RSF (eq.~\ref{RSFeq:constitutivetheta}) and the slip law of eq.~(\ref{eq:defofsliplaw}) intended to cover, in retrospect, only the range $\theta,\theta_{\rm SS}\gg\theta_{\rm X}$ because the high-velocity cutoff and related lower limit of strength (\S\ref{subsec:RSFminimumStrength}), both of which are behaviors important in the range $\theta\lesssim\theta_{\rm X}$, were not considered at the time. 


For $\theta,\theta_{\rm SS}\gg\theta_{\rm X}$, steady state friction is constrained by eq.~(\ref{RSFeq:steadystateconstraint}):
\begin{equation}
    \Omega_{\rm SS}=1
\hspace{10pt}\mbox{(Constraint 1)}.
\label{eq:constraintforsteadystate}
\end{equation}
When eq.~(\ref{eq:constraintforsteadystate}) is met, the constraint by VS tests (eq.~\ref{RSFeq:dotthetaVS}) and that by SHS tests (eq.~\ref{eq:ftNSC_constrained}) become
\begin{flalign}
&f_\delta V= -\frac{L}{D_{\rm c}}\Omega\ln\Omega 
\hspace{10pt}\mbox{(Constraint 2)}
\label{eq:simplifiedslipdepofthetadot}
\\
&f_{t|{\rm NSC}} = 1
\hspace{10pt}\mbox{(Constraint 3)}
\label{eq:simplifiedtimedepofthetadot}
\end{flalign}
We used eq.~(\ref{RSFeq:defofOmega_steadystate}) in deriving eq.~(\ref{eq:simplifiedslipdepofthetadot}).
As far as eq.~(\ref{RSFeq:negligiblehealing}) holds, evolution
laws that satisfy both eqs.~(\ref{eq:simplifiedslipdepofthetadot}) and (\ref{eq:simplifiedtimedepofthetadot}) necessarily satisfy the VS requirement (eq.~\ref{eq:realVScanonical}) and
the SHS$|$NSC requirements (eqs.~\ref{RSFeq:PhiSHSNSCrawtc} and \ref{RSFeq:coincidenceofBs}). 
Thus, the above three equations (eqs.~\ref{eq:constraintforsteadystate}--\ref{eq:simplifiedtimedepofthetadot}) represent a set of principal constraints on the evolution of $\theta$. 

As pointed out earlier (\S\ref{subsec:SHSrequirement}), the simultaneous reproduction of the three canons is substantially equivalent to the reconciliation of the aging law and the slip law. 
In \S\ref{sec:canonicalandside}, we will augment this viewpoint by focusing on the `side effects' of reconciliation seen in
the pioneering attempt of \citet{kato2001composite}. Using insights obtained in \S\ref{sec:canonicalandside}, we seek the necessary conditions for the evolution laws and propose a class of eligible functional forms in \S\ref{RSF322}. In \S\ref{RSF33}, we propose, as examples, specific functional
forms of the evolution law under an assumption of $L\sim D_{\rm c}$. In \S\ref{sec:extensions_DcoverL}, we extend our
proposed evolution laws to cases where $L$ and $D_{\rm c}$ are much different. 
In \S\ref{sec:extensions}, we will check the validity of our proposal for situations including $\theta\lesssim\theta_{\rm X}\cup\theta_{\rm SS}\lesssim\theta_{\rm X}$.

Before proceeding to \S\ref{sec:canonicalandside}, we here point out that eq.~(\ref{eq:simplifiedtimedepofthetadot}) is, in fact, the aging law under the NSC. 
This fact can be proven by showing that, for $\theta,\theta_{\rm SS}\gg\theta_{\rm X}$, the NSC is possible only in a narrow range of $\Omega$. 
Given eqs.~(\ref{eq:simplifiedslipdepofthetadot}) and (\ref{eq:simplifiedtimedepofthetadot}), 
$|f_{\delta}V|\ll |f_t|$ if and only if $\Omega\sim 1\cup\Omega\ll D_{\rm c}/L$. 
[Strictly speaking, for $|\ln (D_{\rm c}/L)|\gg1$, the necessary and sufficient condition for $|f_{\delta}V|\ll f_t$ is $\Omega\sim 1\cup\Omega\ll (D_{\rm c}/L)/|\ln(D_{\rm c}/L)|$. However, unless $D_{\rm c}$ and $L$ differ by very many orders, this difference is negligible, so that basically the former conditions suitable when $|\ln (D_{\rm c}/L)|\lesssim 1$ shall be considered hereafter.] 
Moreover, as VS tests involve situations where $\Omega\sim \Omega_{\rm SS}=1$ (eq.~\ref{eq:constraintforsteadystate}), the VS constraint eq.~(\ref{RSFeq:negligiblehealing}) precludes the
possibility that $\Omega\sim \Omega_{\rm SS}$ satisfies the NSC. 
Therefore, these conclude that, under eqs.~(\ref{eq:constraintforsteadystate})--(\ref{eq:simplifiedtimedepofthetadot}) and (\ref{RSFeq:negligiblehealing}), the NSC can exist only for $\Omega\ll D_{\rm c}/L$. 
Especially, $\Omega\to 0$ corresponds to the ideal SHS$|$NSC tests, where $\tau_{\rm hold}$ is kept much lower than $\tau_{\rm SS}(V_{\rm prior})$ so that ideal stationary contact is almost achieved. As we have adopted the SHS canon (eqs.~\ref{RSFeq:PhiSHSNSCrawtc} and \ref{RSFeq:coincidenceofBs}) from such tests in \S\ref{subsec:SHSrequirement}, we regard $\Omega\to 0$ as the ideal NSC limit. 
For the aging law (eq.~\ref{eq:defofaginglaw}), where $D=D_{\rm c} = L$ is assumed, the above proposition (NSC $\subset \Omega\ll D_{\rm c}/L$) reduces to NSC $\subset \Omega\ll 1$, resulting in that eq.~(\ref{eq:simplifiedtimedepofthetadot}) is the aging law under the NSC. 
Thus, the reconciliation of the aging law and the slip law is equivalent to the reconciliation of the time-driven healing of (static) friction in SHS$|$NSC tests and the slip-driven evolution of (dynamic) friction in VS tests. 


Notice that the above argument regarding the NSC relies on the fact that $f_\delta V$ and $f_{t|{\rm NSC}}$ are constrained to be functions of $\Omega$ only (eqs.~\ref{eq:simplifiedslipdepofthetadot} and \ref{eq:simplifiedtimedepofthetadot}). 
Similarly to \citet{ampuero2008earthquake}, who have noticed and utilized that the aging law and the slip law are both unary functions of $\Omega$, the present development heavily
utilizes this property of $f_\delta V$
and $f_{t|{\rm NSC}}$.

To reconcile the canons, we treat $B$, $L$, and $D_{\rm c}$ as constants. According to \citet{nakatani1996effects} and \citet{ryan2018role}, $B$ and $L$ do not change their orders for arbitrary $\tau=\tau_{\rm hold} >0$, and the $L\sim D_{\rm c}$ case is our main concern. Thus, we can neglect the variations in $B$, $L$, and $D_{\rm c}$ as secondary effects irrelevant to reconciling the two laws, although in applying the developed law, we will need to account for the $\tau$-dependence of $B$, which arises for $\tau_{\rm hold}<0.4\sigma$, where $B_{\rm heal}$ varies by factors; that dependence is a part of thermomechanical properties of the RSF parameters, typified by the temperature dependence of $A$ and $B$. 

\subsection{Strategy to reconcile the aging law and the slip law}
\label{sec:canonicalandside}

The present development aims to construct evolution laws that reproduce all the canonical behaviors from SS, VS, and SHS$|$NSC tests, all of which have been enumerated as experimental requirements in \S\ref{sec:3}. However, it is obvious that the VS constraint $|f_t| \ll |f_\delta V|$ (eq.~\ref{RSFeq:negligiblehealing}) and the SHS$|$NSC constraint $|f_t| \gg |f_\delta V|$ (the NSC itself, eq.~\ref{eq:NSCdef}) cannot hold at the same time. Mutual exclusiveness can also be seen even if we do not presume the TDF. The antinomy is most striking in a strengthening phase of evolution, especially when $\Omega\ll 1$ ($\sim D_{\rm c}/L$ from eq.~\ref{RSFeq:LsimilartoDc}). The canonical behavior in VS tests is the solution of $\dot\theta = f := -(L/D_{\rm c}) \Omega \ln \Omega$ (the slip law), while that in SHS$|$NSC tests is the solution of $\dot \theta = f := 1$ (the aging law for $\Omega\ll1$). 
Their predicted strengthening rates are much different when $\Omega\ll D_{\rm c}/L$; for $\Omega\ll1$, the two laws yield an equal value only when $-(L/D_{\rm c})\Omega\ln\Omega=1$, which is outside the strengthening phase $\Omega \ll  D_{\rm c}/L$. 
Thus, we see that the aging law's misprediction of the strength evolution following a negative velocity step in VS tests (Fig.~\ref{fig:AgingSlipVS}) and the slip law's misprediction of the SHS$|$NSC tests (Fig.~\ref{fig:AgingSlipSHS}) are direct manifestations of this antinomy. 

The above inspection of the strengthening phases of VS and SHS$|$NSC tests has revealed that (i) it is a mathematically impossible claim that the VS and SHS canons hold at the same time throughout the entire range of variables (i.e., any positive $V_{\rm before}$ and $V_{\rm after}$ for VS tests and the NSC in SHS tests, where $\Omega  \ll \min (D_{\rm c}/L,1)$). In other words, by noticing (i), we have sadly proven by contradiction that our original goal of complete simultaneous reproduction of the three canons is impossible in principle. 

Nevertheless, it is also the case that (ii) both $f_{\rm VS}$ and $f_{\rm SHS|NSC}$ canons hold well in experiments done so far. The only possibility to solve this apparent contradiction between facts (i) and (ii) is that there might be mutually exclusive two physical variable subspaces where $f\simeq f_{\rm VS}$ and $f\simeq f_{\rm SHS|NSC}$, respectively. In this case, the aging law and the slip law must switch at the boundary of the two domains within each of which either the VS or SHS$|$NSC canon holds. Furthermore, from (ii), each canon-following domain must include the laboratory-probed range of variables where the canon have been verified. Therefore the aging-slip switch is allowed only outside the experimentally verified canonical ranges.

Thus, we here give up the original goal of simultaneous reproduction of the three canons throughout the entire $(V, \theta)$ range and instead adopt a downgraded goal hereafter. The new (downgraded) goal is the simultaneous reproduction of the three canons throughout the $(V, \theta)$ range well probed experimentally so far. As discussed above, this new goal is equivalent to placing the aging-slip switching boundary outside the experimentally probed range.

Maybe by pure heuristics without deductive arguments like the above, \citet{kato2001composite} introduced a switch between the aging law and the slip law. After all, the present development will conclude a modification of their work, so we here take a detailed look at their idea and predicted behaviors. They have proposed the composite law, in which purely time-dependent healing occurs significantly only at slip rates below a threshold $V_{\rm c}$: 
\begin{equation}
    f= \exp(-V/V_{\rm c}) - \Omega\ln\Omega \hspace{5pt} (\mbox{Composite Law}).
\label{eq:defofcompositelaw}
\end{equation}
Equation~(\ref{eq:defofcompositelaw}) is essentially the slip law at $V \gg V_{\rm c}$. At $V \ll V_{\rm c}$, it behaves similarly to the slip law during the weakening phase and similarly to the aging law ($\dot \theta \sim 1$) during the strengthening phase where $\Omega \ll 1$. The composite law reproduces the symmetric response to positive and negative velocity steps (except at limited conditions mentioned later) while reproducing log-t healing in SHS$|$NSC tests (except at limited conditions mentioned later), almost achieving the quest for the perfect evolution law. 

\begin{figure*}
  \includegraphics[width=150mm]{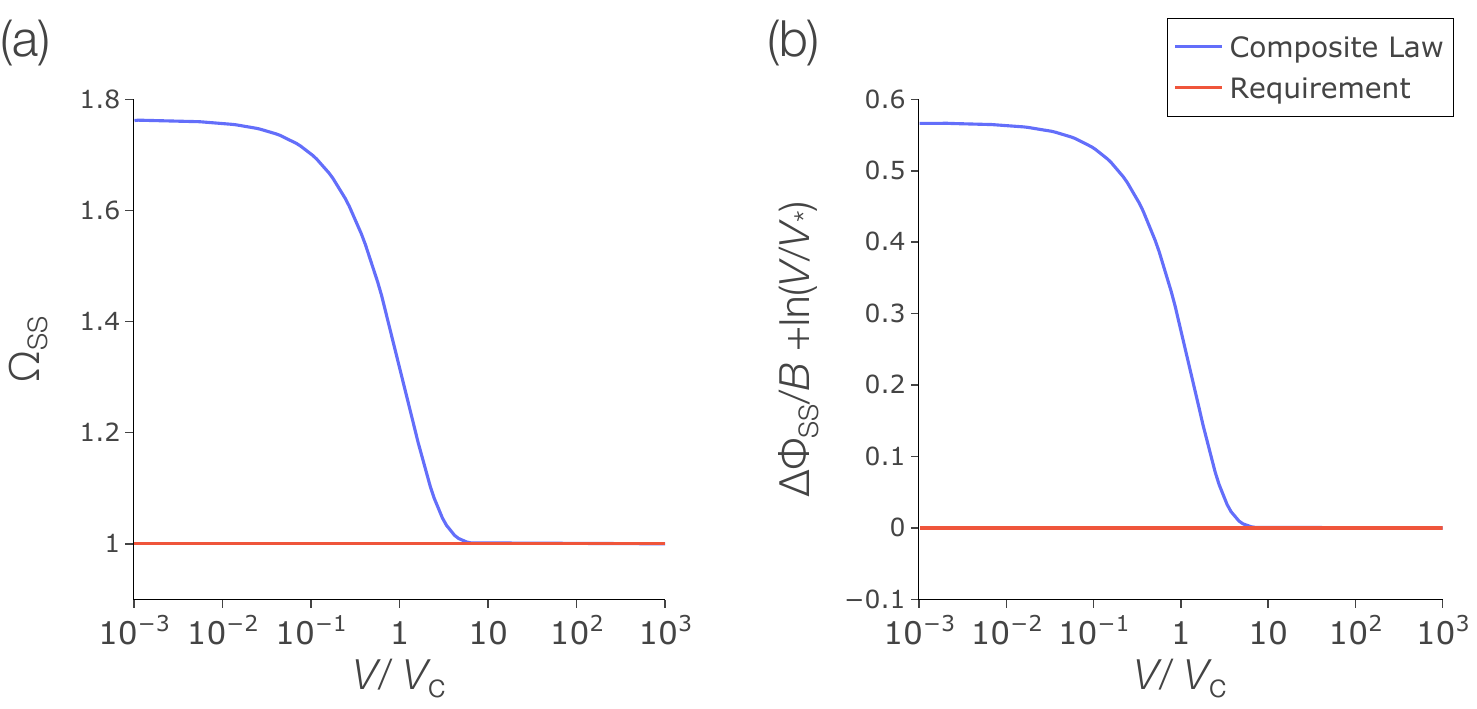}
  \caption{
  Comparison of steady-state values predicted by the composite law (blue) and those given by the SS constraint (eq.~\ref{eq:constraintforsteadystate}; red). Plotted against the slip rate $V$ normalized by $V_{\rm c}$. (a) Steady-state values of $\Omega=V\theta/L$.  (b) Steady-state values of strength $\Phi$. 
  Difference of $\Phi_{\rm SS}(V)$ from the canonical value $\Phi_* - B\ln(V/V_*)$, given by eqs.~(\ref{RSFeq:conversionofthetatophinocut}) and (\ref{eq:constraintforsteadystate}) with the reference slip rate $V_*$ taken so as to satisfy $V_*\gg V_{\rm c}$, is plotted after being normalized by $B$. 
  $\Delta\Phi_{\rm SS}=\Phi_{\rm SS}(V)-\Phi_{\rm SS}(V_*)$ denotes the variation in $\Phi_{\rm SS}$.
  }
  \label{fig:compvsrequirement}
\end{figure*}

However, a well-known side effect exists in its prediction of steady-state $\Phi_{\rm SS}(V)$, as pointed out by \citet{kato2001composite} themselves (Fig.~\ref{fig:compvsrequirement}). As shown in Fig.~\ref{fig:compvsrequirement}b, $\Phi_{\rm SS}(V)$ for $V\ll V_{\rm c}$ is predicted to be higher than the experimental requirement (eq.~\ref{eq:req4ss}) by $W(1)B$, where $W(1)=0.567...$ is the omega constant, contradicting the SS canon. This level shift of $\Phi_{\rm SS}$ around $V_{\rm c}$ results from the extra strengthening mechanism (i.e., the aging law) operating at $V \ll V_{\rm c}$, in addition to the slip law, which alone matches eq.~(\ref{eq:req4ss}) exactly. 
By fitting $\tau_{\rm SS}(V)$ data of \citet{blanpied1998effects}, \citet{kato2001composite} suggested the possible level shift at $V_{\rm c} = 0.01$ $\mu$m/s. However, the data have large scatterings and do not make strong evidence for the reality of the level shift. Furthermore, \citet{kato2001composite} noted that $V_{\rm c} = 0.01$ $\mu$m/s contradicts behaviors in VS tests, as explained in the next paragraph. 

\begin{figure*}
  \includegraphics[width=100mm]{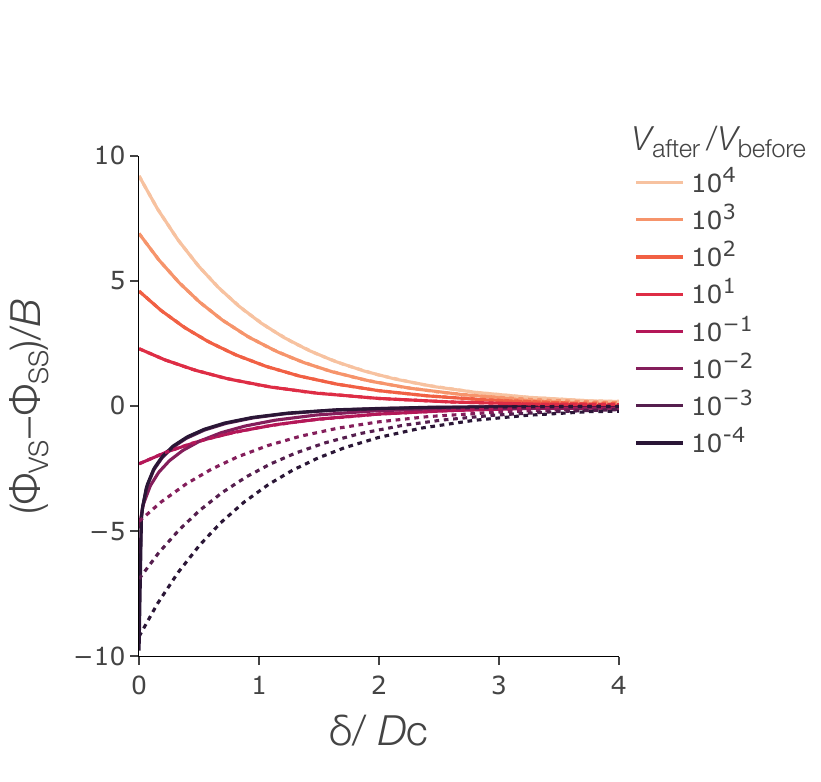}
  \caption{
  Strength evolution in ideal VS tests predicted by the composite law. 
  $L=D_{\rm c}$ and $\theta\gg\theta_{\rm X}$ are assumed. $V_{\rm before}$ is fixed at $100V_{\rm c}$. The visualizing scheme is the same as in Fig.~\ref{fig:AgingSlipVS}. Solid lines of different colors show the strength evolutions for different values of $V_{\rm after}$. 
  The curves of $V_{\rm after}/V_{\rm before}=10^{-3}$ and $10^{-4}$ nearly overlap with each other due to the aging-law-like blowups of the early stage. 
  As a reference, the canonical behavior (eq.~\ref{eq:realVScanonical}) for each $V_{\rm after}$ is shown by the dotted line of the same color, although not visible in cases with $V_{\rm after}\geq 0.1V_{\rm before}$ because it almost coincides with the prediction by the composite law. 
  }
  \label{fig:compositeVS}
\end{figure*}

Though less known, the composite law suffers two other side effects of the switching, rooted in the antinomy between the aging law and the slip law explained earlier in this subsection. The first less-known side effect appears when the hold phase of an SHS test starts with $V\gg V_{\rm c}$; in this case, predicted healing is much less than the canon until $V$ reaches down to $V_{\rm c}$ (\S\ref{sec:SHSanalysiscomposite}, also see Fig.~\ref{fig:SHScomparison}). 
The second less-known side effect appears following a negative velocity step (i.e., in a strengthening phase) when $V_{\rm after} \lesssim V_{\rm c}$ (Fig.~\ref{fig:compositeVS}). In this situation, the composite law is close to the aging law, and hence strengthening follows the log-t healing, exhibiting the transient slip distance much shorter than that predicted for a positive velocity step (i.e., $D_{\rm c}$), as explained for the aging law in \S\ref{subsec:VSrequirement} (Fig.~\ref{fig:AgingSlipVS}). 
\citet{kato2001composite} showed this non-canonical behavior of the composite law by their VS-test simulation assuming $V_{\rm c} = 0.01$ $\mu$m/s, the $V_{\rm c}$-value estimated from the SS tests of \citet{blanpied1998effects} mentioned earlier. 
However, \citet{kato2001composite} also showed that their laboratory VS tests follow canonical behaviors at least down to $V_{\rm after}$ as small as $0.001$ $\mu$m/s~\citep[Fig.~3 of ][]{kato2001composite}, which contradicts their fitting result of SS tests yielding $V_{\rm c} = 0.01$ $\mu$m/s.

Lastly, we briefly describe our strategy to reproduce the three canons. 
Among the three defects of the composite law mentioned above, the level shift of $\Phi_{\rm SS}(V)$, which violates the SS canon (eq.~\ref{eq:req4ss}), is not directly rooted in the antinomy between the VS constraint (eq.~\ref{eq:simplifiedslipdepofthetadot}) and SHS$|$NSC constraint (eq.~\ref{eq:simplifiedtimedepofthetadot}). 
Hence, the present development aims to predict $\Phi_{\rm SS}(V)$ exactly as eq.~(\ref{eq:req4ss}). 
In \S\ref{RSF322}, we will show that it is possible to construct evolution laws that perfectly match the SS canon (eq.~\ref{eq:req4ss}) for all the ranges of $V$. Moreover, we will find functional forms to do so are fairly restricted. 
This finding has facilitated our development. Within such functional forms, we will try to narrow the conditions where the other two unavoidable deviations from canons during the strengthening phase of VS and SHS$|$NSC tests arise. Eventually, we could see that our proposed evolution laws violate canons only at extreme conditions outside the conditions experimentally well verified so far. 
As the constraints eqs.~(\ref{eq:simplifiedslipdepofthetadot}) and (\ref{eq:simplifiedtimedepofthetadot}) are the slip law and the aging law for $\Omega \ll 1$, respectively, the present development, in which we purposefully control the unavoidable side effects of the aging-slip switch, squarely answers the long-standing problem of how to reconcile these two laws~\citep{ampuero2008earthquake}.

\subsection{A class of evolution laws consistent with the three canonical behaviors}\label{RSF322}
Below, we derive a class of evolution laws (of $\theta$) that satisfy thw constraints 1--3. The auxiliary constraint eq.~(\ref{RSFeq:negligiblehealing}) is used as well. The functional form presented below is a sufficient, but not necessary, condition to satisfy these experimental constraints. Hence it represents our ``proposal'' on how to reconcile the experimental requirements. 

\subsubsection{Heuristic derivation}
\label{sec:heuristicderivation}
As discussed in \S\ref{sec:canonicalandside}, the failure of the composite law regarding the reproduction of $\Phi_{\rm SS}(V)$ (Fig.~\ref{fig:compvsrequirement}) is not an unavoidable side effect in reproducing the aging-slip switch implied by the canons.
A simple way to avoid this defect is to adopt functional forms that satisfy the SS constraints eq.~(\ref{eq:constraintforsteadystate}).
Essentially, what functions in eq.~(\ref{eq:constraintforsteadystate}) as a constraint on evolution laws is the invariance of $\Omega$ at steady states: 
\begin{equation}
    \Omega_{\rm SS}=const. (=:C_{\rm SS}).
    \label{RSFeq:Cssconst}
\end{equation}
$C_{\rm SS}$ denotes the constant value of $\Omega_{\rm SS}$, 
which is unity in eq.~(\ref{eq:constraintforsteadystate}): 
\begin{equation}
    C_{\rm SS}=1.
    \label{RSFeq:Cssnormalization}
\end{equation}
$L$ in $\Omega(:=V\theta/L$, eq.~\ref{RSFeq:defofOmega_steadystate}) is a constant defined as the $V$-independent value of $V\theta_{\rm SS}(V)$ at steady state (eq.~\ref{RSFeq:Vthetaconst}). Therefore, eq.~(\ref{RSFeq:Cssnormalization}) is a mere normalization, and the constraint on the evolution law is solely given by eq.~(\ref{RSFeq:Cssconst}). Indeed, $C_{\rm SS}$ can be set to any positive value using the constant-factor indefiniteness of $\theta$. 
[Firstly, set $L$ to $V\theta_{\rm SS}(V)$ based on the $\theta$ defined under a certain unit scale of $c_*$. Secondly, redefine $\theta^\prime$ under a different unit scale $c_*^\prime$. The evolution law in terms of this $\theta^\prime$ has a $V\theta^\prime_{\rm SS}(V)$ different from $L$ and in turn $C_{\rm SS}^\prime :=V\theta^\prime_{\rm SS}(V)$ becomes non-unity.] 
This fact demonstrates that eq.~(\ref{RSFeq:Cssnormalization}) is a mere convention. Considering the above, we keep $C_{\rm SS}$ indefinite for the time being. Nevertheless, we limit our discussion only to $C_{\rm SS}\sim 1$ because the constraints 1--3 are formulated for $C_{\rm SS}=1$. 

Equation~(\ref{RSFeq:Cssconst}) means that $V\theta$ takes a single constant value $C_{\rm SS}$ at fixed points of the evolution law [$f(V, \theta) = 0\Leftrightarrow V\theta=C_{\rm SS}$]. Such $f$ can be constructed by using a unary function $\phi(x)$ such that $\phi(x) = 0$ if and only if $x=C_{\rm SS}L$; substituting $x=V\theta$ into $\phi(x)$, we have an $f$ form: 
\begin{equation}
\dot \theta=f(V,\theta,...)=\phi(V\theta),
\label{RSFeq:fOmega0}
\end{equation}
which necessarily satisfies the SS constraints eq.~(\ref{RSFeq:Cssconst}). 
Then, by limiting our heuristic search for the evolution law only to functions given in the form of eq.~(\ref{RSFeq:fOmega0}), we can automatically preclude the possibility of violating the SS constraint eq.~(\ref{eq:constraintforsteadystate}). 
Incidentally, the composite law eq.~(\ref{eq:defofcompositelaw}) does not conform eq.~(\ref{RSFeq:fOmega0}) as $V$ there appears alone in $\exp(-V/V_{\rm c})$, not as part of the product $V\theta$.

\begin{figure*}
   \includegraphics[width=145mm]{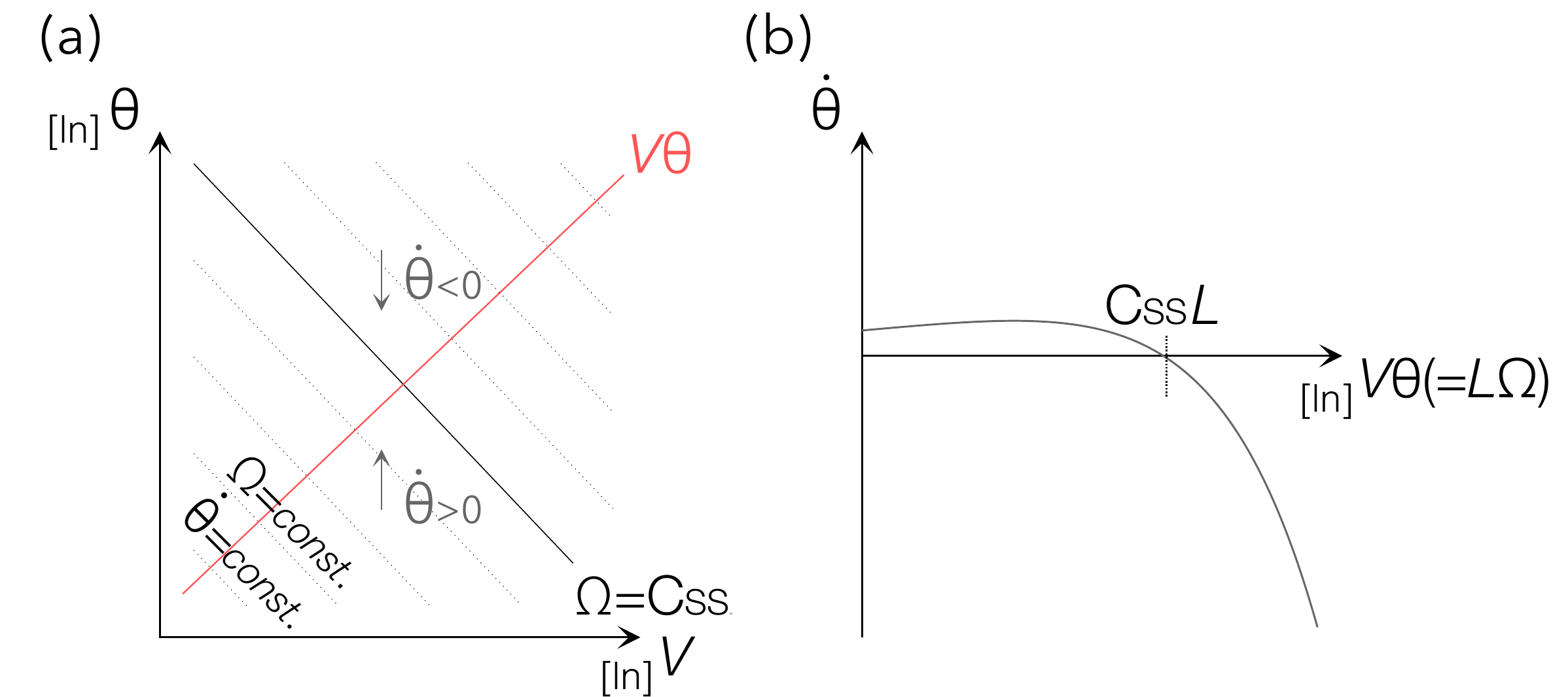}
  \caption{
  The behaviors of the evolution law implied by the function $\phi$ of eq.~(\ref{RSFeq:fOmega0}) (or equivalently, $\psi$ of eq.~\ref{RSFeq:fOmega}). 
  (a) Schematic diagram of the flow field of $\theta$ in $V$-$\theta$ space. Steady states ($\Omega=C_{\rm SS}$) form a straight line, on which $\theta$ is inversely proportional to $V$. The regimes to the upper-right/lower-left of this line are called above/below the steady state, where $\dot\theta$ is negative/positive, respectively. The value of $\dot\theta$ varies along the $V\theta$-axis (the red straight line) and is invariant for the changes in the direction normal to it, shown by the dotted lines of constant $V\theta$.  (b) Schematic diagram of the variation in $\dot\theta$ along the $V\theta$-axis shown in (a).
}
  \label{fig:ReqEvolution}
\end{figure*}

Since $\dot \theta$ is nondimensional, 
we rewrite eq.~(\ref{RSFeq:fOmega0}) in terms of a nondimensional variable. 
Nondimensionalizing $V\theta$ by $L$, we can rewrite the evolution law eq.~(\ref{RSFeq:fOmega0}) with another functional form $\psi$ of $\Omega$ as follows, without loss of generality: 
\begin{equation}
f=\psi(\Omega).
\label{RSFeq:fOmega}
\end{equation}
As $\psi$ is a rewritten form of $\phi$, $\psi$ has a unique zero at $\Omega=C_{\rm SS}$ [$\psi(\Omega)=0 \Leftrightarrow \Omega=C_{\rm SS}$]. Figure~\ref{fig:ReqEvolution} shows the behavior of evolution laws given by eq.~(\ref{RSFeq:fOmega0}) (or equivalently, eq.~\ref{RSFeq:fOmega}). In the $V$-$\theta$ space, $\dot\theta$ varies only along the direction of the $V\theta$-axis while remaining constant along the direction normal to it (dotted lines in Fig.~\ref{fig:ReqEvolution}a). 

Besides the uniqueness of the solution of $\psi=0$, we further impose  
$\psi(\Omega)>0$ for $\Omega<C_{\rm ss}$ and $\psi(\Omega)<0$ for $\Omega>C_{\rm SS}$ in eq.~(\ref{RSFeq:fOmega}) (Fig.~\ref{fig:ReqEvolution}) 
so that $\Omega=C_{\rm SS}$ is an attractor during sliding at a fixed slip rate. $\Omega=C_{\rm SS}$, $\Omega<C_{\rm SS}$, and $\Omega>C_{\rm SS}$ 
represent that the frictional state ($\Phi$ or $\theta$) is at, below, and above the steady state, respectively. 

For $\theta,\theta_{\rm SS}\gg\theta_{\rm X}$, the expressions of $f_{t|{\rm NSC}}$ and $f_\delta V$ (eqs.~\ref{eq:simplifiedslipdepofthetadot} and \ref{eq:simplifiedtimedepofthetadot}) both fit the functional form eq.~(\ref{RSFeq:fOmega}) we consider. As far as eq.~(\ref{RSFeq:fOmega}) holds, the ratio $f_t/f_{t|{\rm NSC}}$ is also a function of $\Omega$, which we denote by $\psi_{\rm c}(\Omega)$ hereafter. Then, under eqs.~(\ref{eq:simplifiedslipdepofthetadot}), (\ref{eq:simplifiedtimedepofthetadot}), and (\ref{RSFeq:fOmega}), the evolution function $f$ under the TDF (eq.~\ref{eq:thermodynamicalf}) can be expressed as
\begin{equation}
    f=\psi_{\rm c}(\Omega)f_{t|{\rm NSC}}+f_\delta V.
    \label{eq:evolutionlaw_compatiblewithfunctionalansatz}
\end{equation}
It is straightforward to show that the corresponding expression in $\Phi$ notation is $F=\psi_{\rm c}F_{t|{\rm NSC}}+F_\delta V$.  Using eqs.~(\ref{eq:simplifiedslipdepofthetadot}) and (\ref{eq:simplifiedtimedepofthetadot}), we have an explicit form of eq.~(\ref{eq:evolutionlaw_compatiblewithfunctionalansatz}): 
\begin{equation}
    f=\psi_{\rm c}(\Omega)-\frac{L}{D_{\rm c}}\Omega\ln\Omega.
    \label{eq:evolutionlaw_compatiblewithfunctionalansatz_substituted}
\end{equation}
The slip-driven evolution $f_\delta$ corresponds to the slip law eq.~(\ref{eq:defofsliplaw}).
The purely time-dependent healing ($f_{t|{\rm NSC}}$) under the NSC corresponds to 
the aging law eq.~(\ref{eq:defofaginglaw}) far below the steady state ($\Omega\ll C_{\rm SS}$).

The experimental requirements enumerated in \S\ref{sec:3} do not include explicit behaviors of the purely time-dependent healing term $f_t$ except at the NSC; hence, the specific functional form of $\psi_{\rm c}$ is undetermined. Nevertheless, we can constrain asymptotic behaviors of $\psi_{\rm c}$ as follows. 
Firstly, since $f_t\to f_{t|{\rm NSC}}$ at the NSC, $\psi_{\rm c}$ is unity at the ideal NSC limit , i.e., $\Omega\to 0$:
\begin{equation}
    \psi_{\rm c}(0)=1. 
    \label{RSFeq:psiat0}
\end{equation}
Since $f_\delta V\to0$ when $\Omega\to0$, eq.~(\ref{RSFeq:psiat0}) guarantees that $f\to f_t$ at the ideal NSC limit, $\Omega\to0$. 

Secondly, another constraint on $\psi_{\rm c}(\Omega)$ is obtained from the symmetry of the transient evolution of $\Phi$ following positive and negative velocity steps. As explained below, the heart of this constraint comes from the requirement that the auxiliary VS constraint eq.~(\ref{RSFeq:negligiblehealing}) be kept in negative-step cases. 
Upon a velocity step $V_{\rm before}\to V_{\rm after}$, $\Omega=V\theta/D_{\rm c}$ instantly changes from unity to $V_{\rm after}/V_{\rm before}$ [precisely speaking, $\Omega$ changes from $C_{\rm SS}(\sim 1)$ to $C_{\rm SS}V_{\rm after}/V_{\rm before}(\sim V_{\rm after}/V_{\rm before}$)]. Therefore, immediately following a large negative step (i.e., $V_{\rm after}/V_{\rm before}\ll1$), there occurs a time duration where $\Omega$ remains much less than unity. For this period, where $\Omega\ll1$, eq.~(\ref{RSFeq:negligiblehealing}) demands that the slip-driven evolution $f_\delta V$ is much greater than the purely time-dependent healing $f_t$ ($=\psi_{\rm c}f_{t|{\rm NSC}}$). 
However, for $\Omega\ll1$, we have already demanded the opposite, that is, the NSC (eq.~\ref{eq:NSCdef}, $f_t \gg f_\delta V$), to reproduce the SHS$|$NSC canon. 
Because $\psi_{\rm c}$ is a unary function of $\Omega$, these two demands almost contradict each other. However, as shown below, we can find a small margin to squeeze them in together.
Applying the above constraint from VS tests (eq.\ref{RSFeq:negligiblehealing}) to eq.~(\ref{eq:evolutionlaw_compatiblewithfunctionalansatz_substituted}), we have 
\begin{equation}
    |\psi_{\rm c}(\Omega)|\ll\frac{L}{D_{\rm c}}|\Omega\ln\Omega|
    \hspace{10pt} (\Omega\gtrsim V_{\rm after}/V_{\rm before}).
    \label{RSFeq:psiinVS}
\end{equation}
In the present development, we demand that eq.~(\ref{RSFeq:psiinVS}) to hold for the range of $V_{\rm after}/V_{\rm before}$ where the VS canon has been confirmed so far. 

Thus, under our functional ansatz that the contributions of all the variables to $\theta$ evolution are combined into the effect of the unary variable $\Omega$ in $f$, we have figured out that, to satisfy both constraints eqs.~(\ref{RSFeq:psiat0}) and (\ref{RSFeq:psiinVS}), $\psi_{\rm c}(\Omega)$ is necessarily constrained by different asymptotic forms above and below a threshold value of $\Omega$, referred to as $\beta$ hereafter:
\begin{equation}
    \psi_{\rm c}(\Omega)=
    \begin{cases}
        o[(L/D_{\rm c})\Omega\ln\Omega] & (\Omega\gg\beta)
        \\
        1 & (\Omega\ll\beta).
    \end{cases}
    \label{RSFeq:reqforpsic}
\end{equation}
Here, $\beta$ is a small positive constant: $\beta\ll1$.
As shown later, the value of $\beta$ is constrained by experiments, and the existence of $\psi_{\rm c}(\Omega)$ conforming eq.~(\ref{RSFeq:reqforpsic}) is guaranteed (Appendix \ref{sec:epsironRange}). We use the expression $x=o(y)$ hereafter to mean $|x|\ll|y|$ [e.g., $x=o(1)\Leftrightarrow |x|\ll 1$].  Since the $f_t$'s behavior of our concern is purely time-dependent healing, we only consider non-negative $\psi_{\rm c}$ ($\psi_{\rm c}\geq0$). 
As a result, evolution functions conforming to eq.~(\ref{eq:evolutionlaw_compatiblewithfunctionalansatz_substituted}) can have stationary points only in the range $\Omega\geq 1(\gg\beta)$, that is, exactly at or above the steady states. Therefore, when both eqs.~(\ref{eq:evolutionlaw_compatiblewithfunctionalansatz_substituted}) and (\ref{RSFeq:reqforpsic}) hold, $f$ has the unique fixed point at $\Omega=1$ because $f=-(L/D_{\rm c})\Omega\ln\Omega+o[(L/D_{\rm c})\Omega\ln\Omega]$ for $\Omega\gg\beta$. 
In summary, we have shown that $\theta$'s evolution laws can be constructed using the functional form $\psi(\Omega)$, which automatically satisfies the SS canon, if we adequately restrain the rate of purely time-dependent healing according to $\Omega$, the distance from the steady state (eq.~\ref{RSFeq:reqforpsic}). Note that what we have shown above is one way of construction; we do not mean that it is a necessary condition. 

So far, eq.~(\ref{RSFeq:fOmega0}) is the only heuristics. Otherwise, the derivation down to eq.~(\ref{RSFeq:reqforpsic}) from eqs.~(\ref{eq:constraintforsteadystate})--(\ref{eq:simplifiedtimedepofthetadot}) plus (\ref{RSFeq:negligiblehealing}) is deductive. Note that eq.~(\ref{RSFeq:reqforpsic}) allows purely time-dependent healing to occur significantly when $\Omega\ll\beta$, even in VS tests. Hence, strength evolution following a large negative step with $V_{\rm after}/V_{\rm before} \ll \beta$ is not symmetric with that following a positive step. This deviation from the canon is a side effect of the aging-slip switch during VS tests discussed in \S\ref{sec:canonicalandside}, also seen in the composite law for $V_{\rm after} < V_{\rm c}$. 
However, currently there are no experimental data for $V_{\rm after}/V_{\rm before}$ less than $10^{-3.5}$~\citep{bhattacharya2022evolution}, so there is no evidence that the asymmetric behavior predicted for such extreme velocity steps is unfactual.

We now look at the roles of $\beta$ in terms of reproducing canonical behaviors. A function $f$ satisfying eq.~(\ref{eq:evolutionlaw_compatiblewithfunctionalansatz}) reduces to $f\simeq \psi_{\rm c}f_{\rm SHS|NSC}+ f_{\rm VS}$ (given eqs.~\ref{eq:fvsisfdeltaV} and \ref{eq:NSCdef}), leading to $f\simeq f_{\rm VS}$ for $\Omega\gg\beta$ and $f\simeq f_{\rm SHS|NSC}$ for $\Omega\ll\beta$ from eq.~(\ref{RSFeq:reqforpsic}). In other words, eqs.~(\ref{eq:evolutionlaw_compatiblewithfunctionalansatz}) and (\ref{RSFeq:reqforpsic}) divide the $\Omega$ range into the two canonical regimes of VS and SHS$|$NSC: 
\begin{equation}
    \psi(\Omega)\simeq
    \begin{cases}
        f_{\rm VS} &(\Omega\gg\beta)
        \\
        f_{\rm SHS|NSC} &(\Omega\ll\beta)
    \end{cases}
    \label{eq:betaisVSSHSboundary}
\end{equation}
Thus, $\beta$ is observable as the scale of $\Omega$ at which 
the state evolution is switched between the VS and SHS$|$NSC canons. 
Because the canon-constrained $f_{\rm VS}$ (eq.~\ref{eq:simplifiedslipdepofthetadot}) is the same as the slip law~\citep{ruina1983slip} and the $f_{\rm SHS|NSC}$ is the same as the aging law for $\Omega\ll1$ ($\supset \Omega\ll\beta$), eq.~(\ref{eq:betaisVSSHSboundary}) also reads as the switching between the aging and slip laws according to $\Omega$.
Moreover, eq.~(\ref{eq:betaisVSSHSboundary}) elucidates how the unavoidable side effects of the aging-slip switch (\S\ref{sec:canonicalandside}) appear in our specific approach using the functional form of $f = \psi(\Omega)$ as well as the TDF $f = f_t + f_\delta V$; sufficiently below the steady state, the side effects harm the reproduction of the VS canon, and in the other ranges of $\Omega$ they harm the reproduction of the SHS canon. 
As will be discussed at the end of \S\ref{RSF322} (and also discussed from another viewpoint in \S\ref{subsec:discussion1}), these side effects are commonly seen in evolution laws trying to reproduce the three canons simultaneously. We may say that these side effects are nearly a logical consequence of those canonical behaviors. 

If eqs.~(\ref{eq:evolutionlaw_compatiblewithfunctionalansatz_substituted}) and (\ref{RSFeq:reqforpsic}) are the case, eq.~(\ref{eq:betaisVSSHSboundary}) suggests two types of experimental information usable to constrain the small positive constant $\beta$ introduced in eq.~(\ref{RSFeq:reqforpsic}). Firstly, eq.~(\ref{RSFeq:reqforpsic}) predicts symmetry breaking for large negative velocity steps such that $V_{\rm after}/V_{\rm before} \lesssim\beta$, in the manner illustrated in Fig.~\ref{fig:predictedmotion_modcomposite}a shown later. The value of $\beta$ is then known from the value of $V_{\rm after}/V_{\rm before}$ below which the VS canon is violated. If this symmetry breaking is not observed down to the lowest $V_{\rm after}/V_{\rm before}$ value experimentally explored so far, that value sets an upper bound of $\beta$. The current situation is closer to the latter; \citet{bhattacharya2015critical} find that the slip law fits VS tests down to $V_{\rm after}/V_{\rm before}$ of 1/100, which constrains $\beta$ to be
\begin{equation}
0< \beta\lesssim 0.01.
\label{RSFeq:betavalueconstraint}
\end{equation}
As mentioned later (\S\ref{subsec:possiblebetadetection}), \citet{bhattacharya2022evolution} concluded $|f-f_t|>|f-f_\delta V|$ for $\Omega$ down to $10^{-3.5}$, based on their VS tests with $V_{\rm after}/V_{\rm before} \geq 10^{-3.5}$. In contrast, also as mentioned there, 
\citet{bhattacharya2015critical} showed that a VS test by \citet{bayart2006evolution} with $V_{\rm after}/V_{\rm before}= 10^{-2}$ indicates that the strength recovery following this large negative step is slightly faster than the slip-law prediction with parameters determined from VS tests with $V_{\rm after}/V_{\rm before} \geq 10^{2}$, $10^{1}$, $10^{-1}$, which suggests the transition from the slip law to the aging law starts kicking in at $\Omega =  10^{-2}$. They concluded that this slight behavior difference for $V_{\rm after}/V_{\rm before}= 10^{-2}$ is statistically significant, though the slight difference may not be sufficient to affirm an extra parameter to express the Aging-Slip switch, as detailed in \S\ref{subsec:possiblebetadetection}. We leave the range $10^{-3.5} < \Omega <  10^{-2}$ as a grey zone for now. It is natural to expect that $\beta$ depends on the details of the experimental conditions.
In the rest of the paper, we assume $\beta\ll1$. 

The second potential source of information to constrain $\beta$ is the change in strength recovery rate during SHS tests. Given that $L\sim D_{\rm c}$ (eq.~\ref{RSFeq:LsimilartoDc}), eqs.~(\ref{eq:evolutionlaw_compatiblewithfunctionalansatz_substituted}) and (\ref{RSFeq:reqforpsic}) predict that the healing rate changes when $\Omega$ crosses $\beta$ during the hold phase. As detailed in Appendix~\ref{sec:SHSHSprinciple}, $\Omega$ increases with time during constant-$\tau$ holds on the interface with $A > B$. Hence, as long as eqs.~(\ref{RSFeq:LsimilartoDc}), (\ref{eq:evolutionlaw_compatiblewithfunctionalansatz_substituted}), and (\ref{RSFeq:reqforpsic}) hold, the strengthening occurs by the aging-law type purely time-dependent healing ($\dot \theta\simeq 1$) early in the hold (where $\Omega\ll \beta$) but switches to the slip-law type slip-driven strengthening ($\dot \theta\sim-\Omega\ln\Omega\ll1$) as $\Omega$ increases past $\beta$ (i.e., $\beta\ll\Omega\ll1$). Consequently, strength recovery slows down at $\Omega\sim \beta$ before saturation around $\Omega\sim \Omega_{\rm SS}(= 1)$ occurs  (\S\ref{sec:SHSanalysismodifiedcomposite}, also see Fig.~\ref{fig:SHScomparison}). This aging-slip crossover time in SHS tests, referred to as $t_{\rm as}$, exhibits a positive dependence on $\beta$ (eq.~\ref{eq:deftas}). If $t_{\rm as}$ is observed, the $\beta$ value is known. If $t_{\rm as}$ is not observed within the hold times tested so far, a lower bound for $\beta$ is known. \citet{dieterich1972time} confirmed steady log-t healing in his constant-$\tau$ holds (at $\tau_{\rm hold}\sim 0.9\Phi_*$) for hold times up to 2 days, which may give a lower bound for $\beta$. However, we do not evaluate it here because this slowdown occurs only on $A > B$ and $D_{\rm c}\sim L$, which are not confirmed for this case.

Lastly, we emphasize the above-discussed symmetry breaking in VS tests and the slowdown of strength recovery in SHS tests, which follow from eqs.~(\ref{RSFeq:LsimilartoDc}), (\ref{eq:evolutionlaw_compatiblewithfunctionalansatz_substituted}), and (\ref{RSFeq:reqforpsic}), are manifestations of unavoidable side effects of the aging-slip switch around $\Omega=\beta$.

\subsubsection{Derivation without heuristics}
\label{sec:generalizingheuristicderivation}

So far, we have taken a strategy to search evolution functions under our heuristic ansatz that presumes unary functions of $\Omega$ (eq.~\ref{RSFeq:fOmega0}, or equivalently, ~\ref{RSFeq:fOmega}) so that the SS canon is secured beforehand. Hence, in deriving eq.~(\ref{RSFeq:reqforpsic}), the constraint on $f_t$ ($= \psi_{\rm c}(\Omega)$ by the ansatz), we only had to consider VS and SHS constraints. In this \S\ref{sec:generalizingheuristicderivation}, we derive constraint on $f_t$ without heuristics. The resultant constraint on $f_t$ is a sufficient and necessary condition to reproduce the three canons, whereas eq.~(\ref{RSFeq:reqforpsic}), derived under the heuristic ansatz, was a sufficient condition. Note that the following argument holds within the standard rate-and-state framework of $\dot\theta=f(V,\theta)=f_t+f_\delta V$, with $f_\delta V$ having the form of eq.~(\ref{eq:simplifiedslipdepofthetadot}) directly constrained by the VS canon (plus part of the SS constraint, strictly speaking); then $f_t$ considered in the following argument is $f_t(V,\theta)$. These are the same circumstances as in the derivation of eq.~(\ref{RSFeq:reqforpsic}), except that the unary ansatz has been removed.

We begin with the VS and SHS canons. 
In the introductory part of \S\ref{sec:development}, we have shown that the NSC  ($|f_t|\gg |f_\delta V|$) holds at $\Omega\to0$, and the NSC can exist only within a right-bounded interval of $\Omega$ [$\Omega\ll D_{\rm c}/L$ for $|\ln D_{\rm c}/L|\lesssim 1$; for $|\ln D_{\rm c}/L|\gg 1$, $\Omega\ll D_{\rm c}/L/|\ln (D_{\rm c}/L)|$]. 
Thus, the existence of the NSC sets a lower bound of the $\Omega$ value sufficiently below which $|f_t|\gg |f_\delta V|$ must hold. 
On the other hand, the VS canon imposes the opposite, namely $|f_t|\ll |f_\delta V|$ (eq.~\ref{RSFeq:negligiblehealing}) within a left-bounded interval of $\Omega$: $\Omega\geq V_{\rm after}/V_{\rm before}(\ll 1)$, which is experienced during the state evolution following a negative velocity step, for any $V=V_{\rm after}$. 
That is, $|f_t|\ll |f_\delta V|$ in VS tests sets an upper bound of the $\Omega$ value sufficiently above which $|f_t|\ll |f_\delta V|$. 
For these $\Omega$ ranges where $|f_t|\gg |f_\delta V|$ and $|f_t|\ll |f_\delta V|$, respectively, to coexist, 
there must exist an $\Omega$ value $\beta$ such that $0<\beta\ll V_{\rm after}/V_{\rm before}(\ll1)$, sufficiently below which $|f_t|\gg |f_\delta V|$ and sufficiently above which $|f_t|\ll |f_\delta V|$. 
Accordingly, $f_t$ is constrained as follows: 
$f_t$ is much smaller than $f_\delta V$ (eq.~\ref{eq:simplifiedslipdepofthetadot}, from VS) for $\Omega\gg \beta$, while it must conform 
$f_{t|{\rm NSC}}$ (eq.~\ref{eq:simplifiedtimedepofthetadot}, from SHS$|$NSC) for $\Omega \ll \beta$:
\begin{equation}
    f_t=
    \begin{cases}
        o[(L/D_{\rm c})\Omega\ln\Omega] & (\Omega\gg\beta)
        \\
        1 & (\Omega\ll\beta).
    \end{cases}
    \label{eq:constraintonftgen}
\end{equation}
Equation~(\ref{eq:constraintonftgen}) differs from eq.~(\ref{RSFeq:reqforpsic}) only in that eq.~(\ref{eq:constraintonftgen}) constrains $f_t$ rather than $\psi_{\rm c}$; the right-hand side, i.e., the required dependency on $V$ and $\theta$, is the same. This stands to reason because the derivation was based on the SHS$|$NSC and VS canons, as in the derivation of eq.~(\ref{RSFeq:reqforpsic}). We may regard eq.~(\ref{eq:constraintonftgen}) as the generalized version of eq.~(\ref{RSFeq:reqforpsic}).

Now, we consider the SS canon, which seems no longer guaranteed beforehand because $f_t$ considered now is not limited to unary functions of $\Omega$. However, as shown below, the SS canon is achieved without additionally imposing $\Omega_{\rm SS}=1$ (eq.~\ref{eq:constraintforsteadystate}); $f_t$ satisfying eq.~(\ref{eq:constraintonftgen}) vanishes near the steady state so that it does not affect the steady-state value $\Omega=1$ set by $f_\delta V$. 
From eqs.~(\ref{eq:simplifiedslipdepofthetadot}) and ~(\ref{eq:constraintonftgen}), the evolution function becomes $f= -(L/D_{\rm c})\Omega\ln\Omega+o[(L/D_{\rm c})\Omega\ln\Omega]$ for $\Omega\gg\beta$, which already guarantees that $\Omega=1$ is a fixed point of the evolution law. Furthermore, if a small additional constraint of non-negativity of $f_t$ is imposed around $\Omega\sim \beta$ (note eq.~\ref{eq:constraintonftgen} does not constrain $f_t$ for this $\Omega$ range):
\begin{equation}
    f_t\geq 0 \hspace{10pt}(\Omega\sim \beta),\label{eq:nonnegativetimehealing}
\end{equation} 
then we have $f>0$ for $\Omega\lesssim\beta(\ll1)$, where $f_t \geq 0$ and $f_\delta V>0$, guaranteeing that $\Omega = 1$ is the unique fixed point for any $V$. 
Equation~(\ref{eq:nonnegativetimehealing}) is virtually trivial as $f_t$ represents purely time-dependent healing, which may not deserve to be recognized as an ``additional'' requirement. 

Thus, the above-derived set of constraints eqs.~(\ref{eq:simplifiedslipdepofthetadot}), (\ref{eq:constraintonftgen}), and (\ref{eq:nonnegativetimehealing}), where the heuristic ansatz has been removed, constitute a necessary and sufficient condition 
to satisfy the experimental constraints 1--3 (eqs.~\ref{eq:constraintforsteadystate}--\ref{eq:simplifiedtimedepofthetadot}) and the auxiliary constraint (eq.~\ref{RSFeq:negligiblehealing}), 
that is, to reproduce all the canonical behaviors (except for $\Omega\sim\beta$, the border regime of the aging-slip switch).

The above argument elucidates an interesting point about the role of eq.~(\ref{RSFeq:reqforpsic}). In deriving eq.~(\ref{RSFeq:reqforpsic}), we intended to secure the SS canon by the unary ansatz. 
However, it has turned out that any function having the asymptotic $V$, $\theta$ dependency conforming eq.~(\ref{eq:constraintonftgen}), practically the same as eq.~(\ref{RSFeq:reqforpsic}), automatically conforms to the SS canon under the trivial assumption of non-negative $f_t$. 
That is, retrospectively speaking, the ansatz was in fact not necessary for our original aim of securing the SS canon.
Both eqs.~(\ref{RSFeq:reqforpsic}) and (\ref{eq:constraintonftgen}) state that $f_t$ must be negligible unless the system is sufficiently below the steady state, 
which is why the steady state is unchanged from the fixed point of $f_\delta V$, $\Omega_{\rm SS}=1$. 
Nevertheless, in the subsequent development (\S\ref{RSF33}--\ref{sec:extensions}), we basically keep the unary ansatz and use eq.~(\ref{RSFeq:reqforpsic}), rather than eq.~(\ref{eq:constraintonftgen}). However, we now have to say that the substantial role of the ansatz is a mere pragmatic convention to avoid considering too many candidates in proposing evolution laws. 

\begin{figure*}
	\includegraphics[width=145mm]{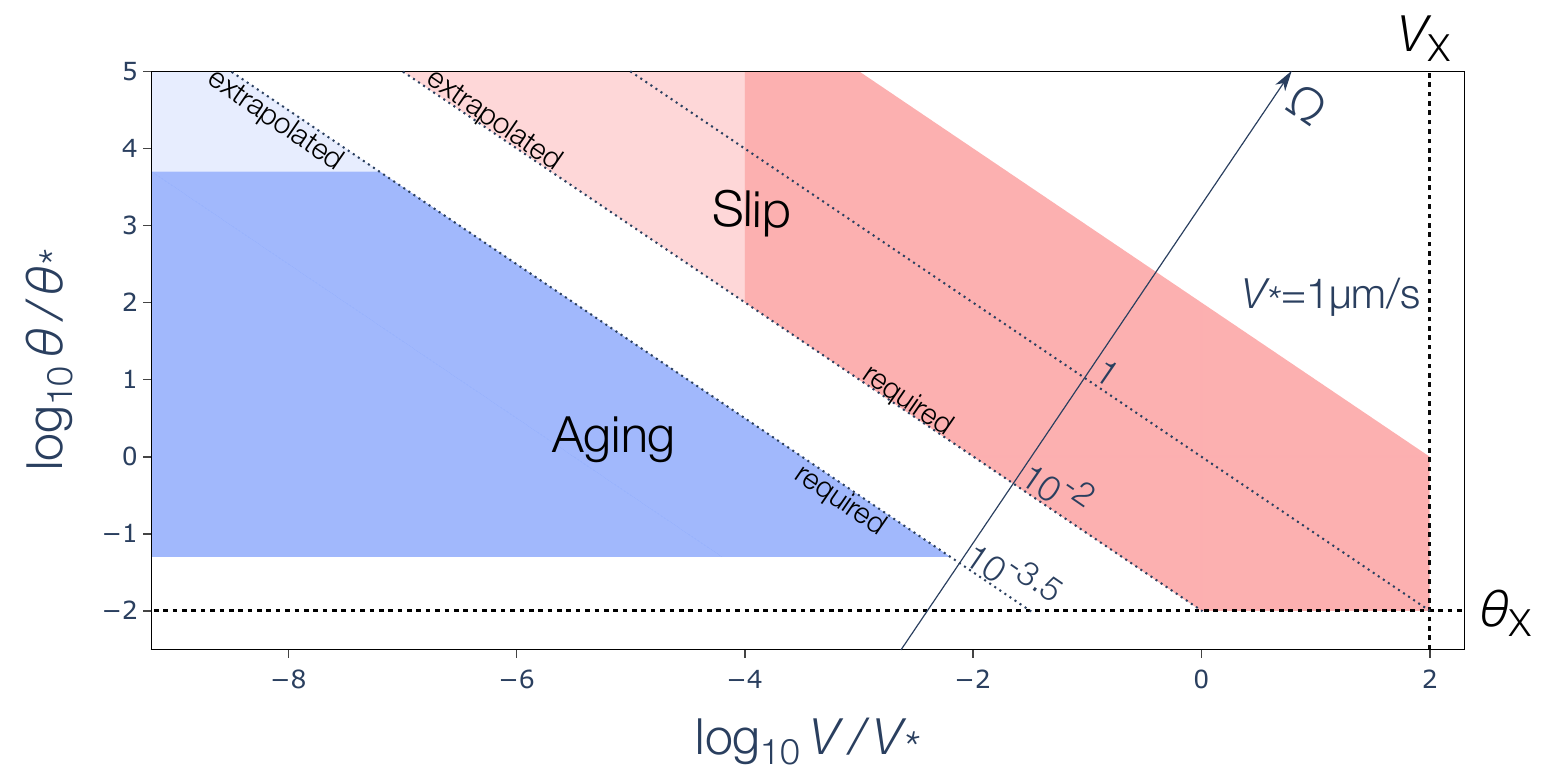}
\caption{
Colored map showing the $(V, \theta)$ regimes where the aging law (blue) is required and the slip law (red) is required based on the compilation of existing experiments. 
$V_*=1$ $\mu$m/s is assumed.
Pale-colored regimes represent where those behaviors are expected; though the experiments have not probed those regimes, we think of no particular reason to expect the behaviors to change. The red regime is based on \citet{kilgore1993velocity} and \citet{bhattacharya2015critical}, while the blue regime is based on \citet{nakatani1996effects}. 
See the text for details. 
}
  \label{fig:zakkuriswitch}
\end{figure*}

In any case, with or without the unary ansatz, the aging-slip switch according to $\Omega$ (eq.~\ref{RSFeq:reqforpsic} or \ref{eq:constraintonftgen}) thus follows from the three canons, as long as $\dot\theta = f (V, \theta)$. At the end of this subsection, we illustrate this logical consequence by compiling relevant VS and SHS$|$NSC tests (Fig.~\ref{fig:zakkuriswitch}). Given the constraints considered here (eqs.~\ref{RSFeq:steadystateconstraint}, \ref{RSFeq:dotthetaVS}, and \ref{eq:ftNSC_constrained}, plus eq.~\ref{RSFeq:negligiblehealing}), the possible functional form of $f$ is limited to one of the following two: $f\simeq f_\delta V=-(L/D_{\rm c})\Omega\ln\Omega$, which reproduces the VS canon, or $f\simeq f_{t|{\rm NSC}}(=1)$, which reproduces the SHS$|$NSC canon. Hence, the remaining task is to identify, in the $(V, \theta)$ space, the regime where evolution agrees with the aging law but contradicts the slip law (blue and pale-blue areas in Fig.~\ref{fig:zakkuriswitch}) and the regime where evolution agrees with the slip law but contradicts the aging law (red and pale-red areas in Fig.~\ref{fig:zakkuriswitch}).

Specifically, we used experiments of \citet{kilgore1993velocity} to define the velocity range ($10^{-4}$--$10^{2}$ $\mu$m/s) of the red area (i.e., slip law required), where both VS and SS canons are required. The range of $\Omega$ where the VS canon is known required is the maximum of $|\log_{10}(V_{\rm after}/V_{\rm before})|$ for which the VS canon has been confirmed. Numerous experiments have confirmed the VS canon for up to 10-fold VS. Recently, \citet{bhattacharya2015critical} have shown that the VS canon is observed even for 100-fold VS. Hence, we have decided that the VS canon has been confirmed up to 100-fold VSs and have adopted $10^{\pm2}$ as the along-$\Omega$ axis ends of the red and pale-red bands. In addition, we have adopted $V_{\rm X}$ of about 100 $\mu$m/s from the high-speed VS tests at $\sigma  = 5$ MPa by \citet{kilgore1993velocity} because we will use SHS$|$NSC tests at $\sigma  = 5$ MPa~\citep{nakatani1996effects} to define the $(V, \theta)$ range where the agreement with the aging-law-like (i.e., the canonical SHS$|$NSC) behavior and the contradiction to the slip-law-like behavior have been confirmed. 
The pale-red area below $V=10^{-4}$ $\mu$m/s, the minimum slip rate of \citet{kato2001composite}, is included by extrapolation given we have no particular observation suggesting the VS behavior to change below a certain slip rate.

As explained in \S\ref{subsec:SHSrequirement}, the SHS$|$NSC canon is required in the $(V, \theta)$ range where the observed log-t healing is far greater than the strengthening predicted by $f_\delta V$, now constrained to be the slip law. Such a range (the blue area in Fig.~\ref{fig:zakkuriswitch}) is constrained by the constant-$\tau$ SHS tests of \citet{nakatani1996effects} with sufficiently low $\tau_{\rm hold}$. Their Fig.~4 b shows that, for SHS tests with $\tau_{\rm hold} > 0.6 \sigma$,  $B_{\rm heal}$ was about $0.01\sigma$. Since their VS tests showed that $A-B$ was about $0.0009 \sigma$, the difference between $A$ and $B$ is less than 10\% of $A$. For such interfaces with $A\simeq B$, $\Omega$ remains close to the initial value $\Omega_{\rm init}$ ($=\exp(-\Delta \tau/A)$) throughout the constant-$\tau$ hold period, and both aging and slip laws predict logarithmic increases of $\Phi$ with $t_{\rm h}$: $\theta(t_{\rm h})=\theta_{\rm SS}(V_{\rm prior})(1+t_{\rm h}/t_{\rm c})$ (eqs.~\ref{eq:agingevolutioninSHS1} and \ref{eq:thetafortime}), though the predicted cutoff time $t_{\rm c}$ is very different between the two evolution laws. Therefore, their series of constant-$\tau$ SHS tests with various levels of $\Delta\tau$ but with the common $V_{\rm prior}$ scan a quadrangular range bounded by oblique constant-$\Omega$ lines and horizontal constant-$\theta$ lines, as detailed below. The slip rate $V$ immediately after the reduction of $\tau$ to the designated $\tau_{\rm hold}$ is given as $V_{\rm init} = V_{\rm prior}\exp(-\Delta \tau/A)$, and then $V$ decreases as $\theta$ increases with $t_{\rm h}$. The blue area in Fig.~\ref{fig:zakkuriswitch} is particularly relying on the seven tests with $\tau_{\rm hold} = 0.75\sigma$ and $t_{\rm h} = 40$ s~\citep[Figure~2 of][]{nakatani1996effects}. 
Corresponding to the natural variability of $\tau_{\rm SS} = 0.83\sigma$--$0.88\sigma$, these seven tests offer a variety of $\Delta\tau = 0.08\sigma$--$0.13\sigma$. For all the seven tests, $V_{\rm prior}$ was about 20 $\mu$m/s, and the strength recovery in the $t_{\rm h}$ of 40 s was about 3$B_{\rm heal}\times \ln10$, which indicates that $t_{\rm h}$ (=40 s) was about $1000t_{\rm c}$, that is, $t_{\rm c}\sim 40$ ms for all the seven tests. In contrast, the slip law predicts that $t_{\rm c}$ for the experimental conditions (i.e., $\Delta \tau/A\in [8, 13]$ and $\Omega\in [-5.5, -3.5]$) ranges $ D_{\rm c}\times$ 19--1700 s/$\mu$m (eq.~\ref{eq:tcslippertcaging}). Assuming the typical laboratory value of $D_{\rm c}\sim 1$ $\mu$m, the predicted $t_{\rm c}$ is $10^2$--$10^4$ times greater than the observed value of 40 ms. Moreover, as pointed out in \S\ref{subsubsec:332}, the slip law predicts strong dependency on $\Delta \tau/A$, contradicting the observation that $t_{\rm c}$ was about constant for $\Delta \tau/A\in  [8, 13]$. Besides, the aging law, which describes the SHS$|$NSC canon well, predicts that $t_{\rm c}$ is constant at $L/V_{\rm prior}$, not depending on $\Delta \tau/A$. 
This prediction is consistent with the observed $t_{\rm c}$ of about 40 ms and typical $D_{\rm c}$ of about 1 $\mu$m if $L\sim D_{\rm c}$ (eq.~\ref{RSFeq:LsimilartoDc}). Thus, these constant-$\tau$ tests contradict the slip law strongly while agreeing with the aging law very well. 
Assuming $A=B$, $\theta/\theta_{\rm SS}(V_{\rm prior})$ increases from approximately 1 to $t_{\rm h}/t_{\rm c}$, and $V$ decreases accordingly along the constant-$\Omega$ contour, as mentioned earlier. 
The blue regime (i.e., aging-law-like behaviors required) in Fig.~\ref{fig:zakkuriswitch} extends up to $t_{\rm h}/t_{\rm c}\sim10^5$ because Figure 4 of \citet{nakatani1996effects} shows the continuation of log-t healing up to such $t_{\rm h}$ (3000 s). 
The blue regime (i.e., aging-law-like behaviors required) in Fig.~\ref{fig:zakkuriswitch} also extends to $V\to 0$, 
since the contradiction to the slip law (and agreement to the aging law) can also be shown for the pale-blue area below $\Omega < 10^{-5.5}$, using the experiments with further lower $\tau_{\rm hold}$ shown in Figure 4 of \citet{nakatani1996effects}. 

In summary, Fig.~\ref{fig:zakkuriswitch} illustrates that, in the velocity range where experiments have confirmed both VS and SHS$|$NSC canons, the aging-slip switch has to occur at some threshold value of $\Omega$, as long as $\dot\theta = f(V, \theta)$ is premised. 

In reading this figure, it must be noted that ``aging-/slip-law-like'' (i.e. $|f-f_t| \lessgtr |f-f_\delta V|$) evolution is distinct from evolution ``conforming the aging/slip canon'' (i.e. $f\simeq f_t$,  $f\simeq f_\delta V$). Rigorous confirmation of the VS-canon conformity of laboratory data is limited to VS tests with $V_{\rm after}/V_{\rm before} > 10^{-2}$. VS tests with further lower $V_{\rm after}/V_{\rm before}$ (down to $10^{-3.5}$ so far) are slip-law-like but can significantly differ from the VS canon \citep{bhattacharya2015critical,bhattacharya2022evolution}. Figure~\ref{fig:zakkuriswitch} leaves this range unpainted. On the other hand, as for SHS, rigorous studies do not exist, to our knowledge, on how closely the laboratory data obeys the aging canon. There is even a possibility that (some of) the results by \citet{nakatani1996effects} are just aging-law-like, not canon-conforming. Under those circumstances, while we painted the whole $(V,\theta)$ space of \citet{nakatani1996effects} down to $\Delta\tau /A =8$ as ``canon-conforming,'' the formal conclusion of the present paper remains $\beta \lesssim0.01$, with the lower bound for $\beta$ unspecified (\S\ref{sec:heuristicderivation}). Nevertheless, finite $\beta$ is instantly concluded mathematically from the SHS tests under $-\log_{10} \Omega\gg1$ and the VS behavior under $|\log_{10}\Omega|=\mathcal O(1)$, not suffering from the current inability to determine the $\beta$ value. Figure~\ref{fig:zakkuriswitch}, though drawn in terms of the concrete values of $\beta$ inferred from experiments probing the currently limited range of $(V, \theta)$ space, is presented only to help readers grasp the general picture of the presently proposed aging-slip switch (i.e., the existence of $\beta$ for $f(V,\theta)$), which has been mathematically proven in \S\ref{sec:generalizingheuristicderivation} by the deduction from the canons (plus the premise of the TDF).

\subsection{Examples of evolution laws consistent with the three canonical behaviors}\label{RSF33}

In \S\ref{RSF322}, we have shown that a class of evolution laws satisfying eqs.(\ref{eq:evolutionlaw_compatiblewithfunctionalansatz_substituted}) and (\ref{RSFeq:reqforpsic}) conform constraints from SS, VS, and SHS experiments. The idea is a simple reconciliation of the slip and aging laws~\citep{kato2001composite,ampuero2008earthquake} by switching the two laws according to the value of $\Omega$ ($:= V\theta/L$), the conventional state variable $\theta$ normalized by its steady-state value for the current slip rate $\theta_{\rm SS}(V)$ [$\Omega= \theta/\theta_{\rm SS}(V)$]. The proposed class of laws approximate $\dot \theta\simeq 1$ (the aging law at the NSC) for $\Omega\ll \beta$, while they approximate the canonical evolution in VS tests, $\dot \theta\simeq-(L/D_{\rm c})\Omega \ln \Omega$ (the slip law). In this subsection \S\ref{RSF33}, we propose concrete examples of this class of evolution laws. 
A stepwise function $\psi_{\rm c}=H(\beta-\Omega)$, where $H(\cdot)$ is the Heaviside function, satisfies eq.~(\ref{RSFeq:reqforpsic}), of course, but it causes an artificial discontinuity in the evolution function. Instead, we seek a procedure to construct smooth functions satisfying eq.~(\ref{RSFeq:reqforpsic}). 
From this subsection onward, we assume $C_{\rm SS}=1$ (eq.~\ref{RSFeq:Cssnormalization}), whereas \S\ref{RSF322} was developed under a less restrictive assumption of $C_{\rm SS}\sim 1$. 
In addition, within this subsection \S\ref{RSF33}, we impose $L\sim D_{\rm c}$ (eq.~\ref{RSFeq:LsimilartoDc}) to simplify the derivation. As discussed in \S\ref{sec:34}, $L\sim D_{\rm c}$ is likely to be the case, if not well confirmed by a wide range of experiments.

\begin{figure*}
	\includegraphics[width=125mm]{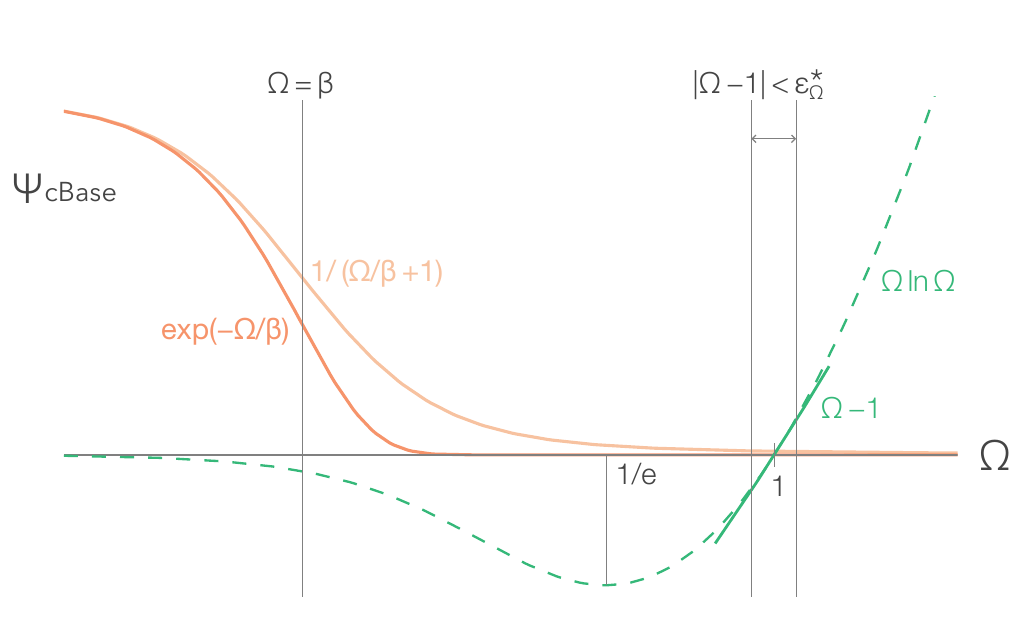}
\caption{
Schematic sketch of the base function $\psi_{\rm cBase}(\Omega)$ of $f_t$, shown for two examples of $\psi_{\rm cBase}=\exp(-\Omega/\beta)$ (orange) and $\psi_{\rm cBase}=1/(\Omega/\beta+1)$ (beige). 
The value of $\Omega$ is indicated by the x-axis with nonuniform stretching of $x=\log_{10}\Omega+\Omega$ for visibility.
For reference, the magnitude of the $f_\delta V$ term ($=\Omega\ln \Omega$) is shown by the dashed line, with its tangent line at $\Omega=1$ (solid line: $\Omega-1$). The gray vertical lines mark the boundaries of the sufficiently below the steady state $\Omega\ll\beta $ and the $\epsilon_\Omega^*$-neighborhood of the steady state, drawn with $\beta=0.01$ and $\epsilon_\Omega^*=0.1$ assumed for illustration. 
}
  \label{fig:mitorizu}
\end{figure*}

Constructing $\psi_{\rm c}$ for $\Omega\gg\beta$ so that it satisfies the first line of eq.~(\ref{RSFeq:reqforpsic}) is not as straightforward as it might sound. The difficulty comes from 
$|f_{t|{\rm VS}}| \ll|f_{\delta|{\rm VS}}V| $ (eq.~\ref{RSFeq:negligiblehealing}) because $f_\delta V$ ($= -(L/D_{\rm c}) \Omega\ln\Omega$) can become infinitesimally small in the vicinity of $\Omega = 1$. To get around, we first make a $\psi_{\rm c}$ that conforms eq.~(\ref{RSFeq:reqforpsic}) except in the vicinity of $\Omega = 1$. Exception handling is next made so that eq.~(\ref{RSFeq:reqforpsic}) is satisfied also in the vicinity of $\Omega = 1$. For this purpose, we introduce here a positive constant $\epsilon_\Omega^*$ and divide the $\Omega$ range into $|\Omega-1|<\epsilon_\Omega^*$ and the rest (Fig.~\ref{fig:mitorizu}). To simplify inequality evaluations in the rest of the derivation, we assume $1-\epsilon_\Omega^* > 1/e$, namely $\epsilon_\Omega^*<1-1/e=0.6321...$, noting that $\Omega\ln\Omega$ takes the unique extremum value at $\Omega = 1/e$ (Fig.~\ref{fig:mitorizu}). Also, we hereafter assume $\beta<1/e^2(=0.135...)$ and $\sqrt \beta\ll1$, again for simplicity of derivation. As discussed in \S\ref{RSF322}, observed symmetry in many existing VS tests has already rejected the possibility that $\beta\geq 1/e^2$. Moreover, recent VS tests employing very large steps by \citet{bhattacharya2015critical} show that $\sqrt{\beta}\lesssim0.1$ (eq.~\ref{RSFeq:betavalueconstraint}).

We now compose a function $\psi_{\rm cBase}(\Omega)$ that satisfies eq.~(\ref{RSFeq:reqforpsic}) outside the $\epsilon_\Omega^*$-vicinity of $\Omega = 1$. Under the considered parameter range of $\beta <1/e<1-\epsilon_\Omega^*$, we obtain the following lower bound of $|\Omega\ln\Omega|$ in the $\Omega$ range such that $\Omega\gg\beta\cap|\Omega-1|\geq\epsilon_\Omega^*$:
\begin{equation}
|\Omega\ln\Omega| \geq
\begin{cases}
|\beta\ln\beta| & (\beta \ll \Omega \leq 1/e)
\\
|(1-\epsilon_\Omega^*)\ln (1-\epsilon_\Omega^*)| & (1/e \leq \Omega \leq 1-\epsilon_\Omega^*)
\\
|(1+\epsilon_\Omega^*)\ln (1+\epsilon_\Omega^*)| & (\Omega \geq1+\epsilon_\Omega^*)
\end{cases} 
\label{eq:boundedxlnxlowerbound}
\end{equation}
To obtain eq.~(\ref{eq:boundedxlnxlowerbound}), we used $\Omega\ln\Omega
\gtrless0$ for $\Omega \gtrless 1$ as well as the fact that the unique extremum of $\Omega\ln\Omega$ is located at $\Omega = 1/e$ (Fig.~\ref{fig:mitorizu}). 
Using eq.~(\ref{eq:boundedxlnxlowerbound}) and $L\sim D_{\rm c}$ (eq.~\ref{RSFeq:LsimilartoDc}), we find that, excepting the $\epsilon_\Omega^*$-vicinity of $\Omega = 1$, eq.~(\ref{RSFeq:reqforpsic}) is satisfied by $\psi_{\rm c}(\Omega)=\psi_{\rm cBase}(\Omega)$ such that
\begin{equation}
    \psi_{\rm cBase}(\Omega)=
    \begin{cases}
        1 & (\Omega\ll\beta)
        \\
        \mathcal O (\beta\ln \beta) & (\beta\ll\Omega\leq 1/e)
        \\
        o[\min_{\pm}(1\pm\epsilon_\Omega^*)\ln (1\pm\epsilon_\Omega^*)]& (\Omega\geq 1/e)       
    \end{cases}
    \label{eq:reqforpsicforO1Omega}
\end{equation}
Note $x=\mathcal O(y)$ denotes $|x|\lesssim|y|$ hereafter. The second line of eq.~(\ref{eq:reqforpsicforO1Omega}) derives from the condition $\psi_{\rm c} = o(\Omega\ln\Omega)$ (eq.~\ref{eq:boundedxlnxlowerbound}), but we put $\mathcal O (\beta\ln \beta)$ instead of $o (\beta\ln \beta)$ straightforwardly expected from eq.~(\ref{eq:boundedxlnxlowerbound}). This is allowed because, for the fairly small $\beta$ assumed here (s.t. $\beta<1/e^2\cap\sqrt{\beta}\ll1$), $\beta\ll\Omega\leq 1/e\Rightarrow |\Omega\ln\Omega|\gg|\beta\ln\beta|$  (eq.~\ref{eq:OmegalnOmegaggbetalnbeta}; see Appendix~\ref{sec:epsironRange} for details).

Equation~(\ref{eq:reqforpsicforO1Omega}) shows a sufficient condition to construct a desired $\psi_{\rm cBase}(\Omega)$ function for a given $\epsilon_\Omega^*$. Next, as $\epsilon_\Omega^*$ has been arbitrary, we seek a way to tune $\epsilon_\Omega^*$ so that $\psi_{\rm cBase}(\Omega)$ satisfies eq.~(\ref{RSFeq:reqforpsic}) outside the $\epsilon_\Omega^*$-vicinity of $\Omega = 1$. 
Recalling the intention of eq.~(\ref{RSFeq:reqforpsic}) explained in \S\ref{RSF322}, one natural choice of $\psi_{\rm c}$ would be a decreasing function having a characteristic $\Omega$ value, $\beta$. For simplicity, we only consider such $\psi_{\rm cBase}$ hereafter. 

From eq.~(\ref{eq:OmegalnOmegaggbetalnbeta}), a monotonically decreasing $\psi_{\rm cBase}$ satisfying eq.~(\ref{RSFeq:reqforpsic}) must conform
\begin{equation}
    \psi_{\rm cBase}(\Omega)=
    \begin{cases}
        1 & (\Omega\ll\beta)
        \\
        \mathcal O (\beta\ln \beta) & (\Omega \gg\beta)
    \end{cases}
    \label{eq:psicBaserequirement}
\end{equation}
Equation~(\ref{eq:psicBaserequirement}) represents a necessary condition for the $\psi_{\rm cBase}$ to be usable as $\psi_{\rm c}$ of eq.~(\ref{RSFeq:reqforpsic}). 
Meanwhile, 
linear approximation of eq.~(\ref{eq:reqforpsicforO1Omega}) with respect to $\epsilon_\Omega^*$ shows that $\psi_{\rm c}=\psi_{\rm cBase}$ satisfying eq.~(\ref{eq:psicBaserequirement}) meets eq.~(\ref{eq:reqforpsicforO1Omega}) up to $\mathcal O(\epsilon_\Omega^*)$ when $|\beta\ln\beta|\ll\epsilon_\Omega^*<1-1/e$ holds, 
suggesting that eq.~(\ref{eq:psicBaserequirement}) is sufficient to satisfy eq.~(\ref{RSFeq:reqforpsic}) outside 
the $\epsilon_\Omega^*$-vicinity of $\Omega = 1$ for a certain value of $\epsilon_\Omega^*$. 
In Appendix~\ref{sec:epsironRange}, we have shown, without approximation, that 
when eq.~(\ref{eq:psicBaserequirement}) is met, eq.~(\ref{RSFeq:reqforpsic}) holds 
in the range $|\Omega-1|\geq\epsilon_\Omega^*$ for $\epsilon_\Omega^*$ such that $|\beta\ln\beta|\ll\epsilon_\Omega^*<1-1/e$ (eq.~\ref{eq:usefulpsicBaseinequality}). 
In summary, as long as eq.~(\ref{eq:psicBaserequirement}) holds, 
tuning $\epsilon_\Omega^*$ is enough to compose the desired $\psi_{\rm c}=\psi_{\rm cBase}$ that satisfies eq.~(\ref{RSFeq:reqforpsic}) in $|\Omega-1|\geq\epsilon_\Omega^*$. 

We now proceed to the remaining $\Omega$ range almost at the steady state ($\Omega_{\rm SS}=1$; eq.~\ref{RSFeq:Cssnormalization}): $|\Omega-1|<\epsilon_\Omega^*$. In this range, eq.~(\ref{RSFeq:reqforpsic}) requires
\begin{equation}
    \psi_{\rm c}(\Omega)=o(\Omega-1).
    \label{eq:smallorderofepsilon}
\end{equation}
A monotonically decreasing $\psi_{\rm cBase}$ constructed above is not enough to satisfy eq.~(\ref{eq:smallorderofepsilon}). 
Notice that eq.~(\ref{eq:smallorderofepsilon}) includes
\begin{equation}
    \psi_{\rm c}(1)=0,
     \label{eq:cutoffatsteadystate}
\end{equation}
which follows eq.~(\ref{eq:evolutionlaw_compatiblewithfunctionalansatz_substituted}) and eq.~(\ref{RSFeq:Cssnormalization}) ($\theta_{\rm SS}=L/V$).
In the first place, eq.~(\ref{RSFeq:RequirementinVS}) derived from the VS canon intends that the slip-driven evolution alone makes $\Omega$ approach exactly to its stationary point $\Omega=1$. Hence, involvement of any purely time-dependent healing within the neighborhood of $\Omega=1$, even minuscule, would shift the steady state from $\Omega=1$, meaning that $C_{\rm SS}$ would deviate from unity. Therefore, $\psi_{\rm c}$ needs to be truncated for $\Omega$ almost at the steady state. A straightforward way to realize eq.~(\ref{eq:smallorderofepsilon}) is to impose $\psi_{\rm c} = 0$ in $\Omega>1-\epsilon_\Omega^*$: 
\begin{equation}
    \psi_{\rm c}(\Omega)=\psi_{\rm cBase} (\Omega)H(1-\epsilon^*_\Omega-\Omega).
    \label{RSFeq:modcompscutofffunction_withHeaviside}
\end{equation}

It is important to note that we need to set an appropriate $\epsilon^*_\Omega$ value so that eq.~(\ref{RSFeq:reqforpsic}) holds for the given form of $\psi_{\rm cBase}$. In Appendix \ref{sec:epsironRange}, we have derived the range of appropriate $\epsilon^*_\Omega$ (eqs.~\ref{eq:epsilonboundgeneral} and \ref{eq:epsilonboundgeneral_improved}). 
Appendix \ref{sec:epsironRange} also shows that such an $\epsilon^*_\Omega$ range necessarily exists for any $\psi_{\rm cBase}$ conforming eq.~(\ref{eq:psicBaserequirement}). 
At least for an example we adopt later in this subsection,  $\psi_{\rm cBase}=\exp(-\Omega/\beta)$, the appropriate range of $\epsilon^*_\Omega$ is relatively wide (eq.~\ref{allowableRangeOfepsilonOmegaStar_exp}), causing no difficulty. 

Thus, we have shown that $\psi_{\rm c}$ satisfying eq.~(\ref{RSFeq:modcompscutofffunction_withHeaviside}) with a suitable $\epsilon_\Omega^*$ value fully satisfies eq.~(\ref{RSFeq:reqforpsic}). Moreover, if we tolerate a very tiny error discussed below, even $\psi_{\rm cBase}$ satisfying eq. (\ref{eq:psicBaserequirement}), without the Heaviside truncation, can be practically adopted as $\psi_{\rm c}$: 
\begin{equation}
    \psi_{\rm c}(\Omega) = \psi_{\rm cBase}(\Omega),
    \label{RSFeq:NoHevisidePsiC_general}
\end{equation}
because the difference made by the truncation in eq.~(\ref{RSFeq:modcompscutofffunction_withHeaviside}) is at most $\psi_{\rm cBase}(1-\epsilon_\Omega^*)$ given the non-negativity and monotonically decreasing nature of $\psi_{\rm cBase}(\Omega)$.
Firstly, excepting the very vicinity of the steady state, where this absolute error of $\mathcal O[\psi_{\rm cBase}(1-\epsilon_\Omega^*)]$ becomes dominant, an evolution law using eq.~(\ref{RSFeq:NoHevisidePsiC_general}) is equivalent to one using eq.~(\ref{RSFeq:modcompscutofffunction_withHeaviside}). 
That is, as long as eq.~(\ref{eq:psicBaserequirement}) holds, 
eq.~(\ref{eq:reqforpsicforO1Omega}) holds under some appropriate $\epsilon_\Omega^*$ value (Appendix \ref{sec:epsironRange}), so that the log-V dependence of the steady-state strength (eq.~\ref{eq:req4ss}) and the log-t healing (eq.~\ref{RSFeq:PhiSHSNSCrawtc}) are reproduced, and the transient strength evolution exponential with slip (eq.~\ref{eq:realVScanonical}) also accurately holds except in the very vicinity of steady states where $|\Omega-1|\ll1$. 
Secondly, although the use of eq.~(\ref{RSFeq:NoHevisidePsiC_general}) certainly makes $\psi_{\rm c}(1)$ nonzero and violates eq.~(\ref{eq:cutoffatsteadystate}), the resulting $C_{\rm SS}$ is only very slightly different from unity, not making any recognizable difference in observable quantities. Since $\Omega\ln\Omega|_{\Omega=1+\epsilon_\Omega}= \epsilon_\Omega+\mathcal O(\epsilon_\Omega^2)$ for $\epsilon_\Omega=o(1)$ (see Fig.~\ref{fig:mitorizu}), $C_{\rm SS}=1+\mathcal O[\psi_{\rm cBase}(1-\epsilon_\Omega^*)]$ when using eq.~(\ref{RSFeq:NoHevisidePsiC_general}). 
In the case of a specific form of $\psi_{\rm cBase}$ later proposed, the $\mathcal O[\psi_{\rm cBase}(1-\epsilon_\Omega^*)]$ absolute error can be below the rounding errors of floating-point arithmetic in computing the strength evolution. Thus, eq.~(\ref{RSFeq:NoHevisidePsiC_general}) can be made numerically equivalent to eq.~(\ref{RSFeq:modcompscutofffunction_withHeaviside}) by using a suitable $\psi_{\rm cBase}$. 

Following the above considerations, we propose two example evolution laws where $\exp(-\Omega/\beta)$ is adopted as an example of $\psi_{\rm cBase}$ satisfying eq.~(\ref{eq:psicBaserequirement}). 
The first proposal is 
\begin{equation}
f =\exp(-\Omega/\beta)H(1-\epsilon^*_\Omega-\Omega)
-
\frac L{D_{\rm c}}\Omega\ln\Omega,
\label{RSFeq:exmodcomplaw}
\end{equation}
obtained by substituting $\psi_{\rm c}$ of eq.~(\ref{RSFeq:modcompscutofffunction_withHeaviside}) into eq.~(\ref{eq:evolutionlaw_compatiblewithfunctionalansatz_substituted}). 
The second proposal is 
\begin{equation}
f =\exp(-\Omega/\beta) -
\frac L{D_{\rm c}}\Omega\ln\Omega
\hspace{10pt} \mbox{(Modified Composite Law)},
\label{RSFeq:modcomplaw}
\end{equation}
obtained by substituting $\psi_{\rm c}$ of eq.~(\ref{RSFeq:NoHevisidePsiC_general}) into eq.~(\ref{eq:evolutionlaw_compatiblewithfunctionalansatz_substituted}). 

Given that $\epsilon_\Omega^*$ is set following eq. (\ref{allowableRangeOfepsilonOmegaStar_exp}), eq.~(\ref{RSFeq:exmodcomplaw}) reproduces all the canonical behaviors throughout the $(V, \theta)$ range verified experimentally so far. Although eq.~(\ref{RSFeq:exmodcomplaw}) does suffer the two side effects of the aging-slip switch explained in \S\ref{RSF322}, they occur outside the experimentally verified range. Although $\psi_{\rm cBase}=\exp[-(\Omega/\beta)^n]$ with appropriate $n$ is more faithful to eq.~(\ref{eq:psicBaserequirement}) (see eq.~\ref{eq:exppsicbasenrange}), the $\psi_{\rm cBase}=\exp(-\Omega/\beta)$ proposed above is regarded as such when $\beta\gtrsim 10^{-10}$ (\S\ref{subsec:cutoffmodcomplawdetail}). In the present paper, we assume $\beta\gtrsim 10^{-10}$ when $\psi_{\rm cBase}=\exp(-\Omega/\beta)$ is adopted. 

In the case of eq.~(\ref{RSFeq:modcomplaw}), error quantification is necessary because, as mentioned earlier, it violates eq.~(\ref{eq:smallorderofepsilon}) in the close neighborhood of the steady state. 
The absolute difference of eq.~(\ref{RSFeq:modcomplaw}) 
from eq.~(\ref{RSFeq:exmodcomplaw}) is at most $\exp[-(1-\epsilon_\Omega^*)/\beta]$, which is $\psi_{\rm cBase}(1-\epsilon_\Omega^*)$ for $\psi_{\rm cBase}=\exp(-\Omega/\beta)$.
This upper-bound value depends positively on $\beta$ and $\epsilon_\Omega^*$. 
The range of $\beta$ and $\epsilon_\Omega^*$ worth consideration would be $\beta\leq 0.01$ (given eq.~\ref{RSFeq:betavalueconstraint}) and $\epsilon_\Omega^*\leq 1/10$, where the absolute error does not exceed $\psi_{\rm cBase}(1-\epsilon_\Omega^*-0) < 10^{-39}$, negligible in 64-bits floating-point computation of the strength $\Phi$.

Hereafter, we refer to eq.~(\ref{RSFeq:modcomplaw}) as the `modified composite law' because its form under $L=D_{\rm c}$ is exactly the composite law (eq.~\ref{eq:defofcompositelaw}) by \citet{kato2001composite}, with the constant $V_{\rm c}$ in eq.~(\ref{eq:defofcompositelaw}) replaced by $\beta L/\theta$. To sum up, by turning on/off the purely time-dependent healing according to $\Omega$ (i.e., the $\theta$ normalized by its steady-state value $\theta_{\rm SS}$), instead of the composite law's  switching according to the slip rate $V$, we could avoid violating the SS canon eq.~(\ref{eq:req4ss}), which the composite law violates around $V=V_{\rm c}$. This mechanism applies to any evolution law conforming eqs.~(\ref{eq:evolutionlaw_compatiblewithfunctionalansatz_substituted}) and (\ref{RSFeq:reqforpsic}), not limited to the particular examples of eqs.~(\ref{RSFeq:exmodcomplaw}) and (\ref{RSFeq:modcomplaw}).

\begin{figure*}
	\includegraphics[width=150mm]{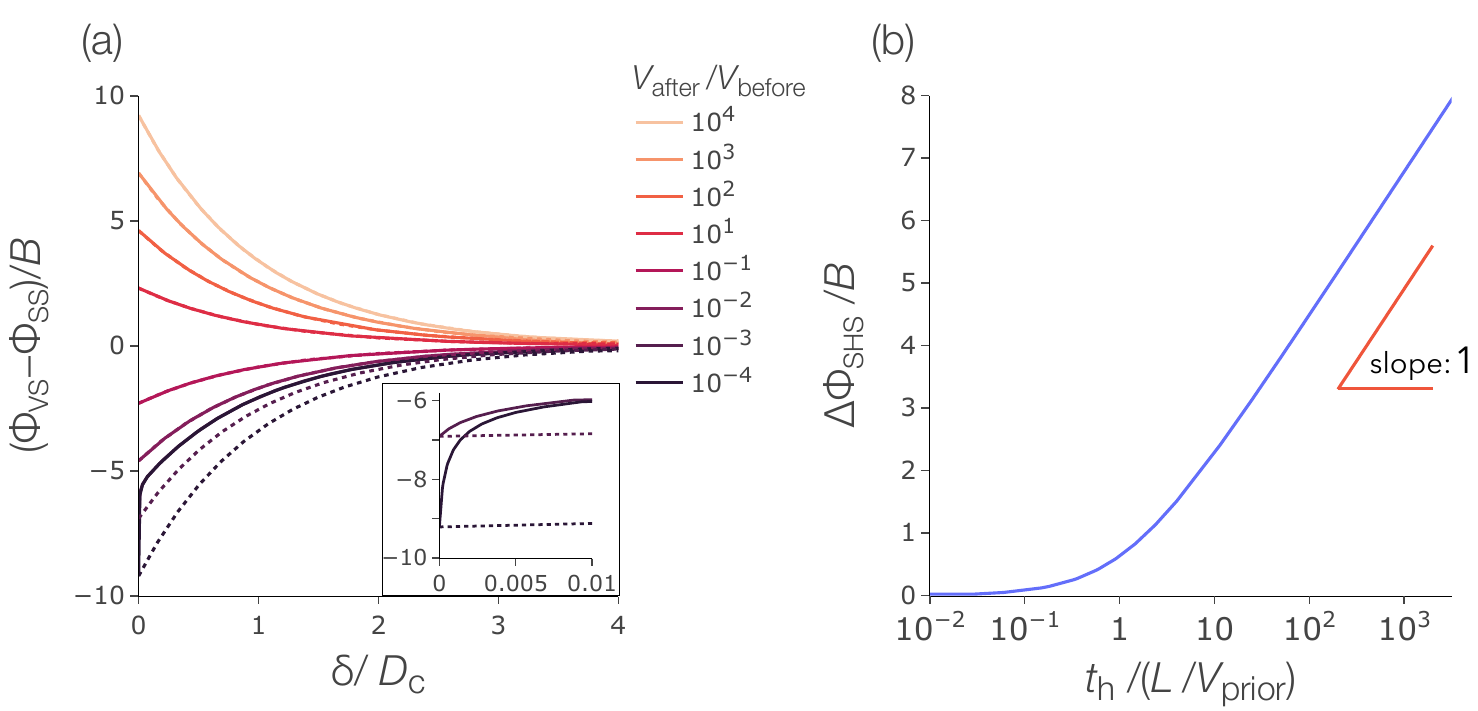}
\caption{
Strength evolution predicted by the modified composite law. 
$A=B$, $L=D_{\rm c}$, $\theta\gg\theta_{\rm X}$, and $\beta = 0.001$ are assumed. 
(a) Predicted strength evolution for ideal VS tests. 
The visualizing scheme follows that of Fig.~\ref{fig:compositeVS}. Differing from the composite law, the prediction (solid curves) depends only on the ratio $V_{\rm after}/V_{\rm before}$, not affected by their absolute values because the modified composite law does not have a characteristic speed. Dotted curves show the canonical behaviors (eq.~\ref{eq:realVScanonical}), which are not visible for cases with  $V_{\rm after}\geq 0.01V_{\rm before}$ as they almost coincide with the prediction by the modified composite law. 
The inset shows the blowup of the early stage of the evolution for $V_{\rm after}/V_{\rm before}=10^{-3}$, $10^{-4}$. 
(b) Predicted strength evolution for constant-$\tau$ SHS tests. 
Following a steady-state sliding at $V_{\rm prior}$, shear stress is abruptly reduced by $\Delta\tau = 9A$ and kept constant. The visualizing scheme follows that of Fig.~\ref{fig:AgingSlipSHS}. The red wedge indicates the slope of the $\log t_{\rm h}$ asymptote. 
}
  \label{fig:predictedmotion_modcomposite}
\end{figure*}

The rest of this subsection is devoted to inspecting the behaviors (Fig.~\ref{fig:predictedmotion_modcomposite}) of the modified composite law (eq.~\ref{RSFeq:modcomplaw}), which is simpler and easier to implement than eq.~(\ref{RSFeq:exmodcomplaw}) (as $\epsilon_\Omega^*$ specification is unnecessary). Note that, as shown above, the strength evolution of eq.~(\ref{RSFeq:exmodcomplaw}) is numerically indistinguishable from that of eq.~(\ref{RSFeq:modcomplaw}) as long as $\beta\ll1$. 

Figures~\ref{fig:predictedmotion_modcomposite}a and \ref{fig:predictedmotion_modcomposite}b show the predictions for VS tests and SHS tests (the former assumes an infinitely stiff apparatus; for the latter, hold periods are under the NSC). Both assumed $\beta=0.001$, $A=B$, and $L=D_{\rm c}$. Sufficiently small $\theta_{\rm X}$ was assumed so that $\theta$, $\theta_{\rm SS}\gg\theta_{\rm X}$ is kept throughout the simulations. As a result, the predicted behaviors shown in Fig.~\ref{fig:predictedmotion_modcomposite} do not include the possible effects of $\theta_{\rm X}$. We do not include the predicted velocity dependence of the steady-state strength in Fig.~\ref{fig:predictedmotion_modcomposite} because eq.~(\ref{RSFeq:modcomplaw}) has the $f=\psi(\Omega)$ form, which guarantees complete reproduction of eq.~(\ref{eq:req4ss}) (\S\ref{RSF322}).

Figure~\ref{fig:predictedmotion_modcomposite}a shows the transient evolution following velocity steps with different $V_{\rm after}/V_{\rm before}$. The $\theta$ value is initially $\theta_{\rm SS}(V_{\rm before})$ $(=D_{\rm c}/V_{\rm before})$ and evolves under a constant $V$ $(=V_{\rm after})$. 
As in \citet{ampuero2008earthquake}, we plotted the difference of evolving strength $\Phi$ from the eventual steady-state strength $\Phi_{\rm SS}(V_{\rm after})$, after normalization by $B$. 
When $V_{\rm after}/V_{\rm before}\gg\beta$ holds, we see that the VS canon (eq.~\ref{eq:realVScanonical}, dotted lines in Fig.~\ref{fig:predictedmotion_modcomposite}a) is met. 

By contrast, when $V_{\rm after}/V_{\rm before}\lesssim\beta$ ($V_{\rm after}/V_{\rm before} = 10^{-3}, 10^{-4}$), the predicted evolution deviates significantly from the VS canon (eq.~\ref{eq:realVScanonical}), especially in the early stage in the $\delta$-$\Phi$ plot. As shown in the blowup of the early stage of the two cases (Fig.~\ref{fig:predictedmotion_modcomposite}a, inset), the strengthening is rapid per unit slip. This behavior comes from the purely time-dependent healing term $f_t$, which is the dominant term when $\Omega<\beta$; when $V_{\rm after}$ is low, much time elapses while earning a small slip displacement so that the time-dependent healing appears rapid when plotted against $\delta$. Hence, this gross violation of the VS canon (eq.~\ref{eq:realVScanonical}) for $V_{\rm after}/V_{\rm before}\ll\beta$ is the inevitable side effect of the purely time-dependent $B\ln t$ healing at the NSC explained in \S\ref{sec:canonicalandside}. This mechanism is the same as that causing the similar behavior in the composite-law prediction of VS tests with $V_{\rm after} < V_{\rm c}$ \citep[also see Fig.~\ref{fig:compositeVS}]{kato2001composite} and as that causing the too quick strengthening in the aging-law prediction of VS tests with $V_{\rm after}/V_{\rm before}\ll1$ (Fig.~\ref{fig:AgingSlipVS}b). 

The above-mentioned aging-driven rapid strengthening in the earliest part of the transient evolution following a large negative step is a gross violation of the VS canon. On the other hand, Fig.~\ref{fig:predictedmotion_modcomposite}a indicates that the subsequent part of the predicted evolution traces the canonical VS behavior following a negative step from $V^\prime_{\rm before}$ such that $V^\prime_{\rm before}\sim V_{\rm after}/\beta$. 
In other words, excepting the very early stage where rapid, time-driven strengthening dominates the prediction, the modified composite law predicts a slip-driven exponential evolution with the fixed characteristic distance specified by the law’s parameter $D_{\rm c}$. 
This transition is because the modified composite law mutes $f_t$ according to $\Omega$; hence, the slip-law-ike behavior takes over in the later stage where $\Omega$ exceeds $\beta$. Contrastingly, the aging law does not have a mute mechanism of $f_t$, and the composite law mutes $f_t$ according to the absolute level of $V$. Therefore, these two laws do not predict slip-law-like behaviors even later in the transient evolution, and the steady state following a negative step is achieved in a much shorter distance than the parameter $D_{\rm c}$. In this regard, it might be arguable that the modified composite law is still better because it at least reproduces the independence of the characteristic evolution distance $D_{\rm c}$ from the sign and magnitude of the step~\citep{ruina1980friction,ruina1983slip}. In any case, we emphasize that this switch of the evolution from the log-time healing to the exponential slip relaxation, predicted by the modified composite law, is a direct logical consequence of the general character of our idea eq.~(\ref{RSFeq:reqforpsic}), or the generalized version, eq.~(\ref{eq:constraintonftgen}). 

Figure~\ref{fig:predictedmotion_modcomposite}b shows the prediction on constant-$\tau$ SHS tests, where, following the initialization by the steady-state sliding at $V_{\rm prior}$, shear stress is abruptly dropped by $\Delta\tau$ at the beginning of hold and is then held constant at $\tau_{\rm hold}$ during the hold (\S\ref{subsubsec:331}). The strength $\Phi$ during the hold is computed using the modified composite law eq.~(\ref{RSFeq:modcomplaw}). $V$ equals $V_{\rm prior} \exp(-\Delta\tau/A)$ immediately after the beginning of the hold $t_{\rm h} = +0$, and then gradually decreases according to the constitutive law eq.~(\ref{RSFeq:constitutivePhiB}) as the strength $\Phi$ increases. During the hold, from eq.~(\ref{RSFeq:constitutivetheta}) with $\theta(0)=D_{\rm c}/V_{\rm prior}$, we see 
\begin{equation}
  \Omega=\exp(-\Delta \tau/A)(\theta V_{\rm prior}/D_{\rm c})^{1-B/A}. 
  \label{eq:stressstepSHS}
\end{equation}
In Fig.~\ref{fig:predictedmotion_modcomposite}b, we set 
$V_{\rm prior}=V_*$, $\Delta \tau=9A$, and $A=B$.
Under this condition, $\Omega\ll\beta$ holds throughout $t_{\rm h} > 0$, and the modified composite law, as expected, reproduces the log-time healing eq.~(\ref{RSFeq:PhiSHSNSCrawtc}) (Fig.~\ref{fig:predictedmotion_modcomposite}b), as the aging law does. Also, the predicted cutoff time $t_{\rm c}$ coincides with $D_{\rm c}/V_{\rm prior}$, consistent with the confirmation by \citet{nakatani2006intrinsic} using the data of stationary-loadpoint SHS tests by \citet{marone1998effect}.

For interfaces with $A< B$ (Fig.~\ref{fig:SHScomparison}, d3), $\Omega$ decreases as $\theta$ increases, as implied by eq.~(\ref{eq:stressstepSHS}). Hence, if $\Omega\ll\beta$ holds at $t_{\rm h} = +0$, the modified composite law, which satisfies eq.~(\ref{RSFeq:reqforpsic}), predicts log-time healing throughout $t_{\rm h} >0$, again conforming to the SHS canon. However, for $A > B$  (Fig.~\ref{fig:SHScomparison}, d1), the modified composite law predicts deviation from the canon; although the canonical log-time healing is predicted in the early stage, strengthening will slow down at the time when $\Omega$ comes close to $\beta$ because $\Omega$ increases with $\theta$ on interfaces with $A > B$ (see eq.~\ref{eq:stressstepSHS}). This canon-violating behavior derives from eq.~(\ref{RSFeq:reqforpsic}), where $f$ drastically drops from $f \simeq f_{t|{\rm NSC}} = 1$ to $f \simeq f_\delta V = -(L/D_{\rm c})\Omega\ln\Omega$ (see Appendix \ref{sec:SHSHSprinciple} for details). Therefore, this is another side effect of the modified composite law, again a direct consequence of the aging-slip switch that has brought the aforementioned side effect in large negative VS tests. As far as $L\sim D_{\rm c}$ (eq.~\ref{RSFeq:LsimilartoDc}), this side effect on SHS tests is again a direct logical consequence of the general character of our idea (\ref{RSFeq:reqforpsic}) or, more generally, from eq.~(\ref{eq:constraintonftgen}) (see \S\ref{sec:SHSanalysismodifiedcomposite} for details).

At the end of this subsection, we emphasize that the assumed exponential form $\exp(-\Omega/\beta)$ for $\psi_{\rm cBase}(\Omega)$ is not a unique choice, just like \citet{kato2001composite} introduced the exponential function as an example in proposing the composite law. It can be replaced by any other function that can satisfy eq.~(\ref{eq:psicBaserequirement}), such as $1/[1+(\Omega/\beta)^n/\beta]$, with a sufficiently large positive number $n$. 
This function with a larger $n$ resembles $\exp(-\Omega/\beta)$ and converges to a step function when $n \to \infty$.
This non-uniqueness of the $\psi_{\rm cBase}(\Omega)$ function matters especially regarding the stringentness of the constraint eq.~(\ref{RSFeq:reqforpsic}) on $\psi_{\rm c}$. In the above derivations, we implicitly assumed that the ``$\gg$'' and ``$\ll$'' signs in the branch condition (i.e., $\Omega \gg \beta$ and $\Omega \ll \beta$) of eq.~(\ref{RSFeq:reqforpsic}) means the difference by the same order of magnitude as that meant by those signs in $|f_t| \gg |f_\delta V|$ (eq.~\ref{eq:NSCdef} converted into $\theta$-notation) and $|f_t| \ll |f_\delta V|$ (eq.~\ref{RSFeq:negligiblehealing}). However, there actually is no reason to believe so, and we cannot preclude the possibility that the correct (i.e., matches closely to the experimental results) $\psi_{\rm cBase}$ is a function that changes much more slowly with $\Omega$ than the $\exp(-\Omega/\beta)$ adopted by the example laws eqs.~(\ref{RSFeq:exmodcomplaw}) and (\ref{RSFeq:modcomplaw}). The $\psi_{\rm cBase}=1/(1+\Omega/\beta)$, drawn in beige together with $\exp(-\Omega/\beta)$ (orange) in Fig.~\ref{fig:mitorizu}, exemplifies such a mild muting function.

\subsection{Supplemental consideration on $L/D_{\rm c}$}
\label{sec:extensions_DcoverL}
In the derivation of the modified composite law from eqs.~(\ref{eq:evolutionlaw_compatiblewithfunctionalansatz_substituted}) and (\ref{RSFeq:reqforpsic}) in \S\ref{RSF33}, 
we searched for the desired form of $\psi_{\rm c}$ with $L\sim D_{\rm c}$ (eq.~\ref{RSFeq:LsimilartoDc}) premised. This premise facilitated the derivation, especially in obtaining eqs.~(\ref{eq:reqforpsicforO1Omega}) and (\ref{eq:smallorderofepsilon}) from eq.~(\ref{RSFeq:reqforpsic}). However, as mentioned in \S\ref{subsec:VSrequirement}, there is no a priori reason to think that $L\sim D_{\rm c}$. 
We here generalize the constraints eqs.~(\ref{eq:reqforpsicforO1Omega}) and (\ref{eq:smallorderofepsilon}) on the evolution law so that they apply also to $L\ll D_{\rm c}$ and $L\gg D_{\rm c}$, by repeating their derivations in \S\ref{RSF33} but without using $L\sim D_{\rm c}$ this time. For $|\Omega-1|\geq\epsilon_\Omega^*$, eq.~(\ref{eq:reqforpsicforO1Omega}) is replaced by
\begin{equation}
    \psi_{\rm cBase}(\Omega)=
    \begin{cases}
        1 & (\Omega\ll\beta)
        \\
        \mathcal O [(L/D_{\rm c})\beta\ln \beta] & (\beta\ll\Omega\leq 1/e)
        \\
        o[\min_{\pm}(L/D_{\rm c})(1\pm\epsilon_\Omega^*)\ln (1\pm\epsilon_\Omega^*)]& (\Omega\geq 1/e)       
    \end{cases}
    \label{eq:reqforpsicforO1Omega_generalized}
\end{equation}
For $|\Omega-1|<\epsilon_\Omega^*$, eq.~(\ref{eq:smallorderofepsilon}) is replaced by
\begin{equation}
    \psi_{\rm c}(\Omega)|_{\Omega=1+\epsilon_\Omega}=o[(L/D_{\rm c})\epsilon_\Omega].
    \label{eq:smallorderofepsilon_generalized}
\end{equation}
We can also have a generalized version of eq.~(\ref{eq:psicBaserequirement}) for any $L/D_{\rm c}$ value: 
\begin{equation}
    \psi_{\rm cBase}(\Omega)=
    \begin{cases}
        1 & (\Omega\ll\beta)
        \\
        \mathcal O [(L/D_{\rm c})\beta\ln \beta] & (\Omega \gg\beta)
    \end{cases}
    \label{eq:psicBaserequirement_arbitraryLoverDc}
\end{equation}
By multiplying eq.~(\ref{eq:psicBaserequirement_arbitraryLoverDc}) by the Heaviside function $H(1-\epsilon^*_\Omega-\Omega)$, we can compose $\psi_{\rm c}$, as we derived eq.~(\ref{RSFeq:modcompscutofffunction_withHeaviside}) in \S\ref{RSF33}. See Appendix~\ref{sec:epsironRange} for the allowable range of $\epsilon_\Omega^*$. Also, as in eq.~(\ref{RSFeq:NoHevisidePsiC_general}), we can drop the Heaviside truncation as long as we do not mind the slight deviation of $C_{\rm SS}$ from unity. 
One example to satisfy eq.~(\ref{eq:psicBaserequirement_arbitraryLoverDc}) is the rapidly decaying exponential $\psi_{\rm c}$ (eqs.~\ref{RSFeq:exmodcomplaw} and \ref{RSFeq:modcomplaw}) proposed for $L \sim D_{\rm c}$ in \S\ref{RSF33}, which satisfies eq.~(\ref{eq:psicBaserequirement_arbitraryLoverDc}) for quite a wide range of conditions of $\ln(L/D_{\rm c})=\mathcal O(1)$ (\S\ref{subsec:cutoffmodcomplawdetail}). 

Thus, the effect of $L/D_{\rm c}$ is minor in deriving evolution laws that meet the experimental requirements. However, it may be worth noting that the decay rate of $\psi_{\rm cBase}$ required by eq.~(\ref{eq:reqforpsicforO1Omega_generalized}) depends on $L/D_{\rm c}$. For $L\ll D_{\rm c}$, the constraint of eq.~(\ref{eq:reqforpsicforO1Omega_generalized}) is qualitatively the same as that for $L \sim D_{\rm c}$ although more rapid decay is required. 
For $L\lesssim D_{\rm c}$, $\psi_{\rm cBase}$ satisfying eq.~(\ref{eq:reqforpsicforO1Omega_generalized}) needs to mute the purely time-dependent healing to virtually zero in the range $\Omega\gg\beta$: 
\begin{equation}
    \psi_{\rm cBase}(\Omega)\ll \psi_{\rm cBase}(0) \hspace{10pt}(\Omega\gg\beta),
    \label{eq:decayingpuretimehealing}
\end{equation}
which is essentially the same constraint on $\psi_{\rm cBase}$ as in \S\ref{RSF33}. 

By contrast, the constraint of eq.~(\ref{eq:reqforpsicforO1Omega_generalized}) becomes very tolerant for $L\gg D_{\rm c}$. In addition to the rapidly decaying $\psi_{\rm cBase}$ (eq.~\ref{eq:decayingpuretimehealing}), which of course works fine in this case as well, we can even use $\psi_{\rm cBase}$ not muting the purely time-dependent healing at all even for $\Omega\gg\beta$ because, in this case, $f_\delta V$ is very ``loud'' for $\Omega\gg\beta$. 
Namely, for $L\gg D_{\rm c}$, $|f_t| \ll |f_\delta V|$ can be secured also by a $\psi_{\rm cBase}$ such that  
\begin{equation}
    \psi_{\rm cBase}(\Omega)\sim \psi_{\rm cBase}(0) \hspace{10pt}(\Omega\gg \beta).
    \label{eq:constantsmallpuretimehealing}
\end{equation}
Rigorously stated, for $\beta\,(\ll\sqrt{\beta}<1/e)$ such that $|\beta\ln\beta|\sim D_{\rm c}/L$, 
$\psi_{\rm cBase}$ conforming eq.~(\ref{eq:constantsmallpuretimehealing}) satisfies eq.~(\ref{eq:reqforpsicforO1Omega_generalized}), if we select $\epsilon_\Omega^*(<1-1/e)$ so that $\epsilon_\Omega^*\gg D_{\rm c}/L$ (Appendix~\ref{sec:betavalue4unitypsicbase}). 

The end member of functions conforming eq.~(\ref{eq:constantsmallpuretimehealing}) is the least decaying decreasing function: a constant function $\psi_{\rm cBase}=1$. Adopting it for $\psi_{\rm c}$ (eq.~\ref{RSFeq:NoHevisidePsiC_general}), we obtain 
\begin{equation}
\dot \theta= 1- \frac L{D_{\rm c}}\Omega\ln\Omega.
\label{eq:prevmodcomplaw_ft1}
\end{equation}
If the exact satisfaction of eq.~(\ref{RSFeq:reqforpsic}) is desired, it is enough to multiply $\psi_{\rm cBase}$ by the Heaviside function as done in eq.~(\ref{RSFeq:modcompscutofffunction_withHeaviside}). Appendix~\ref{sec:betavalue4unitypsicbase} shows the allowable range of $\epsilon_\Omega^*$ for this example. 
In such cases as eqs.~(\ref{eq:constantsmallpuretimehealing}) and (\ref{eq:prevmodcomplaw_ft1}), where the adopted $\psi_{\rm cBase}$ does not involve $\beta$ as an explicit parameter of the function, appropriate values of $\beta$ must be sought before choosing $\epsilon_\Omega^*$ accordingly. Appendix~\ref{sec:betavalue4unitypsicbase} also shows a general procedure to construct appropriate $\beta$ for a given $\psi_{\rm cBase}$, including application cases to eqs.~(\ref{eq:constantsmallpuretimehealing}) and (\ref{eq:prevmodcomplaw_ft1}). 
As discussed there, a certain range of values is considered appropriate as $\beta$, but the order of $\beta$ is fixed when we adopt a weakly decaying $\psi_{\rm cBase}$ (eq.~\ref{eq:constantsmallpuretimehealing}). 

By introducing another state variable $\theta^\prime:=(D_{\rm c}/L)\theta$, scaled so that $\theta^\prime_{\rm SS}=D_{\rm c}/V$ instead of $L/V$, eq.~(\ref{eq:prevmodcomplaw_ft1}) is converted to
\begin{equation}
\dot \theta^\prime= (D_{\rm c}/L) -\Omega^\prime\ln\Omega^\prime,
\label{eq:prevmodcomplaw}
\end{equation}
where $\Omega^\prime:=V\theta^\prime/D_{\rm c}$. Note $\dot \theta^\prime=(D_{\rm c}/L)\dot\theta$ and $\Omega^\prime=\Omega$. 
Equation~(\ref{eq:prevmodcomplaw}) can be read as the sum of the slip law and the purely time-dependent healing of the magnitude $D_{\rm c}/L$ (instead of unity).

When $L \sim D_{\rm c}$, eq.~(\ref{eq:decayingpuretimehealing}), which mutes $f_t$ to substantially zero in the non-NSC regime, has been deduced as the unique form of constraint on $f_t$ (\S\ref{RSF33}). By contrast, as shown above, when $L\gg D_{\rm c}$, eq.~(\ref{eq:constantsmallpuretimehealing}) ($f_t=const.$ except in the close vicinity of
steady state) is also acceptable as the constraint on $f_t$ alternative to eq.~(\ref{eq:decayingpuretimehealing}). 
Then, if $L\gg D_{\rm c}$, rather than $L \sim D_{\rm c}$ implicitly presumed in most studies, turns out to be the case, we have to determine, in order to write down evolution laws, which of eq.~(\ref{eq:decayingpuretimehealing}) or (\ref{eq:constantsmallpuretimehealing}) is the case. 
Two evolution laws resulting from eqs.~(\ref{eq:decayingpuretimehealing}) and (\ref{eq:constantsmallpuretimehealing}) predict significantly different behaviors, distinguishable in experiments. 
For example, if eq.~(\ref{eq:decayingpuretimehealing}) is adopted, SHS$|$NSC tests on interfaces with $A>B$ are expected to show the slowdown of log-t healing (Fig.~\ref{fig:SHScomparison}, d1) as we detailed for the modified composite law (\S\ref{RSF33}; also see \S\ref{sec:SHSanalysismodifiedcomposite}). In contrast, if eq.~(\ref{eq:constantsmallpuretimehealing}) is adopted, the resulting evolution laws do not predict such a slowdown even for interfaces with $A>B$. Thus, provided that the relation between $L$ and $D_{\rm c}$ has not been established, we are ready to write down evolution laws under whatever relation between $L$ and $D_{\rm c}$. 

Incidentally, in this \S\ref{sec:extensions_DcoverL}, development for any $L/D_{\rm c}$ has been under the unary ansatz of $\psi_{\rm c}$ (eq.~\ref{RSFeq:fOmega0}, or equivalently, \ref{RSFeq:fOmega}), and accordingly, we sought 
satisfaction of eq.~(\ref{RSFeq:reqforpsic}). However, all the arguments in \S\ref{sec:extensions_DcoverL} hold without this ansatz by replacing eq.~(\ref{RSFeq:reqforpsic}) with eq.~(\ref{eq:constraintonftgen}), which applies to any $f_t(V,\theta)$, not limited to of $\psi_{\rm c}(\Omega)$ form. 

\subsection{Incorporation of the strength minimum}
\label{sec:extensions}

So far, our derivation has only targeted the canonical behaviors in the range $\theta,\theta_{\rm SS}\gg\theta_{\rm X}$, where the canonical behaviors well constrain the state evolution law. In this subsection \S\ref{sec:extensions}, we extend the evolution laws so that they can also reproduce known features in the range $\theta\lesssim\theta_{\rm X}\cup\theta_{\rm SS}\lesssim\theta_{\rm X}$. The derivation in this subsection applies to any $L/D_{\rm c}$ value.

The validity of evolution laws becomes quite a complicated issue when strength minimum is incorporated, so we first clarify this point. The target of our analysis so far is the $\mathcal O(B)$ variation in strength $\Phi$ seen in canonical behaviors. Hence, the developed evolution laws are meant to describe $\mathcal O(B)$ variations in $\Phi$.
For $\theta\gg\theta_{\rm X}$, the evolution of $\theta$ is an exponential magnification of the variation in strength (eq.~\ref{RSFeq:conversionofphitothetanocut}), and we may say that the $\Phi/B$ variation, measured as the leading order of $\mathcal O(1)$, allows (indirect) determination of $\mathcal O(1)$ variation in $\ln\theta$. 
By contrast, for $\theta\ll\theta_{\rm X}$, a considerable change in $\ln\theta$ accompanies little [$o(B)$] change in strength, and hence $\theta$ is not well constrained even in the log scale (eq.~\ref{RSFeq:conversionofphitotheta}; also, see Fig.~\ref{fig:Xstate}). However, this only means that the variable $\theta$ excessively zooms small variations in strength when $\theta\ll\theta_{\rm X}$. In other words, such incapability to constrain $\theta$ little affecting the strength does not prevent the derivation of evolution laws reproducing the evolution of strength seen in canonical behaviors. In fact, we can avoid the above problem by simply defining a new state variable $\tilde \theta$, which includes the $\theta_{\rm X}$ as an offset:
\begin{equation}
    \tilde \theta:=\theta+\theta_{\rm X}.\label{eq:thetatilde}
\end{equation}
This state variable $\tilde \theta$ always has the same $\Phi$ dependency (eq.~\ref{RSFeq:conversionofphitotheta}) as that of $\theta$ for the range $\theta\gg\theta_{\rm X}$ (eq.~\ref{RSFeq:conversionofphitothetanocut}) mentioned above, and even for $\theta\ll\theta_{\rm X}$, the value of $\tilde \theta$ can be well constrained. Hence, we are able to derive meaningful evolution laws for $\tilde \theta$ from the canons. In fact, though counter-intuitive, the varying rate of $\theta$ (i.e., the evolution law in $\theta$ notation) is as well constrained as that of $\tilde \theta$ through $\dot \theta=\dot{\tilde \theta}$, despite the weak constraint on the value of $\theta$ for $\theta\ll\theta_{\rm X}$. 


Now, we constrain the evolution law according to the canonical behaviors, this time considering the behaviors in the range $\theta\lesssim\theta_{\rm X}\cup \theta_{\rm SS}\lesssim\theta_{\rm X}$ as well. We begin with the SS canon, as we did so in the development for $\theta,\theta_{\rm SS}\gg\theta_{\rm X}$. 
For the entire range of $V$, including $V\gg L/\theta_{\rm X}$, $V\sim L/\theta_{\rm X}$, and $V\ll L/\theta_{\rm X}$, \citet{weeks1993constitutive} has shown that their experimental data of the steady-state shear stress $\tau_{\rm SS}$ are well fitted by eq.~(\ref{eq:N1}), namely the direct-effect term plus the steady-state strength, if the steady-state strength has the following velocity dependence involving a high-speed cutoff:
\begin{flalign}
    \Phi_{\rm SS}(V)&=\Phi_{\rm X}+B\ln \left(\frac{L}{V\theta_{\rm X}}+1\right)\label{eq:consequentialstrengthwithXstate}
    \\
    &\simeq
    \left\{
    \begin{array}{cc}
    \Phi_{\rm X}-B\ln(V\theta_{\rm X}/L)     & (V\ll L/\theta_{\rm X})\\
    \Phi_{\rm X}    & (V\gg L/\theta_{\rm X})
    \end{array}
    \right.
\end{flalign}
This satisfies both the experimental requirement (eq.~\ref{eq:req4ss}) for the SS test when $\theta_{\rm SS}\gg \theta_{\rm X}$ and
the strength minimum (eq.~\ref{RSFeq:xstatePhi}). 
Through eq.~(\ref{RSFeq:conversionofthetatophi}), eq.~(\ref{eq:consequentialstrengthwithXstate}) is equivalent to $\Omega_{\rm SS}=1$ (eq.~\ref{RSFeq:steadystateconstraint}). 
Therefore, considering the above result by \citet{weeks1993constitutive}, we regard eq.~(\ref{RSFeq:steadystateconstraint}) as the constraint from the SS canon for the entire range of $\theta$ and $\theta_{\rm SS}$.

The VS constraint (eq.~\ref{RSFeq:dotthetaVS}) holds for any $\Phi$ and $\Phi_{\rm SS}$. Thus, no modification is necessary to incorporate the strength minimum. 
This eq.~(\ref{RSFeq:dotthetaVS}) is the only constraint on $f_{\delta}$, except that it must be negligible at the NSC, which is automatically met as long as eq.~(\ref{RSFeq:dotthetaVS}) on $f_{\delta}$ and eq.~(\ref{eq:ftNSC_constrained}) on $f_{t|{\rm NSC}}$ hold. 
Hence, we can adopt the slip-dependent evolution function $f_{\delta{\rm |VS}}$ of eq.~(\ref{RSFeq:dotthetaVS}), which involves the effect of $\theta_{\rm X}$,  as the evolution law $f_{\delta}$ satisfying the VS canon (eq.~\ref{eq:relaxeddifferentialVSrequirement}) for the entire range of $\theta$ and $\theta_{\rm SS}$. 

For convenience in subsequent derivations, we introduce the following variable:
\begin{equation}
        \tilde\Omega  :=\tilde\theta V/L.\label{eq:Omegatilde}
\end{equation}
Under $\Omega_{\rm SS}=1$ (eq.~\ref{RSFeq:steadystateconstraint}), the steady-state value of $\tilde \Omega$, $\tilde \Omega_{\rm SS}$, can be written as 
\begin{equation}
    \tilde\Omega_{\rm SS}=1+V\theta_{\rm X}/L.
    \label{eq:tildeOmegassspecified}
\end{equation}
Using eqs.~(\ref{eq:Omegatilde}) and (\ref{eq:tildeOmegassspecified}), $f_\delta V$ equated to $f_{\delta{\rm |VS}}V$ of eq.~(\ref{RSFeq:dotthetaVS}) is expressed as
\begin{equation}
    f_\delta V=-(L/D_{\rm c})\tilde\Omega\ln (\tilde\Omega/\tilde \Omega_{\rm SS}).
    \label{eq:dotthetaVS_rewritten}
\end{equation}

Also, purely time-dependent healing under the NSC has already been constrained by eq.~(\ref{eq:ftNSC_constrained}), which already considers the effect of $\theta_{\rm X}$ and is therefore valid for any $\theta$ and $\theta_{\rm SS}$:
\begin{equation}
    f_{t|{\rm NSC}}= 1. \tag{\ref{eq:ftNSC_constrained} again}
\end{equation}

Thus, all that is missing is $f_t$ under the non-NSC. 
Similar to our derivation of eq.~(\ref{RSFeq:reqforpsic}) for $\theta,\theta_{\rm SS}\gg\theta_{\rm X}$, the asymptotic form of $f_t$ under the non-NSC is constrained by eq.~(\ref{RSFeq:negligiblehealing}) of VS tests. 
With eq.~(\ref{eq:dotthetaVS_rewritten}), eq.~(\ref{RSFeq:negligiblehealing}) can be rewritten into
\begin{equation}
    f_{t|{\rm VS}}=o[(L/D_{\rm c})\tilde\Omega\ln (\tilde\Omega/\tilde \Omega_{\rm SS})].
    \label{RSFeq:negligiblehealing_4fdeltagen}
\end{equation}
When we derive eq.~(\ref{eq:dotthetaVS_rewritten}) from eq.~(\ref{RSFeq:dotthetaVS}), we have removed the conditional subscript $|$VS on the left-hand side of eq.~(\ref{eq:dotthetaVS_rewritten}) to indicate that eq.~(\ref{eq:dotthetaVS_rewritten}) represents the $f_\delta$ part of the evolution law being developed rather than the description of constraints in a specific type of experiment. Unlike eq.~(\ref{eq:dotthetaVS_rewritten}), eq.~(\ref{RSFeq:negligiblehealing_4fdeltagen}) is still an experimental constraint on the evolution law, so we keep $|$VS on the left-hand side of eq.~(\ref{RSFeq:negligiblehealing_4fdeltagen}). 
This is for the pragmatic reason of distinguishing it from another constraint but for the NSC (eq.~\ref{eq:ftNSC_constrained}).

Thus, regarding the constraint on $f_t$, the NSC ($|f_t|\gg |f_\delta V|$) gives the range where the SHS canon applies while eq.~(\ref{RSFeq:negligiblehealing}) ($|f_t|\ll |f_\delta V|$) gives the range where the VS canon applies. 
On the other hand, regarding $f_\delta$, since the definition of the NSC ($|f_t|\gg |f_\delta V|$) means that $f_\delta$ is negligible, the explicit form of $f_\delta$ under the NSC is poorly constrained [from $\mathcal O(B)$ variations in $\Phi$]. However, this conversely means that we can assume eq.~(\ref{eq:dotthetaVS_rewritten}) for the entire range, including the NSC, without causing any relevant differences in the resultant behavior of $\Phi$. For this reason, in the present derivation, we can apply eq.~(\ref{eq:dotthetaVS_rewritten}) for the entire range. 
Similarly, noting that $f_t$ does not affect the evolution as long as $|f_t|\ll |f_\delta V|$ (eq.~\ref{RSFeq:negligiblehealing}) holds, we can modify the constraint of eq.~(\ref{RSFeq:negligiblehealing_4fdeltagen}) to the following stricter one: 
\begin{equation}
    f_{t|{\rm VS}}=o[(L/D_{\rm c})(\tilde\Omega/\tilde \Omega_{\rm SS})\ln (\tilde\Omega/\tilde \Omega_{\rm SS})],
    \label{RSFeq:negligiblehealing_4fdeltagen_sufficient}
\end{equation}
where $f_t$ is more strongly muted than in eq.~(\ref{RSFeq:negligiblehealing_4fdeltagen}) because $\tilde \Omega_{\rm SS} \geq \Omega_{\rm SS}=1$ as seen from eqs.~(\ref{RSFeq:steadystateconstraint}) and (\ref{eq:tildeOmegassspecified}). 
Equation~(\ref{RSFeq:negligiblehealing_4fdeltagen_sufficient}) is a sufficient condition of eq.~(\ref{RSFeq:negligiblehealing_4fdeltagen}), and the use of eq.~(\ref{RSFeq:negligiblehealing_4fdeltagen_sufficient}) facilitates subsequent derivation.

Using eq.~(\ref{RSFeq:negligiblehealing_4fdeltagen_sufficient}), we can derive the evolution law by a logic parallel to that in \S\ref{RSF322}. 
Specifically, by noting that $\tilde\Omega/\tilde \Omega_{\rm SS}$ connects to $\Omega$ when $\theta,\theta_{\rm SS}\gg \theta_{\rm X}$ (eqs.~\ref{RSFeq:defofOmega_steadystate} and \ref{eq:thetatilde}--\ref{eq:tildeOmegassspecified}), we can regard eq.~(\ref{RSFeq:negligiblehealing_4fdeltagen_sufficient}) as a generalized form of $f_{t|{\rm VS}}=o[(L/D_{\rm c})\Omega\ln \Omega]$ derived from eqs.~(\ref{RSFeq:negligiblehealing}) and (\ref{eq:simplifiedslipdepofthetadot}), where $\tilde\Omega/\tilde \Omega_{\rm SS}$ replaces $\Omega$ (i.e., $\Omega/\Omega_{\rm SS}$ given eq.~\ref{RSFeq:steadystateconstraint}).
Therefore, $f_t$ satisfying eqs.~(\ref{eq:ftNSC_constrained}) and (\ref{RSFeq:negligiblehealing_4fdeltagen_sufficient}) can be derived by repeating the same procedure as in \S\ref{RSF322}, with $\Omega$ replaced by $\tilde\Omega/\tilde \Omega_{\rm SS}$.

Since the right-hand side of eq.~(\ref{RSFeq:negligiblehealing_4fdeltagen_sufficient}) is a unary function of $\tilde\Omega/\tilde \Omega_{\rm SS}$, eq.~(\ref{RSFeq:negligiblehealing_4fdeltagen_sufficient}) is equivalent to that $f_t$ equals the right-hand side of eq.~(\ref{RSFeq:negligiblehealing_4fdeltagen_sufficient}) in the range of $\tilde\Omega/\tilde \Omega_{\rm SS}$ that can occur in VS tests. 
In ideal VS tests, where the interface slip rate instantly changes from $V_{\rm before}$ to $V_{\rm after}$, the farthest deviation of the strength from the steady state occurs immediately after the step. Hence, we see that $|\ln (\tilde \Omega/\tilde \Omega_{\rm SS})|\leq |\ln [(L/V_{\rm before}+\theta_{\rm X})/(L/V_{\rm after}+\theta_{\rm X})|$ (eqs.~\ref{eq:thetatilde}, \ref{eq:Omegatilde}, and \ref{eq:tildeOmegassspecified}). 
This inequality obtained for ideal VS tests is approximately valid for realistic (non-ideal) VS tests. 
Thus, eq.~(\ref{RSFeq:negligiblehealing_4fdeltagen_sufficient}) can be rewritten into
\begin{equation}
    f_t=o\left[\frac{L}{D_{\rm c}}\frac{\tilde\Omega}{\tilde \Omega_{\rm SS}}\ln\frac{\tilde\Omega}{\tilde \Omega_{\rm SS}}\right] 
    \hspace{10pt} \left(\left(\ln \frac{ \min(V_{\rm after},V_{\rm X})}{\min(V_{\rm before},V_{\rm X})}\right)\left/\left(\ln\frac{\tilde \Omega}{\tilde \Omega_{\rm SS}}\right)\gtrsim\right.1\right),
    \label{eq:sufficientft_gen}
\end{equation}
where we have introduced $V_{\rm X}:=L/\theta_{\rm X}$. We used $(1/x+1/y)\sim 1/\min(x,y)$ for $x,y>0$. 
Note eq.~(\ref{eq:sufficientft_gen}) constrains $f_t$ for above ($\tilde \Omega>\tilde \Omega_{\rm SS}$) and below ($\tilde \Omega<\tilde \Omega_{\rm SS}$) the steady state in the transient following positive ($V_{\rm after}>V_{\rm before}$) and negative ($V_{\rm after}<V_{\rm before}$) steps, respectively.

Lastly, we look for $f_t$ satisfying both eq.~(\ref{eq:ftNSC_constrained}) (constraint for the NSC) and eq.~(\ref{eq:sufficientft_gen}) (constraint for $|f_t|\ll |f_\delta V|$). 
Since the right-hand side of eq.~(\ref{eq:sufficientft_gen}) is a unary function of $\tilde \Omega/ \tilde \Omega_{\rm SS}$, we can construct the desired $f_t$ by using unary functions of $\tilde \Omega/ \tilde \Omega_{\rm SS}$, denoted by $\tilde \psi_{\rm c}$:
\begin{equation}
    f_t= \tilde \psi_{\rm c} (\tilde \Omega/ \tilde \Omega_{\rm SS}).
    \label{eq:psicprime}
\end{equation}
In the derivation below, we additionally assume $\tilde\psi_{\rm c}$ to be a non-negative, monotonically decreasing function. Then, the monotonic increase of $x\ln x$ for $x\geq 1$, taken together with the monotonic decrease of $\tilde\psi_{\rm c}$, guarantees that eq.~(\ref{eq:sufficientft_gen}) holds during the transient evolution following positive steps, where $\tilde \Omega\geq \tilde \Omega_{\rm SS}$, as long as eq.~(\ref{eq:sufficientft_gen}) holds at steady states, where $\tilde \Omega=\tilde \Omega_{\rm SS}$. Hence, we only have to request $f_t$ of eq.~(\ref{eq:psicprime}) to conform eq.~(\ref{eq:sufficientft_gen}) during the transient evolution following negative steps:
\begin{equation}
    f_t=o\left[\frac{L}{D_{\rm c}}\frac{\tilde\Omega}{\tilde \Omega_{\rm SS}}\ln\frac{\tilde\Omega}{\tilde \Omega_{\rm SS}}\right] 
    \hspace{10pt} \left(\frac{\tilde \Omega}{\tilde \Omega_{\rm SS}}\gtrsim \frac{ \min(V_{\rm after},V_{\rm X})}{\min(V_{\rm before},V_{\rm X})}\right).
    \label{eq:sufficientft_gen_ftdecreasing}
\end{equation}
In summary, we can construct $f_t$ that works fine under both NSC and non-NSC by imposing  eqs.~(\ref{eq:ftNSC_constrained}), (\ref{eq:psicprime}), and (\ref{eq:sufficientft_gen_ftdecreasing}). We see 
that the use of the following $f_t=\tilde \psi_{\rm c}$ is sufficient to satisfy these equations:
\begin{equation}
    \tilde\psi_{\rm c} (\tilde\Omega/\tilde \Omega_{\rm SS})= 
    \begin{cases}
        1 & (\tilde \Omega/\tilde\Omega_{\rm SS}\ll \tilde\beta)
        \\
        o[(L/D_{\rm c})(\tilde\Omega/\tilde \Omega_{\rm SS})\ln (\tilde\Omega/\tilde \Omega_{\rm SS})] & (\tilde \Omega/\tilde\Omega_{\rm SS}\gg \tilde\beta)
    \end{cases}
    \label{eq:psicprimeconstraint}
\end{equation}
where $\tilde \beta$ is a small positive constant. 

Now, we consider constructing a full evolution law as $f=f_t+f_\delta V$, using $f_t$ of eqs.~(\ref{eq:psicprime}) and (\ref{eq:psicprimeconstraint}) and $f_\delta V$ of eq.~(\ref{eq:dotthetaVS_rewritten}). Then, $f$ becomes a function of two variables $V$ and $\theta$. In the latter half of \S\ref{RSF322}, we have shown that $f_t$ in such forms must satisfy eq.~(\ref{eq:constraintonftgen}) for $\theta,\theta_{\rm SS}\gg\theta_{\rm X}$. Therefore, for $\tilde\psi_{\rm c}$ conforming eq.~(\ref{eq:psicprimeconstraint}) to be usable in this construction, $f_t$ conforming eqs.~(\ref{eq:psicprime}) and (\ref{eq:psicprimeconstraint}) must also satisfy eq.~(\ref{eq:constraintonftgen}) for $\theta,\theta_{\rm SS}\gg\theta_{\rm X}$. Recalling $\tilde \Omega/\tilde \Omega_{\rm SS}\simeq \Omega$ for $\theta,\theta_{\rm SS}\gg\theta_{\rm X}$, we see that such is possible only if 
\begin{equation}
    \tilde\beta=\beta.
    \label{eq:tildebetaisbeta}
\end{equation}
Under eq.~(\ref{eq:tildebetaisbeta}), eq.~(\ref{eq:psicprimeconstraint}) coincides with eq.~(\ref{RSFeq:reqforpsic}), with $\Omega$ replaced by $\tilde \Omega/\tilde \Omega_{\rm SS}$. Hence, by adopting the same function
as $\psi_{\rm c}$ that satisfies eq.~(\ref{RSFeq:reqforpsic}), 
\begin{equation}
    \tilde\psi_{\rm c}(x)=\psi_{\rm c} (x),
    \label{eq:psiprimeispsi}
\end{equation}
$\tilde \psi_{\rm c}$ satisfies eq.~(\ref{eq:psicprimeconstraint}). Examples of such $\psi_{\rm c}$ include those obtained in \S\ref{RSF33} and \S\ref{sec:extensions_DcoverL}, such as, for $L \sim D_{\rm c}$, eq.~(\ref{RSFeq:modcompscutofffunction_withHeaviside}) and eq.~(\ref{RSFeq:NoHevisidePsiC_general}), though the latter causes a slight deviation of $C_{\rm SS}$. 
Note that we here consider only decreasing functions for $\psi_{\rm c}$ so that eq.~(\ref{eq:sufficientft_gen}) can be replaced by eq.~(\ref{eq:sufficientft_gen_ftdecreasing}). 

Thus, we have shown that evolution laws incorporating the strength minimum can be constructed as $f=f_t+f_\delta V$, using the $f_\delta$ of eq.~(\ref{eq:dotthetaVS_rewritten}) and the $f_t= \psi_{\rm c} (\tilde \Omega/ \tilde \Omega_{\rm SS})$ with $\psi_{\rm c}$ of eq.~(\ref{RSFeq:reqforpsic}), which satisfies eq.~(\ref{eq:psicprimeconstraint}), a sufficient condition of eqs.~(\ref{eq:ftNSC_constrained}) and (\ref{RSFeq:negligiblehealing_4fdeltagen}): 
\begin{equation}
    \dot\theta(=\dot{\tilde\theta})=\psi_{\rm c}(\tilde\Omega/\tilde\Omega_{\rm SS})-\frac{L}{D_{\rm c}}\tilde\Omega\ln(\tilde\Omega/\tilde\Omega_{\rm SS}).
    \label{eq:evolutionlaw_compatiblewithfunctionalansatz_substituted2}
\end{equation}
Because $V$ appears alone in eq.~(\ref{eq:evolutionlaw_compatiblewithfunctionalansatz_substituted2}) through the $V\theta_{\rm X}/L$ term involved in the variable $\tilde\Omega/\tilde\Omega_{\rm SS}$, this evolution law is no longer a unary function of $\Omega$ (or equivalently, $V\theta$). Nevertheless, $\theta_{\rm SS}=L/V$ (eq.~\ref{eq:steadystateform_theta}; 
i.e., $\Omega_{\rm SS}=1$, eq.~\ref{RSFeq:steadystateconstraint}) is strictly kept through eq.~(\ref{eq:tildeOmegassspecified}). 

Notice that what fixes the $\tilde \Omega_{\rm SS}$ value is eq.~(\ref{eq:tildeOmegassspecified}), not eq.~(\ref{eq:evolutionlaw_compatiblewithfunctionalansatz_substituted2}). 
The evolution law for $\theta$ is given as the combination of eqs.~(\ref{eq:tildeOmegassspecified}) and (\ref{eq:evolutionlaw_compatiblewithfunctionalansatz_substituted2}). 
In other words, transient evolution and value assignment to the steady state (as a function of $V$) are separable. This distinct characteristic is a direct consequence of part of the VS constraints, where $f_\delta $ (eq.~\ref{RSFeq:dotthetaVS}) drives the system toward the steady state while $f_t$ disappears at the steady state, that is, 
$
    f_{t}=o[(L/D_{\rm c})\tilde\Omega\ln (\tilde\Omega/\tilde \Omega_{\rm SS})]
    $ for $
    (\tilde\Omega\sim \tilde \Omega_{\rm SS}),
$
which is a special case of eq.~(\ref{RSFeq:negligiblehealing_4fdeltagen}) in the vicinity of steady states.

We now check if the evolution law eq.~(\ref{eq:evolutionlaw_compatiblewithfunctionalansatz_substituted2}) has a unique stationary point, one of the desired properties as discussed in \S\ref{RSF322}. Because $\psi_{\rm c}$ is non-negative, the stationary points of eq.~(\ref{eq:evolutionlaw_compatiblewithfunctionalansatz_substituted2}) can exist only in $\tilde \Omega/\tilde\Omega_{\rm SS}\geq 1$. Further recalling that $f_t$-part of eq.~(\ref{eq:evolutionlaw_compatiblewithfunctionalansatz_substituted2}), $\psi_{\rm c}(\tilde \Omega/\tilde\Omega_{\rm SS})$, has been assumed to be monotonically decreasing, and also noticing that its $f_\delta V$-part, $-(\tilde \Omega/\tilde\Omega_{\rm SS})\ln (\tilde \Omega/\tilde\Omega_{\rm SS})$, is also monotonically decreasing in the range $\tilde \Omega/\tilde\Omega_{\rm SS}\geq 1$, we see that the trivial stationary point $\tilde \Omega/\tilde \Omega_{\rm SS}=1$, that is, $\Omega=\Omega_{\rm SS}$, is the only stationary point of eq.~(\ref{eq:evolutionlaw_compatiblewithfunctionalansatz_substituted2}).

At the end of this subsection, we visually compare eq.~(\ref{eq:evolutionlaw_compatiblewithfunctionalansatz_substituted2}) with eq.~(\ref{eq:evolutionlaw_compatiblewithfunctionalansatz_substituted}) derived in \S\ref{RSF33}. Both are the sum of the slip-driven state evolution $f_\delta V$ and the purely time-dependent healing $f_t$ that disappears when strength comes to the close vicinity of steady states. We notice that the replacement of $\Omega$ in eq.~(\ref{eq:evolutionlaw_compatiblewithfunctionalansatz_substituted}) with $\tilde \Omega$ and another replacement of its steady-state value $C_{\rm SS}=1$ (eq.~\ref{RSFeq:Cssnormalization}) in $\ln\Omega(=\ln(\Omega/C_{\rm SS}))$ with $\tilde \Omega_{\rm SS}$ (eq.~\ref{eq:tildeOmegassspecified}) yields eq.~(\ref{eq:evolutionlaw_compatiblewithfunctionalansatz_substituted2}). The former replacement corresponds to $\theta\to\tilde \theta$, and the latter corresponds to 
$L/V\to L/V+\theta_{\rm X}$. As expected from such relationships between them, it is easy to confirm that eq.~(\ref{eq:evolutionlaw_compatiblewithfunctionalansatz_substituted2}) reproduces eq.~(\ref{eq:evolutionlaw_compatiblewithfunctionalansatz_substituted}) for $\theta,\theta_{\rm SS}\gg\theta_{\rm X}$, as it ought to. Thus, the constructed eq.~(\ref{eq:evolutionlaw_compatiblewithfunctionalansatz_substituted2}) (plus eq.~\ref{eq:tildeOmegassspecified}) is the modification of eq.~(\ref{eq:evolutionlaw_compatiblewithfunctionalansatz_substituted}) so that experimental requirements for $\theta\lesssim\theta_{\rm X}\cup\theta_{\rm SS}\lesssim\theta_{\rm X}$ are satisfied as well. In conclusion, the evolution law consisting of eqs.~(\ref{RSFeq:reqforpsic}),  (\ref{eq:tildeOmegassspecified}), and (\ref{eq:evolutionlaw_compatiblewithfunctionalansatz_substituted2}) satisfies all the experimental requirements for the entire range of $\theta$ and $\theta_{\rm SS}$, including $\theta,\theta_{\rm SS}\gg\theta_{\rm X}$, $\theta\lesssim\theta_{\rm X}$, and $\theta_{\rm SS}\lesssim\theta_{\rm X}$, except the unavoidable side effects of aging-slip switch predicted outside the conditions experimentally probed so far.

\section{Discussion}
\label{sec:discussion}
As shown in \S\ref{sec:3}, the slip law is the very requirement from VS and SS tests, while the aging law is one of the natural functional forms satisfying the requirements from SHS and SS tests. However, neither law can simultaneously satisfy requirements from VS, SHS (SHS$|$NSC, where behaviors contradict the VS canon), and SS tests. In fact, no differential equation can simultaneously reproduce the strengthening rate observed in VS tests and that observed in SHS$|$NSC tests (\S\ref{sec:canonicalandside}). This simple mathematical fact is, we would say, the fundamental reason why we have been struggling for so long to reconcile the slip and aging laws~\citep[e.g.,][]{kato2001composite}. This issue has critical geophysical implications because the two laws predict quite different behaviors for many aspects of earthquake cycles that may entail both slip weakening during fault rupture and time-dependent healing of locked zones. Examples include the recurrence interval~\citep{kato1999model}, the earthquake nucleation~\citep{ampuero2008earthquake}, 
and the aseismic interaction between megathrust earthquakes and repeaters~\citep{ariyoshi2014trial}. 

The present work has theoretically shown that simultaneous reproduction of the VS and SHS canons necessarily requires switching between the aging-law-like evolution and the slip-law-like evolution, while evolution laws involving such a switch are to predict behaviors violating the VS canon and the SHS canon under the conditions where the switch happens. Accepting such non-desired behaviors predicted as inevitable side effects, we carried on our development because behaviors violating the VS and SHS canons near the
Slip/Aging switch are expected to occur in the $(V,\theta)$ range not well probed experimentally so far and hence can be hidden there. However, given this acceptance of the yet-to-be-observed side effects, we still must be aware that only certain forms of aging-slip switches are consistent with the existing experimental results. For example, the aging-slip switch introduced in the composite law~\citep{kato2001composite} violates the SS canon in the range already confirmed by experiments, not only having the side effects (\S\ref{sec:canonicalandside}) violating the VS and SHS canons at the transient regime. 

With the above understandings in mind, we have decided to go back to the starting point of the RSF law, the compilation of experimental facts; the present paper has deduced evolution laws by recompiling the three canons, avoiding the heuristic approach usually taken in earlier studies. During the present development along this general, deductive approach, we have realized that simultaneous, complete reproduction of VS and SHS canons is impossible in principle. This understanding has become a big turning point in the present development; we have discarded the original goal of a complete reproduction of the three canons, which has now turned out to be a mathematical impossibility, and have instead set a new downgraded goal of reproducing the three canons throughout the $(V, \theta)$ range in which experiments have so far confirmed the canons well. In that we have narrowed the range of phenomena to be reproduced, our downgrading of the goal resembles the case of the composite law~\citep{kato2001composite}, who have given up the SS canon. However, thanks to the present theoretical identification of the reason for the impossibility, we have additionally realized that the composite law's violation of the SS canon is not an inevitable consequence of the three canons. As a result, we could derive, in a reasonably general form, a class of evolution laws that reproduce all the three canons, except the inevitable side effects of the aging-slip switch, namely the deviation from the canonical behaviors predicted for the strengthening phases of VS and SHS$|$non-NSC tests. 
We managed to make those canon-violating side effects appear only in the $(V,\theta)$ range beyond the range where the existing experiments have well confirmed the three canons. 
Furthermore, with reasonable generality, we could deduce that the switch of the strengthening mechanisms must be controlled by the value of $\theta/\theta_{\rm SS}(=\Omega)$. Thus, the presently achieved elucidation of the inevitable side effects consequential of the three canons is, we believe, a fundamental answer to the long-standing problem of reconciling the aging law and the slip law.

In this \S\ref{sec:discussion}, we work on several issues arising from the above-summarized insights obtained by \S\ref{sec:development}. The first issue is whether the control factor of the canons-required aging-slip switch must be $\Omega$. In \S\ref{subsec:discussion1}, we further generalize the derivations of \S\ref{sec:generalizingheuristicderivation} to find that the $\Omega$-dependent switch is nearly an inevitable consequence of the three canons. The second issue is the physical meaning of the $\Omega$-dependent switch of the evolution mechanism. In \S\ref{subsec:macrophysicsOmegadepswitch}, we discuss this point from a macrophysical viewpoint. The third issue is experimental validation. Although the presently proposed class of evolution laws involving the $\Omega$-dependent switch, deduced from the compilation of the three canons without heuristics, is nearly the unique solution, this does not guarantee the predicted behaviors, especially the side effects of the switching, are the case. 
In \S\ref{subsec:possiblebetadetection}, we will validate the modified composite law (eq.~\ref{RSFeq:modcomplaw}) using an existing experiment that seems to include the $(V, \theta)$ range where the predicted side effects of the aging-slip switch begin to appear.

\subsection{Expression of evolution laws consistent with the three canons derived without presuming the thermodynamic formulation}
\label{subsec:discussion1}

If a class of evolution laws is derived without assumptions other than the canons, agreement of their predictions with experiments means that it is the unique correct answer (i.e., the correct physical law), at least within the conditions under which the canons hold. In contrast, if assumptions other than the canons are used in derivation, their agreement with experiments does not guarantee it is the unique correct answer; other laws outside the scope limited by the (a priori) assumptions may also reproduce all the canons. For this reason, the present work aims to derive evolution laws with minimal assumptions to secure as much generality as possible. Below we reiterate, from the above viewpoint of generality, our derivations in \S\ref{RSF322}, and then, for further generality, we will redo the derivation without the assumption of the TDF (eq.~\ref{RSFeq:thermodynamicformPhi}, $d\Phi=...dt+...d\delta$), which was adopted both in \S\ref{sec:heuristicderivation} and \S\ref{sec:generalizingheuristicderivation}.

Under the recognition of the necessity of the aging-slip switch (\S\ref{sec:canonicalandside}), we have developed evolution laws in \S\ref{RSF322}. Our first proposal was eq.~(\ref{RSFeq:reqforpsic}) in \S\ref{sec:heuristicderivation}. It was derived under the ansatz that $f$ is a unary function of $\Omega$, which is a heuristic assumption. Then, in \S\ref{sec:generalizingheuristicderivation}, we redid the derivation without the unary ansatz and obtained eq.~(\ref{eq:constraintonftgen}), a generalized version of eq.~(\ref{RSFeq:reqforpsic}). Both eqs.~(\ref{RSFeq:reqforpsic}) and (\ref{eq:constraintonftgen}), combined with eq.~(\ref{eq:simplifiedslipdepofthetadot}), reproduce all three canons throughout the ($V, \theta$) range well probed experimentally so far. Although eq.~(\ref{eq:constraintonftgen}) was derived without the unary ansatz, the right-hand side of eqs.~(\ref{RSFeq:reqforpsic}) and (\ref{eq:constraintonftgen}) are the same, expressing that the aging-slip switch occurs according to the value of $\Omega$. Therefore, we see an $\Omega$-dependent aging-slip switch is necessary even if functions not being unary of $\Omega$ are included in candidates of the evolution function $f$.

However, even the generalized version (eq.~\ref{eq:constraintonftgen}) was derived under two a priori assumptions besides the canons (\S\ref{sec:generalizingheuristicderivation}). The first assumption was the TDF, as mentioned already. It represents an a priori belief (as a working hypothesis) that variation in strength consists of a response to the time-lapse ($F_t$) and that to the slip ($F_\delta$). In the derivation, the TDF contributed substantially to assigning the VS and SHS canons to the slip response ($f_\delta$) and the time response ($f_t$), respectively. 
By contrast, the derivation below shall be done without the TDF assumption. The second assumption was the functional limitation $\dot\theta=f(V,\theta)$. We will keep this. One reason is to forbid using an almighty oracle variable for switching evolution functions between VS and SHS$|$NSC tests; the use of an oracle would be mere tautology of experimental constraints shown in \S\ref{sec:3}. Although oracle exclusion may be achievable in other ways, we use $\dot\theta=f(V,\theta)$, which has been standard since \citet{ruina1983slip}, for simplicity of derivation. 
The functional space $\dot \theta =f(V,\theta)$ contains $\dot\theta = f(V, \theta, \tau)$ as $\tau$ can be expressed in terms of $V$ and $\theta$ (eq.~\ref{RSFeq:constitutivetheta} or \ref{RSFeq:constitutivethetacutoff}). 
Still, we have to admit that the $f(V,\theta)$ 
presumption has precluded some plausible effects, such as the shear stress dependence~\citep{nagata2012revised}, written as $\dot\theta = f(V, \theta, \dot \tau)$ [or equivalently, $\dot\theta = f(V, \theta, \dot V)$]. 
Considerations of evolution functions not being of $f(V,\theta)$ are left to future studies. 

Now, we start the derivation of a further generalized form of eq.~(\ref{eq:constraintonftgen}) with the TDF assumption removed, whereas we still assume $\dot\theta=f(V,\theta)$ as mentioned above. We only consider $\theta,\theta_{\rm SS}(V)\gg \theta_{\rm X}$ because the strength minimum can be incorporated by the procedure shown in \S\ref{sec:extensions}. 
We redo the derivation of experimental constraints from the canonical behaviors without assuming the TDF this time. Constraints from the SS canon remain eq.~(\ref{RSFeq:steadystateconstraint}) because the original derivation of eq.~(\ref{RSFeq:steadystateconstraint}) in \S\ref{subsec:SSrequirement} did not utilize the TDF. 

Under eq.~(\ref{RSFeq:steadystateconstraint}), 
we can obtain VS constraints simply by rewriting eq.~(\ref{eq:sliplawPhiform}) into $\theta$-notation:
\begin{equation}
f_{\rm VS} = -(L/D_{\rm c})\Omega\ln \Omega .   
\label{eq:VSconstraint4nonTDF}
\end{equation}
Again, this derivation does not utilize the TDF, although the resulting eq.~(\ref{eq:VSconstraint4nonTDF}) has the same form as the constraint on $f_\delta V$ (eq.~\ref{eq:simplifiedslipdepofthetadot}), obtained with the TDF assumption.

To express constraints from SHS tests under stationary contact, we introduced the NSC in \S\ref{subsec:SHSrequirement}. However, the definition of the NSC (eq.~\ref{eq:NSCdef}) assumes the TDF; hence, we here need to express the stationary contact in other ways. We do this by replacing `NSC' with the literal definition of stationary contact, `$\delta\to0$'. 
Repeating \S\ref{subsubsec:332} with `NSC' replaced by `$\delta\to0$',
we have 
\begin{equation}
f_{\rm SHS|\delta \to 0}=1.
\label{eq:fSHS_constrained}
\end{equation}
Although derivation this time did not presume the TDF, the resulting eq.~(\ref{eq:fSHS_constrained}) represents the log-t healing, the same form as the constraint on the purely time-dependent term under the NSC (eq.~\ref{eq:ftNSC_constrained}) obtained under the TDF assumption. 

Next, we need to specify the regimes where each constraint applies. Again, we cannot use expressions that necessitate the TDF. As discussed in \S\ref{RSF322}, VS tests require (i) $f=f_{\rm VS}$ for $\Omega\gtrsim V_{\rm after}/V_{\rm before}$ probed by existing VS tests. Note that this requirement (i) is not conditioned on $V_{\rm after}$ itself because the VS canon has been confirmed over a wide range of absolute slip rates. Because of (i), the canonical behavior of SHS tests ($f=1$, eq.~\ref{eq:fSHS_constrained}) is allowed only in the $V$-$\theta$ regime sufficiently below the steady state that excludes the $\Omega$ range in the strengthening phase following the negative velocity step. Meanwhile, the SHS canon requires (ii) $f=f_{\rm SHS|\delta \to 0}$ for stationary contact. This time, as mentioned earlier, we cannot express stationary contact as the NSC (eq.~\ref{eq:NSCdef}), which is only definable under the TDF. 
Instead, we here express stationary contact by invoking constant-$\tau$ SHS tests. If we, upon the start of the hold period, drop the shear stress by much more than $A$ from the level [$\tau_{\rm SS}(V_{\rm prior})$] just before the start of hold, stationary contact is realized and kept for a long time throughout the hold period (\S\ref{sec:SHSsaturation}). When we consider a large-stress-drop limit of $e^{-\Delta \tau/A}\to 0$, $\Omega\to 0$ holds for $0<t_{\rm h}<\infty$. Therefore, from (ii), we can state $f=f_{\rm SHS|\delta \to 0}$ when $\Omega\to 0$. 
Thus, from (i) and (ii), we have 
\begin{equation}
        f\simeq
    \begin{cases}
        f_{\rm VS} & (\Omega \gtrsim  V_{\rm after}/V_{\rm before})
        \\
        f_{\rm SHS|\delta \to 0} & (\Omega\to 0)
    \end{cases}
    \label{eq:constraintonftgen_withoutTF}
\end{equation}
Equation~(\ref{eq:constraintonftgen_withoutTF}) is a generalized version of eq.~(\ref{eq:constraintonftgen}), deduced without the TDF assumption. Note that the constraining behaviors (i.e., the right-hand side) for low $\Omega$ and high $\Omega$ are the same between eqs.~(\ref{eq:constraintonftgen}) and (\ref{eq:constraintonftgen_withoutTF}), although eq.~(\ref{eq:constraintonftgen}) expressed the aging-slip switch as tuning of $f_t$ because the VS canon was described elsewhere (eq.~\ref{eq:simplifiedslipdepofthetadot}) as the constraint on $f_\delta V$.

The same discussion as that for eq.~(\ref{eq:constraintonftgen}) (+ eq.~\ref{eq:simplifiedslipdepofthetadot}) in \S\ref{sec:generalizingheuristicderivation} guarantees evolution laws conforming eq.~(\ref{eq:constraintonftgen_withoutTF}) automatically satisfies the SS constraint (eq.~\ref{RSFeq:steadystateconstraint}). 
Hence, like eq.~(\ref{eq:constraintonftgen}) (+ eq.~\ref{eq:simplifiedslipdepofthetadot}) under the TDF, but without the TDF assumption this time, eq.~(\ref{eq:constraintonftgen_withoutTF}) is a necessary and sufficient condition to reproduce the three canons simultaneously, except for the range not experimentally probed so far. 

Below is a cautious remark regarding the difference between the low-velocity limit $V\to 0$ and the far-below-the-steady-state limit $\Omega\to 0$.
Because $V\to0$ and $\Omega\to0$ are equivalent unless $\theta\to\infty$, we cannot distinguish $V\to0$ from $\Omega\to0$ when we consider constant-$\tau$ SHS tests for finite hold time. 
Therefore, the SHS condition (ii) used in the above derivation of eq.~(\ref{eq:constraintonftgen_withoutTF}), which we took as a constraint at the low-$\Omega$ limit, might be a condition at the low-$V$ limit in fact. 
This difference may matter in describing the state evolution in loadpoint-hold SHS tests where both $\theta$ and $V$ change gradually, and the change in $\Omega$ is far slower so that the limits represented by $\Omega\to0$ and $V\to0$ can be different. 
Hence, some ambiguity is left in the above derivation and eq.~(\ref{eq:constraintonftgen_withoutTF}). 
Such ambiguity did not exist in the derivation of eq.~(\ref{eq:constraintonftgen}) in \S\ref{sec:generalizingheuristicderivation} because the NSC cannot exist outside the low-$\Omega$ regime (proven at the beginning of \S\ref{sec:development}). This ambiguity is part of the reason why we first showed eq.~(\ref{eq:constraintonftgen}). 
Yet, experiments so far confirm that the VS canon holds independently of $V$, which singly rejects the emergence of the conditional branch according to the value of $V$, supporting the $\Omega$-dependent switch as expressed in eq.~(\ref{eq:constraintonftgen_withoutTF}). Equation (\ref{eq:constraintonftgen}) expresses this with a single constant $\beta$. 
Such $\beta$, as the scale of the aging-slip switch (cf. Appendix~\ref{sec:betavalue4unitypsicbase}), also exists for eq~(\ref{eq:constraintonftgen_withoutTF}), not limited to the TDF (though $\dot \theta=f(V,\theta)$ is still assumed), but $f$ in eq~(\ref{eq:constraintonftgen_withoutTF}) is largely unconstrained over a transitional $\Omega$ range.

To sum up, the aging-slip switch according to $\Omega$ (eq.~\ref{eq:constraintonftgen_withoutTF}), namely, the existence of a characteristic value of $\Omega$, $\beta$, in eq.~(\ref{eq:constraintonftgen}), is a general conclusion deduced from the three canons as far as $\dot\theta=f(V,\theta)$. 
Excepting \citet{bhattacharya2015critical,bhattacharya2022evolution} discussed later in \S\ref{subsec:possiblebetadetection}, nobody has observed this switch, but there is no evidence denying it so far. Given the generality kept in the derivations, if future experiments reject the existence of $\beta$, it will logically mean that the three canons are not the case somewhere in the regimes experimentally probed so far (again, if we ignore the possibility $\dot\theta\neq f(\theta,V)$). Simultaneous reproduction of the three canons is the target of the present work and the historical development of the RSF. As has turned out in the present work, the difficulty of reconciling the aging and slip laws is not such a superficial matter that just nobody has yet thought of a clever formula to reproduce the three canons simultaneously. Thence, we believe the general conclusion of this work, that is, aging-slip switch around $\Omega\sim\beta$, has fundamental importance, warranting future experimental and microphysical studies to verify or falsify it.

\subsection{Macrophysical implications of the $\Omega$-controlled aging-slip switch}
\label{subsec:macrophysicsOmegadepswitch}
In \S\ref{subsec:discussion1}, we have shown that the simultaneous reproduction of the three canons (throughout the range probed by experiments so far) is essentially equivalent to the aging-slip switch according to $\Omega$. How, then, would the $\Omega$- rather than $V$-control of the switch be interpreted?

Noting $\Omega=\theta/\theta_{\rm SS}$, the $\Omega$-dependent nature of the aging-slip switch means that the switch occurs according to the relative difference of instantaneous $\theta$ from its steady-state value for the instantaneous slip rate. When the strength minimum is incorporated, the switch occurs according to $\tilde \Omega/\tilde\Omega_{\rm SS}=\tilde \theta/\tilde \theta_{\rm SS}$, as in $\tilde \psi_{\rm c}(\tilde \Omega/\tilde \Omega_{\rm SS})$ (eq.~\ref{eq:psicprime}), so that the control factor of the switch becomes the relative difference of $\tilde \theta=\theta+\theta_{\rm X}$ from its steady-state value. Then, under \citet{dieterich1978time}'s interpretation of $\theta$ as the average contact time, it is implied the aging-slip switch is controlled by the contact time for $\theta\gg\theta_{\rm X}$ while it becomes almost independent of the contact time for $\theta\ll\theta_{\rm X}$. Such clumsiness, however, disappears in the $\Phi$-notation. 
From eqs.~(\ref{RSFeq:conversionofthetatophi}) and (\ref{eq:thetatilde})--(\ref{eq:tildeOmegassspecified}), we have
\begin{equation}
    \ln \frac{\tilde \Omega}{\tilde \Omega_{\rm SS}}=\frac{\Phi-\Phi_{\rm SS}}B.
    \label{eq:Omega2strength}
\end{equation}
For $\theta,\theta_{\rm SS}\gg\theta_{\rm X}$, 
from eqs.~(\ref{RSFeq:conversionofthetatophinocut_taustar}), (\ref{RSFeq:constitutivePhiA}), and (\ref{RSFeq:defofOmega_steadystate}),
we have
\begin{equation}
    \ln\Omega=\frac{\tau-\tau_{\rm SS}}B = \frac{\Phi-\Phi_{\rm SS}}B.
    \label{eq:Omega2strength_uncut}
\end{equation}
Notice that the direct effect does not contribute to the difference of the shear stress from its steady-state value (eq.~\ref{RSFeq:constitutivePhiA}), that is, $\tau-\tau_{\rm SS} =  \Phi - \Phi_{\rm SS}$.
Hence, for any values of $\theta$, our evolution laws set the aging-slip switch according to $\Phi-\Phi_{\rm SS}$ ($\tau-\tau_{\rm SS}$). 
The threshold $\Phi-\Phi_{\rm SS}$ value of the aging-slip switch is $B\ln \beta$ (eqs.~\ref{eq:constraintonftgen} and \ref{eq:Omega2strength}). 
Note that eq.~(\ref{eq:Omega2strength}) also indicates that $\tilde \Omega/\tilde \Omega_{\rm SS}$ affects strength change only through its logarithm. 
Then, the crossover regime of the aging-slip switch around $\Omega\sim \beta$ in eq.~(\ref{eq:constraintonftgen}) is narrowed in the $\Phi$ scale so that eq.~(\ref{eq:constraintonftgen}) switches the constraints on the evolution depending on whether the strength $\Phi$ is significantly higher or lower than the threshold value $\Phi_{\rm SS}+B\ln \beta$. 

The above-mentioned $\tau-\tau_{\rm SS} =  \Phi - \Phi_{\rm SS}$ rewrites eq.~(\ref{eq:Omega2strength}) into 
\begin{equation}
    \ln \frac{\tilde \Omega}{\tilde \Omega_{\rm SS}} = \frac{\tau-\tau_{\rm SS}}B \label{eq:Omega2stresss}.
\end{equation}
Therefore, the aging-slip switch controlled by $\tilde \Omega/\tilde \Omega_{\rm SS}$ can also be regarded as a switch according to the difference of the shear stress from its steady-state value.

By using eq.~(\ref{eq:Omega2stresss}), for interfaces with $|A-B|/B\ll1$ as typically observed in experiments [e.g., $|A-B|/B\simeq 1/30$ reported by \citet{bhattacharya2015critical}], we can propose a more physically intelligible interpretation of the $\Omega$-dependent aging-slip switch as below. 
In the remaining part of this subsection \S\ref{subsec:macrophysicsOmegadepswitch}, we assume $|A-B|/B\ll1$ unless otherwise noted. 
Then, the steady-state frictional resistance $\tau_{\rm SS}$ remains close to the reference value $\tau_*$ unless hugely extensive range of velocities is considered; we obtain the following relation from the SS canon (eq.~\ref{eq:req4ss}), using the identity eq.~(\ref{eq:Omega2stresss}):
\begin{equation}
    \ln\frac{\tilde \Omega}{\tilde \Omega_{\rm SS}}=\frac{\tau-\tau_*}B+\mathcal O[(|A-B|/B)\ln (V/V_*)],
    \label{eq:Omega2stressexcess}
\end{equation}
which corresponds to the observation that dynamic friction depends little on velocity, as presumed in classical tribology. 
Note eq.~(\ref{eq:Omega2stressexcess}) is written with the reference of frictional resistance $\tau_*$, instead of the reference strength $\Phi_*$, given their numerical coincidence (eq.~\ref{eq:steadystatestrengthstressequality}).  
As long as the second term of eq.~(\ref{eq:Omega2stressexcess}) is negligible, the aging-slip switch according to $\Omega$ (precisely, $\tilde \Omega/\tilde\Omega_{\rm SS}$) can be approximately interpreted as a switch according to $\tau$.
Thus, we argue that the essence of the presently proposed evolution law may be the stress-dependent switch of the aging and slip laws, in contrast to the velocity-dependent switching proposed by \citet{kato2001composite}.

As mentioned in \S\ref{sec:canonicalandside}, \citet{kato2001composite} proposed to switch the aging and slip laws according to the slip rate $V$, under the expectation that there must be a critical slip rate $V_{\rm c}$ where the switch occurs. 
Upon introducing the composite law, \citet{kato2001composite} said, ``To accommodate a friction law to experimental observations of both slide-hold-slide tests and velocity stepping tests, the evolution equation of $\theta$ must be similar to the slip law near the steady state and must approach unity at very small velocities.'' [Note: for readability in the present discussion, we have replaced some expressions from the original: state variable $\to \theta$; (3) $\to$ the slip law]. We can then summarize their requirement on the evolution law, here denoted by $f_{\rm KT}$, in a form analogous to eq.~(\ref{eq:constraintonftgen_withoutTF}): 
\begin{equation}
        f_{\rm KT}\simeq
    \begin{cases}
        f_{\rm VS} & (\Omega \sim \Omega_{\rm SS})
        \\
        f_{\rm SHS|NSC} & (V\to 0)
    \end{cases}
    \hspace{10pt}
    \label{eq:ideaofKato}
\end{equation}

Here, $f_{\rm VS}$ intends, as in eq.~(\ref{eq:constraintonftgen_withoutTF}), the exponential convergence toward the steady state over a constant slip distance (e.g., eq.~\ref{RSFeq:dotthetaVS}), but does not include the enforcement of steady-state value following the SS canon. We think this is what \citet{kato2001composite} intended when they mentioned `similar to the slip law' in the above excerpt. Although `the slip law' (eq.~\ref{eq:defofsliplaw}) enforces both behaviors of VS and SS, it is clear that the latter SS-canon enforcement was not intended by \citet{kato2001composite}, given that their proposed composite law does not follow the SS canon. 

Because $V\to 0$ and $\Omega\to 0$ cannot be distinguished for finite $\theta$ in our derivation of eq.~(\ref{eq:constraintonftgen_withoutTF}), 
we may say that Kato and Tullis's (2001) idea expressed by eq.~(\ref{eq:ideaofKato}) is correct from the conceptual viewpoint of the aging-slip switch. 
The only major failure of the composite law (eq.~\ref{eq:defofcompositelaw}) is the deviation from the SS canon. [Strictly speaking, eq.~(\ref{eq:ideaofKato}) has one more problem: constraint by $f_{\rm VS}$ needs to be imposed for a broader $\Omega$ range than said in the first line of eq.~(\ref{eq:ideaofKato}). This problem can be understood if one notices that the aging law, which does not reproduce the VS canon, actually satisfies eq.~(\ref{eq:ideaofKato}). Nevertheless, we here ignore this shortcoming for this discussion at an upper conceptual level.] 

Thus, when expressed in the form of eq.~(\ref{eq:ideaofKato}), the composite law can be interpreted as the switch of the dominant evolution mechanism depending on the kinematic state of the interface, i.e., (nearly) sticking or (sufficiently) slipping, similar to the pre-RSF refinement of friction laws. Thence, with the aid of the present abstraction of their essence (eq.~\ref{eq:ideaofKato}), we may retrospectively say that, upon their recognition of the violation of the SS canon by their composite law, \citet{kato2001composite} have already noticed the difficulty of explaining the SS canon under the traditional kinetic interpretation of the switch depending on stick or slip. In this context, our solution is to change the control factor of the switching; the present work proposes that the switch is controlled by the level of the applied stress $\tau$ as argued earlier in this subsection (eq.~\ref{eq:constraintonftgen_withoutTF} + eq.~\ref{eq:Omega2stressexcess}). When $\tau$ is significantly below the dynamic frictional resistance ($\tau\ll \tau_{\rm SS}\simeq \tau_*$), purely time-dependent healing dominates the evolution. In contrast, when $\tau$ is close to the dynamic frictional resistance ($\tau\sim \tau_{\rm SS}\simeq \tau_*$), slip-driven evolution (i.e., the slip law) dominates the strength variation no matter how slow the slip is.

Evolution laws adopting $\tau$-dependent switch and $V$-dependent switch tend to have similar forms because $V$ and $\tau$ are related by the constitutive relation (eq.~\ref{RSFeq:constitutivePhiB}). 
An example of the similarity is that, as pointed out in \S\ref{RSF33}, the modified composite law (eq.~\ref{RSFeq:modcomplaw}) we have proposed can be obtained by merely replacing the constant $V_{\rm c}$ of the composite law (eq.~\ref{eq:defofcompositelaw}) with a variable of velocity dimension inversely proportional to $\theta$. Nevertheless, their predictions are qualitatively different because $V_*\exp(-\Phi/A)$, the proportional constant between the $V$ and $\exp(\tau/A)$ in eq.~(\ref{RSFeq:constitutivePhiB}), drastically changes with $\mathcal O(B)$ variations in $\Phi$. These two laws, namely the $\tau$-controlled switch and $V$-controlled switch, can be distinguished in both the healing rate in SHS tests and the strengthening rate following a negative VS. In \S\ref{subsec:possiblebetadetection}, we analyze the VS test data of \citet{bayart2006evolution} to find 
that the modified composite law better explains the data than the composite law. 

Besides, when we start looking at third-order $\mathcal O(A-B)$ variations in frictional resistance, $\Omega$-controlled switching, which is no longer equivalent to $\tau$-controlled switching, is required to reproduce the three canons, as shown in the present full development including $\mathcal O(A-B)$ terms up to \S\ref{subsec:discussion1}.

\subsection{Potential detection of $\beta$ from an experiment}
\label{subsec:possiblebetadetection}
As far as $\dot \theta=f(V,\theta)$, we have theoretically shown that the aging-slip switch must occur around $\Omega=\beta$ to reproduce the three canonical behaviors confirmed by experiments so far. Such evolution laws involving an aging-slip switch are necessarily accompanied by some side effects violating the canons (\S\ref{sec:canonicalandside}; also, see \S\ref{sec:heuristicderivation}). Among them, relatively accessible by experiments may be the behavior following a large negative velocity step ($V_{\rm after}/V_{\rm before}\ll\beta$), which is expected to exhibit too rapid strengthening compared to the prediction by the slip law, that is, the VS canon (e.g., Fig.~\ref{fig:predictedmotion_modcomposite}a). 
Although existing VS tests generally obey the slip law as we presumed as the VS canon throughout the present development, a slight deviation has been reported in fact. \citet{bhattacharya2015critical} realized that an interesting behavior of the Nagata law~\citep{nagata2012revised}, which is the aging law plus a stress-dependent term $-c d \tau$ representing an experimentally detected instantaneous effect of the applied shear stress to weaken the interface; that is, the Nagata law changes from the aging law to the slip law for large velocity steps. They further showed that some VS datasets involving multiple steps with various levels of $V_{\rm after}/V_{\rm before}$ (ranging $10^{-2}$ to $10^{2}$) are fit well by the Nagata law with $c = 50$, contrasting that the slip law cannot fit the large negative VS. This finding of \citet{bhattacharya2015critical} implies that the involvement of the aging-slip switch improves fitting to the experiments. 

As \citet{bhattacharya2015critical} themselves point out, assuming $c$ as large as 50 is unreasonable as the shear-stress effect of the Nagata law, given that \citet{nagata2012revised} obtained $c = $2 by direct measurement of $d\Phi/d\tau$. However, from the viewpoint of the aging-slip switch, their Nagata law fits suggest the aging-slip switch according to $\Omega$ we deduced from the canons. Below, we try fitting the data shown in \citet{bhattacharya2015critical} using the proposed modified composite law. Of course, making a formal statistical conclusion on the aging-slip switch according to $\Omega$ requires to demonstrate the increase in likelihoods sufficient to counter the increase in the number of parameters. The slight deviation found in the $V_{\rm after}/V_{\rm before}$ range of the data used in \citet{bhattacharya2015critical} is probably not enough to conclude the existence of the additional parameter (excessively large $c$, or reasonably, $\beta$)~\citep{bhattacharya2015critical}. Therefore, a definitive conclusion must be postponed until more suitable data becomes available. Further detailed discussion shall be shown at the end of this subsection, where we also consider \citet{bhattacharya2022evolution} involving exploration of a further lower range of $\Omega$.

\begin{figure*}
	\includegraphics[width=145mm]{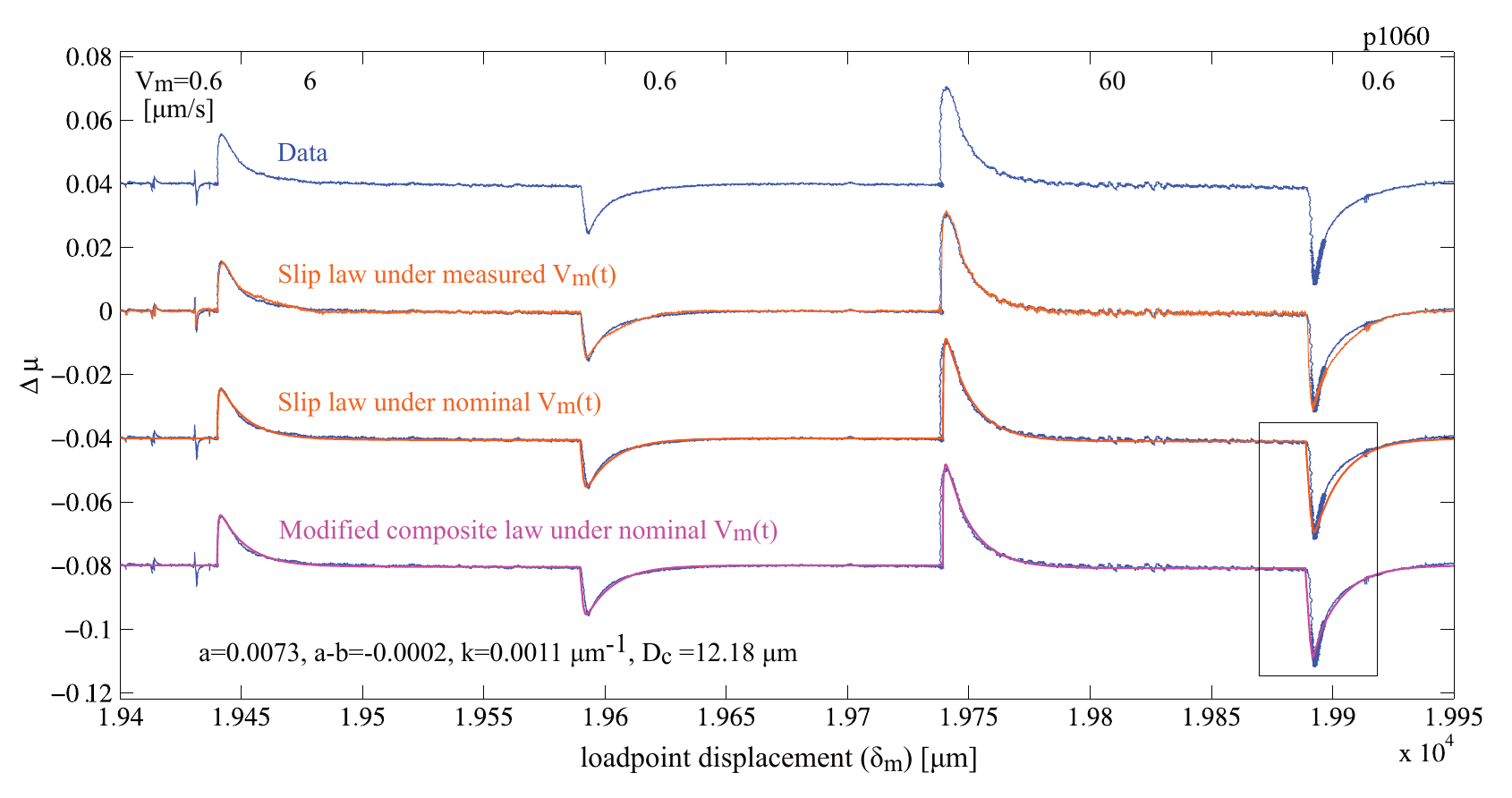}
\caption{
Comparison of VS tests by \citet{bayart2006evolution} with predictions by the slip law and the modified composite law.
Variations in normalized frictional stress ($\Delta \mu$ in $\tau/\sigma$) are plotted against the loadpoint displacement. The same experimental data, reproduced from Fig.~2(A) of \citet{bhattacharya2015critical}, are plotted in all four rows in blue. First row: Data only. Second row: The orange line shows the prediction by the slip law under the measured loadpoint velocity $V_{\rm m}(t)$. Parameter values estimated by \citet{bhattacharya2015critical} are shown at the lower left corner of the figure, where $k$ is normalized by $\sigma$, and are used for all the simulations shown in the second to fourth rows. Third row: The orange line shows the prediction by the slip law under the nominal $V_{\rm m}(t)$ taken from the terminal load-point velocities at respective steps. Fourth row: The pink line shows the prediction by the modified composite law with $\beta = 0.01$ under the nominal $V_{\rm m}(t)$ taken from the terminal load-point velocities at respective steps.
}
  \label{fig:bhatfit}
\end{figure*}

The VS test data ($\tau/\sigma$ versus the loadpoint displacement) are shown by blue curves in all four rows of Figure~\ref{fig:bhatfit}. The second row is the comparison with the slip-law prediction (orange) with parameter values (shown in the figure) estimated by \citet{bhattacharya2015critical}. 
All this figure's simulations (the second to the fourth rows) employ this set of parameter values. 
In the simulation, finite stiffness $k$ of the loading apparatus was considered through the spring slider model $\dot\tau=k(V_{\rm m}-V)$ (eq.~\ref{eq:springslidereq}), where $V_{\rm m}$ is the loadpoint velocity. For $V_{\rm m}(t)$, \citet{bhattacharya2015critical} used the measured movement of the load point rather than the values commanded to the servo system because the actual movement does not necessarily follow the command perfectly. 
The shown experiment includes four VS tests. The first three tests with $V_{\rm after}/V_{\rm before} = 10$, $1/10$, and $100$ are very well reproduced by the slip law, but in the fourth test done with $V_{\rm after}/V_{\rm before}=1/100$, actual strengthening is significantly more rapid than the slip law prediction~\citep{bhattacharya2015critical}.

The fourth row of Fig.~\ref{fig:bhatfit} compares the same data with the prediction (pink) by the modified composite law (eq.~\ref{RSFeq:modcomplaw}) with $\beta=0.01$. 
The modified composite law fits the data very well, including the fourth VS with $V_{\rm after}/V_{\rm before} = 1/100$, which was not well reproduced by the slip law (second row) as mentioned above. Thus, the data for $V_{\rm after}/V_{\rm before} = 1/100$ seem to support the reality of the aging-slip switch by observing the expected side effect. 
However, in the above simulation shown in the fourth row, $V_{\rm m}(t)$ was taken from the terminal load-point velocities at respective steps, $0.6 \to 6 \to 0.6 \to 60 \to 0.6$ $\mu$m/s because we do not have actual $V_{\rm m}(t)$ record and servo commands. 
Hence, there remains a possibility that the improvement from the second row to the fourth row resulted from the incorrectness of the nominal $V_{\rm m}(t)$ we assumed in the simulation of the fourth row. 
To clear this doubt, we did another simulation, where the slip-law prediction was calculated under the nominal $V_{\rm m}(t)$ series used in the fourth row. The result is shown in the third row (orange), showing almost the same result as the accurate slip-law simulation employing the measured $V_{\rm m}(t)$ (second row); only the fourth VS is mispredicted in the same sense as in the second row. Therefore, we argue that the fitting improvement seen in the fourth row comes from the side effect of the aging-slip switch around $\Omega\sim \beta (= 0.01)$.

Besides, we should point out that it seems unlikely for the composite law~\citep{kato2001composite} to explain the VS test data in Figure~\ref{fig:bhatfit}. The data have two negative velocity steps: the first one with $V_{\rm m} = 6 \to 0.6\mu$m/s and the second one with $V_{\rm m} = 60 \to 0.6\mu$m/s. The first one followed slip-law prediction, but the second one did not. The second one is in between the slip law's and the aging law's predictions. 
As shown above, the modified composite law could explain both by turning on the aging effect only for the second one where $V_{\rm after}/V_{\rm before} \lesssim \beta(= 0.01)$. In contrast, the composite law turns on and off the aging effect depending on the absolute value of $V_{\rm after}$. Because $V_{\rm after}$ is the same (0.6 $\mu$m/s) in the two negative steps, it seems unlikely for the composite law to turn on the aging effect only for one of the two. If $V_{\rm c} \ll V_{\rm after}$(=0.6 $\mu$m/s), the composite law predicts that both tests exhibit slip-law-like behaviors. If $V_{\rm c} \gg V_{\rm after}$(=0.6 $\mu$m/s), aging-law-like behaviors are predicted for the both. Hence, this data set of \citet{bayart2006evolution} suggests the superiority of the modified composite law compared to the composite law, though aging-slip switch is implemented in both laws (in different ways). However, again, this argument is based on only one experiment, and further evaluation by more experiments is necessary for this point.

In the above, we have pointed out that 
the modified-composite law with $\beta=0.01$ explains the data of \citet{bayart2006evolution} for $V_{\rm after}/V_{\rm before} = 0.01$, which \citet{bhattacharya2015critical} have found to be basically slip-law-like but somewhat skewed toward aging-law-like behaviors, which are approximated by the Nagata law with excessively large $c$-values. 
At the same time, we admit that some datasets do not support the aging-slip switch. 
First, \citet{bhattacharya2015critical} note that the above-mentioned better fit of the Nagata law than the slip law, which supports the aging-slip switch, was not evident in one of the three datasets they examined. 
Second, \citet{bhattacharya2022evolution} have performed VS tests, planned for capturing the composite-law-like aging-slip switch, strategically including quite low  nm/s-order slip rates and small and large steps.
Although they have conducted VS tests with $V_{\rm after}/V_{\rm before}$ down to $10^{-3.5}$, they have not reported the aging-slip switch. It implies $\beta$ for their samples is below $10^{-3.5}$ if exits. 
Still, this observation of \citet{bhattacharya2022evolution} may originate from the long-term trend~\citep[similar to that in][]{blanpied1998effects} in their data, not fully taken into their analysis. 
They reported that the observed slip-strengthening behaviors are even slower than the slip law, which would presumably indicate the above-mentioned long-term trend. 
However, when the long-term trend is removed by a linear fit, we find that the detrended data for $V_{\rm after}/V_{\rm before}=10^{-3.5}$ are faster than the slip law, and the modified composite law with $\beta=0.01$ well fits those detrended data, although the statistical significance of our fitting is yet to be scrutinized. 
Thus, we deem $10^{-3.5}<\Omega < 10^{-2}$ still under debate.

After all, we do not know much about the $\beta$ value, because  VS tests with $V_{\rm after}/V_{\rm before}\leq 10^{-2}$ is still scarce. 
However, we emphasize that the existence of $\beta$ has been deduced from the reconciliation of the three canons under the premise of $f(V, \theta)$, and our SHS canon (significant log-t healing in low-$\tau_{\rm hold}$ SHS) is set at $\Omega$ far-far below steady states ($\Omega_{\rm SS}=1$), that is, $\Omega\to0$, much lower than $10^{-3.5}$. As the proposers, we reckon that VS tests with further small $V_{\rm after}/V_{\rm before}$ will find the expected switch. 
Also, note that the value of $\beta$, like the value of $A$ and $B$, can depend on various experimental conditions other than the variables of friction law ($\tau$, $V$, $\Phi$), including rock types, temperature, normal stress, and gouge conditions. Hence, $\beta$ must be considered a fitting parameter, which should be experimentally determined under each condition, as $A$ and $B$ are. For prudence, we still regard the aging-slip switch and its side effects as an unconfirmed theoretical prediction of the present work, and we would like to present the data analysis shown in Fig.~\ref{fig:bhatfit} as cursory support to our proposal.

\section{Conclusion}
It has been an outstanding problem that no evolution laws ever proposed can simultaneously reproduce canonical behaviors in the three standard types of low-velocity friction experiments (VS, SHS, and SS). 
Especially relevant to the present work are the flaws of two representative evolution laws; the aging law fails badly in reproducing the VS canon, while the slip law does so in reproducing the SHS canon. In the present study, by recompiling the experimental requirements from the VS, SHS, and SS tests, we have proposed a class of evolution laws reproducing the three canons simultaneously throughout the $(V,\theta)$ range experimentally probed so far.

In order to put the experimental requirements in a clear mathematical perspective, we first tidied up the RSF formulation so that the roles of the two constituent equations are separated; the constitutive law describes the slip rate induced by the shear stress applied on the interface with a given strength while the variation in strength is exclusively taken care of by the evolution law (\S\ref{subsec:RSFseparation}).
Then, we have derived the constraints on the evolution law from the three canons expressed in the total differential form (\S\ref{sec:3}). In formulating the experimental requirements, we introduced a new concept of NSC, negligible slip condition, akin to the traditional concept of quasi-stationary contact, to clearly define the SHS canon (\S\ref{subsec:SHSrequirement}). Evolution laws thus derived from the VS and SHS$|$NSC canons have been found to coincide with the slip law and the aging law under the NSC, respectively. At the same time, however, we have mathematically proven that there is a range of variables where the VS canon contradicts the SHS canon under the NSC (\S\ref{sec:canonicalandside}). In other words, the very canonical behaviors in VS and SHS$|$NSC tests lead to a logical conclusion that, by reduction to absurdity, evolution laws consistent with both canons are only possible for a limited range of the laws' variables. Hence, the dominant mechanism of evolution must switch between the aging- and slip-driven ones outside this range. This existence proof of the switch of the VS and SHS canons is one crucial finding of the present study. 

Thus-recognized impossibility of simultaneous reproduction of the three canons throughout the entire range of variables has forced us to downgrade the goal of the present development (\S\ref{sec:canonicalandside}). We have decided to seek evolution laws satisfying the three canons over the $(V,\theta)$ range where existing experiments have confirmed the canons, and we could find a general precedence to construct such laws. The derivation was done in several steps, with an increasing degree of generality. In the first derivation (\S\ref{sec:heuristicderivation}), to prevent the violation of the SS canon, which was a major problem of the composite law~\citep{kato2001composite}, the first evolution law involving the aging-slip switch, we started with the heuristic ansatz of $f=\psi(\Omega)$, which is a sufficient condition for the SS canon to hold. For $f_\delta V$, the slip law has been adopted as it is the unique solution to produce the symmetric evolution following positive and negative velocity steps. Since $f_\delta V$ is thus constrained, the aging-slip switch becomes a matter of tuning $f_t$. 
We implemented the aging-slip switch by tuning $f_t$ so that it is significantly less than $|f_\delta V|$ except when $\theta$ is sufficiently below the steady state for the instantaneous $V$ where the $f_t$ becomes $f_t$-term of the aging law (\S\ref{sec:heuristicderivation}). Thus-constructed class of evolution laws (eq.~\ref{eq:evolutionlaw_compatiblewithfunctionalansatz_substituted} with eq.~\ref{RSFeq:reqforpsic}) satisfy the three canons under the $(V,\theta)$ range of the existing experiments. 
A general procedure to obtain concrete expressions belonging to 
this class of evolution laws is presented in \S\ref{RSF33}, with specific examples (eqs.~\ref{RSFeq:exmodcomplaw} and \ref{RSFeq:modcomplaw})
employing an exponential function to mute $f_t$. Interestingly, eq.~(\ref{RSFeq:modcomplaw}) has a form analogous to the composite law; when its constant parameter $V_{\rm c}$, the upper $V$ threshold for the purely time-dependent healing, is replaced with a variable inversely proportional to $\theta$, it becomes eq.~(\ref{RSFeq:modcomplaw}).

In the second step (\S\ref{sec:generalizingheuristicderivation}), we have further shown that the aging-slip switch occurring around $\Omega=\beta$ is an inevitable consequence of the three canons even when evolution laws are sought in a more general scope of $\dot \theta=f(V,\theta)=f_t+f_\delta V$ with the $f=\psi(\Omega)$ ansatz removed (eq.~\ref{eq:constraintonftgen}).

The above-proposed class of evolution laws (eq.~\ref{eq:evolutionlaw_compatiblewithfunctionalansatz_substituted} with eq.~\ref{eq:constraintonftgen}) have two side effects of the aging-slip switch. The first is the too-rapid strengthening following a large negative velocity step ($V_{\rm after}/V_{\rm before}\ll\beta$) in the early stage where $\Omega \ll \beta$. 
The second is the slowdown of healing around $\Omega= \beta$ in SHS tests on the interfaces with $A > B$ (\S\ref{sec:SHSanalysismodifiedcomposite}). 
To see if these predicted side effects are the case, we suggest VS tests with $V_{\rm after}/V_{\rm before}\ll\beta$ for the former, and long-$t_{\rm hold}$ SHS tests on interfaces with $A>B$ for the latter. We have found one existing VS test~\citep{bayart2006evolution,bhattacharya2015critical} exhibiting behavior that appears to be the former, with $\beta\sim 0.01$ (\S\ref{subsec:possiblebetadetection}). 

In the third step (\S\ref{subsec:discussion1}), we have shown that similar constraints on the evolution laws can be deduced without presuming the thermodynamic formulation. This most general derivation in the present paper shows that the aging-slip switch according to $\Omega$ is equivalent to simultaneous satisfaction of the three canons observed by the existing experiments if the evolution function depends only on $V$ and $\theta$. 
Furthermore, for interfaces with $|A-B|\ll B$, the $\Omega$-dependent aging-slip switch can be approximately regarded to be controlled by the applied shear stress; the dominant evolution mechanism (aging or slip) switches depending on whether the applied stress is significantly lower than the dynamic frictional resistance. This macrophysical interpretation contrasts with classical tribology, where the switch occurs depending on whether the interface is sufficiently slipping (\S\ref{subsec:macrophysicsOmegadepswitch}).

\begin{acknowledgments}
We would like to thank Dr. Hugo Perfettini, Dr. Paul Antony Selvadurai, Dr. So Ozawa, Dr. Bunichiro Shibazaki, Dr. Yukitoshi Fukahata, Dr. Takeshi Iinuma, and Dr. Takane Hori for valuable discussions and encouragement. 
We also acknowledge careful reviews and insightful comments from Dr. Allan Rubin, Dr. Shiqing Xu, Dr. Baoning Wu, and an anonymous reviewer. 
This study was supported in part by JSPS KAKENHI Grant Numbers 22KJ1658, 23K19082, and 21H04507 and by the Ministry of Education, Culture, Sports, Science and Technology (MEXT) of Japan, under its Earthquake and Volcano Hazards Observation and Research Program.
\end{acknowledgments}

\section*{Data availability}

The data underlying this article are available from \citet{bhattacharya2015critical}. 

\bibliographystyle{gji}


\appendix

\renewcommand{\theequation}{\thesection.\arabic{equation}}
\renewcommand{\thefigure}{\thesection.\arabic{figure}}
\section{Predictions of different evolution laws during quasi-stationary contact}
\label{sec:SHSHSprinciple}
\setcounter{equation}{0}
\setcounter{figure}{0}

This appendix \ref{sec:SHSHSprinciple} presents the behaviors expected of different evolution laws discussed in the present paper (i.e., the aging, slip, composite, and modified composite laws) during constant-$\tau$ SHS tests explained below. 
In this type of SHS test, following a steady-state sliding at $V=V_{\rm prior}$, shear stress is abruptly reduced by $\Delta \tau$ ($>0$) to a designated level $\tau_{\rm hold}$ ($< \tau_{\rm SS}(V_{\rm prior})$) and held constant for a designated time duration $t_{\rm h}$. 
Regarding the log-t healing behavior, we focus on constant-$\tau$ SHS tests because the NSC can be ensured throughout the hold period by setting the $\tau_{\rm hold}$ sufficiently low. 

For all the four evolution laws considered here (unless $V\lesssim V_{\rm c}$ for the composite law), initial conditions at the beginning of a quasi-stationary hold period (defined as $t=+0$) are given as follows: 
\begin{flalign}
    V|_{\rm t=+0}&=:V_{\rm init}=V_{\rm prior}e^{-\Delta\tau/A}
    \label{eq:initialV}
    \\
    \theta|_{\rm t=+0}&=:\theta_{\rm init}=L/V_{\rm prior}
    \label{eq:initialtheta}
\end{flalign}
Equation~(\ref{eq:initialV}) derives from eq.~(\ref{RSFeq:constitutivetheta}), irrespective of the evolution law here considered. 
Note that eq.~(\ref{eq:initialtheta}) assumes each evolution law satisfies $\Omega_{\rm SS}=1$ (eq.~\ref{RSFeq:steadystateconstraint}). 
This is not always the case, for instance, for the composite law, where $\Omega_{\rm SS}$ is given by $\Omega_{\rm SS}\ln\Omega_{\rm SS}=\exp(-V/V_{\rm c})$. Associated remarks will appear in \S\ref{sec:SHSsaturation} and \S\ref{sec:SHSanalysiscomposite}. Throughout Appendix \ref{sec:SHSHSprinciple}, we also assume $D=L=D_{\rm c}$ and $\theta\gg \theta_{\rm X}$. 

By integrating respective evolution laws, coupled with the RSF constitutive law (eq.~\ref{RSFeq:constitutivetheta}), we will examine how the state evolves during the quasi-stationary hold period. 
The constant-stress condition ($\dot\tau=0$) reduces eq.~(\ref{RSFeq:constitutivetheta}) to the following total differential form, 
\begin{equation}
    A d\ln V= -B d\ln \theta. 
    \label{eq:stressholdVthetarelation}
\end{equation}
Given eq.~(\ref{eq:stressholdVthetarelation}), the solution of the above initial value problem for any given evolution law satisfies
\begin{equation}
    V/V_{\rm init}= (\theta/\theta_{\rm init})^{-B/A}
    \label{eq:Vasthetafunc}
\end{equation}
\begin{equation}
    \Omega/\Omega_{\rm init}=(\theta/\theta_{\rm init})^{1-B/A}
    \label{eq:Omegaasthetafunc}
\end{equation}
where $\Omega_{\rm init}:=\Omega|_{t=+0}$. 
Equation (\ref{eq:Vasthetafunc}) shows that $V$ decreases as the interface strengthens under the constant $\tau_{\rm hold}$ imposed. 
Equation (\ref{eq:Omegaasthetafunc}) shows that $\Omega$ remains constant if $A=B$, monotonically increases if $A>B$, and monotonically decreases if $A<B$. Using eqs.~(\ref{eq:Vasthetafunc}) and (\ref{eq:Omegaasthetafunc}), we will analyze the evolution of $\theta$ (and $\Phi$ via eq.~\ref{RSFeq:conversionofthetatophinocut}) for each evolution law in \S\ref{sec:SHSaginglaw}--\ref{sec:SHSanalysismodifiedcomposite}, after cautioning a problem of constant-$\tau$ SHS tests for interfaces with $A>B$ in \S\ref{sec:SHSsaturation}. We focus on behaviors under the NSC by setting $\tau_{\rm hold}$ sufficiently low such that
\begin{equation}
    \Omega_{\rm init}=e^{-\Delta\tau/A}< \beta,
    \label{eq:initialomega}
\end{equation}
excepting eqs.~(\ref{eq:compthetalast}) and (\ref{eq:modcompthetalast}). 

\begin{figure*}
   \includegraphics[width=133mm]{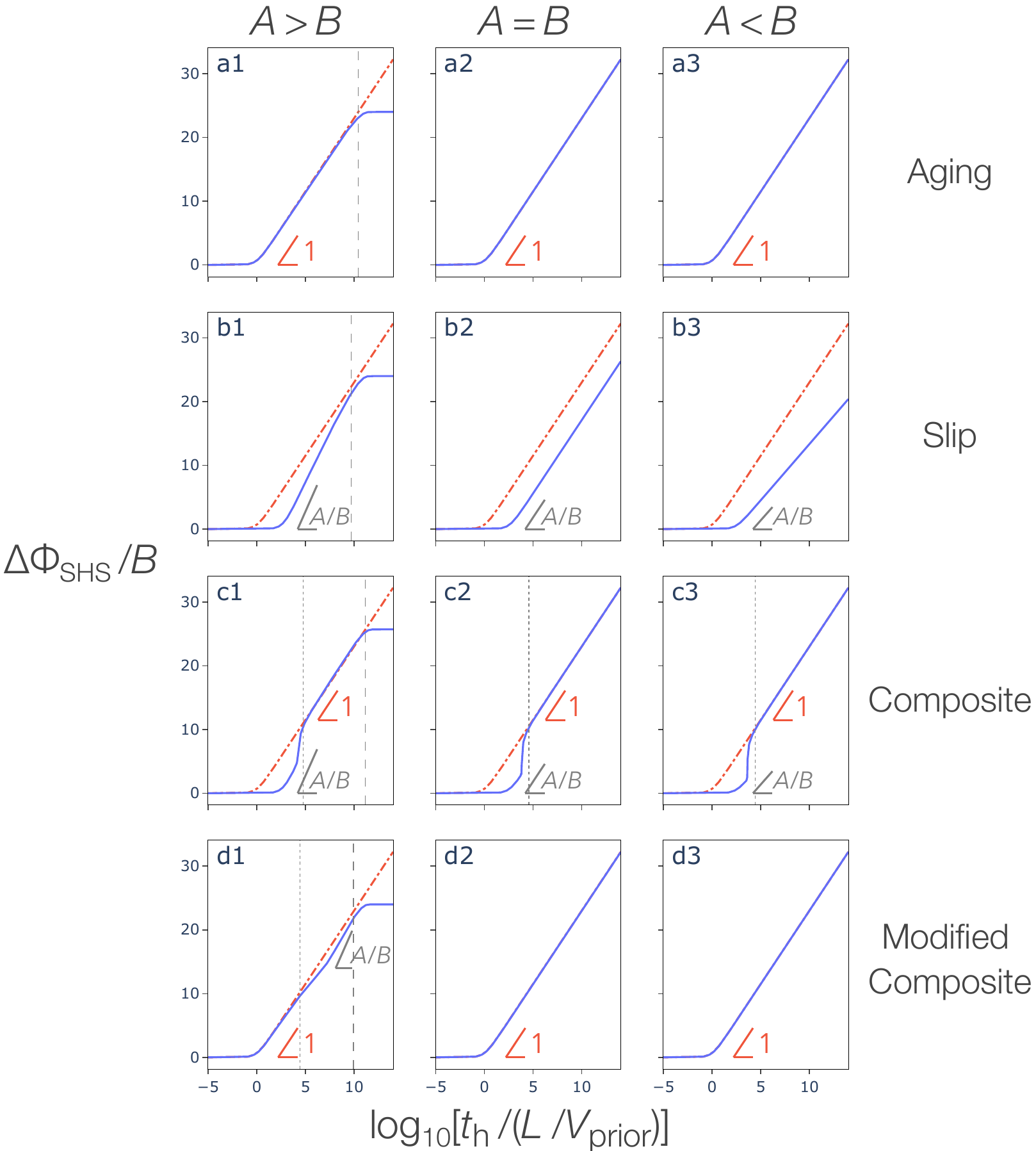}
  \caption{Predictions of the aging, slip, composite, and modified-composite laws in the constant-$\tau$ hold SHS test (blue lines), compared to the canonical behavior (corresponding to eq.~\ref{eq:fSHS_constrained}, red dash-dots). 
  Strength increase $\Delta \Phi_{\rm SHS}=\Phi-\Phi_{\rm init}$ normalized by $B$ is shown in semilog plots versus normalized hold time $t_{\rm h}/(L/V_{\rm prior})$.
  $B/A$ is set to $2/3$, $1$, and $4/3$ to represent $A>B$, $A=B$, and $A<B$ cases, respectively. 
  $\Delta \tau/A$ is set to $8$, which suffices the NSC. 
  Composite-law panels use $V_{\rm init}/V_{\rm c}=100$ to compute the $V_{\rm init}\gg V_{\rm c}$ cases, which remarkably differ from the aging-law patterns. 
  Modified-composite-law panels use $\beta=0.01$. 
  Vertical lines and wedges indicate 
  analytic estimates of
  the transient times (saturation: dashed lines; switch: dotted lines) and slopes for the predicted healing curves, derived in Appendix~\ref{sec:SHSHSprinciple}.
}
  \label{fig:SHScomparison}
\end{figure*}

Figure~\ref{fig:SHScomparison} summarizes the exact evolutions obtained by numerical integration for all cases of the present concern. We will explain each case through analytical evaluations and discuss how it agrees or disagrees with the canonical behavior of log-t healing (\S\ref{subsec:SHSrequirement} and \S\ref{sec:34}).

\subsection{Healing saturation---a flaw of constant-$\tau_{\rm hold}$ tests for frictional interfaces with $A>B$}
\label{sec:SHSsaturation}

Figure~\ref{fig:SHScomparison} suggests that, for interfaces with $A > B$, all of the four evolution laws predict the eventual saturation of healing after a certain $t_{\rm h}$. As explained below, this saturation occurs in constant-$\tau$ holds on interfaces with $A > B$, irrespective of the assumed evolution law. 

In general, $\Phi_{\rm SS}(V)$ increases during a hold period because $V$ decreases as healing proceeds (i.e., $\theta$ increases, eq.~\ref{eq:Vasthetafunc}). When $A \leq B$, while both $\Phi$ and $\Phi_{\rm SS}$ increases with time, $\Phi$ never catches up with $\Phi_{\rm SS}$; in fact, eq.~(\ref{eq:Omegaasthetafunc}) guarantees that $\Phi$ more and more overwhelms $\Phi_{\rm SS}$. For $A > B$, the situation is the opposite. As healing proceeds, $\Phi$ increases and comes closer to $\Phi_{\rm SS}$, which also increases but at a slower pace. Hence, $\Phi$ eventually catches up, achieving a steady state, whereby healing saturates by the convergence to a steady state suitable for the imposed $\tau_{\rm hold}$. Afterward, $\Phi$ remains at that plateau level. 
This mechanism has been discussed by \citet{ruina1983slip} in terms of the stability of the RSF behavior under a constant $\tau$. 

We here estimate how long it takes to reach such saturation for $A>B$. For a quick estimate, we assume that the canonical $B\ln t$ healing (eq.~\ref{eq:fSHS_constrained}) holds until the time of saturation. Then, eq.~(\ref{eq:Omegaasthetafunc}) becomes 
\begin{equation}
    \Omega/\Omega_{\rm init}=(1+t/t_{\rm c})^{1-B/A}.
    \label{eq:SHScanonicalOmega}
\end{equation}
For $A > B$, eq.~(\ref{eq:SHScanonicalOmega}) predicts that $\Omega = \Omega_{\rm SS}$ ($=1$, eq.~\ref{RSFeq:steadystateconstraint}) will take place at a positive finite time $t_*$: 
\begin{equation}
    t_*= t_{\rm c}\left(\Omega_{\rm init}^{-1/(1-B/A)}-1\right),
    \label{eq:tstar}
\end{equation}
which is an estimate of the saturation time. 
Note that this estimate does not apply to the slip law, which includes no purely time-dependent healing, that is, $f_t = 0$ (or equivalently, $F_t = 0$). Although the slip law does predict log-t healing \citep[e.g.,][also see Fig.~\ref{fig:SHScomparison}]{ruina1983slip}, its mechanism is slip-dependent strengthening, where $\Phi$ chases the moving target $\Phi_{\rm SS}(V)$ floating above (cf. eq.~\ref{eq:sliplawinddelta}). 
Thus, as detailed in \S\ref{sec:SHSsliplaw}, the `apparent' healing predicted by the slip law is $A\ln t$ instead of the canonical $B\ln t$ predictable by the other three laws, all of which involve purely time-dependent healing. The estimate of the saturation time $t_*^\prime$ for the slip law will be given in \S\ref{sec:SHSsliplaw}. 

We now flip our attention to the strength in the plateau stage after healing saturation (see Fig.~\ref{fig:SHScomparison}, left column). As long as the canonical SS behavior (eq.~\ref{eq:req4ss}) holds, eq.~(\ref{eq:Omegaasthetafunc}) tells that at the saturation plateau,
\begin{equation}
    \theta/\theta_{\rm init}=(\Omega_{\rm SS}/\Omega_{\rm init})^{1/(1-B/A)},
    \label{eq:plateaulevelrelation}
\end{equation}
which holds even when the canonical SHS behavior (eq.~\ref{eq:fSHS_constrained}) is violated. 
Consequently, the strength increase $\Delta\Phi_{\rm SHS}$ (in Fig.~\ref{fig:SHScomparison}) maintained in the plateau stage is 
\begin{equation}
B\ln(\theta/\theta_{\rm init})=[B/(1-B/A)]\ln(\Omega_{\rm SS}/\Omega_{\rm init}).   
\label{eq:netstrengthvariation_SHS}
\end{equation}
The right hand side of eq.~(\ref{eq:netstrengthvariation_SHS}) takes the same value as long as $\Omega_{\rm SS}(V)=const.$ (eq.~\ref{RSFeq:Cssconst}) holds. 
Therefore, the predicted value of the saturated strength enables us to tell if the assumed evolution law obeys the canonical SS behavior. 
In contrast, the composite law gives 
$\Omega_{\rm SS}\simeq 1$ for $V\gg V_{\rm c}$ but gives $\Omega_{\rm SS}\simeq 1/W(1)(=1.763...)$ for $V\ll V_{\rm c}$, where $W(\cdot)$ denotes the Lambert W function. Hence, when $V$ crosses $V_{\rm c}$ during the hold, 
the composite law violates eq.~(\ref{RSFeq:Cssconst}) and predicts 
the strength at the saturation plateau to be greater than expected from the canonical SS behavior. 

Lastly, we emphasize that the long-time saturation predicted for $A > B$ (Fig.~\ref{fig:SHScomparison}, left column) should not be taken as a flaw of the evolution law; rather, this healing saturation simply represents a flaw of the constant-$\tau$ healing tests. As explained earlier in \S\ref{sec:SHSsaturation}, for interfaces with $A > B$, this testing scheme is fated to end up with the steady state expected of the designated $\tau_{\rm hold}$, no longer functioning as a healing test. Nevertheless, we also note that, even for $A > B$, constant-$\tau$ SHS tests with $t_{\rm h} \ll t_*$ (or $\ll t_*^\prime$ for the slip law) can test laws' ability to reproduce the canonical healing behavior. Note $t_*$ (and $t_*^\prime$) can be extended as desired by employing a lower $\tau_{\rm hold}$.

\subsection{Aging law}
\label{sec:SHSaginglaw}
For $A\leq B $, $\Omega$ keeps decreasing when $\dot \theta>0$ (eq.~\ref{eq:Omegaasthetafunc}). Then, $\Omega\ll1$ holds forever if we set $\Omega_{\rm init} \ll 1$ by imposing a sufficiently low $\tau_{\rm hold}$. In this situation, $f$ of the aging law (eq.~\ref{eq:defofaginglaw}) is reduced to approximately unity throughout the hold period. Hence
\begin{equation}
    \theta_{\rm aging}\simeq 
     t+t_{\rm c|aging} \hspace{10pt}(A\leq B),
     \label{eq:agingevolutioninSHS1}
\end{equation}
where $t_{\rm c|aging}$ denotes the following approximate representation of the cutoff time for the aging law:
\begin{equation}
    t_{\rm c|aging}=\theta_{\rm init}=L/\rm V_{\rm prior},
    \label{eq:tcagingistcexpected}
\end{equation}
which agrees with the empirical relation eq.~(\ref{eq:tcdetection});  
in Appendix~\ref{sec:SHSHSprinciple}, variables conditioned by the name of an evolution law denote predictions by the evolution law. 
We see that the canonical $B\ln t$ healing (eqs.~\ref{RSFeq:PhiSHSNSCrawtc} and \ref{RSFeq:coincidenceofBs}) 
by converting eq.~(\ref{eq:agingevolutioninSHS1}) into its $\Phi$-notation form via eq.~(\ref{RSFeq:conversionofthetatophinocut}). 
The predicted healing can be found in subpanels a2 and a3 in Fig.~\ref{fig:SHScomparison}. 

By contrast, for $A>B$, $\Omega$ is strictly monotonically increasing, and the healing will saturate within a finite hold time $t_{*|{\rm aging}}$ as explained in \S\ref{sec:SHSsaturation}. Nonetheless, as long as $\tau_{\rm hold}$ is chosen so that $\Omega_{\rm init} \ll 1$, the earlier part of the hold (i.e., $t \ll t_{*|{\rm aging}}$) observes the same canonical $B\ln t$ healing as seen for $A\leq B$ interfaces. Thus, the prediction of the aging law for $A>B$ is summarized as
\begin{equation}
    \theta_{\rm aging}\sim 
    \begin{cases}
     t+t_{\rm c|aging} &(A> B\cap t\ll t_{*|\rm aging})
     \\
     t_*+t_{\rm c|aging} & (A> B\cap t\gtrsim t_{*|\rm aging})
     \end{cases}
     \label{eq:agingevolutioninSHS2}
\end{equation}
Plugging eq.~(\ref{eq:tcagingistcexpected}) into eq.~(\ref{eq:tstar}), we obtain an analytic expression of $t_{*|{\rm aging}}$: 
\begin{equation}
    t_{\rm *|aging}=\theta_{\rm init}(\Omega_{\rm init}^{1/(B/A-1)}-1).
    \label{eq:tstaraging}
\end{equation}
Note that eq.~(\ref{eq:agingevolutioninSHS2}) has only an order-level accuracy for $t\sim t_{*|\rm aging}$ because our derivation assumed the continuation of the canonical healing (i.e., $\dot\theta = 1$) up to the eventual steady state, whereas actual $\dot\theta$ gradually decreases (but remains to be of $\mathcal O(1)$). 
This approximation error does not propagate to the later phase of $t\gg t_{*|\rm aging}$ because $t_{\rm *|aging}$ is defined as the time when the canonical log-t healing reaches the ceiling $\Omega=\Omega_{\rm SS}$, given that the strength increment after saturation depends on $\Omega_{\rm init}/\Omega_{\rm SS}$ only (\S\ref{sec:SHSsaturation}, eq.~\ref{eq:plateaulevelrelation}).
The symbol $\sim$ in Appendix~\ref{sec:SHSHSprinciple} denotes similar rough evaluations.

\subsection{Slip law}
\label{sec:SHSsliplaw}

The slip law includes no purely time-dependent healing (i.e., $f_{t|{\rm slip}} = 0$). Nevertheless, it has been known that the slip law does predict, as mentioned in \S\ref{subsubsec:331}, `apparent' healing for conditions expected during the hold periods of many conventional SHS tests~\citep{ruina1983slip,beeler1994roles}, where the motion of the shear-loading piston is halted. As seen from eq.~(\ref{eq:sliplawinddelta}) below, this apparent healing is strictly a slip-driven increase of the strength towards $\Phi_{\rm SS}(V)$, which is higher than $\Phi_{\rm init}$ because $V$ during a hold period is lower than $V_{\rm prior}$. 
However, as mentioned in the introductory part of \S\ref{subsec:SHSrequirement}, the existence of purely time-dependent healing has been unambiguously proven~\cite[e.g.,][]{nakatani1996effects} by the observation of significant log-t healing in SHS$|$NSC tests employing holds during which the shear stress was kept sufficiently low (\S\ref{subsubsec:331}). In this subsection \S\ref{sec:SHSsliplaw}, we analyze the behavior of the slip law during holds at a constant $\tau_{\rm hold}$ sufficiently lower than $\tau_{\rm SS}(V_{\rm prior})$. 

The slip law is often easier to analyze when worked in the $\Phi$-notation. 
Using the conversion formula eq.~(\ref{RSFeq:conversionofthetatophinocut}), eq.~(\ref{eq:defofsliplaw}) is rewritten as follows: 
\begin{equation}
    \frac{d\Phi}{d\delta } = -D_{\rm c}^{-1}[\Phi-\Phi_{\rm SS}(V)].
    \label{eq:sliplawinddelta}
\end{equation}
Besides, when $d\tau=0$, eq.~(\ref{eq:stressholdVthetarelation}) and 
the logarithmic $V$ dependence of the steady-state strength
(eq.~\ref{eq:req4ss}, satisfied by the slip law) lead to 
\begin{equation}
    d\Phi_{\rm SS}(V) =-B d\ln V = (B/A)d\Phi.
    \label{eq:PhiversusPhiSS_constanttau}
\end{equation}
Therefore, during a constant-$\tau$ hold, where $d\tau=0$, 
we find the following from eq.~(\ref{eq:PhiversusPhiSS_constanttau}) and
the differentiation of eq.~(\ref{eq:sliplawinddelta}) with respect to the slip $\delta$: 
\begin{equation}
    \frac{d^2 \Phi}{d\delta^2 }= -(1-B/A)D_{\rm c}^{-1} \frac{d \Phi}{d\delta }.
    \label{eq:dphiODE}
\end{equation}
Equation~(\ref{eq:dphiODE}) is a differential equation of $d\Phi/d\delta$ and has the following solution for the initial condition given through eq.~(\ref{eq:sliplawinddelta}): 
\begin{equation}
    \frac{d\Phi}{d\delta}= -\frac{\Phi_{\rm init}-\Phi_{\rm SS}(V_{\rm init})}{D_{\rm c}}e^{-(1-B/A)\delta/D_{\rm c}},
    \label{eq:dphiddeltasolution}
\end{equation}
where $\Phi_{\rm init}$ (eq.~\ref{eq:Phiinitvalue}) is the $\Phi$ value that corresponds to $\theta_{\rm init}$ via eq.~(\ref{RSFeq:conversionofthetatophinocut}).
Integrating eq.~(\ref{eq:dphiddeltasolution}) once more and noticing $\Phi-\Phi_{\rm SS}=B\ln \Omega$, we have, for $A\neq B$, 
\begin{equation}
    \Phi= \Phi_{\rm init}-\frac{B\ln \Omega_{\rm init}}{1-B/A}\left(1-e^{-(1-B/A)\delta/D_{\rm c}}\right).
    \label{eq:phiforslip}
\end{equation}
The solution for $A=B$ is equal to the $A/B\to1$ limit of eq.~(\ref{eq:phiforslip}) as long as $\delta/D_{\rm c}$ is finite. 
Thus, for brevity, the subsequent analysis in this section \S\ref{sec:SHSsliplaw} treats the $A=B$ case as the $A/B\to1$ limit. 

\subsubsection{Predicted behaviors for $t\ll t_{\rm c|slip}(\Omega_{\rm init}^{-1/|A/B-1|}-1)$}
\label{subsubsec:moderateholdforslip}
When converted into the $\Phi$-notation via eq.~(\ref{RSFeq:conversionofthetatophinocut}), eq.~(\ref{eq:Vasthetafunc}) becomes  
\begin{equation}
    \frac{d\delta}{dt}= V_{\rm init} e^{-(\Phi-\Phi_{\rm init})/A},
    \label{eq:VisfuncofPhi}
\end{equation}
Equation~(\ref{eq:VisfuncofPhi}), combined with eq.~(\ref{eq:phiforslip}), fully determines the time evolution of $\Phi(t)$ during a constant-$\tau$ hold period. 
Its exact solution is not in a closed form (see eq.~\ref{eq:whatdirtysolution}), 
but as below we can obtain a simple closed-form solution as long as 
\begin{equation}
|1-B/A|\delta/D_{\rm c}\ll1,  
\label{eq:smallslipdistanceforsliplaw}
\end{equation}
that is, when the slip during the hold is relatively small. 
As shown in \S\ref{subsubsec:longholdforslip},
$\Omega\sim 1$ can be realized outside this range so that eq.~(\ref{eq:smallslipdistanceforsliplaw}) lies in $\Omega\ll1$, where the canonical $f_{t|{\rm NSC}}$ (eq.~\ref{eq:ftNSC_constrained}, \S\ref{subsec:SHSrequirement} in the text) and $f_\delta V=-\Omega\ln\Omega$ of the slip law contradict each other. 
Therefore, subsequent predictions in this subsection are all we need to compare against the canonical healing behavior to (in)validate the slip law. 
Taking the lowest order of $(1-B/A)\delta/D_{\rm c}$, eq.~(\ref{eq:phiforslip}) reduces to 
\begin{equation}
    \Phi-\Phi_{\rm init}\simeq -(B\ln \Omega_{\rm init})\delta/D_{\rm c}.
    \label{eq:phiforslip2}
\end{equation}
Substituting eq.~(\ref{eq:phiforslip2}) into eq.~(\ref{eq:VisfuncofPhi}), we have
\begin{equation}
    \frac{d\delta}{dt}\simeq V_{\rm init} e^{-[(B/A)\ln(1/\Omega_{\rm init})] \delta /D_{\rm c}}.
\end{equation}
Its solution is
\begin{equation}
    \delta \simeq \frac{AD_{\rm c}}{B\ln (1/\Omega_{\rm init})} \ln(1+t/t_{\rm c|slip}),
    \label{eq:slipfortime}
\end{equation}
which and eq.~(\ref{eq:phiforslip2}) yield
\begin{equation}
    \Phi\simeq \Phi_{\rm init}+ A\ln (1+t/t_{\rm c|slip});
    \label{eq:phifortime}
\end{equation}
$t_{\rm c|slip}$, appearing both in eqs.~(\ref{eq:slipfortime}) and (\ref{eq:phifortime}), is the lower cutoff time for the log-t evolution of the slip and strength. In the course of derivation above, the following approximate expression of $t_{\rm c|slip}$ is obtained as well: 
\begin{equation}
    t_{\rm c|slip}=\frac{D_{\rm c}}{V_{\rm init}}\frac A{B\ln (1/\Omega_{\rm init})}.
    \label{eq:tcsliplaw}
\end{equation} 

Equation~(\ref{eq:phifortime}) demonstrates that, at least when the slip that has occurred during the hold is still small enough, the slip law does predict a healing phenomenon linear with the logarithm of the time elapsed since the beginning of the constant-$\tau$ hold. However, the predicted slope of the log-t healing is $A$, instead of $B$, and clearly disagrees with the canonical SHS behavior (see Fig.~\ref{fig:SHScomparison}, b1-b3 for confirmation by the exact numerical solutions as well). 
Furthermore, eq.~(\ref{eq:tcsliplaw}), combined with eqs.~(\ref{eq:initialV}) and (\ref{eq:initialomega}), tells that the slip law overestimates the cutoff time by a factor of $A/[B\Omega_{\rm init}\ln (1/\Omega_{\rm init})]$ compared with the canonical value (eq.~\ref{eq:tcdetection}), which is correctly reproduced by the aging law: 
\begin{equation}
    t_{\rm c|slip}/t_{\rm c|aging}=\frac AB \frac{1}{\Omega_{\rm init}\ln (1/\Omega_{\rm init})}\left(=\frac {A^2} {B\Delta \tau}e^{\Delta \tau/A}\right). 
    \label{eq:tcslippertcaging}
\end{equation}
This overprediction of the cutoff time is visible in Fig.~\ref{fig:SHScomparison}, where the slip-law prediction (blue curve) is shifted to the right by approximately 100 times compared with the canonical healing (red dash-dotted curve).

We present a few more useful relations regarding the slip law for the convenience of call-ups from the other sections. 
Differentiation of eq.~(\ref{eq:slipfortime}) leads to 
\begin{equation}
    V\simeq V_{\rm init}/(1+t/t_{\rm c|slip}).
    \label{eq:Vfortime}
\end{equation}
Equation~(\ref{RSFeq:conversionofthetatophinocut}) converts eq.~(\ref{eq:phifortime}) into the $\theta$-notation as 
\begin{equation}
    \theta_{\rm slip}\simeq \theta_{\rm init}(1+t/t_{\rm c|slip})^{A/B}.
    \label{eq:thetafortime}
\end{equation}
From eqs.~(\ref{eq:Vfortime}) and (\ref{eq:thetafortime}), we have 
\begin{equation}
    \Omega\simeq \Omega_{\rm init}/(1+t/t_{\rm c|slip})^{1-A/B}.
    \label{eq:Omegafortime}
\end{equation}

Equations~(\ref{eq:slipfortime}), (\ref{eq:phifortime}), and (\ref{eq:Vfortime})--(\ref{eq:Omegafortime}) are valid for $|1-B/A|\delta/D_{\rm c}\ll1$, inheriting the validity range of eq.~(\ref{eq:phiforslip2}), which has been the lowest-order approximation with respect to $|1-B/A|\delta/D_{\rm c}$. 
Because $|1-B/A|\delta/D_{\rm c}\ll1$ holds forever when $A/B\to 1$, these solutions become exact without limitation for $A=B$.  
On the other hand, their approximation capability has an upper time limit for $A\neq B$.
If we extrapolate eq.~(\ref{eq:slipfortime}) beyond its validity range, $|1-B/A|\delta/D_{\rm c}$ reaches unity at time $t_*^\prime$ such that 
\begin{equation}
     \ln (1+t_*^\prime/t_{\rm c|slip})=\frac{\ln(1/\Omega_{\rm init})}{|A/B-1|},
\end{equation}
or in an explicit form,
\begin{equation}
     t_*^\prime=t_{\rm c|slip}\left(\Omega_{\rm init}^{-1/|A/B-1|}-1\right).
     \label{eq:tstarprimefromtcslip}
\end{equation}
It is safe to say that the 1st-order approximations eqs.~(\ref{eq:slipfortime}), (\ref{eq:phifortime}), and (\ref{eq:Vfortime})--(\ref{eq:Omegafortime}) are valid for $t\ll t_*^\prime$. 
Equations~(\ref{eq:tcagingistcexpected}), (\ref{eq:tcslippertcaging}), and (\ref{eq:tstarprimefromtcslip}) yield a convenient expression of $t_*^\prime$ in terms of the initial values of $\theta$ and $\Omega$: 
\begin{equation}
     t_*^\prime=\frac AB \frac{\Omega_{\rm init}^{-1/|A/B-1|}-1}{\Omega_{\rm init}\ln (1/\Omega_{\rm init})}\theta_{\rm init}.
     \label{eq:tstarprime}
\end{equation}

\subsubsection{Predicted behaviors for $t\gtrsim t_{\rm c|slip}(\Omega_{\rm init}^{-1/|A/B-1|}-1)$}
\label{subsubsec:longholdforslip}

Now we show the prediction of the slip law for $t \gtrsim t_*^\prime$, separately for $A > B$ and $A < B$. 
For $A=B$, the $A\ln t$ healing (eq.~\ref{eq:phifortime}) is predicted for any positive $t$, since $t_*^\prime$ goes infinity when $A\to B$. 

For $A > B$, extrapolations of eqs.~(\ref{eq:Omegafortime}) and (\ref{eq:tstarprimefromtcslip}) predict 
$\Omega$ reaches $\Omega_{\rm SS} = 1$ at $t_*^\prime$, 
telling that the $A\ln t$ healing under the slip law saturates around $t_*^\prime$, as illustrated by the numerical result Fig.~\ref{fig:SHScomparison} (b1). 
Therefore, in the same sense as that $t_*$ (eq.~\ref{eq:tstar}), defined with the dare extrapolation of the canonical $B\ln t$ healing, works as a rough estimate of the saturation time for the aging law (\S\ref{sec:SHSaginglaw}), we can say that $t_*^\prime$ works as a rough estimate of the saturation time predicted by the slip law on the interfaces with $A > B$. Consequently, via derivation similar to that in \S\ref{sec:SHSaginglaw}, which has led to eq.~(\ref{eq:agingevolutioninSHS2}) for the aging law, we can conclude 
\begin{equation}
    \theta_{\rm slip}(t)\sim \theta_{\rm init}[1+\min(t,t_*^\prime)/t_{\rm c|slip}]^{A/B} \hspace{10pt}(A> B).
    \label{eq:SHSsliplawstablecase}
\end{equation}
Equation~(\ref{eq:SHSsliplawstablecase}) represents $A\ln t$ healing and its saturation at the plateau level of 
eq.~(\ref{eq:plateaulevelrelation}). Since the slope $B_{\rm heal}$ of the log-t healing predicted by the slip law is $A$ (eq.~\ref{eq:phifortime}) instead of the canonical $B$, $t_*^\prime$ (eq.~\ref{eq:tstarprime}) is not identical to $t_*$ (eq.~\ref{eq:tstar}). 
Nonetheless, a logical parallel is present between eqs.~(\ref{eq:tstar}) and (\ref{eq:tstarprime}); 
$t_*^\prime$ in eq.~(\ref{eq:tstarprime}) is equal to $t_*$ in eq.~(\ref{eq:tstar}) after $A$ and $B$ are interchanged, and then eqs.~(\ref{eq:tstarprime}) and (\ref{eq:tstar}) are consistent in the lowest-order approximation with respect to $|A-B|/A$. 

Now, we move to the more challenging case of $A < B$. For $t \gg t_*^\prime$,  there is no a priori reason to expect that $\theta(t)$ does not deviate significantly from the $A\ln t$ healing (eq.~\ref{eq:phifortime}). However, the numerical solution shown in Fig.~\ref{fig:SHScomparison} (b3) indicates no appreciable deviation, hinting that $A\ln t$ may hold forever if $A < B$:
\begin{equation}
    B\ln[\theta_{\rm slip}(t)/\theta_{\rm init}]\sim A\ln(1+t/t_{\rm c|slip}) \hspace{10pt}(A< B).
    \label{eq:SHSsliplawunstablecase}
\end{equation}
Using the parameter sets the simulation of Fig.~\ref{fig:SHScomparison} adopts, we have confirmed that eq.~(\ref{eq:SHSsliplawunstablecase}) agrees with the numerical results within a \%-order accuracy up to $t = 10^{100} \times (L/V_{\rm prior})$, the maximum simulation time we tried. In fact, we can provide a proof of eq.~(\ref{eq:SHSsliplawunstablecase}), which occupies the rest of this subsection.

Equation~(\ref{eq:SHSsliplawunstablecase}) is an order estimate, and thus eq.~(\ref{eq:SHSsliplawunstablecase}) has already been proven for $t \lesssim t_*^\prime$ through the derivation of eq.~(\ref{eq:phifortime}). The following calculation then focuses on the validity of eq.~(\ref{eq:SHSsliplawunstablecase}) for $t \gg t_*^\prime$. 
Throughout the remainder of the proof, we suppose $A < B$ and $t \gg t_*^\prime$, where we derive eq.~(\ref{eq:SHSsliplawunstablecase}) via a perturbation theory by utilizing the fact that healing has proceeded well when $t\gg t_*^\prime$ (eqs.~\ref{eq:thetafortime} and \ref{eq:tstarprimefromtcslip}). 

When eq.~(\ref{eq:Omegaasthetafunc}) holds, 
the slip law (eq.~\ref{eq:defofsliplaw}) reduces to
\begin{equation}
    \dot\theta=-\Omega_{\rm init}(\theta/\theta_{\rm init})^{1-B/A}\ln[\Omega_{\rm init}(\theta/\theta_{\rm init})^{1-B/A}].
    \label{eq:constantstresssliplaw}
\end{equation}
As is the case in eq.~(\ref{eq:constantstresssliplaw}), 
it has recently been pointed out by \citet{noda2023tertiary} that the constant stress condition reduces an evolution law of the $\dot\theta=f(V,\theta)$ form to a unary function of $\theta$ via eq.~(\ref{eq:Vasthetafunc}).
The variable separation integral of eq.~(\ref{eq:constantstresssliplaw}) is 
\begin{equation}
    -\Omega_{\rm init}t=\theta_{\rm init}
    \int^{\theta/\theta_{\rm init}}_1 \frac{dx}{x^{1-B/A}\ln(\Omega_{\rm init}x^{1-B/A})},
\end{equation}
which is solved by using the exponential integral ${\rm Ei}(\cdot)$ as
\begin{equation}
    -\Omega_{\rm init}t=\theta_{\rm init}\frac{\Omega_{\rm init}^{1/(1-A/B)}}{1-B/A}{\rm Ei}\left[
    \ln\left(
    \Omega_{\rm init}^{-1/(1-A/B)}x^{B/A}
    \right)
    \right]^{\theta/\theta_{\rm init}}_1.
    \label{eq:whatdirtysolution}
\end{equation}
Further, 
by using 
$\Omega_{\rm init}^{-1/(1-A/B)}(\theta/\theta_{\rm init})^{B/A}\gg1$ for $t\gg t_*^\prime$ (from eqs.~\ref{eq:thetafortime} and \ref{eq:tstarprimefromtcslip})
and the following lowest order representation of ${\rm Ei}(\ln(x))$ for $|\ln x|\gg1$, 
\begin{equation}
    {\rm Ei}(\ln(x))=x/\ln(x)+\mathcal O[x/\ln^2 (x)],
\end{equation}
we can reduce eq.~(\ref{eq:whatdirtysolution}) to
\begin{equation}
    -\Omega_{\rm init}t\sim \frac{\theta_{\rm init}\Omega_{\rm init}^{1/(1-A/B)}}{1-B/A}
    \left\{
    \frac{\Omega_{\rm init}^{-1/(1-A/B)}(\theta/\theta_{\rm init})^{B/A}}{\ln\left(\Omega_{\rm init}^{-1/(1-A/B)}(\theta/\theta_{\rm init})^{B/A}\right)}
    -
    {\rm Ei}\left[
    \frac{\ln(1/\Omega_{\rm init})}{(1-A/B)}
    \right]
    \right\}.
    \label{eq:unstablesliplawasymptote}
\end{equation}
Canceling out the signs on both sides in eq.~(\ref{eq:unstablesliplawasymptote}) given $A<B$ ($1-B/A<0$), 
then taking the logarithm, 
and finally utilizing the representation of $t_{\rm c|slip}$ (from eqs.~\ref{eq:tcagingistcexpected} and \ref{eq:tcslippertcaging}), 
we find the leading order of eq.~(\ref{eq:unstablesliplawasymptote}) satisfies
\begin{equation}
    \ln(t/t_{\rm c|slip})\sim (B/A)\ln(\theta/\theta_{\rm init})+
    \mathcal O[\ln\ln(\theta/\theta_{\rm init})]+\mathcal O(\ln\ln\Omega_{\rm init}).
\end{equation}
It shows that the slip law keeps $A\ln t$ healing even for $t\gg t_*^\prime$ when $A<B$:
\begin{equation}
    \frac{B\ln(\theta/\theta_{\rm init})}{A\ln(t/t_{\rm c|slip}+1)}\sim 1.
\end{equation}

\subsection{Composite law}
\label{sec:SHSanalysiscomposite}

The behavior of the composite law (eq.~\ref{eq:defofcompositelaw}) differs a lot depending on $V_{\rm init}/V_{\rm c}$. 
The numerical solutions in Fig.~\ref{fig:SHScomparison} (c1-c3) are for $V_{\rm init}/V_{\rm c}\gg 1$, where the predicted healing significantly differs from the canonical log-t healing. 
We start the analysis below with the other, more straightforward case of $V_{\rm init}/V_{\rm c}\ll 1$, where the composite law reproduces the SHS$|$NSC canon. 

If $V_{\rm init}/V_{\rm c}\ll 1$, $V/V_{\rm c}\ll 1$ holds forever as $V$ monotonically decreases during constant-$\tau$ healing tests (eq.~\ref{eq:Vasthetafunc}). Therefore, the composite law is reduced to 
\begin{equation}
    \dot \theta \simeq 1-\Omega\ln\Omega \hspace{10pt} (V_{\rm init}\ll V_{\rm c}).
    \label{eq:lowVcomposite}
\end{equation}
Under the presently concerned NSC, $\Omega\ll 1$ so that eq.~(\ref{eq:lowVcomposite}) is approximated by $\dot \theta\simeq 1$, reproducing the canonical log-t healing. Note that, for $A > B$, the healing saturation explained in \S\ref{sec:SHSsaturation} occurs given the constancy of $\Omega_{\rm SS}$ (eq.~\ref{RSFeq:Cssconst}) throughout the hold period; the composite-law prediction of the plateaued strength value may be different from that of the aging law because the $\Omega_{\rm SS}$ value expected of eq.~(\ref{eq:lowVcomposite}) is $1/W(1)(=1.763...)$, rather than unity expected of the aging law, while the composite law predicts $\Omega_{\rm init}=e^{-\Delta \tau/A}$ as in the aging law for ordinary cases of $V_{\rm prior}\gg V_{\rm c}$. 
After all, we can conclude
\begin{equation}
    \theta_{\rm composite}\simeq \theta_{\rm aging} \hspace{10pt}(V_{\rm init}\ll V_{\rm c}).
\end{equation}

If $V_{\rm init}\gtrsim V_{\rm c}$, the composite law behaves almost like the slip law until the time $t_{\rm sa}$ when $V$ decelerates to $V_{\rm c}$, while it behaves almost like the aging law afterward. We can estimate this switching time $t_{\rm sa}$ by using the slip-law prediction of $V$ (eq.~\ref{eq:Vfortime}): 
\begin{equation}
    t_{\rm sa}= t_{\rm c|slip}(V_{\rm init}/V_{\rm c}-1).
    \label{eq:defoftsa}
\end{equation}
In cases where the healing saturation predicted by the slip law begins earlier than this switch from the slip law to the aging law, the evolution remains to follow the slip law, and thus healing saturation occurs while still in the slip-law regime; $\Phi$ stops increasing, and in turn, $V$ stops decreasing and stays at a value higher than $V_{\rm c}$. 
That is, when $A>B\cap t_{\rm sa}\gtrsim t_{*}^\prime$, the composite law behaves like the slip law over the entire time range: 
\begin{equation}
    \theta_{\rm composite }\simeq \theta_{\rm slip} \hspace{10pt}(V_{\rm init}\gtrsim  V_{\rm c}\cap A>B\cap t_{\rm sa}\gtrsim t_{*|\rm slip}).
\end{equation}
In the remaining cases [Fig.~\ref{fig:SHScomparison} (c1--c3) all deal with such cases], the evolution is slip-law-like until $t_{\rm sa}$, after which the evolution is the aging-law-like:
\begin{equation}
    \theta_{\rm composite}(t)\sim 
    \begin{cases}
    \theta_{\rm slip}(t) &(V_{\rm init}\gtrsim V_{\rm c}\cap t\lesssim t_{\rm sa})
    \\
    \theta_{\rm aging}(t-t_{\rm sa})|_{\theta(0)=\theta_{\rm slip}(t_{\rm sa}),\Omega(0)/\Omega_{\rm SS}=W(1)\theta_{\rm slip}(t_{\rm sa})V_{\rm c}/D_{\rm c}} &(V_{\rm init}\gtrsim V_{\rm c}\cap t\gg t_{\rm sa})
    \end{cases}
    \label{eq:compthetalast}
\end{equation}
where $\theta_{\rm slip}(t)$ before the crossover is evaluated with eq.~(\ref{eq:thetafortime}). 
For brevity, eq.~(\ref{eq:compthetalast}) specifies initial ($\theta$, $\Omega$) values in the lower part of the right-hand side (instead of eqs.~\ref{eq:initialtheta} and \ref{eq:initialomega} consistently used in this appendix section including the left-hand side of eq.~\ref{eq:compthetalast}); eq.~(\ref{eq:compthetalast}) for $t\gg t_{\rm sa}$ ($V\ll V_{\rm c}$) accounts for the fact that $\Omega_{\rm SS}$ changes by a factor of $1/W(1)$ as the decreasing $V$ crosses $V_{\rm c}$ from above. 
Note that because $t_{\rm sa}$ is defined as the time when $V=V_{\rm c}$, eq.~(\ref{eq:stressholdVthetarelation}) yields
\begin{equation}
    \theta_{\rm slip}(t_{\rm sa})= \theta_{\rm init}(V_{\rm init}/V_{\rm c})^{A/B},
    \label{eq:thetasliptsa}
\end{equation}
which is certainly satisfied by eqs.~(\ref{eq:thetafortime}) and (\ref{eq:defoftsa}). 

Equation~(\ref{eq:compthetalast}) also predicts the healing saturation after a long hold for $A>B$. 
Because the saturation time $t_{\rm *|aging}$ for $\theta_{\rm aging}$ is given by eq.~(\ref{eq:tstaraging}) when the initial condition is $(\theta(0),\Omega(0))=(\theta_{\rm init},\Omega_{\rm init})$, 
the saturation time of $\theta_{\rm composite}$ in eq.~(\ref{eq:compthetalast}) (here denoted by $t_{\rm *|composite}$) is expressed by eq.~(\ref{eq:tstaraging}) after the initial-value replacement and time translation as per eq.~(\ref{eq:compthetalast}):
\begin{flalign}
    t_{\rm *|composite}&=t_{\rm sa}+ \theta_{\rm slip}(t_{\rm sa})\left[\left(\frac{\theta_{\rm slip}(t_{\rm sa})V_{\rm c}}{D_{\rm c}}W(1)\right)^{1/(B/A-1)}-1\right]
    \\
    &=
    t_{\rm sa}+ \theta_{\rm init}(V_{\rm init}/V_{\rm c})^{A/B}\left\{(V_{\rm init}/V_{\rm c})^{\frac{A/B-1}{B/A-1}}[W(1)\Omega_{\rm init}]^{1/(B/A-1)}-1\right\}
\end{flalign}
where $\theta_{\rm slip}(t_{\rm sa})$ is evaluated by eq.~(\ref{eq:thetasliptsa}). 

\subsection{Modified composite law}
\label{sec:SHSanalysismodifiedcomposite}
For $A\leq B$, $\Omega$ does not increase during healing under a constant $\tau$ (eq.~\ref{eq:Omegaasthetafunc}). Hence, if the initial condition meets $\Omega <\beta$ as assumed throughout Appendix \ref{sec:SHSHSprinciple}, $\Omega$ remains below $\beta$ forever, and the modified composite law remains to approximate the aging law over the entire time range: 
\begin{equation}
    \theta_{\rm mod.composite}\sim \theta_{\rm aging} \hspace{10pt}(A\leq B).
    \label{eq:modcompsimilartoaging}
\end{equation}
Therefore, the modified-composite-law predictions in constant-$\tau$ healing tests agree with the canonical $B\ln t$ healing for $A\leq B$, as seen in the numerical example Fig.~\ref{fig:SHScomparison} (d2 \& d3). 
Note, however, that eq.~(\ref{eq:modcompsimilartoaging}) is only an order estimate because the condition $\Omega <\beta$ includes the neighborhood of $\Omega =  \beta$ as well as $\Omega \ll \beta$, thus only guaranteeing that $f_t = \exp(-\Omega/\beta) > 1/e$, whereas the aging law holds that $f_t = 1$. 

For $A > B$, the modified composite law approximates the aging law in the early stage where $\Omega < \beta$ and the slip law in the later stage where $\Omega \gg \beta$. As shown in \S\ref{sec:SHSanalysiscomposite}, the composite law can saturate before it switches from the slip law to the aging law. In contrast, the modified composite law, which approximates the aging law only for $\Omega < \beta \ll 1$, predicts that saturation ($\Omega$ approaches unity from below) is necessarily preceded by the switch from aging-law-like to slip-law-like evolution (Fig.~\ref{fig:SHScomparison}, d1). 
The switch time $t_{\rm as}$ when $\Omega$ reaches $\beta$ is estimated as follows by using eq.~(\ref{eq:Omegaasthetafunc})
with eqs.~(\ref{eq:tcagingistcexpected}), (\ref{eq:agingevolutioninSHS2}), and (\ref{eq:modcompsimilartoaging}):
\begin{equation}
    t_{\rm as}= t_{\rm c|aging}[(\beta/\Omega_{\rm init})^{1/(1-B/A)}-1].
    \label{eq:deftas}
\end{equation}
Before $t_{\rm as}$, the evolution is in the early stage approximated by the aging law: 
\begin{equation}
    \theta_{\rm mod.composite}\sim \theta_{\rm aging} \hspace{10pt}(A>B\cap t\lesssim t_{\rm as}).
\end{equation}
The approximate evolution for the later stage after $t_{\rm as}$ is obtained from the slip-law prediction in \S\ref{sec:SHSsliplaw}, by replacing ($t$, $\theta_{\rm init}$, $\Omega_{\rm init}$) with ($t-t_{\rm as}$, $\theta_{\rm aging}(t_{\rm as})$, $\beta$): 
\begin{equation}
    \theta_{\rm mod.composite}(t)\sim \theta_{\rm slip}(t-t_{\rm as})|_{\theta(0)=\theta_{\rm aging}(t_{\rm as}),\Omega(0)=\beta} \hspace{10pt}(A>B\cap t \gg t_{\rm as}),
    \label{eq:modcompthetalast}
\end{equation}
where the same substitution notation as in eq.~(\ref{eq:compthetalast}) is used. 
Because evolution after $t_{\rm as}$ is approximated by the slip law, 
the saturation time $t_{\rm *|mod.composite}$ of $\theta_{\rm mod.composite}$ can be estimated from eq.~(\ref{eq:tstarprime}) with the initial-value replacement and time translation as per eq.~(\ref{eq:modcompthetalast}), considering the saturation time $t_*^\prime$ for $\theta_{\rm slip}$ is given by eq.~(\ref{eq:tstarprime}) with the initial condition of $(\theta(0),\Omega(0))=(\theta_{\rm init},\Omega_{\rm init})$:
\begin{equation}
    t_{\rm *|mod.composite}=\frac AB \frac{\beta^{-1/(A/B-1)}-1}{\beta\ln (1/\beta)}(t_{\rm as}+t_{\rm c|aging})+t_{\rm as},
\end{equation}
where we used $\theta_{\rm aging}(t_{\rm as})=t_{\rm as}+t_{\rm c|aging}$ from eq.~(\ref{eq:agingevolutioninSHS2}).

Lastly, we note that any evolution laws satisfying eqs.~(\ref{RSFeq:LsimilartoDc}), (\ref{eq:evolutionlaw_compatiblewithfunctionalansatz_substituted}), and (\ref{RSFeq:reqforpsic}) are described by the same asymptotics [$\dot \theta \sim 1-\Omega\ln\Omega$ ($\Omega\ll\beta$) and
$\dot \theta \sim -\Omega\ln\Omega$ ($\Omega\gg\beta$)] as the modified composite law. 
Therefore, all the above-derived behaviors in \S\ref{sec:SHSanalysismodifiedcomposite} should apply to any evolution laws satisfying eqs.~(\ref{RSFeq:LsimilartoDc}), (\ref{eq:evolutionlaw_compatiblewithfunctionalansatz_substituted}), and (\ref{RSFeq:reqforpsic}), which define an eligible class of evolution laws (\S\ref{RSF322}). 

\section{Predictions of the slip law during stationary-loadpoint SHS tests}
\label{sec:slip_law_pred_lph}
\setcounter{figure}{0}
In the present paper, we have taken up SHS$|$NSC tests, where slip during the hold phase is too small for the slip law to cause significant healing, as the experiments requiring a correction of the slip law. On the other hand, in the main text, we did not consider stationary-loadpoint SHS tests at all because observations in the other types of experiments are enough to constrain the evolution law (\S\ref{subsec:SHSrequirement}).
However, although we have to emphasize that no existing evolution laws can reproduce stationary-loadpoint SHS tests with high machine stiffness~\citep{bhattacharya2017does}, it is well known that loadpoint-SHS tests under typical (i.e., not so high) stiffness $k$ of usual apparatus are well reproduced by the slip law not involving purely time-dependent healing~\citep{bhattacharya2017does}. 
One puzzling factor in the aging-versus-slip problem is that some healing experiments require purely time-dependent healing (the aging law) while other healing tests are well described by the slip-driven state evolution (the slip law).
We here answer it by examining how the constraints on the evolution law would be modified if stationary-loadpoint SHS tests with typical $k$ are taken as an additional requirement.

\begin{figure*}
\centering
\includegraphics[width=130mm]{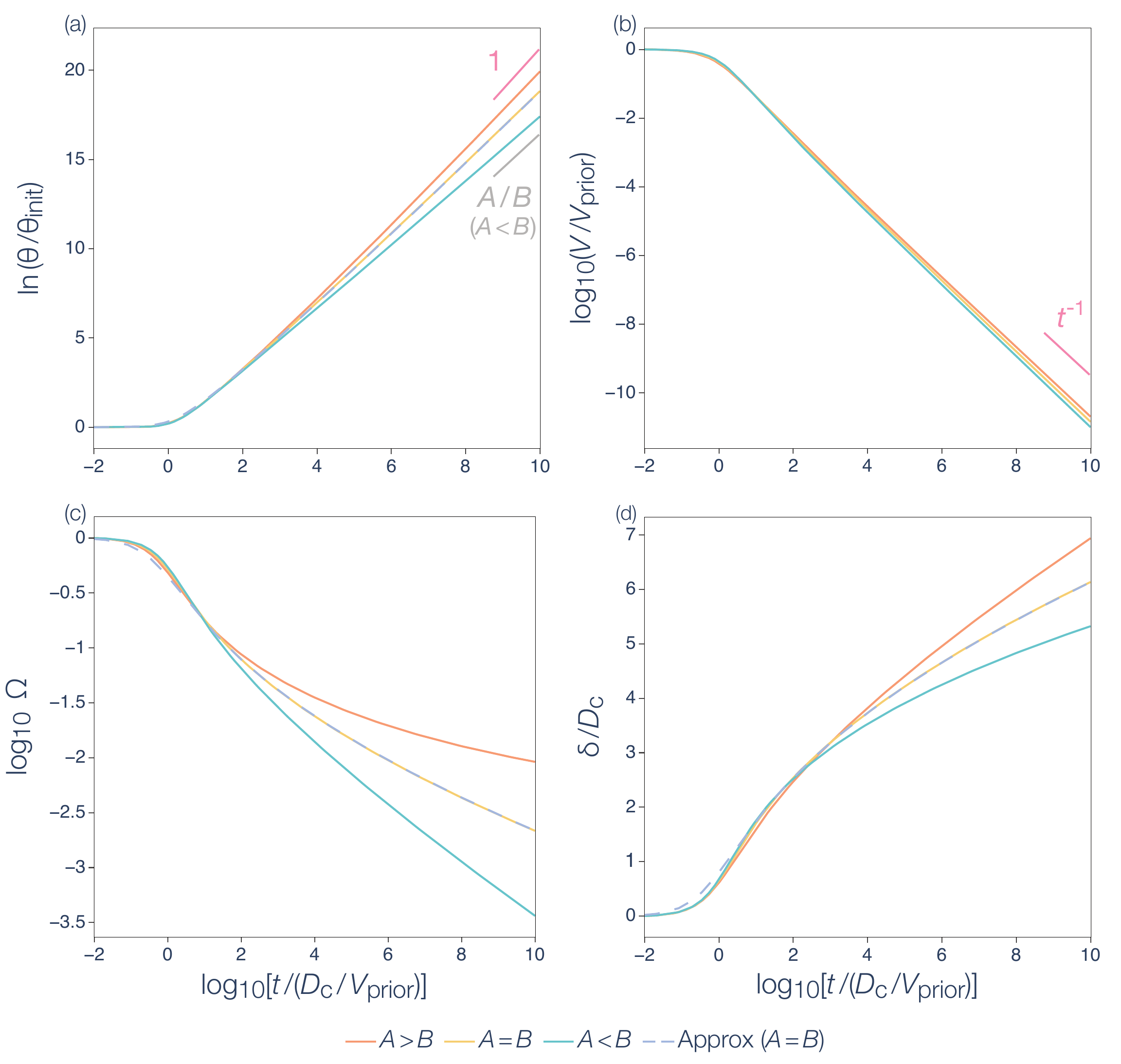}
\caption{
Predictions of the slip law for typical stationary-loadpoint SHS tests, where $kD_{\rm c}\sim A\sim B$ (numerical: solid lines; analytical: broken lines). 
$B/A = 1.2, 1, 0.8$ are assumed for $A>B$, $A=B$, $A<B$ cases, respectively. The other non-dimensionalized parameter, $kD_{\rm c}/A$, was set to $A/B$. 
Approximate analytical solutions (eqs.~\ref{eq:LPHSHSslip_deltaWRTtime}--\ref{eq:LPHSHSslip_thetaWRTtime}) for $\delta\lesssim D_{\rm c}/|1-B/A|$ are shown for the $A=B$ case only, as the illustrated solutions for $A\neq B$ are almost collapsed to that for $A=B$ when $\delta\ll D_{\rm c}/|1-B/A|$. 
For $\ln(V/V_{\rm prior})$ and
$\ln(\theta/\theta_{\rm init}$), slopes expected from the analytical solutions obtained in \S\ref{subsec:LPH_solution4superlongtime} are also indicated.
}
\label{fig:loadpointhold_summary}
\end{figure*}

Our conclusions can be grasped from Fig.~\ref{fig:loadpointhold_summary}, which shows the evolution of slip, slip rate, strength, and $\Omega$ against time $t$ during the hold phase of stationary-loadpoint SHS tests when the slip law holds. Three cases with $A > B$, $A = B$, and $A < B$ are simulated by assuming $k \sim A/D_{\rm c}$, which is typical in existing experiments. Although $V$ keeps decreasing as $1/t$, $\Omega$ does not decay much; $\Omega$ remains above $10^{-2}$ for $t < 1000(V_{\rm prior}/D_{\rm c})$. Because in most stationary-loadpoint SHS tests, $V_{\rm prior}\sim1$ $\mu$m/s, $D_{\rm c} \sim 1$ $\mu$m, and hold time is up to $1000$ s, this result means that $\Omega$ has been probed only above 0.01. To achieve $\Omega =10^{-2}$ for typical values $A\simeq B$, $t \sim 4\times 10^{6} (V_{\rm prior}/D_{\rm c})$ is necessary. 
Hence, the range of $\Omega$ probed by stationary-loadpoint SHS tests with typical $k$ ($kD_{\rm c}/\sigma$) and $t$ ($V_{\rm prior}t/D_{\rm c}$), where the slip law is shown to hold, is a mere subset of the range probed in VS tests (discussed in \S\ref{sec:heuristicderivation}). Therefore, requirement from stationary-loadpoint SHS tests with typical $k$ is automatically satisfied by satisfying the requirement from VS tests. In other words, the slip law's success in reproducing stationary-loadpoint SHS tests with typical $k$ means nothing more than the slip law's success in reproducing VS tests. 

In the main text, we have pointed out the necessity of aging-slip switch across the border between the regime covered by existing VS tests and that covered by constant-$\tau$-hold SHS tests. The above argument based on Fig.~\ref{fig:loadpointhold_summary} additionally shows that this aging-slip switch occurs across the border between the regime covered by existing stationary-loadpoint SHS tests with typical $k$ and that covered by constant-$\tau$-hold SHS tests with sufficiently low $\tau_{\rm hold}$.

Besides, Fig.~\ref{fig:loadpointhold_summary}b shows that $V$ decreases to $V_{\rm prior}/1000$ in $t\sim 1000(V_{\rm prior}/D_{\rm c})$, a realistic experimental time. Hence, typical stationary-loadpoint SHS tests can reject the aging-slip switch at $V_{\rm c} \sim 10^{-8}$ m/s suggested by \citet{kato2001composite}. On the other hand, because of the limited range of reachable $\Omega$ (Fig.~\ref{fig:loadpointhold_summary}), typical stationary-loadpoint SHS tests are difficult to access our proposed condition for the aging-slip switch at $\Omega\sim \beta \lesssim 10^{-2}$. 

The rest of this appendix shows analytical derivations to support the prediction of the slip law for stationary-loadpoint SHS tests illustrated in Fig.~\ref{fig:loadpointhold_summary}, that is, $V$ slows down as $1/t$, while $\ln\Omega$ changes much more slowly, roughly as $-\sqrt{\ln t}$ as shown below. 

Under the slip law, governing equations during the hold stage (i.e, $V_{\rm m} = 0$) of stationary-loadpoint SHS tests are the spring-slider relation (eq.~\ref{eq:springslidereq}), the RSF constitutive law (eq.~\ref{RSFeq:constitutivePhiA}), and the slip law (eq.~\ref{eq:sliplawinddelta}). For ease of derivation, we convert these relations into differential equations with respect to slip $\delta$, as we did in \S\ref{sec:SHSsliplaw}. 
\begin{flalign}
    &\frac{d\tau}{d\delta} =-k.
    \label{eq:LPHSHS_SlipLaw_goveq1}
    \\
    &\frac{d\tau}{d\delta}=A\frac{d\ln V}{d\delta}+\frac{d\Phi}{d\delta}
    \label{eq:LPHSHS_SlipLaw_goveq2}
    \\
    &\frac{d\Phi}{d\delta}=-\frac{\Phi-\Phi_{\rm SS}}{D_{\rm c}}
    \label{eq:LPHSHS_SlipLaw_goveq3}
\end{flalign}
Initial conditions are $V = V_{\rm prior}$ and $\Omega=1$ [i.e., $\Phi=\Phi_{\rm SS}(V_{\rm prior})$]. Appendix \ref{sec:slip_law_pred_lph} assumes $D_{\rm c} = L$ and $\Omega_{\rm SS}=1$ as presumed in the slip law.

In the below, we first solve the coupled differential equations (\ref{eq:LPHSHS_SlipLaw_goveq1})--(\ref{eq:LPHSHS_SlipLaw_goveq3}) and obtain the solutions as functions of $\delta $ (\S\ref{subsec:LPH_solution4delta}). Then, in \S\ref{subsec:LPH_solution4AeqB}, we will obtain approximate expressions of those solutions as functions of time, for $\delta \lesssim D_{\rm c}/|1-B/A|$, which holds very well for typical frictional properties $A\simeq B$. \S\ref{subsec:LPH_solution4superlongtime} will discuss behaviors for $\delta \gg D_{\rm c}/|1-B/A|$. However, as will be shown in \S\ref{subsec:LPH_solution4AeqB}, such large $\delta$ is unlikely to be involved in existing stationary-loadpoint SHS tests, and hence their reproducibility is not a concern of the present paper, which aims at reproduction of consensual canons based on large body of experiments. Nevertheless, we include \S\ref{subsec:LPH_solution4superlongtime} to point out that the slip law predicts $A/B$ dependence of $B_{\rm heal}$ in the stationary-loadpoint SHS tests (Fig.~\ref{fig:loadpointhold_summary}), which are partly visible even in the $\delta \lesssim D_{\rm c}/|1-B/A|$ regime. The other aspects of \S\ref{subsec:LPH_solution4superlongtime} have already been discussed by \citet{bhattacharya2017does}, but included here for completeness of derivation. 

\subsection{Exact solutions as functions of slip $\delta$}
\label{subsec:LPH_solution4delta}

Equations~(\ref{eq:LPHSHS_SlipLaw_goveq1})--(\ref{eq:LPHSHS_SlipLaw_goveq3}) are linear equations in terms of $\tau$, $\ln V$, and $\Phi$ ($\ln\theta$). Therefore, it is easy to obtain the solutions of the coupled differential equations (\ref{eq:LPHSHS_SlipLaw_goveq1})--(\ref{eq:LPHSHS_SlipLaw_goveq3}) as functions of $\delta$. In the following, we use $\ln \Omega$ instead of $\Phi$ to simplify calculations.

By eliminating $\tau$ from eqs.~(\ref{eq:LPHSHS_SlipLaw_goveq1})--(\ref{eq:LPHSHS_SlipLaw_goveq3}), we obtain 
\begin{equation}
\left(
    \begin{array}{cc}
    \frac{A-B}{B} &1\\
    -1 &1
    \end{array}
\right)
\left(
    \begin{array}{c}
    \frac{d\ln V}{d\delta }\\
    \frac{d\ln \Omega}{d\delta }
    \end{array}
\right)
=
-\left(
    \begin{array}{c}
    k/B\\
    D_{\rm c}^{-1}\ln\Omega
    \end{array}
\right)
\label{eq:LPHSHS_SlipLaw_goveqmat}
\end{equation}
Equation~(\ref{eq:LPHSHS_SlipLaw_goveqmat}) is reduced to two differential equations: 
\begin{flalign}
\frac{d\ln V}{d\delta }&=-\frac k A+\frac{B}{AD_{\rm c}}\ln\Omega \label{eq:EOMs_VOmega_lph1}
\\
\frac{d\ln \Omega}{d\delta }&=-\frac k A-\frac{A-B}{AD_{\rm c}}\ln\Omega \label{eq:EOMs_VOmega_lph2}
\end{flalign}
The solution of eq.~(\ref{eq:EOMs_VOmega_lph2}) with the initial condition $\ln\Omega = 0$ is
\begin{equation} 
\ln \Omega=\frac{kD_{\rm c}}{A-B}
\left(
e^{-\frac{A-B}{AD_{\rm c}}\delta}-1
\right).
\label{eq:LPHSHSslip_OmegaWRTdelta}
\end{equation}
By using eq.~(\ref{eq:LPHSHSslip_OmegaWRTdelta}),
the solution of eq.~(\ref{eq:EOMs_VOmega_lph1}) with the initial condition $\ln V=\ln V_{\rm prior}$ is found to be
\begin{equation} 
\ln (V/V_{\rm prior})=
-\left(\frac k A +\frac{kB/A}{A-B}
\right)\delta 
-\frac{kD_{\rm c}B}{(A-B)^2}
\left(
e^{-\frac{A-B}{AD_{\rm c}}\delta}-1
\right).
\label{eq:LPHSHSslip_VWRTdelta}
\end{equation}
As above, the slip-law prediction during the stationary-loadpoint hold has been obtained in closed-form functions of slip $\delta$. However, in time $t$ domain, those exact solutions are not expressed in closed forms (exemplified by eq.~\ref{eq:coresolution_lph_slip} for $A=B$ using a special function). Thus, in the following, we will derive approximate closed-form solutions (\S\ref{subsec:LPH_solution4AeqB} for moderate slip; \S\ref{subsec:LPH_solution4superlongtime} for much large slip). 

\subsection{Approximate solutions for $\delta \lesssim D_{\rm c}/|1-B/A|$ as functions of time $t$}
\label{subsec:LPH_solution4AeqB}
For $\delta \lesssim D_{\rm c}/|1-B/A|$, 
a series expansion in $\delta |1-B/A|/D_{\rm c}$ is helpful to solve eqs.~(\ref{eq:LPHSHSslip_OmegaWRTdelta}) and (\ref{eq:LPHSHSslip_VWRTdelta}) as functions of time.
The range $\delta \lesssim D_{\rm c}/|1-B/A|$ becomes wider when $A$ and $B$ become closer to each other. For $A = B$, any $\delta <\infty$ is included in this range. 

The lowest-order approximations of eqs.~(\ref{eq:LPHSHSslip_OmegaWRTdelta}) and (\ref{eq:LPHSHSslip_VWRTdelta}) with respect to $\delta |1-B/A|/D_{\rm c}$ are
\begin{flalign}
\ln\Omega &\simeq -\frac{k}{A}\delta
\label{eq:LPHSHSslip_OmegaWRTdelta_approx}
\\
\ln \frac V {V_{\rm prior}}&\simeq-\frac{k}{A}\delta -\frac 1 2 \frac{Bk}{A^2D_{\rm c}}\delta ^2.
\label{eq:LPHSHSslip_VWRTdelta_approx}
\end{flalign}
In \S\ref{subsec:LPH_solution4AeqB}, $\simeq$ denotes the lowest-order approximation in terms of $|1-B/A|\delta /D_{\rm c}$ (exact for $A=B$). 
Because $V=d\delta /dt$, eq.~(\ref{eq:LPHSHSslip_VWRTdelta_approx}) is an ordinary differential equation of slip $\delta$ with respect to time $t$, which can be integrated by separation of variables: 
\begin{equation}
    V_{\rm prior}t
    \simeq
    \sqrt{\frac{\pi A^2D_{\rm c}}{2Bk}}e^{-kD_{\rm c}/(2B)}
    {\rm erfi}\left(
    \sqrt{\frac{Bk}{2A^2 D_{\rm c}}}(\delta+AD_{\rm c}/B)\right)+C,
    \label{eq:coresolution_lph_slip}
\end{equation}
where ${\rm erfi}(x):={\rm erf}(ix)/i$ denotes the imaginary error function, and $C$ is a constant of integration.

Further using series expansion ${\rm erfi}(x)=-i+\exp(x^2)/(x\sqrt{\pi})[1+\mathcal O(x^{-2})]$, 
we simplify eq.~(\ref{eq:coresolution_lph_slip}) to
\begin{equation}
    V_{\rm prior}t
    \simeq
    \frac{A}{k}\left\{\left(\frac{B\delta}{AD_{\rm c}}+1\right)^{-1}
    e^{\frac{kD_{\rm c}}{2B}\left[\left(\frac{B\delta}{AD_{\rm c}}+1\right)^2-1\right]+...}
    -1
    \right\},
    \label{eq:LPHSHSslip_deltaWRTtime_before}
\end{equation}
where the $...$ term denotes non-leading orders. The constant of integration has been determined by the initial condition of $\delta=0$ at $t=0$. 

Equation~(\ref{eq:LPHSHSslip_deltaWRTtime_before}) can be implicitly solved for $(B\delta)/(AD_{\rm c})+1$ as
\begin{equation}
    \left(\frac{B\delta}{AD_{\rm c}}+1\right)\simeq\sqrt{1+\frac{2B}{kD_{\rm c}}\left[\ln (t/t_{\rm c,lph}+1)+\ln\left(\frac{B\delta}{AD_{\rm c}}+1\right)\right]+...},
\end{equation}
or explicitly, using recursive substitution of the left-hand side to the right-hand side,  
\begin{equation}
    \left(\frac{B\delta}{AD_{\rm c}}+1\right)\simeq\sqrt{1+\frac{2B}{kD_{\rm c}}\left[\ln \left(\frac {t}{t_{\rm c,lph}}+1\right)+ \ln\sqrt{1+\frac{2B}{kD_{\rm c}}\ln \left(\frac {t}{t_{\rm c,lph}}+1\right)}\right]+...},
    \label{eq:LPHSHSslip_deltaWRTtime_verymess}
\end{equation}
where 
\begin{equation}
t_{\rm c,lph}:=  \frac{A}{k} V_{\rm prior}^{-1}.
\end{equation}
Finally, collecting the non-leading orders in eq.~(\ref{eq:LPHSHSslip_deltaWRTtime_verymess}),
we can arrive at the solution of $\delta$ as a function of time as 
\begin{equation}
\delta\simeq 
\frac{AD_{\rm c}}{B}
\left\{-1+
\sqrt{
1+\frac{2B}{kD_{\rm c}}\ln\left(t/t_{\rm c,lph}+1\right)
+\lambda(t)}
\right\},
\label{eq:LPHSHSslip_deltaWRTtime}
\end{equation}
where $\lambda$ denotes a set of correction terms: 
\begin{equation}
    \lambda(t):=\frac{B}{kD_{\rm c}}\ln\left[1+\frac{2B}{kD_{\rm c}}\ln \left(\frac {t}{t_{\rm c,lph}}+1\right)\right]+...
\label{eq:LPHSHSslip_WRTtime_correction}
\end{equation}
The first term on the right-hand side of eq.~(\ref{eq:LPHSHSslip_WRTtime_correction}) represents the lowest-order correction.
The right-hand side of eq.~(\ref{eq:LPHSHSslip_deltaWRTtime}), neglecting the $...$ term in $\lambda(t)$, serves as an approximate closed-form solution of $\delta (t)$ for $\delta \lesssim D_{\rm c}/|1-B/A|$, 
the accuracy of which is evident in Fig.~\ref{fig:loadpointhold_summary} for $\delta \ll D_{\rm c}/|1-B/A|$.

Approximate sketches of the evolution of $\ln \Omega$ and $\ln V$ as functions of $t$ are obtained by plugging eq.~(\ref{eq:LPHSHSslip_deltaWRTtime}) into eqs.~(\ref{eq:LPHSHSslip_OmegaWRTdelta_approx}) and (\ref{eq:LPHSHSslip_VWRTdelta_approx}):
\begin{flalign}
    \ln\Omega&\simeq 
-\frac{kD_{\rm c}}{B}
\left\{-1+
\sqrt{
1+\frac{2B}{kD_{\rm c}}
\ln\left(t/t_{\rm c,lph}+1\right)
+\lambda(t)
}
\right\}
\label{eq:LPHSHSslip_OmegaWRTtime}
\\
\ln \frac V {V_{\rm prior}}&\simeq \ln\Omega -\frac 1 2 \frac{kD_{\rm c}}{B}\left\{-1+
\sqrt{
1+\frac{2B}{kD_{\rm c}}\ln\left(t/t_{\rm c,lph}+1\right)
+\lambda(t)
}
\right\}^2
\label{eq:LPHSHSslip_rateWRTtime}
\end{flalign}
Noting $d\ln\Omega = d\ln V + d\ln\theta$, the corresponding strength recovery is
\begin{equation}
    B\ln(\theta/\theta_{\rm init})\simeq\frac{kD_{\rm c}}{2}\left\{-1+
\sqrt{
1+\frac{2B}{kD_{\rm c}}\ln\left(t/t_{\rm c,lph}+1\right)
+\lambda(t)
}
\right\}^2.
\label{eq:LPHSHSslip_thetaWRTtime}
\end{equation}
The closed-form approximate solutions eqs.~(\ref{eq:LPHSHSslip_deltaWRTtime})--(\ref{eq:LPHSHSslip_thetaWRTtime}) [with the $...$ term in eq.~(\ref{eq:LPHSHSslip_WRTtime_correction}) neglected for $\lambda(t)$] are plotted in Fig.~\ref{fig:loadpointhold_summary} for $A = B$. 
They agree well with the numerical solutions, demonstrating accuracy of approximation for $\delta \ll D_{\rm c}/|1-B/A|$. 

Incidentally, notice Fig.~\ref{fig:loadpointhold_summary}a indicates that the slope $B_{\rm heal}$ is below $B$ for $tV_{\rm prior}/D_{\rm c}<10^{10}$, although
eq.~(\ref{eq:LPHSHSslip_thetaWRTtime}) suggests $B_{\rm heal}$ converges to $B$ when $t\to\infty$. Therefore, in loadpoint-hold SHS tests with realistic range of hold time, the slip law predicts that, when $A = B$, then $B_{\rm heal}<B$. 

For stationary-loadpoint SHS tests with typical $k$ and frictional properties $A$ and $B$, where $kD_{\rm c} \sim A\sim B$, these approximate expressions of the solutions (i.e., eqs.~\ref{eq:LPHSHSslip_deltaWRTtime}--\ref{eq:LPHSHSslip_thetaWRTtime}) reduce to $\delta/D_{\rm c}\sim\ln \Omega\sim \sqrt{\ln t/t_{\rm c,lph}}$, where $t_{\rm c,lph}\sim D_{\rm c}/V_{\rm prior}$. Hence, the decrease of $\Omega$ is much slower than the decrease of $V$, which slows down in inverse proportion to $t$. Thus, the conclusion we described at the beginning of Appendix \ref{sec:slip_law_pred_lph} using Fig.~\ref{fig:loadpointhold_summary} has been proved analytically. 
We also notice that the slip amount nondimensionalized by $D_{\rm c}$ is comparable to the drop of $\ln\Omega$. 

The above approximate solutions are valid until the time when $\delta$ reaches $D_{\rm c}/|1-B/A|$. This timing, denoted by $t^*_{\rm lph}$, is roughly estimated by plugging $\delta = D_{\rm c}/|1-B/A|$ into eq.~(\ref{eq:LPHSHSslip_deltaWRTtime}):
\begin{equation}
    t_{\rm lph}^*\sim t_{\rm c,lph} e^{\frac{kD_{\rm c}}{2B}|A/B-1|^{-2}}.
\end{equation}
We see that $\ln t^*_{\rm lph}$ is inversely proportional to the square of $(A/B-1)$. For typical rocks with $|A/B-1| \lesssim 1/5$, for example, stationary-loadpoint SHS tests with typical $k$, which is on the order of $B/D_{\rm c}$, $t^*_{\rm lph}$ is as much as $10^5t_{\rm c,lph}$ or greater. Thus, most existing experiments are done within the time range where $\delta \lesssim D_{\rm c}/|1-B/A|$.

Lastly, we calculate the values of $\theta$ and $\Omega$ at $t^*_{\rm lph}$, partly for use in \S\ref{subsec:LPH_solution4superlongtime}. Plugging $\delta (t^*_{\rm lph}) = D_{\rm c}/|1-B/A|$ into (\ref{eq:LPHSHSslip_OmegaWRTdelta_approx}) and (\ref{eq:LPHSHSslip_VWRTdelta_approx}), we obtain
\begin{equation}
    B\ln [\theta(t_{\rm lph}^*)/\theta_{\rm init})]\simeq \frac{kD_{\rm c}}{2|A/B-1|^2}.
    \label{eq:junction_thetavalue_lph}
\end{equation}
Here, we used $d\ln \Omega=d\ln V+d\ln \theta$. 
The approximate solutions (eqs.~\ref{eq:LPHSHSslip_deltaWRTtime}--\ref{eq:LPHSHSslip_thetaWRTtime}) are valid until the strength recovery reaches the level of eq.~(\ref{eq:junction_thetavalue_lph}). Similarly, by plugging $\delta (t^*_{\rm lph}) = D_{\rm c}/|1-B/A|$ into (\ref{eq:LPHSHSslip_OmegaWRTdelta_approx}), we obtain
\begin{equation}
    \ln\Omega(t_{\rm lph}^*)\simeq -\frac{kD_{\rm c}}{|A-B|}.
    \label{eq:junction_Omegavalue_lph}
\end{equation}
In the main text, we have discussed the aging-slip switch is expected to occur around $\Omega\sim\beta \in [10^{-4}$,$10^{-2}$]. Hence, for stationary-loadpoint SHS tests with typical $k (\sim A/D_{\rm c})$ and on typical rocks ($|A-B| \ll A$), eq.~(\ref{eq:junction_Omegavalue_lph}) suggests that the aging-slip switch is expected to occur within the condition examined in \S\ref{subsec:LPH_solution4AeqB}, that is, $\delta\lesssim D_{\rm c}/|1-B/A|$. 

\subsection{Approximate solutions for $\delta \gg D_{\rm c}/|1-B/A|$  as functions of time $t$}
\label{subsec:LPH_solution4superlongtime}
As shown in \S\ref{subsec:LPH_solution4AeqB}, most of the existing stationary-loadpoint SHS tests were done only
for $\delta\lesssim D_{\rm c} /|1-B/A|$. 
However, the analytic approximate expressions derived in
\S\ref{subsec:LPH_solution4AeqB} do deviate significantly from the numerical solutions as $\delta$  approaches
$D_{\rm c} /|1-B/A|$ (Fig.~\ref{fig:loadpointhold_summary}). For example, $\Omega(t)$ and $\delta(t)$ are relatively sensitive to $B/A$,
but the analyses in \S\ref{subsec:LPH_solution4AeqB} do not capture this feature. Those deviations in fact reflects
behaviors for $\delta\gg D_{\rm c} /|1-B/A|$ we examine in this section.

The strength recovery can be decomposed as
\begin{equation}
    B\ln(\theta/\theta_{\rm init})=B\ln [\theta(t^*_{\rm lph})/\theta_{\rm init}]+B\ln [\theta(t)/\theta(t^*_{\rm lph})].
\end{equation}
The first term is already given by eq.~(\ref{eq:junction_thetavalue_lph}). In the below, we evaluate the second term. The behavior for large $t$ differs depending on the sign of $A - B$, so we work on each case separately. 

\subsubsection{$A>B$ cases}
\label{subsec:LPH_solution4superlongtime_stable}

For $A > B$, eq.~(\ref{eq:LPHSHSslip_OmegaWRTdelta}) indicates that $\Omega$ stops decreasing at $\delta \gtrsim D_{\rm c}A/|A-B|$, reaming at the level
\begin{equation} 
\ln \Omega\sim -\frac{kD_{\rm c}}{A-B}
\hspace{10pt}(A>B\cap \delta \gtrsim D_{\rm c}/|1-B/A|).\label{eq:largeslip_lph_AgB}
\end{equation}
Under this condition of nearly constant $\ln\Omega$, strength recovery is known if $\ln V$ is constrained because $d\ln\Omega=d\ln\theta+d\ln V$. By taking the difference of eq.~(\ref{eq:LPHSHSslip_VWRTdelta}), where the second term is proportional to $\ln\Omega$ (eq.~\ref{eq:LPHSHSslip_OmegaWRTdelta}), from its value at $t = t^*_{\rm lph}$, we have
\begin{equation}
\ln [V(t)/V(t^*_{\rm lph})]\sim 
-\frac k {A-B}[\delta(t)-\delta(t^*_{\rm lph})]
\hspace{10pt}(A>B\cap \delta \gtrsim D_{\rm c}/|1-B/A|).
\end{equation}
This is a differential equation for $\delta$ using $V=d\delta/dt$, 
which is solved as follows as in \S\ref{sec:SHSsliplaw}, 
\begin{equation}
\begin{array}{l}
\delta(t)\sim\delta(t^*_{\rm lph})+k^{-1}(A-B)\ln [(t-t^*_{\rm lph})/t_{\rm c2+}+1]
\\
V(t)\sim V(t^*_{\rm lph})/[(t-t^*_{\rm lph})/t_{\rm c2+}+1]
\end{array}
\hspace{10pt}(A>B\cap \delta \gtrsim D_{\rm c}/|1-B/A|)
\label{eq:asymptoticdelta_positiveAminB}
\end{equation}
with
\begin{equation}
    t_{c2+}:=\frac{A-B}{kV(t^*_{\rm lph})}.
\end{equation}
Now we can calculate the strength recovery as $d\ln \theta\sim -d\ln V$ via eqs.~(\ref{eq:largeslip_lph_AgB}) and (\ref{eq:asymptoticdelta_positiveAminB}):
\begin{equation}
    B\ln [\theta(t)/\theta(t^*_{\rm lph})]\sim B\ln [(t-t^*_{\rm lph})/t_{\rm c2+}+1]
\hspace{10pt}(A>B\cap \delta \gg D_{\rm c}/|1-B/A|).
\label{eq:asymptotictheta_positiveAminB}
\end{equation}
Equation~(\ref{eq:asymptotictheta_positiveAminB}) indicates that the slip law with $A > B$ predicts $B\ln t$ healing for stationary-loadpoint SHS tests with a very long duration. Our numerical experiments support this conclusion (Fig.~\ref{fig:loadpointhold_summary}).

\subsubsection{$A<B$ cases}
\label{subsec:LPH_solution4superlongtime_unstable}
We here derive that the slip law with $A < B$ predicts $A\ln t$ healing (Fig.~\ref{fig:loadpointhold_summary}) for stationary-loadpoint SHS tests with a very long duration, just as we saw for constant-$\tau_{\rm hold}$ SHS tests (\S\ref{sec:SHSsliplaw}). 
Throughout the remaining of \S\ref{subsec:LPH_solution4superlongtime_unstable}, $A < B$ and $t > t^*_{\rm lph}$ are assumed. 
Under this condition, from eq.~(\ref{eq:LPHSHSslip_OmegaWRTdelta}),
\begin{equation}
\ln\Omega= \mathcal O\{\exp[(B/A-1)\delta /D_{\rm c}]\},    
\label{eq:OmegaOrder_negativeAminB}
\end{equation}
and thus from eqs.~(\ref{eq:LPHSHSslip_OmegaWRTdelta}) and (\ref{eq:LPHSHSslip_VWRTdelta}),
\begin{equation}
\ln V/V_{\rm prior}=\frac{B}{B-A}\ln\Omega+...,
\label{eq:VperOmegaRatio_negativeAminB}
\end{equation}
where the $...$ term denotes non-leading orders.

With the aid of eqs.~(\ref{eq:OmegaOrder_negativeAminB}) and (\ref{eq:VperOmegaRatio_negativeAminB}), 
we can obtain the lowest-order solution by dropping non-leading-order terms. Derivation is easier when we first solve for $d\ln\Omega/dt$, rather than starting with $d\delta/dt$ as we did in \S\ref{subsec:LPH_solution4AeqB} and \S\ref{subsec:LPH_solution4superlongtime_stable}. By noting $d\ln\Omega/dt=(d\ln\Omega/d\delta)(d\delta/dt)$, 
eqs.~(\ref{eq:EOMs_VOmega_lph2}), (\ref{eq:OmegaOrder_negativeAminB}) and (\ref{eq:VperOmegaRatio_negativeAminB}) yield
\begin{equation} 
\frac{d\ln\Omega}{dt}=\frac{ (B-A)V_{\rm prior}\ln\Omega}{AD_{\rm c}} e^{\frac{B}{B-A}\ln\Omega+...}.
\end{equation}
Its solution is 
\begin{equation}
\frac{B-A}{AD_{\rm c}} V_{\rm prior} t + C^\prime= {\rm Ei}\left(-\frac{B}{B-A}\ln\Omega\right)+...,
\label{eq:expfunc_negativeAminB}
\end{equation}
where ${\rm Ei}(\cdot )$ is the exponential integral, 
and $C^\prime$ is a constant of integration.
Using $\ln {\rm Ei}(x)=x +\mathcal O(\ln x)$, 
eq.~(\ref{eq:expfunc_negativeAminB}) is reduced to 
\begin{equation}
\ln\left[\frac{B-A}{AD_{\rm c}} V_{\rm prior}(t-t^*_{\rm lph})+\Omega(t^*_{\rm lph})^{-B/(B-A)}\right]= -\frac{B}{B-A}\ln \Omega(t)+...
\label{eq:expfunc_negativeAminB_approx}
\end{equation}
$C^\prime$ has been determined by $\Omega=\Omega(t^*_{\rm lph})$, the boundary condition at $t=t^*_{\rm lph}$. 
Using (\ref{eq:junction_Omegavalue_lph}), eq.~(\ref{eq:expfunc_negativeAminB_approx}) is reduced to 
\begin{equation}
\ln [\Omega(t)/\Omega(t^*_{\rm lph})]= -\frac{B-A}{B}
\ln[(t-t^*_{\rm lph})/t_{\rm c2-}+1]+...,
\label{eq:Omegaasymptotic_negativeAminB}
\end{equation}
where 
\begin{equation}
    t_{\rm c2-}:=\frac{A}{B-A}\frac{D_{\rm c}}{V_{\rm prior}}e^{kD_{\rm c}B/(B-A)^2}.
    \label{eq:secondcutofftime_afterswitch_negativeAminB}
\end{equation}
Equation~(\ref{eq:Omegaasymptotic_negativeAminB}) yields $\ln V\sim -\ln t$ via eq.~(\ref{eq:VperOmegaRatio_negativeAminB}).

Lastly, using the time-domain solution of $\ln\Omega$ (eq.~\ref{eq:Omegaasymptotic_negativeAminB}), we obtain the time-domain solution of $\ln \theta$. 
Because $d\ln \Omega=d\ln V+d\ln \theta$, eq.~(\ref{eq:VperOmegaRatio_negativeAminB}) yields  
\begin{equation}
    \ln [\theta/\theta(t^*_{\rm lph})]=-\frac{A}{B-A}\ln\Omega+...\label{eq:lph_asymptotic_negativeAminB_thetaOmegarelation}
\end{equation}
Thus, we see that ln-t healing occurs for $\delta \gg D_{\rm c}/|1-B/A|$ on interfaces with $A < B$. 
From eqs.~(\ref{eq:Omegaasymptotic_negativeAminB}) and (\ref{eq:lph_asymptotic_negativeAminB_thetaOmegarelation}), we have 
\begin{equation}
    B\ln [\theta/\theta(t^*_{\rm lph})]= A\ln[(t-t^*_{\rm lph})/t_{\rm c2-}+1]+...
    \label{eq:asymptotictheta_negativeAminB}
\end{equation}
Namely, the slip law predicts that the slope $B_{\rm heal}$ of the ln-t healing on $A< B$ interfaces is $A$, in agreement with the numerical solution (Fig.~\ref{fig:loadpointhold_summary}). This is in contrast with that $B_{\rm heal}$ was predicted to be $B$ (eq.~\ref{eq:asymptotictheta_positiveAminB}) for $A > B$ interfaces.



\section{Predictions of the aging law during VS tests with sufficiently high machine stiffness}
\label{sec:VSagingderivation}

Consider ideal VS tests where $\delta=t V_{\rm after}$. For such tests, the early stage following a sufficiently large positive $V$ step (i.e., $V_{\rm after}/V_{\rm before}\gg1$) achieves $\Omega \gg 1$, where the aging law (eq.~\ref{eq:defofaginglaw}) predicts $\dot\theta\simeq -\Omega$. In the early stage following a sufficiently large negative $V$ step ($V_{\rm after}/V_{\rm before}\ll1$), $\Omega\ll1$ holds, and eq.~(\ref{eq:defofaginglaw}) becomes $\dot\theta\simeq1$. Then, further using eq.~(\ref{RSFeq:conversionofthetatophinocut}), we obtain
\begin{equation}
    \frac{\Phi_{\rm VS|aging}(\delta)-\Phi_{\rm SS}(V_{\rm after})}{B}\simeq
    \begin{cases}
        \ln (V_{\rm after}/V_{\rm before})-\delta/D_{\rm c} & (\Omega\gg1)
        \\
        \ln (\delta/D_{\rm c}+V_{\rm after}/V_{\rm before}) &(\Omega\ll 1)
    \end{cases}
    \label{eq:agingVStransient}
\end{equation}
which indicates strong asymmetry~\citep{ampuero2008earthquake}. 

The early frictional evolution following a positive step (eq.~\ref{eq:agingVStransient} top) approximates linear slip weakening with a constant slope $B/D_{\rm c}$ independent of the $V$-step magnitude, and hence the convergence slip distance to the steady-state increases with the log magnitude $\ln(V_{\rm after}/V_{\rm before})$ of the positive $V$ step, another disagreement with the canonical behavior~\citep{nakatani2001conceptual}. 

In contrast, the early part following a negative $V$ step (eq.~\ref{eq:agingVStransient} bottom) approximates purely time-dependent healing ($\dot\theta\simeq f_t = 1$, dotted lines in Fig.~\ref{fig:AgingSlipVS}b, inset). 
This explains rapid strengthening in the early stage of $\delta$-$\Phi$ plot, as in Fig.~\ref{fig:AgingSlipVS}b, 
and the convergence distance anti-correlates to the magnitude of the negative step; if we define the relaxation distance as the distance required for $|\Phi_{\rm VS|aging}(\delta)-\Phi_{\rm SS}(V_{\rm after})|$ to decrease by a factor of $1/e$, it is estimated to be $(V_{\rm after}/V_{\rm before})^{1/e}D_{\rm c}$ from eq.~(\ref{eq:agingVStransient}, bottom), which predicts a significant dependence of the relaxation distance on the magnitude of the negative step, in disagreement with the VS canon (eq.~\ref{eq:realVScanonical}).

\section{Derivation of $B=B_{\rm heal}$ from other empirical formulae}
\label{sec:SHSderivationdetail}

As with the relationship between $L$ and $D_{\rm c}$, there is no a priori reason that the value of $B$, which appears in eq.~(\ref{eq:req4ss}) describing the log-V dependence of the steady-state strength, coincides with the value of $B_{\rm heal}$ in eq.~(\ref{RSFeq:PhiSHSNSCrawtc}) for the log-t healing. 
Contrasting with that some laboratory evidence exists for $L \sim D_{\rm c}$ (eq.~\ref{RSFeq:LsimilartoDc}), including \citet{nakatani2006intrinsic} who concluded eq.~(\ref{RSFeq:LsimilartoDc}) by comparing the stationary-loadpoint SHS and VS tests on the same frictional interface shown in \citet{marone1998effect}, serious comparison of $B$ and $B_{\rm heal}$ is difficult because, as $D_{\rm c}$ is often very short around $\mathcal O(1)$ $\mu$m, reliable estimate of the absolute value of $B$ (and $A$) requires nearly ideal VS tests on an unworldly stiff loading apparatus~\citep[e.g.,][]{blanpied1998quantitative,nagata2012revised,kame2015earthquake}, whereas $A-B$ can be easily constrained from SS tests unless the machine stiffness is so low that disastrous stick-slip occurs to mess up the experiment. The above consideration makes it questionable that we required $B=B_{\rm heal}$ (eq.~\ref{RSFeq:coincidenceofBs}) as part of the canonical behaviors to be reproduced by good evolution laws. 

However, to the authors' surprise, the proposition $B=B_{\rm heal}$, in fact, can be derived as a logical consequence of some other empirical formulae (eqs. \ref{eq:req4ss}, \ref{RSFeq:PhiSHSNSCrawtc}, and \ref{eq:tcdetection}), as long as we presume $\dot\theta = f(\theta, V)$ as we do when constructing the specific functional form of $f$. 
We show the proof below. 

For simplicity, we consider $\theta\gg\theta_{\rm X}$, and associated approximations are denoted by $\simeq$. Note that neither $B$ nor $B_{\rm heal}$ matters for $\theta\ll\theta_{\rm X}$ because the strength $\Phi$, the observable, varies little with $\theta$, being approximately $\Phi_{\rm X}$. When $B\neq B_{\rm heal}$, $c_*$ is not an arbitrary constant, and then the convention $c_* = 1$ (eq.~\ref{eq:normalizationofcstar}) is not applicable. Hence, we keep $c_*$ as a variable in the subsequent derivation. 
The starting points are eq.~(\ref{eq:steadystateform_theta}), which follows the SS requirement of eq.~(\ref{eq:req4ss}), eq.~(\ref{eq:thetaSHSNSC}), which is an identity, eqs.~(\ref{eq:tc_depon_Bheal_cstar}) and (\ref{eq:thetanscwithbheal}), which follow eq.~(\ref{eq:thetaSHSNSC}) and the SHS requirement of eq.~(\ref{RSFeq:PhiSHSNSCrawtc}), and eq.~(\ref{eq:tcdetection}), which is another empirical formula of SHS tests. 

To begin with, we substitute 
$\theta_{\rm init}=\theta_{\rm SS}(V_{\rm prior})$ ($=L/V_{\rm prior}$, from eq.~\ref{eq:steadystateform_theta}) into eq.~(\ref{eq:tc_depon_Bheal_cstar}). Then, comparing it to eq.~(\ref{eq:tcdetection}), we find
\begin{equation}
    c_*^{-1}\simeq V_{\rm prior}^{1-B_{\rm heal}/B}L_{\rm heal}^{B_{\rm heal}/B}/L.
    \label{eq:cstarwithbheal}
\end{equation}
In the meantime, from eqs.~(\ref{eq:thetaSHSNSC}) and (\ref{eq:thetanscwithbheal}), 
\begin{equation}
    c_*^{-1}
    \theta_{\rm SHS|NSC}\simeq (t_{\rm h}+t_{\rm c})^{B_{\rm heal}/B}.
    \label{eq:holdtimedependentcstar_theta}
\end{equation}
Differentiation of eq.~(\ref{eq:holdtimedependentcstar_theta}) with respect to $t_{\rm h}$ yields
\begin{equation}
    c_*^{-1} f_{\rm SHS|NSC}\simeq (B_{\rm heal}/B)(t_{\rm h}+t_{\rm c})^{B_{\rm heal}/B-1}.
    \label{eq:holdtimedependentcstar_f}
\end{equation}
Eliminating $t_{\rm h}+t_{\rm c}$ from eqs.~(\ref{eq:holdtimedependentcstar_theta}) and (\ref{eq:holdtimedependentcstar_f}), we have
\begin{equation}
    c_*^{-1} f_{\rm SHS|NSC}\simeq (B_{\rm heal}/B)(c_*^{-1} \theta_{\rm SHS|NSC})^{1-B/B_{\rm heal}}.
    \label{eq:ftnscwithbhealcstar}
\end{equation}
Finally canceling $c_*$ in eqs.~(\ref{eq:cstarwithbheal}) and (\ref{eq:ftnscwithbhealcstar}), we arrive at 
\begin{equation}
    f_{\rm SHS|NSC}=\frac{B_{\rm heal}}{B}\frac{L}{L_{\rm heal}}\left(\frac{V_{\rm prior}\theta_{\rm SHS|NSC}}L\right)^{1-B/B_{\rm heal}}.
    \label{eq:ftnscwithbhealwithoutcstar}
\end{equation}
For $f_{\rm SHS|NSC}$ to be a function only of the instantaneous $V$ and $\theta$($=\theta_{\rm SHS|NSC}$) values for any given history of $V$, $f_{\rm SHS|NSC}$ must be independent of $V_{\rm prior}$. Equation~(\ref{eq:ftnscwithbhealwithoutcstar}) shows that such is possible only when $B=B_{\rm heal}$. 

Thus, even if we do not set forth $B=B_{\rm heal}$ as an experimental requirement (eq.~\ref{RSFeq:coincidenceofBs}), the development of the evolution law from the canons and $t_{\rm c}=L/V_{\rm prior}$ (eq.~\ref{eq:tcdetection}) is not affected, as long as it presumes an evolution law depending only on the rate $V$ and state $\theta$ with temporal locality, $\dot\theta = f(\theta, V)$. 
However, we note that 
eq.~(\ref{eq:tcdetection}) is yet to be scrutinized under the NSC, 
and the above analysis does not exclude the possibility that temporally local evolution laws with $B\neq B_{\rm heal}$ become possible if the evolution laws include responses to factors other than $V$ and $\theta$~\citep[e.g., normal and shear stresses; ][]{linker1992effects,nagata2012revised}.

\section{Tolerance of $\epsilon_\Omega^*$ in eqs.~(93) and (95)}
\label{sec:epsironRange}
\setcounter{equation}{0}

To construct the function $\psi_{\rm c}$ using the strategy of eq.~(\ref{RSFeq:modcompscutofffunction_withHeaviside}), 
the value of $\epsilon^*_\Omega$ must be chosen appropriately so that eq.~(\ref{RSFeq:reqforpsic}) holds.  In \S\ref{subsec:AppD1}, for a decreasing function $\psi_{\rm cBase}$ that satisfies eq.~(\ref{eq:psicBaserequirement_arbitraryLoverDc}), we show
eq.~(\ref{RSFeq:reqforpsic}) is met if
\begin{equation}
    |\beta \ln\beta| \ll\epsilon^*_\Omega(<1-1/e).
    \label{eq:epsilonboundgeneral}
\end{equation}

In \S\ref{subsec:AppD2}, we further show that the tolerance of $\epsilon^*_\Omega$ relaxes when $\psi_{\rm cBase}$ has some favorable properties. Specifically, we will prove that the tolerance of $\epsilon^*_\Omega$ relaxes to
\begin{equation}
    (D_{\rm c}/L)\psi_{\rm cBase}(1)\ll\epsilon^*_\Omega(<1-1/e)
    \label{eq:epsilonboundgeneral_improved}
\end{equation}
when $\psi_{\rm cBase}(\Omega)$ satisfies all the following additional conditions, atop eq.~(\ref{eq:psicBaserequirement_arbitraryLoverDc}):
\begin{align}
    \psi_{\rm cBase}(1/2)\ll (L/D_{\rm c})\beta \label{eq:add1}
    \\
    \psi_{\rm cBase}=\Psi(\Omega/\beta) \label{eq:add2}
    \\
    d^n\Psi/dx^n=\mathcal O(\Psi) \label{eq:add3}
\end{align}

Throughout this Appendix~\ref{sec:epsironRange}, we assume $\beta\ll\sqrt{\beta}<1/e$ ($\sqrt{\beta}\ll1\cap\beta<1/e^2$) and $0<\epsilon_\Omega^*<1-1/e$ as we do in the main text. 
The following proofs are valid with any value of 
$L/D_{\rm c}$. 

\subsection{Proof for eq.~(\ref{eq:epsilonboundgeneral})}
\label{subsec:AppD1}

Equation~(\ref{eq:epsilonboundgeneral}) represents the tolerance of $\epsilon^*_\Omega$ 
if the following relation holds for $\psi_{\rm cBase}$ that satisfies eq.~(\ref{eq:psicBaserequirement_arbitraryLoverDc}):
\begin{equation}
    \psi_{\rm cBase}(\Omega)=
    \begin{cases}
        1 & (\Omega\ll\beta)
        \\
        o [(L/D_{\rm c})\Omega \ln \Omega] & (\Omega \gg\beta \cap |\Omega-1|\gg |\beta\ln\beta|)
    \end{cases}
    \label{eq:usefulpsicBaseinequality}
\end{equation} 
Therefore, what we aim to prove here is the proposition that 
$\psi_{\rm cBase}$ decaying as strongly as eq.~(\ref{eq:psicBaserequirement_arbitraryLoverDc}) tells already satisfies eq.~(\ref{RSFeq:reqforpsic}) except for the range $|\Omega-1|\lesssim |\beta\ln\beta|$. 
Since eq.~(\ref{eq:usefulpsicBaseinequality}) is the same as eq.~(\ref{eq:psicBaserequirement_arbitraryLoverDc}) for $\Omega \ll\beta$, we only need to prove it for $\Omega \gg\beta$, where the concerned proposition (if eq.~\ref{eq:psicBaserequirement_arbitraryLoverDc} is true then eq.~\ref{eq:usefulpsicBaseinequality} is true) is reduced to the following:
if $\psi_{\rm cBase}=\mathcal O[(L/D_{\rm c})\beta\ln\beta]$ (eq.~\ref{eq:psicBaserequirement_arbitraryLoverDc}), then $\psi_{\rm cBase}=o[(L/D_{\rm c})\Omega\ln\Omega]$ except for $|\Omega-1|\lesssim |\beta\ln\beta|$ (eq.~\ref{eq:usefulpsicBaseinequality}). Namely, 
\begin{equation}
    |\Omega\ln\Omega|\gg |\beta\ln\beta|\hspace{10pt} (\Omega \gg\beta \cap |\Omega-1|\gg |\beta\ln\beta|).
    \label{eq:C1subject}
\end{equation}
In the subsequent proof of eq.~(\ref{eq:C1subject}), we split the relevant $\Omega$ range into three intervals separated at $\Omega=1/e,1$ and evaluate $|\Omega\ln\Omega|$ in each interval. 

We begin with the range $\beta\ll\Omega<1/e$. 
Note, in this appendix \ref{sec:epsironRange}, $\beta\ll\sqrt\beta<1/e$ holds because $\sqrt\beta\ll1$ and $\beta<1/e^2$ are assumed. Then we can consider an $\Omega$ range $\beta\ll\Omega<\sqrt\beta$ within $\Omega<1/e$. 
For $\beta\ll\Omega<\sqrt\beta$, $\Omega/\beta<1/\sqrt\beta$ holds. Then, taking its logarithm, we see $\ln(\Omega/\beta)<(1/2)|\ln\beta|$, which is followed by
\begin{flalign}
\frac{\Omega\ln\Omega}{\beta\ln\beta}&=\frac{\Omega}{\beta}\frac{-\ln\beta-\ln(\Omega/\beta)}{-\ln\beta}    
\\
&=\frac{\Omega}{\beta}\left(1-\frac{\ln(\Omega/\beta)}{|\ln\beta|}\right)\\
&>\frac12\frac\Omega\beta
\end{flalign}
From the inverse of the above, for $\beta\ll\Omega<\sqrt\beta$, 
\begin{equation}
    \frac{\beta\ln\beta}{\Omega\ln\Omega}<2\beta/\Omega=o(1).
\end{equation}
We thus have
\begin{equation}
    |\Omega \ln\Omega |\gg |\beta \ln \beta| \hspace{10pt} (\beta \ll \Omega < \sqrt\beta).
    \label{eq:OmegalnOmegaggbetalnbetanarrower}
\end{equation}
Further noting that $|\Omega\ln\Omega|$ monotonically increases within $\Omega\leq 1/e$, we see the same inequality as eq.~(\ref{eq:OmegalnOmegaggbetalnbetanarrower}) holds up to $\Omega=1/e$:
\begin{equation}
    |\Omega \ln\Omega |\gg |\beta \ln \beta| \hspace{10pt} (\beta \ll \Omega \leq  1/e).
    \label{eq:OmegalnOmegaggbetalnbeta}
\end{equation}
We remark here that the proof of eq.~(\ref{eq:OmegalnOmegaggbetalnbeta}) does not necessarily require $\sqrt\beta\ll1$ but generally holds when $\beta^n\ll1$ for $1/(1-n)=\mathcal O(1)$. 

Next, we consider $1/e\leq \Omega \leq 1$. 
In this range, $|\Omega \ln \Omega |$ is convex upwards and hence never falls below the line connecting its endpoints, $(e-1)^{-1}(\Omega-1)$: 
\begin{equation}
    |\Omega \ln\Omega |\geq (e-1)^{-1}(1-\Omega).
\end{equation}
By utilizing this, for the target range of eq.~(\ref{eq:C1subject}) within $1/e\leq \Omega<1$, namely for $(1/e\leq \Omega<1)\cap (|\Omega-1|\gg|\beta\ln\beta|)=(|\beta \ln\beta|  \ll  1- \Omega \leq 1-1/e)$, we have
\begin{equation}
    |\Omega \ln \Omega | \gg (e-1)^{-1} |\beta \ln \beta|\sim |\beta\ln\beta| \hspace{10pt} (|\beta \ln\beta|  \ll  1- \Omega \leq 1-1/e),
    \label{eq:D1lemma2}
\end{equation}
where an order evaluation $1/(e-1)=0.58...\sim 1$ is used. 

Lastly, for $\Omega \geq 1$, $|\Omega \ln \Omega |$ is convex downwards and hence never falls below the tangent line at its starting point, $\Omega-1$. Therefore,
\begin{equation}
    |\Omega \ln\Omega |\geq \Omega -1.
\end{equation}
This leads to
\begin{equation}
    |\Omega \ln \Omega | \gg |\beta \ln \beta| \hspace{10pt} (\Omega-1 \gg|\beta \ln\beta|).
    \label{eq:D1lemma3}
\end{equation}

Equations~(\ref{eq:OmegalnOmegaggbetalnbeta}), (\ref{eq:D1lemma2}), and (\ref{eq:D1lemma3}) combined are equivalent to eq.~(\ref{eq:C1subject}), and hence 
eq.~(\ref{eq:usefulpsicBaseinequality}) has been proven for $\psi_{\rm cBase}$ that satisfies eq.~(\ref{eq:psicBaserequirement_arbitraryLoverDc}). 
[Note: Strictly speaking, what the calculus leading to eqs.~(\ref{eq:OmegalnOmegaggbetalnbetanarrower}) and (\ref{eq:D1lemma2}) shows is $|\Omega \ln \Omega | \gg |\beta \ln \beta|$  for $\Omega\gg2\beta$, or $|\Omega \ln \Omega | \gg |\beta \ln \beta|/2$ for $\Omega\gg\beta$, not being $|\Omega \ln \Omega | \gg |\beta \ln \beta|$ for $\Omega\gg\beta$. However, this is not a problem for the present purpose because we do not distinguish $\Omega\sim\beta$ and $\Omega\sim2\beta$ in constructing the evolution law from eq.~(\ref{eq:psicBaserequirement_arbitraryLoverDc}).]

\subsection{Proof for eq.~(\ref{eq:epsilonboundgeneral_improved})}
\label{subsec:AppD2}
In this subsection \S\ref{subsec:AppD2}, we consider a special case where the decreasing function $\psi_{\rm cBase}$ additionally satisfies all the conditions of eqs.~(\ref{eq:add1})--(\ref{eq:add3}). 
Equation~(\ref{eq:add1}) constrains the decreasing function $\psi_{\rm cBase}$ to decay sufficiently below $(L/D_{\rm c})\beta$ by $\Omega=1/2$. 
This stipulation does not necessarily have to be made at $\Omega=$``$1/2$''.
In fact, the following proof of eq.~(\ref{eq:epsilonboundgeneral_improved}) works even when eq.~(\ref{eq:add1}) is replaced by the same constraint at 
$\Omega=1-|\tilde\beta\ln\tilde\beta|$, where $\tilde\beta$ is a constant such that $\beta \ll\tilde\beta \ll1$: 
\begin{equation}
    \psi_{\rm cBase}(1-|\tilde\beta\ln\tilde\beta|)\ll (L/D_{\rm c})\beta,
\end{equation}
because the role of eq.~(\ref{eq:add1}) in the proof is to guarantee that $\psi_{\rm cBase}$ is sufficiently small at the minimum $\Omega$ in $\beta\lesssim |\Omega-1|\lesssim |\beta\ln\beta|$ (in eq.~\ref{eq:psicBaseineequality_athalf}). In the proof below, we adopt eq.~(\ref{eq:add1}) just for brevity. 

Proving eq.~(\ref{eq:epsilonboundgeneral_improved}) as (a subset of) the tolerance of $\epsilon_\Omega^*$ is equivalent to showing that eq.~(\ref{RSFeq:reqforpsic}) holds outside the $\epsilon_\Omega^*$-neighborhood of $\Omega=1$ with $\epsilon_\Omega^*$ satisfying eq.~(\ref{eq:epsilonboundgeneral_improved}). 
Thus, we below show that eq.~(\ref{RSFeq:reqforpsic} holds for $|\Omega-1| \gg (D_{\rm c}/L)\psi_{\rm cBase}(1)$ given that $\psi_{\rm cBase}$ satisfies the additional assumptions eqs.~(\ref{eq:add1})--(\ref{eq:add3}), atop eq.~(\ref{eq:psicBaserequirement_arbitraryLoverDc}). 

Equation~(\ref{eq:usefulpsicBaseinequality}) already guarantees that eq.~(\ref{RSFeq:reqforpsic}) holds for $|\Omega-1|\gg|\beta\ln\beta|$. Therefore, if there exists no $\Omega$ such that $(D_{\rm c}/L)\psi_{\rm cBase}(1)\ll |\Omega-1|\lesssim |\beta\ln\beta|$, eq.~(\ref{eq:usefulpsicBaseinequality}) guarantees eq.~(\ref{eq:epsilonboundgeneral_improved}), and the proof is done here. Thence, in the below, we show that eq.~(\ref{RSFeq:reqforpsic}) holds in the range $(D_{\rm c}/L)\psi_{\rm cBase}(1)\ll |\Omega-1|\lesssim |\beta\ln\beta|$, presuming that $\psi_{\rm cBase}(1)$ is so small that such an $\Omega$ range exists.

We evaluate eq.~(\ref{RSFeq:reqforpsic}) firstly for the range $\beta\lesssim |\Omega-1|\lesssim |\beta\ln\beta|$ and secondly for the remaining range $(D_{\rm c}/L)\psi_{\rm cBase} (1)\ll
|\Omega -1|\ll \beta$ (if exists). The assumption of eq.~ (\ref{eq:add1}), together with the monotonic decrease of $\psi_{\rm cBase}$, guarantees the following inequality for $\Omega\geq 1/2$, which includes the currently examined range $\beta\lesssim |\Omega-1|\lesssim  |\beta\ln\beta|$:
\begin{equation}
    \psi_{\rm cBase}\leq \psi_{\rm cBase}(1/2)\ll (L/D_{\rm c})\beta.
    \label{eq:psicBaseineequality_athalf}
\end{equation}
Combining this and the following relation valid for $\beta\lesssim |\Omega-1|\lesssim |\beta\ln\beta|\ll1$ (note $\beta<|\beta\ln \beta|$ for $\beta<1/e$),
\begin{equation}
    |\Omega\ln\Omega|=|\Omega-1|+\mathcal O[(\Omega-1)^2]\gtrsim \beta,
\end{equation}
we have
\begin{equation}
    \psi_{\rm cBase} (\Omega) \ll (L/D_{\rm c})|\Omega\ln\Omega| \hspace{10pt}(\beta\lesssim |\Omega-1|\lesssim |\beta\ln\beta|), \label{eq:strongdecayingpsi_smallerthanfdelta}
\end{equation}
which shows eq.~(\ref{RSFeq:reqforpsic}) holds in this range of $\Omega$. 

Next, we work on the other range $(D_{\rm c}/L)\psi_{\rm cBase}(1)\ll |\Omega-1|\ll \beta$. 
Equations~(\ref{eq:add2}) and (\ref{eq:add3}) guarantee 
\begin{equation}
    \psi_{\rm cBase}(\Omega)= \Psi(1/\beta)[1+\mathcal O(|\Omega-1|/\beta)].
    \label{eq:D2improvingprocess2}
\end{equation}
From eq.~(\ref{eq:D2improvingprocess2}), for $\Omega$ such that 
$(D_{\rm c}/L)\psi_{\rm cBase} (1)\ll
|\Omega -1|\ll \beta$, if exists, we have 
\begin{equation}
    |\Omega \ln \Omega| =|\Omega -1|+\mathcal O[(\Omega -1)^2]\gg (D_{\rm c}/L)\psi_{\rm cBase} (1)= (D_{\rm c}/L)\psi_{\rm cBase} (\Omega)[1+\mathcal O(|\Omega-1|/\beta)],
\end{equation}
that is
\begin{equation}
\psi_{\rm cBase}\ll (L/D_{\rm c})|\Omega \ln \Omega| 
\hspace{10pt} ((D_{\rm c}/L)\psi_{\rm cBase} (1)\ll
|\Omega -1|\ll \beta). 
\label{eq:strongdecayingpsi_fdeltaintersection}
\end{equation}
Equations (\ref{eq:usefulpsicBaseinequality}), (\ref{eq:strongdecayingpsi_smallerthanfdelta}), and (\ref{eq:strongdecayingpsi_fdeltaintersection}) lead to 
\begin{equation}
    \psi_{\rm cBase}(\Omega)=
    \begin{cases}
        1 & (\Omega\ll\beta)
        \\
        o [(L/D_{\rm c})\Omega \ln \Omega] & (\Omega \gg\beta \cap |\Omega-1|\gg (D_{\rm c}/L)\psi_{\rm cBase}(1))
    \end{cases}
\end{equation}
We have thus obtained eq.~(\ref{eq:epsilonboundgeneral_improved}) as the tolerance of $\epsilon_\Omega^*$ such that eq.~(\ref{RSFeq:modcompscutofffunction_withHeaviside}) satisfies eq.~(\ref{RSFeq:reqforpsic}) when using such $\psi_{\rm cBase}$ that conforms eqs.~(\ref{eq:psicBaserequirement_arbitraryLoverDc}) and (\ref{eq:add1})--(\ref{eq:add3}). 

Besides, we note that even without the assumption of eqs.~(\ref{eq:add2}) and (\ref{eq:add3}), the assumption of eq.~(\ref{eq:add1}) is enough to relax the lower bound of $\epsilon_\Omega^*$ to $\beta$ from $|\beta\ln\beta|$ in eq.~(\ref{eq:epsilonboundgeneral}) (via eq.~\ref{eq:strongdecayingpsi_smallerthanfdelta}), although $\beta$ and $|\beta\ln\beta|$ are of almost the same order unless $\beta$ is very small.

\subsection{Tolerance of $\epsilon_\Omega^*$ in eq.~(\ref{RSFeq:exmodcomplaw})}
\label{subsec:cutoffmodcomplawdetail}

In this \S\ref{subsec:cutoffmodcomplawdetail}, we evaluate the tolerance for $\epsilon_\Omega^*$ given by eq.~(\ref{eq:epsilonboundgeneral_improved}), especially for the evolution law of eq.~(\ref{RSFeq:exmodcomplaw}), namely eq.~(\ref{RSFeq:modcompscutofffunction_withHeaviside}) that adopts $\exp(-\Omega/\beta)$ as $\psi_{\rm cBase}$. We assume $L \sim D_{\rm c}$, throughout \S\ref{subsec:cutoffmodcomplawdetail} unless otherwise noted.

Before using eq.~(\ref{eq:epsilonboundgeneral_improved}), we need to confirm that it is applicable to this case with $\psi_{\rm cBase}=\exp(-\Omega/\beta)$. From $\exp(-0.5/\beta)\ll\beta$ and $d\exp(-x)/dx=\mathcal O[\exp(-x)]$, it is easy to see that $\psi_{\rm cBase}=\exp(-\Omega/\beta)$ satisfies eqs.~(\ref{eq:add1})--(\ref{eq:add3}), the additional requirements on $\psi_{\rm cBase}$ for eq.~(\ref{eq:epsilonboundgeneral_improved}) to be applicable. The remaining job is to evaluate if eq.~(\ref{eq:psicBaserequirement})  (eq.~\ref{eq:psicBaserequirement_arbitraryLoverDc} for $L\sim D_{\rm c}$) holds for $\psi_{\rm cBase}=\exp(-\Omega/\beta)$. For this purpose, we consider a more general case where $\psi_{\rm cBase}=\exp[-(\Omega/\beta)^n]$ and determine the range of $n (>0)$ satisfying eq.~(\ref{eq:psicBaserequirement}). Or equivalently, we determine the range of $n >0$ such that
\begin{equation}
    \exp[-(\Omega/\beta)^n]\lesssim|\beta\ln\beta| \hspace{10pt} (\Omega\gg\beta).
    \label{eq:exponentialpsicbaseconstraint}
\end{equation}
By taking the logarithm of both sides, and then raised to the power of $1/n$, we obtain 
\begin{equation}
    \Omega/\beta\gtrapprox |\ln|\beta\ln\beta||^{1/n},
    \label{eq:exppsicbaserange}
\end{equation} 
where (and throughout \S\ref{subsec:cutoffmodcomplawdetail}) we define the symbol $\gtrapprox$ such that $x \gtrapprox y \Leftrightarrow \exp(x^n) \gtrsim \exp(y^n)$. 
In the considered range of $\Omega\gg\beta$, eq.~(\ref{eq:exppsicbaserange}) holds unless the right-hand side is extremely large. When the order evaluation is made in the decimal system, $|\ln|\beta\ln\beta||^{1/n}\lessapprox 10$ is sufficient for eq.~(\ref{eq:exppsicbaserange}) to hold for $\Omega\gg\beta$. Then, by taking its logarithm, we obtain a bound of $n$ for $\psi_{\rm cBase}=\exp[-(\Omega/\beta)^n]$ to satisfy eq.~(\ref{eq:psicBaserequirement}): 
\begin{equation}
n\geq \frac{1}{\ln10}\ln|\ln|\beta\ln\beta||-corr.
\label{eq:exppsicbasenrange}
\end{equation}
where the term $corr$ on the right-hand side is the correction associated with the replacement of $\gtrapprox$ with $\geq$ upon taking the logarithm. The first term of the right-hand side is the leading order of this bound, which is a monotonically decreasing function of $\beta (<1/e^2)$ and takes values of $1.29...$ and $0.48...$ at $\beta=10^{-10}$ and $\beta=10^{-2}$, respectively. For this range of $\beta$, as the leading order is close to unity, $corr$ is about $\ln(\ln(4)) = 0.32...$. 
Here, we again considered the order evaluation in the decimal system, where numbers differing by up to a factor of $4$ from each other may be regarded as of the same order. Thus, we see $n = 1$ satisfies eq.~(\ref{eq:exppsicbasenrange}) for $\beta\gtrsim 10^{-10}$, and hence eq.~(\ref{eq:epsilonboundgeneral_improved}) is applicable to the considered case with $\psi_{\rm cBase}=\exp(-\Omega/\beta)$ when $\beta\gtrsim 10^{-10}$. We remark here that this conclusion, derived above under $L\sim D_{\rm c}$, also holds for somewhat wider conditions of $\ln L/D_{\rm c}=\mathcal O(1)$ because the evaluation of eq.~(\ref{eq:exppsicbasenrange}) deals with the logarithm of the logarithm of the quantities being compared.

Now we use eq.~(\ref{eq:epsilonboundgeneral_improved}) to evaluate the lower bound of $\epsilon^*_\Omega$ in the considered case of $\psi_{\rm cBase}=\exp(-\Omega/\beta)$. 
Given $L \sim D_{\rm c}$, eq.~(\ref{eq:epsilonboundgeneral_improved}) becomes 
\begin{equation}
    \exp(-1/\beta)\ll\epsilon^*_\Omega\ll1.
    \label{allowableRangeOfepsilonOmegaStar_exp}
\end{equation}
Since $\beta \lesssim 0.01$ has been constrained by existing experiments (eq.~\ref{RSFeq:betavalueconstraint}), $\exp(-1/\beta)$ is very small, being $10^{-43.24...}$ for $\beta = 0.01$. Hence, we can use very small $\epsilon^*_\Omega$ so that the $\epsilon^*_\Omega$-neighborhood of the steady state, namely, the range $|\Omega - 1| < \epsilon^*_\Omega$ in which $\psi_{\rm c}$ has to satisfy eq.~(\ref{eq:smallorderofepsilon}), can be taken so narrow that this range becomes numerically null, which is good news. Nevertheless, it is still necessary to set $\epsilon_\Omega^*$ appropriately for given $\beta$, according to eq.~(\ref{allowableRangeOfepsilonOmegaStar_exp}). A convenient setting is $\epsilon_\Omega^*=\sqrt{e^{-1/\beta}}$, which necessarily satisfies eq.~(\ref{allowableRangeOfepsilonOmegaStar_exp}) as far as $\beta\ll1$. 

When $\beta$ is considerably less than $10^{-10}$, eq.~(\ref{eq:exppsicbasenrange}) no longer guarantees the appropriateness of $n=1$ [i.e., $\psi_{\rm cBase}=\exp(-\Omega/\beta)$]. This stands to reason, as can be understood by invoking a simple fact that any $n<\infty$ violates eq.~(\ref{eq:exponentialpsicbaseconstraint}) when $\beta\to0$ (exceptional is $n\to \infty$, a simple step function, which is applicable even when $\beta\to0$). The considered functional form $\psi_{\rm cBase}=\exp(-\Omega/\beta)$ is useful only for an appropriate range of $\beta$. It would be possible to widen the applicability of $\psi_{\rm cBase}=\exp(-\Omega/\beta)$ for $\beta$ from the above-proven range of $\beta$, by replacing the order evaluation using $\ll$ and $\gg$ in the text with another multi-scale order evaluation also using $\lll$ and $\ggg$ (and possibly, with more $<$ and $>$). However, we do not attempt that in the present paper as we do not see any acute reason justifying the efforts. 

\section{Construction of $\beta$ for a given $\psi_{\rm c}$}
\label{sec:betavalue4unitypsicbase}
\setcounter{equation}{0}

In \S\ref{RSF322}, we have derived eq.~(\ref{RSFeq:reqforpsic}) as a constraint that $\psi_{\rm c}(\Omega)$ ($=f_t$) must conform. When compared with eq.~(\ref{eq:betaisVSSHSboundary}), eq.~(\ref{RSFeq:reqforpsic}) can also be read as the expression of the scale of $\Omega$ around which the aging-slip switch must occur. 
When we constructed $\psi_{\rm c}$ for a given $\beta$ in \S\ref{RSF33}, we implemented $\psi_{\rm c}$ for give $\beta$, which worked as the scale parameter in a unary function of $\Omega/\beta$. Alternatively, eq.~(\ref{RSFeq:reqforpsic}) allows the opposite approach, introducing $\beta$ for given $\psi_{\rm c}$, where we adjust the value of $\beta$ so that the canons can be reproduced with a given $\psi_{\rm c}(\Omega)$. Obviously, such an approach is necessary when we consider $\psi_{\rm c}$ forms that do not involve $\beta$ as an explicit parameter, as in the case of eq.~(\ref{eq:prevmodcomplaw_ft1}) discussed in \S\ref{sec:extensions_DcoverL}. Also, the inversion of $\beta$ from experimentally obtained state evolution (or from its time derivative, $f = f_t + f_\delta V$) belongs to the latter approach. In the below, we present a general procedure to construct $\beta$ for a given $\psi_{\rm c}$, followed by application to examples of $\psi_{\rm c}(\Omega)$ discussed in \S\ref{sec:extensions_DcoverL}, where this procedure is necessary. 

As discussed above, the value of $\beta$ in eq.~(\ref{RSFeq:reqforpsic}) is the scale of $\Omega$ around which the aging-slip switch occurs (eq.~\ref{eq:betaisVSSHSboundary}). Since eq.~(\ref{eq:betaisVSSHSboundary}) does not specify the behavior for $\Omega\sim\beta$, the steepness of the aging-slip switch---namely, whether the switch is a crossover or a transition---is unknown. Then, some 1-order ambiguity is intended in the concept of $\beta$. Nevertheless, from the viewpoint of eventually writing down a concrete evolution law, it is desirable to assign a single value to $\beta$. In this sense, we propose, as a practical procedure, to define a representative value $\beta^\prime$ to be the value of $\Omega$ such that 
\begin{equation}
    |f_{t|{\rm NSC}}|=|f_\delta V| \hspace{10pt} (\Omega\ll 1). \label{eq:defofbetaprime}
\end{equation}

As mentioned above, somewhat different values also function as $\beta$ due to the unconstrained transient behavior of the aging-slip switch, but $\beta^\prime$ obtained by the above procedure necessarily satisfies eq.~(\ref{RSFeq:reqforpsic}) whatever the switch is, implying that the proposed procedure is robust. 
Actually, the value of $\beta$ appears as $\ln \beta$ in $\Phi/B$ so that this 1-order ambiguity in the $\beta$ value is just within an $\mathcal O(B)$ factor in the strength evolution law. 

As seen from eq.~(\ref{eq:defofbetaprime}), $\beta^\prime$ is determined by the behavior of $f_t$ and $f_\delta V$ in $\Omega\ll1$, and hence the behavior of $f_t$ in the vicinity of steady states ($\epsilon_\Omega^*$-vicinity in \S\ref{RSF33}) is irrelevant. 
Then, using $\psi_{\rm c}$ for $\Omega\ll1$, namely the $\psi_{\rm cBase}$ introduced in \S\ref{RSF33}, we can rewrite the definition of $\beta^\prime$ (eq.~\ref{eq:defofbetaprime}) into 
\begin{equation}
    |\psi_{\rm cBase}|=|f_\delta V| \hspace{10pt} (\beta^\prime\ll1). \label{eq:defofbetaprime_2}
\end{equation}

As an example, we below determine $\beta^\prime$ for the case of constant $\psi_{\rm cBase}$, which can satisfy eq.~(\ref{RSFeq:reqforpsic}) when $L\gg D_{\rm c}$ (\S\ref{sec:extensions_DcoverL}). 
Following the main text (eq.~\ref{eq:prevmodcomplaw_ft1}), we adopt 
\begin{equation}
    \psi_{\rm cBase}=1.\label{eq:constantpsicBase}
\end{equation}
Since this $\psi_{\rm cBase}$ does not involve the parameter $\beta$, this is an obvious case where we have to determine $\beta$ for the given $\psi_{\rm c}$. The $f_\delta V$ term in this example is already fixed by eq.~(\ref{eq:simplifiedslipdepofthetadot}). Then, eq.~(\ref{eq:defofbetaprime_2}) becomes
\begin{equation}
    -\beta^\prime\ln\beta^\prime=D_{\rm c}/L.
    \label{eq:betaforsmallconstantft}
\end{equation}
Here we presume $\beta\ll\sqrt{\beta}<1/e$ for $\beta=\beta^\prime$ as in the main text. 
Solving eq.~(\ref{eq:betaforsmallconstantft}), we obtain $\beta^\prime=\exp[W_{-1}(-D_{\rm c}/L)]$ for $\psi_{\rm cBase}=1$ given $\beta^\prime\ll1$, using  the analytic continuation of Lambert W function $W_{-1}(\cdot)$. 
Thus, $\beta^\prime$ has been obtained for this example case, but for completion of writing down the evolution law, we carry on to specify appropriate $\epsilon_\Omega^*$, following Appendix~\ref{sec:epsironRange}. As long as eq.~(\ref{eq:betaforsmallconstantft}) holds, eq.~(\ref{eq:psicBaserequirement_arbitraryLoverDc}) becomes
\begin{equation}
    \psi_{\rm cBase}(\Omega)=
    \begin{cases}
        1 & (\Omega\ll\beta)
        \\
        \mathcal O (1) & (\Omega \gg\beta)
    \end{cases}
    \label{eq:psicBaserequirement_arbitraryLoverDc_constantft}
\end{equation}
The said $\psi_{\rm cBase}$ of eq.~(\ref{eq:constantpsicBase}) satisfies eq.~(\ref{eq:psicBaserequirement_arbitraryLoverDc_constantft}) (i.e., eq.~\ref{eq:psicBaserequirement_arbitraryLoverDc} for eq.~\ref{eq:betaforsmallconstantft}), so we can use eq.~(\ref{eq:epsilonboundgeneral}) to obtain the associated tolerance of $\epsilon_\Omega^*$. Using the $\beta=\beta^\prime$ value of eq.~(\ref{eq:betaforsmallconstantft}), eq.~(\ref{eq:epsilonboundgeneral}) reduces to 
\begin{equation}
    D_{\rm c}/L\ll \epsilon_\Omega^*<1-1/e, \label{eq:epsilonOmegarange4constantpsicBase}
\end{equation}
where we presume $\epsilon_\Omega^*<1-1/e$ as in the main text. 
Thus, given $L\gg D_{\rm c}$, we can construct an appropriate evolution law, with $\psi_{\rm cBase}=1$, $\beta=\exp[W_{-1}(-D_{\rm c}/L)]$, and $\epsilon_\Omega^*$ of eq.~(\ref{eq:epsilonOmegarange4constantpsicBase}). 

As another example, we remark that a similar conclusion is obtained when we adopt $\psi_{\rm cBase}\sim1$ (eq.~\ref{eq:constantsmallpuretimehealing}) instead of $\psi_{\rm cBase}=1$. In this case, $\beta=\beta^\prime$ is constrained by $-\beta^\prime\ln\beta^\prime\sim D_{\rm c}/L$, while the tolerance of $\epsilon_\Omega^*$ again becomes eq.~(\ref{eq:epsilonOmegarange4constantpsicBase}). 

\label{lastpage}
\end{document}